\documentclass[a4paper,12pt]{book}

\usepackage[utf8]{inputenc}
\usepackage[left=2cm,right=2cm,top=2cm,bottom=3cm]{geometry}

\usepackage{xcolor}
\usepackage[thref]{ntheorem}
\theoremstyle{break}
\newtheorem{definition}{Definition}[section]

\usepackage[linktoc=all,hidelinks]{hyperref}
\usepackage{physics}
\usepackage{amsfonts}    
\usepackage{amssymb}     
\usepackage{stmaryrd}    
\usepackage[toc]{appendix}

\usepackage{etoolbox}
\AtBeginEnvironment{appendices}{%
  \let\clearpage\relax
  
  \setcounter{section}{0}%
}
\usepackage[dvipsnames]{xcolor}
\usepackage[mathscr]{eucal}
\usepackage{tikz-cd}
\usepackage{physics}
\usepackage{cancel}
\usepackage{framed}
\usepackage{bbm}
\usepackage{cite}
\usepackage{amsmath}
\usepackage{amsfonts}
\usepackage{amssymb}
\usepackage{appendix}
\usepackage{graphicx}
\usepackage{afterpage}
\usepackage[nottoc,numbib]{tocbibind}
\usepackage{ytableau}
\usepackage{enumitem} 
\usepackage{ytableau}
\usepackage{genyoungtabtikz}
\Yboxdim{8pt}
\Ylinethick{1pt}
\newcommand{\R}{\mathbb{R}}
\newcommand{\Enveloping}{\mathcal{U}}
\newcommand{\Ideal}{\mathcal{I}}

\newcommand{\Casimir}{\mathcal{C}}
\newcommand{\hs}{\mathfrak{hs}}
\newcommand{\hsdeformed}{{\ensuremath{\pmb{\mathrm{A}}_u}}}
\newcommand{\AlgInd}[1]{\mathsf{#1}}
\newcommand{\1}{\mathbf{1}}
\newcommand{\Centraliser}{\mathcal{Z}}
\newcommand{\Weyl}{\mathcal{A}}
\newcommand{\WeylClifford}{\mathcal{A}}
\newcommand{\bosOsc}{\mathsf{a}}
\newcommand{\ferOsc}{\mathsf{c}}

\newcommand{\Maxwell}{\mathcal{M}}
\newcommand{\Joseph}{\mathcal{J}}

\newcommand{\Base}{\mathscr{X}}
\newcommand{\langlecket}[1]{\pmb\{#1\pmb\}}
\newcommand{\Completion}{\pmb\gamma}
\newcommand{\Curvature}{\mathcal{R}}
\newcommand{\dR}{\mathrm{d}}
\newcommand{\Fedosov}{\mathfrak{D}}
\newcommand{\Fock}{\mathfrak{F}}
\newcommand{\Functions}{\mathscr{C}^\infty}
\newcommand{\Rep}[2]{#1 \pmb{\triangleright} #2}
\newcommand{\SymDer}{\partial_{\scriptscriptstyle\nabla}}
\newcommand{\Taylor}{j_{\scriptstyle\nabla}}
\newcommand{\WeylAlg}{\mathcal{A}}
\newcommand{\WeylBundle}{\mathscr{W}}
\newcommand{\Wigner}{\mathcal{W}}

\newcommand{\fud}[2]{{}^{#1}{}_{#2}\,}
\newcommand{\fdu}[2]{{}_{#1}{}^{#2}\,}

\makeatletter
\newcommand{\subalign}[1]{%
  \vcenter{%
    \Let@ \restore@math@cr \default@tag
    \baselineskip\fontdimen10 \scriptfont\tw@
    \advance\baselineskip\fontdimen12 \scriptfont\tw@
    \lineskip\thr@@\fontdimen8 \scriptfont\thr@@
    \lineskiplimit\lineskip
    \ialign{\hfil$\m@th\scriptstyle##$&$\m@th\scriptstyle{}##$\crcr
      #1\crcr
    }%
  }
}
\makeatother

\newcommand{\TikzRect}[2]{\filldraw[color=black,fill=red]  (#1-\R,#2-\R) rectangle (#1+\R,#2+\R);}
\newcommand{\TikzRectB}[2]{\filldraw[color=black,fill=blue]  (#1-\R,#2-\R) rectangle (#1+\R,#2+\R);}

\newcommand{\FIELDS}{
\begin{tikzpicture}[scale=0.7]
\tikzset{%
  >=latex, 
  inner sep=0pt,%
  outer sep=2pt,%
  mark coordinate/.style={inner sep=0pt,outer sep=0pt,minimum size=3pt,
    fill=black,circle}%
}
\def\R{0.15}
\def\mar{0.15}

\draw[->,black,thick] (0,0) -- (0,11) node[left]{$\# A'$ };
\draw[->,black,thick] (0,0) -- (15,0) node[right]{$\# A$} ;

\filldraw[color=black,fill=green]  (14,6) circle (\R);
\node[right] at (14.3,6) {one-forms, $\omega$};

\TikzRect{14}{7}  
\node[right] at (14.3,7) {zero-forms, $C$};

\filldraw[color=black,fill=green]  (3,0) circle (\R);
\filldraw[color=black,fill=green]  (2,1) circle (\R);
\filldraw[color=black,fill=green]  (1,2) circle (\R);
\filldraw[color=black,fill=green]  (0,3) circle (\R);

\filldraw[color=black,fill=green]  (3,1) circle (\R);
\filldraw[color=black,fill=green]  (2,2) circle (\R);
\filldraw[color=black,fill=green]  (1,3) circle (\R);

\filldraw[color=black,fill=green]  (4,1) circle (\R);
\filldraw[color=black,fill=green]  (3,2) circle (\R);
\filldraw[color=black,fill=green]  (2,3) circle (\R);
\filldraw[color=black,fill=green]  (1,4) circle (\R);

\filldraw[color=black,fill=green]  (4,2) circle (\R);
\filldraw[color=black,fill=green]  (3,3) circle (\R);
\filldraw[color=black,fill=green]  (2,4) circle (\R);

\filldraw[color=black,fill=green]  (5,2) circle (\R);
\filldraw[color=black,fill=green]  (4,3) circle (\R);
\filldraw[color=black,fill=green]  (3,4) circle (\R);
\filldraw[color=black,fill=green]  (2,5) circle (\R);

\draw[green,thick] (3,0) -- (0,3) -- (2,5) -- (5,2) -- (3,0);

    \draw[-latex,thick] (4.5,6) node[right,scale=1.0]{$\omega^{A(t-1),A'(2s-t-1)}$}
        to[out=-180,in=50] (2.15, 5.15);
        
    \draw[-latex,thick] (5,8) node[right,scale=1.0]{$C^{A(t-1),A'(2s-t+1)}$}
        to[out=-180,in=0] (2.15,7);

    \draw[-latex,thick] (9,2.5) node[right,scale=1.0]{$C^{A(2s-t+1),A'(t-1)}$}
        to[out=180,in=-30] (7.2, 2);

    \draw[-latex,thick] (4.5,4) node[right,scale=1.0]{$\omega^{A(2s-t-1),A'(t-1)}$}
        to[out=180,in=90] (5, 2.15);

    \draw[red,thick] (2,11) -- (0,9) -- (2,7) -- (5,10);
    \TikzRect{2}{7}

    \draw[red,thick] (11,2) -- (9,0) -- (7,2) -- (10,5);
    \TikzRect{7}{2}
    
    \draw[rounded corners] ( 2-0.5 , 5-0.5 ) rectangle (2.5,7.5) {}; 
    \draw[rounded corners] ( 5-0.5 , 2-0.5 ) rectangle (7.5,2.5) {}; 
    
    \draw (-0.1,9) -- (0.1,9);
    \node[left] at (-0.1,9) {\footnotesize$2s$};
    \draw[dashed] (-0.1,7) -- (2,7);
    \node[left] at (-0.1,7) {\footnotesize$2s-t+1$};
    \draw[dashed] (-0.1,5) -- (2,5);
    \node[left] at (-0.1,5) {\footnotesize$2s-t-1$};
    \node[left] at (-0.1,3) {\footnotesize$2s-2t$};
    
    \draw[dashed] (7,2) -- (7,-0.1);
    \node[below] at (7,-0.1) {\footnotesize$2s-t+1$};
    \draw (9,0.1) -- (9,-0.1);
    \node[below] at (9,-0.1) {\footnotesize$2s$};
    \draw[dashed] (5,2) -- (5,-0.7);
    \node[below] at (5,-0.7) {\footnotesize$2s-t-1$};
    \node[below] at (3,-0.1) {\footnotesize$2s-2t$};
\end{tikzpicture}}

\newcommand{\FIELDSDescr}{
\begin{tikzpicture}[scale=0.7]
\tikzset{%
  >=latex, 
  inner sep=0pt,%
  outer sep=2pt,%
  mark coordinate/.style={inner sep=0pt,outer sep=0pt,minimum size=3pt,
    fill=black,circle}%
}
\def\R{0.15}
\def\mar{0.15}

\draw[->,black,thick] (0,0) -- (0,11) node[left]{$\# A'$ };
\draw[->,black,thick] (0,0) -- (15,0) node[right]{$\# A$} ;

\filldraw[color=black,fill=green]  (14,6) circle (\R);
\node[right] at (14.3,6) {one-forms, $\omega$};

\TikzRect{14}{7}  
\node[right] at (14.3,7) {zero-forms, $C$};

\filldraw[color=black,fill=green]  (7,0) circle (\R);
\filldraw[color=black,fill=green]  (6,1) circle (\R);
\filldraw[color=black,fill=green]  (5,2) circle (\R);
\filldraw[color=black,fill=green]  (4,3) circle (\R);
\filldraw[color=black,fill=green]  (3,4) circle (\R);
\filldraw[color=black,fill=green]  (2,5) circle (\R);
\filldraw[color=black,fill=green]  (1,6) circle (\R);
\filldraw[color=black,fill=green]  (0,7) circle (\R);

\draw[green,thick] (7,0) -- (0,7);

    \draw[-latex,thick] (4.5,6) node[right,scale=1.0]{$\omega^{A(t-1),A'(2s-t-1)}$}
        to[out=-180,in=50] (2.15, 5.15);
        
    \draw[-latex,thick] (5,8) node[right,scale=1.0]{$C^{A(t-1),A'(2s-t+1)}$}
        to[out=-180,in=0] (2.15,7);

    \draw[-latex,thick] (9,2.5) node[right,scale=1.0]{$C^{A(2s-t+1),A'(t-1)}$}
        to[out=180,in=-30] (7.2, 2);

    \draw[-latex,thick] (5,-1) node[left,scale=1.0]{$\omega^{A(2s-t-1),A'(t-1)}$}
        to[out=0,in=-90] (5, 1.85);

    \draw[red,thick] (0,9) -- (9,0);
    \TikzRect{2}{7}
    \TikzRect{1}{8}
    \TikzRect{0}{9}
    \TikzRect{3}{6}
    \TikzRect{4}{5}
    \TikzRect{5}{4}
    \TikzRect{6}{3}
    \TikzRect{7}{2}
    \TikzRect{8}{1}
    \TikzRect{9}{0}

    \TikzRect{7}{2}
    
    \draw[rounded corners] ( 7-0.5 , -0.5 ) rectangle (9.5,.5) {};
     \draw[rounded corners] ( -0.5 , 7-0.5 ) rectangle (.5,9.5) {};    
    
    \draw[rounded corners] ( 2-0.5 , 5-0.5 ) rectangle (2.5,7.5) {}; 
    \draw[rounded corners] ( 5-0.5 , 2-0.5 ) rectangle (7.5,2.5) {}; 
\end{tikzpicture}}

\usepackage{tikz}
\usetikzlibrary{patterns}
\newcommand{\FIELDSBeyond}{
\begin{tikzpicture}[scale=0.7]
\tikzset{%
  >=latex, 
  inner sep=0pt,%
  outer sep=2pt,%
  mark coordinate/.style={inner sep=0pt,outer sep=0pt,minimum size=3pt,
    fill=black,circle}%
}
\def\R{0.15}
\def\mar{0.15}

\draw[->,black,thick] (0,0) -- (0,11) node[left]{$\# A'$ };
\draw[->,black,thick] (0,0) -- (15,0) node[right]{$\# A$} ;

    \draw[red,thick] (2,11) -- (0,9);
    \draw[red,dashed] (0,9) -- (6,3);
    \draw[red, thick] (6,3) -- (11,8);
    \TikzRect{6}{3}

    \draw[blue,thick] (11,2) -- (9,0);
    \draw[blue,dashed] (9,0) -- (3,6);
    \draw[blue,thick] (3,6) -- (8,11);
    \TikzRectB{3}{6}
    
    \draw[pattern=north west lines, pattern color=gray]
    (8,11) -- (3,6) -- (6,3) -- (11,8);
    
    \draw (-0.1,9) -- (0.1,9);
    \node[left] at (-0.1,9) {\footnotesize$2s$};
    
    \draw (9,0.1) -- (9,-0.1);
    \node[below] at (9,-0.1) {\footnotesize$2s$};
    
    \draw[dashed] (6,3) -- (6,-0.1);
    \node[below] at (6,-0.1) {\footnotesize$s+k+1$};
    \draw[dashed] (3,6) -- (3,-0.1);
    \node[below] at (3,-0.1) {\footnotesize$s-k+1$};
\end{tikzpicture}}

\numberwithin{equation}{section}
\begin{document}

\thispagestyle{empty}

\begin{figure}[h]
\centering
\includegraphics[width=30mm]{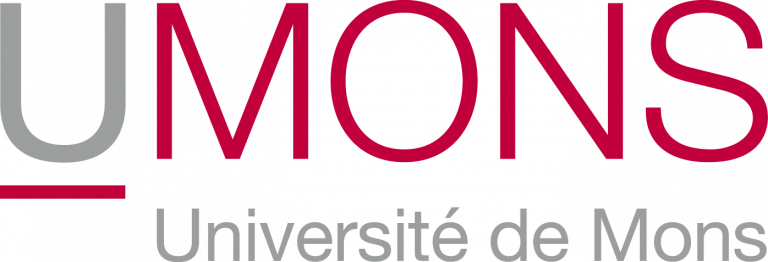}%
\hfill
\includegraphics[width=30mm]{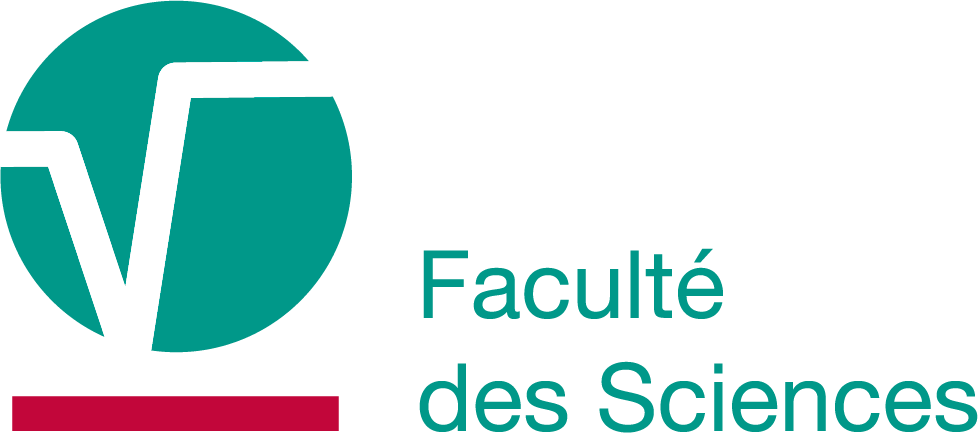}
\end{figure}

\begin{center}
\vspace*{20mm}
\rule{\textwidth}{0.4pt} 
\vspace{10mm}
{\Large \scshape Interactions of conformal and partially massless higher spin fields} 
\vspace{10mm}
\rule{\textwidth}{0.4pt} 

\vspace{15mm}
{\large Shailesh Dhasmana}

\vspace{5mm}
Service de Physique de l'Univers, Champs et Gravitation\\
Facult\'e des Sciences\\
Universit\'e de Mons

\vspace{4mm}
July 2025 

\vspace{15mm}
{\scshape Th\`ese pr\'esent\'ee en vue de l'obtention\\
du grade de Doctorat en sciences}

\vspace{15mm}
{\scshape Promoteur de th\`ese}\\
Prof. Evgeny Skvortsov, \quad Universit\'e de Mons, Belgique

{\scshape Co-promoteur de th\`ese}\\
Dr. Thomas Basile (Postdoc),\quad Universit\'e de Mons, Belgique

\vspace{5mm}

{\scshape Membres du jury}\\[7pt]
\begin{tabular}{l c l}
Prof. Xavier Bekaert, && Institut Denis Poisson, Tours, France\\
Dr. Dmitry Ponomarev, && ITMP, MSU, Moscou, Russie\\
Prof. Nicolas Boulanger, && Universit\'e de Mons, Belgique\\
Prof. Andrea Campoleoni, && Universit\'e de Mons, Belgique\\
\end{tabular}
\end{center}

\chapter*{Acknowledgments}
\addcontentsline{toc}{chapter}{Acknowledgments}
\pagenumbering{roman}

First and foremost, I am profoundly indebted to my supervisor, Evgeny Skvortsov,
for his unwavering patience and tailored guidance, which allowed me to grow at my own pace. This thesis would not have been possible without his constant encouragement and support during this period. From learning higher spin theory to discovering my particular research interest around it is an important journey for me, and none of this would have been possible without his guidance. I am thankful to him for giving me the projects that cover a wide range of topics. It opens up my horizon and will help me in my future research endeavor. Also, coming to Mons from India during the pandemic was not very easy at first for me personally; even after that, I struggled a lot health-wise. I sincerely appreciate his understanding during these times and later, whether health-related or family matters, as he always allowed me the flexibility to resolve challenges in my own time. This empathy is what I remain profoundly thankful for.

I am equally grateful to my co-supervisor, Thomas Basile, whose generosity with time and wisdom went far beyond academia. From our initial discussions through to finalizing this manuscript, his consistent support and belief in my progress were invaluable. Whether deciphering theoretical puzzles or untangling bureaucratic knots, Thomas was my steadfast ally. To any in need of a "civic counselor,": you know the person to call.

I extend my sincere thanks to Thomas, Arghya, Chrysoula, and Ismael for their invaluable support, both to me and my wife, on numerous personal and administrative matters. Their willingness to help made navigating challenges far easier, and I deeply appreciate their kindness.

A PhD doesn't mean putting life on hold. My four years in Mons were made enjoyable thanks to a wonderful group of colleagues - the "weird" crew from the ground floor and Antwerp: Richard Van Dongen, Mattia Serrani, Arsenii Sukhanov, Josh O'Connor, and Ismael Ahlouche.

Beyond the laughter and good times, I've learned so much from them - from the first "principles of holography" to other "insights" during the Modave summer school. Our regular discussions about work progress were invaluable for staying motivated and managing the pressures of research. I'm truly grateful for all the stimulating conversations, helpful feedback, and much-needed moments of levity we shared along the way.  

In addition, I would like to thank the full staff of the unit for providing support, facilities, and an environment to work without worrying about anything other than research.

I owe my deepest gratitude to my family, Shri. Nirmal Dhasmana, Smt. Shakuntala Dhasmana, Abhilasha, Nitesh, Neha, and Swati for their unwavering support throughout this journey. Their constant encouragement and sacrifices gave me the strength to persevere.  

Above all, I owe my deepest thanks to my wife, Swati. She has been my constant companion, standing with me through every challenge and celebrating every small victory. She has contributed to every aspect of my life in one way or another and today, with the completion of this thesis, the sense of joy and achievement belongs as much to her as it does to me.

\vfill

\paragraph{Funding:} This research project was supported by the European Research Council (ERC) under the European Union’s Horizon 2020 research and innovation programme (grant agreement No 101002551)

\chapter*{Abstract}

This thesis investigates the interactions of partially massless (PM) fields in 4-dimensional (anti)de Sitter spaces, along with conformal higher spin fields and their coupling to matter in arbitrary dimensions. The first part of the thesis deals with PM fields and PM algebras. A reformulation of PM fields is proposed and studied using a novel chiral formulation, inspired by Penrose's twistor approach to massless fields in Minkowski space. This reformulation enables explicit construction of Yang-Mills-type interactions and current couplings. Next, an oscillator realisation for PM higher spin algebras is given in terms of bosonic and fermionic oscillators. The construction is based on the Weyl-Clifford algebra. 

The second part of the thesis derives the coupling between a massless scalar field and a background of higher spin fields within a manifestly covariant framework, employing Fedosov quantization techniques, called the "parent formulation". This formalism yields, in particular, an explicit covariant expression for the coupling between scalar fields and higher spin conformal gravity.

\setcounter{tocdepth}{2}
\tableofcontents
\pagenumbering{arabic}
\chapter*{\emph{An Invitation}}
\addcontentsline{toc}{chapter}{Introduction}

One of the most enduring challenges in modern theoretical physics is to formulate a consistent theory of quantum gravity that is, a framework unifying Einstein’s General Relativity (GR) with the quantum field theoretic description of the other fundamental forces encapsulated in the Standard Model (SM).  GR describes gravity as the geometry of spacetime, while the SM describes the electromagnetic, weak, and strong interactions.  In their current forms, these two theories are entirely independent: each is brilliantly successful within its domain, yet fundamentally incompatible with the other.

The SM’s unification of three forces suggests that gravity, the lone force outside its remit, should also fit into the same framework.  Pursuing this idea broadly termed “quantum gravity” requires both pushing experimental probes beyond the SM’s energy reach and challenging the foundational assumptions of our existing theories.  Unfortunately, the Planck scale at which quantum-gravitational effects become significant lies far beyond the capabilities of any foreseeable collider, so progress must rely primarily on theoretical innovation.

Over the past half-century, physicists have explored a variety of radical approaches, such as string theory and loop quantum gravity, that replace one or both standard frameworks with entirely new structures and then attempted to recover GR and the SM as low-energy limits.  These efforts have yielded remarkable insights, including the AdS/CFT correspondence, but the overarching goal of a complete quantum gravity remains elusive.

In parallel, more conservative extensions adhere closely to the SM’s and GR's established principles, adding new symmetries or fields without discarding the underlying framework.  Examples include emergent-gravity scenarios motivated by black-hole thermodynamics, Holographic duality and higher spin gauge theories.  The latter one investigates whether a theory of fundamental particles of spin greater than two might play a role in our universe or not. This offers a promising avenue for extending the SM and probing the deep connections between geometry, quantum mechanics, and fundamental interactions.

Another compelling motivation for studying higher spin particles comes from string theory\cite{Bengtsson:1986nk}, see also \cite{Francia:2002pt, Francia:2010qp, Francia:2006hp}. String theory contains an infinite tower of massive higher spin fields that interact consistently. By taking the low-tension limit, in which these masses vanish, one can extract hints about their interactions. Conversely, a deeper understanding of higher spin dynamics may shed new light on string theory itself, which has so far focused primarily on its low-spin, massless sector and its low-energy interactions. Thus, in this low-tension limit, there exists an underlying symmetry called higher spin symmetry, which leads many to conjecture that string theory could be a broken phase of a higher spin symmetric theory\cite{Gross:1988ue}.

Although higher spin theory in its modern form has developed over the last 20–30 years, its roots trace back to E. Majorana’s 1932 work \cite{Majorana1932} and Dirac’s 1936 equations for arbitrary spin \cite{Dirac:1936RWE}. These were the first instances of Lorentz-covariant equations of motion for particles of arbitrary spin. A proper mathematical foundation arrived in 1939 with E. Wigner’s group-theoretic classification of one-particle states as unitary irreducible representations of the Poincaré group \(ISO(3,1)\) \cite{Wigner:1939cj}.  One fixes the invariant \(p^2=m^2\ge0\) (massive vs.\ massless), and then classifies states by their little-group representations: \(SO(3)\) for \(m>0\) (spin \(s\), dimension \(2s+1\)) and \(SO(2)\) for \(m=0\) (helicity \(\pm\lambda\)).

Fierz’s seminal 1939 paper \cite{Fierz:1939ix} then gave the first comprehensive treatment of free higher spin fields in tensor language, and Frønsdal \cite{Fronsdal:1978rb} introduced a consistent free gauge theory of the massless higher spins in flat spacetime via symmetric tensors with trace constraints. The central question of whether one can construct consistent interacting theories that include at least one massless field of spin \(s>2\) remains notoriously difficult.  Various no-go theorems (see \cite{Bekaert:2010hw} for details) demonstrate severe obstructions to coupling such fields to gravity or to themselves while preserving both gauge invariance and locality. Briefly, a few of them are as follows (see\cite{Bekaert:2010hw} for more detail),  
\begin{enumerate}
  \item \textbf{Weinberg’s Low-Energy Theorem (1964)\cite{Weinberg:1964ew}}\\
    In the soft limit of emitting a massless spin-\(s\) particle, one finds the conservation condition
    \[
      \sum_{i} g^{(s)}_{i}\,p_{i}^{\mu_{1}}\cdots p_{i}^{\mu_{s-1}} \;=\; 0.
    \]
    The symbol \( g_i^{(s)} \) represents the coupling constant associated with the emission of a soft massless particle of spin \( s \) from the \( i \)-th external hard particle in a scattering amplitude. For \(s=1\), \(g_i^{(1)}\) is the electric charge of particle \(i\), and the soft factor \(\sum_i g_i^{(1)}\) enforces charge conservation.  For \(s=2\), \(g_i^{(2)}\) is the gravitational coupling. Hence, soft theorem encodes both momentum conservation and the equivalence principle.  
    For \(s>2\), this forces all couplings \(g^{(s)}_{i}\) to vanish, forbidding most of the nontrivial interactions.  

  \item \textbf{Coleman–Mandula Theorem (1967)\cite{Coleman:1967ad}}\\
    Any nontrivial, analytic S-matrix in four dimensions with finitely many particle species can only realize a direct product of the Poincaré algebra with internal symmetries.  No extra higher spin conserved charges are allowed.  

    \item \textbf{Weinberg–Witten Theorem (1980)\cite{Weinberg:1980kq}}\\
    Massless particles of spin \(>1\) cannot carry a Lorentz-covariant, gauge-invariant stress-energy tensor.  Equivalently, no universal two-derivative coupling to gravity exists for \(s\geq2\) in flat space.  
\end{enumerate}
These classic no-go theorems for interacting massless higher spin fields rest on flat-space assumptions, a Poincaré-invariant S-matrix, minimal two-derivative couplings, and well-defined asymptotic states.  Of course, one may relax any of these hypotheses as long as no physical or mathematical inconsistency arises.  In particular, a nonzero cosmological constant \(\Lambda\) eliminates the very notion of an S-matrix, but, as later turned out, the appropriate observables are the boundary correlation functions\footnote{In ordinary flat‐space QFT, the S-matrix is the set of transition amplitudes between `in' and `out' states defined on asymptotic Minkowski null infinity.  In $AdS$ there is no such notion (particles can never “escape” to infinity), so the role of scattering amplitudes is played instead by boundary correlators in the dual CFT.  The “holographic S-matrix” is then recovered by taking an appropriate flat-space limit of those correlators.}. Before AdS/CFT was established, this argument presented a possibility to evade one or more of the flat-space conditions and thus motivates the study of higher spin theories in \((A)dS\).  Indeed, Fradkin and Vasiliev \cite{Fradkin:1987ks, Fradkin:1986qy} showed that \(\Lambda\neq0\) permits nontrivial cubic vertices.\footnote{It is worth mentioning that there are nontrivial cubic interactions of massless higher-spin fields in flat space as well, including the gravitational ones, as was shown around the same time in \cite{Bengtsson:1983pd,Bengtsson:1986kh}.}  At first sight, their construction in \((A)dS_{d}\) appears to “circumvent” the flat-space no-go theorems simply by exploiting the dimensionful scale \(\Lambda\).  However, full consistency demands quartic and higher order interactions.  Although Vasiliev later formulated a non-linear system of interacting higher spin equations in \((A)dS\) \cite{Vasiliev:1988sa, Vasiliev:1990en, Vasiliev:1995dn}, those equations inevitably generate infinite-derivative (i.e.\ non-local) interactions and contain infinitely many free coefficients that crucially affect physical observables, hence no local quartic and higher-order interactions can be obtained from Vasiliev's equations. Due to this, one can't interpret the equations as providing a (local) higher-spin field theory in $AdS$ and it remains an open question whether and how one can extract physical observables from this system, for details see \cite{Boulanger:2015ova}. 

 Also, many of the arguments of the no-go theorems are based on manifest Lorentz invariance, but it turns out that in the light cone approach, one directly deals with physical degrees of freedom and avoids the dependence on a particular description of fields. Such an approach was used to classify cubic interactions \cite{Bengtsson:1983pd,Bengtsson:1986kh,Metsaev:1991mt, Metsaev:1991nb} and in \cite{Ponomarev:2016lrm} it was finally shown how to avoid no-go theorems and establish consistent non-trivial higher spin gravity in flat spacetime. Based on these results, it was further shown that there exists a complete and consistent theory, called \emph{Chiral higher spin theory}, in four dimensions\cite{Ponomarev:2016lrm}.

Chiral Higher spin gravity (HiSGRA) is of huge interest as this is the only perturbatively local field theory with propagating massless fields. It is at least one loop finite \cite{Skvortsov:2018jea,Skvortsov:2020wtf,Skvortsov:2020gpn}. Also, recently, chiral higher spin theory has been fully developed in a covariant manner in a series of works \cite{Skvortsov:2022syz, Sharapov:2022wpz, Sharapov:2022nps}. This description is inspired by the pure-connection formalism of gravity by Plebanski \cite{Plebanski:1977zz} and \cite{Krasnov:1970bpz}. Chiral HiSGRA admits two simple, consistent truncations which are to be seen as higher spin extensions of self-dual Yang-Mills (HS-SDYM) and self-dual gravity (HS-SDGRA) \cite{Ponomarev:2017nrr, Krasnov:2021nsq}. These theories are also the subject of huge interest from the twistor front, as this description is very twistor-friendly. Indeed, in \cite{Herfray:2022prf}, a twistor description of HS-SDYM and HS-SDGRA is given, extending the famous Ward correspondence and Non-linear graviton theorem for the low spin case. This naturally suggests that the twistor formulation for full chiral higher spin theory must exist, see \cite{Mason:2025hs} for recent progress on this front.  All of these make Chiral HiSGRA a very interesting, important tool to explore higher spin gravity in general\footnote{Very recently, \cite{Serrani:2025owx} presented a complete classification of chiral higher spin theories with one- and two-derivative vertices. Remarkably, this gives the first examples of gravitational (self-dual) theories with only a finite number of interacting higher spin fields and greatly expands the known landscape. Such truncations also appear to be realizable within covariant chiral higher spin actions \cite{Krasnov:2021nsq} and developed in  \cite{Skvortsov:2022syz, Sharapov:2022nps}.}.

Now, from the point of view of the AdS/CFT correspondence \cite{Maldacena:1997re, Gubser:1998bc, Witten:1998qj}, it became apparent that a consistent theory of massless higher spin fields is very natural.  In particular, these fields possess precisely the structure required to be holographically dual to a free or critical vector model, a correspondence first conjectured in \cite{ref9, Klebanov:2002ja}, and admit fermionic generalizations, as described in \cite{Sezgin:2003pt, Leigh:2003gk}. 
One of the crucial results in this context is that in three-dimensional CFTs, the existence of a stress tensor and a single conserved higher spin current forces the theory to be free \cite{Maldacena:2011jn}. It was proved that such a current generates an infinite tower of conserved currents, and the resulting Ward identities uniquely fix all stress-tensor and higher spin correlators to coincide with those of either a free boson or a free fermion.  Hence, no nontrivial interacting CFT can possess exact higher spin symmetry. Subsequently, such investigation is extended to four dimensions in \cite{Alba:2013yda} and then to $d$ dimensions in \cite{Alba:2015upa}. This result, from the point of view of $AdS/CFT$, must be dual to some bulk higher spin theory. Indeed, in \cite{Boulanger:2013zza}, a classification of higher‐spin algebras in $\mathrm{AdS}_d$ for $d>3$ is done.  Under mild assumptions, they obtain a complete solution in $d=4$ and $d>7$, demonstrating that the algebra governing symmetric higher spin fields in the bulk and, equivalently, the algebra of exactly conserved, totally symmetric higher‐spin currents in the boundary CFT$_{d-1}$ is unique.

In this context, a foundational result in representation theory is the \emph{Flato-Frønsdal theorem} \cite{Flato:1978qz}, which lies at the heart of higher spin holography by demonstrating how bulk higher spin gauge fields emerge from simpler boundary degrees of freedom.  This relationship is encapsulated in the following group‐theoretic statement:
\paragraph{Flato-Frønsdal Theorem:} \textit{`` In $AdS_{4}$, the tensor product of two singletons decomposes into a tower of massless bulk fields.''}

This is understood as follows:
\begin{itemize}
  \item A scalar \emph{singleton} (or “Rac”) is a lowest‐weight module \(D\bigl(\tfrac12,0\bigr)\) of \(\mathfrak{so}(3,2)\) that has no propagating bulk degrees of freedom, but corresponds to a free conformal field on the AdS\(_4\) boundary. The tensor product of two such singletons decomposes into an infinite direct sum of  \(\mathfrak{so}(3,2)\) irreducible modules:
    \[
      D\bigl(\tfrac12,0\bigr)\;\otimes\;D\bigl(\tfrac12,0\bigr)
      \;=\;\bigoplus_{s=0}^{\infty}D\bigl(s+1,s\bigr)\,,
    \]
    where each \(D(s+1,s)\) is the lowest‐weight module describing a massless spin-\(s\) field in AdS\(_4\). Thus, although an individual singleton does not propagate in the bulk, two singletons together generate the full spectrum of massless higher spin fields.
  \item \emph{On boundary side:} Conserved spin-$s$ currents $J^{(s)}(x)$ in the free CFT and \\\emph{On bulk side:} Free massless spin-$s$ gauge fields $\varphi^{(s)}$ in AdS.
\end{itemize}

\noindent The Flato–Frønsdal theorem guarantees the correspondence
so that the infinite tower of boundary currents matches exactly the infinite tower of bulk higher spin fields.

Thus, it is evident that understanding the physics of (Anti-)de Sitter space is of particular importance for higher spin theories and vice versa.  From a physical standpoint, including a cosmological constant also aligns with observations: our universe’s accelerated expansion \cite{Planck2015XX} implies a small but positive \(\Lambda\), realizing de Sitter spacetime. These backgrounds exhibit features without flat-space counterparts.  In Minkowski space, for example, the Poincaré group (via Wigner’s classification) admits only strictly massless gauge fields.  By contrast, the isometry algebra with \(\Lambda\neq0\), \(\mathfrak{so}(d,2)\) in $AdS_{d+1}$ or \(\mathfrak{so}(d+1,1)\) in $dS_{d+1}$, supports a much richer spectrum of irreducible representations \cite{Newton1950,Thomas1941,Deser:2003gw,Deser:2001us} which leads bigger set of gauge fields then just massless gauge fields.  Besides the familiar massive and massless cases, one also finds what is known as \emph{partially massless} (PM) gauge fields, whose gauge symmetries have no analogue in flat space and are thus a distinctive hallmark of \((A)dS\)\cite{Deser:1983mm, Deser:1983tm}. For a geometrical formulation of such fields, see \cite {Skvortsov:2006at}.

More precisely, a partially massless field of spin \(s\) and depth \(t\) is described by a symmetric tensor \(\phi^{(s,t)}_{\mu_{1}\dots\mu_{s}}\) in \(AdS\).  Its gauge invariance involves \(t\) derivatives on the parameter \(\xi_{\mu_{t+1}\dots\mu_{s}}\):
\[
\delta\phi^{(s,t)}_{\mu_{1}\dots\mu_{s}}
\;\propto\;
\nabla_{(\mu_{1}}\!\nabla_{\mu_{2}}\cdots\nabla_{\mu_{t}}\,
\xi_{\mu_{t+1}\dots\mu_{s})}\,.
\]
Here, the integer \(t\) (with \(1\le t\le s\)) is the \emph{depth} of the partially massless gauge symmetry\footnote{Note that in the literature, there are different conventions used for depth $t$ depending upon for what value $t$ one recovers the massless case. In this thesis, we will always make the convention clear and explicit whenever we discuss it.}: it counts the number of covariant derivatives acting on the gauge parameter.  Compared to the single‐derivative gauge transformation of a strictly massless spin-\(s\) field (which has \(t=1\)), this higher-derivative structure in the gauge transformation is responsible for keeping the number of physical degrees of freedom to an intermediate value between the massless case and the massive case. It is in this sense that PM fields are a generalization of the massless case. It is important to note that the partially massless fields are not unitary in $(A)dS$, but they are unitary in $dS$, thus making them more physical and important in the latter case\cite{Baumann:2017jvh}. 

The global symmetry, corresponding to the above PM higher spin local gauge symmetry, is called the PM higher spin algebra. Remarkably, these PM higher–spin algebras, with odd depth $t=1,3,\dots ,2\ell-1$, are isomorphic to the higher symmetries of the so–called “higher–order singleton” (satisfying \(\Box^{\ell}\phi=0\)).  In this way, the well–known correspondence between higher–spin algebras and singleton symmetries generalizes directly to a bijection between PM higher–spin algebras and the symmetries of higher–order singletons. (Note that even though a higher power of $\Box$ may seem unphysical from a unitarity point of view, they may have a chance to arise in a real physical situation. For example $\Box^2=0$ operator, which is called the "Biharmonic equation", describes many interesting physical situations in elasticity, fluids, etc., see for details \cite{Selvadurai2000, Chau2018, LandauLifshitz1970, mak2018solving}).

From the holographic point of view, these fields again play an important role. A generalized holographic dictionary between boundary conserved currents and bulk gauge fields was proposed in \cite{Dolan:2001ih}. It turns out that the boundary dual of a tower of \text{partially massless} fields is the higher‐order scalar singleton ($\square^{\ell}\,\phi(x)=0$)\cite{Bekaert:2013zya}. This leads to a generalization of \emph{Flato-Frønsdal} theorem for massless fields \cite{Basile:2014wua}.

If $D\bigl(\tfrac{d-2\ell}{2},0\bigr)$ is the space of solutions of $\square^{\ell}\,\phi(x)=0$ in d dimensions then the tensor product of two higher‐order scalar singletons decomposes into a direct sum of \(\mathfrak{so}(d,2)\) irreducible modules describing partially conserved currents (or massless fields) of all integer ranks \(s\ge0\) and all odd depths \(t=2k-1\) with \(k=1,\dots,\ell\):
\[
D\bigl(\tfrac{d-2\ell}{2},0\bigr)\;\otimes\;D\bigl(\tfrac{d-2\ell}{2},0\bigr)
\;=\;\bigoplus_{s=0}^{\infty}\;\bigoplus_{k=1}^{\ell}
D\bigl(d+s-2k\,,\,s\bigr)\,.
\]
This formula generalises the original Flato–Frønsdal theorem to higher order \(\ell>1\) and to depths \(t>1\).  In the holographic duality, the left‐hand side corresponds to the singlet sector of the \(U(N)\) (or for \(O(N)\), the symmetric) vector model, while the right‐hand side can be viewed either as the spectrum of composite primary operators in the CFT or as the spectrum of fields in $(A)dS_{d+1}$ (for details, see\cite{Basile:2014wua}). Just like the massless higher spin algebra, the PM higher spin algebra is also an infinite-dimensional algebra generalizing the massless case by adding more generators corresponding to non-trivial depth $t$. 

The above theorem is very promising and provides a huge motivation to study PM fields seriously. It turns out that PM fields are more difficult to handle. One of the major problems in the context of PM higher spin theory is constructing an interacting theory. Even for the simplest case of spin $s=2$ and $t=2$ (PM graviton), where we expect its cosmological relevance (see \cite{Baumann:2017jvh}), fully consistent and realistic models of interacting partially massless fields remain elusive. The central obstacle (or a part of it) is enforcing gauge invariance beyond the linearized level: this requirement alone suffices to forbid any theory of a single, self-interacting PM spin-2 field (see, e.g.,\cite{deRham:2013wv, Joung:2014aba, Garcia-Saenz:2015mqi}). Such no-go results naturally raise the question of whether a deeper obstruction prevents nontrivial PM dynamics altogether. As of today, there is no fully satisfactory, nontrivial interacting theory of a single partially massless higher‐spin field that is both local and unitary.  What does exist falls into two categories:
\begin{itemize}
  \item \textbf{Cubic and higher‐derivative vertices in \((A)dS\) \cite{Joung:2012rv,Boulanger:2012dx}.}  
    For any spin \(s>2\), one constructs and classifies a large, explicitly solvable sector of PM cubic interactions in (A)dS, identifies a sharp mass/depth selection rule, governing when PM couplings exist, and provides explicit PM vertices. These cubic results are expected to be further constrained by the higher-order consistency, but no general quartic completion is known. Specifically for the spin-2 case in \cite{Joung:2014aba}, it was concluded that there cannot exist a unitary theory of a PM spin-two field coupled to Einstein gravity
    with a perturbatively local Lagrangian.  Hence, the problem of PM interaction is still wide open\footnote{In this thesis, we will provide examples of the interacting theory of partially massless field via chiral approach.}.

  \item \textbf{Multi‐field, non‐unitary constructions \cite{Boulanger:2019zic}.} It was shown that one can indeed build a complete interacting theory for a multiplet of PM gravitons in \((A)dS_{4}\), provided that the internal metric is not positive definite (thus relaxing classical unitarity). This construction shows that gauge invariance per se does not fundamentally rule out nontrivial PM spin-2 interactions; only the combined insistence on unitarity and single-field self-couplings does (see also \cite{Joung:2019wwf}).  
\end{itemize}

Therefore, even though partially massless fields are a natural extension of the strictly massless case, they present formidable obstacles to constructing a fully interacting theory. Achieving such a theory would not only deepen our understanding of \((A)dS\) physics but also shed new light on higher spin holography and higher spin gravity in general. 

Motivated by these challenges and successes of Chiral massless higher spin gravity, as discussed above, it is natural to seek an analogous chiral, twistor-inspired description of PM fields, with the hope that such a framework might finally unlock the problem of consistent PM interactions. With these goals in mind, the first part of this thesis is therefore devoted to exploring chiral formulation of partially massless higher spin fields and partially massless algebra in four dimensions, and we will see that indeed this allows us to write some non-trivial interactions (see chapter 2). 

There is another important class of higher spin theory called \emph{Conformal Higher Spin} (CHS) theory. Just as higher spin theory is an extension of Einstein gravity, CHS theory is a higher spin extension of conformal gravity. CHS theory provides a remarkable example of a consistent interacting higher spin theory in a flat space background, which is local to all orders, and its generalization to arbitrary curved spacetime has been achieved. These theories have, in addition to diffeomorphism symmetry, the Weyl symmetry. This is first suggested by Fradkin-Tseytlin's \cite{Fradkin:1985am}, and developed in \cite{Tseytlin:2002gz,Segal:2002gd, Bekaert:2010ky,Basile:2022nou}.

These CHS models appear to be important for various reasons.
\begin{enumerate}
     \item In the framework of the AdS/CFT correspondence~\cite{Maldacena:1997re, LiuTseytlin1998, Sundborg2001} conformal higher‐spin fields in $d$ dimensions arise precisely as the boundary values of massless higher‐spin fields propagating in $\mathrm{AdS}_{d+1}$.  This construction directly generalizes the standard $AdS/CFT$ to its higher spin extension. Thus, a detailed study of conformal higher‐spin theories in $d$ dimensions can shed light on the properties and interactions of their bulk $AdS$ counterparts.

    \item Just as conformal gravity leads to Cartan geometry, when seen as a gauge theory of $\mathfrak{so}(d, 2)$ in terms of a Cartan-connection one-form valued in the (conformal) isometry algebra, conformal higher spin gravity offers a broader laboratory for studying “higher spin geometry". The geometries probed by conformal higher spin theories are generally parameterized by two data: by the type of matter fields we start with and by the conformally-invariant equation they obey. Also, the fact that CHS is a generalization of Conformal gravity leads to a variety of conformal invariants \cite{Nutma:2014pua, Grigoriev:2016bzl, Beccaria:2017nco, Kuzenko:2019eni, Kuzenko:2019ill}.

    \item Another motivation for examining conformal higher spin theories in four dimensions stems from the availability of twistor methods\cite{Penrose:1968me, Eastwood:1981jy}. Since twistors are inherently adapted to conformal structures, a twistor description of CHS could yield deeper geometric insight into it. Conformal gravity in twistor space was first written in \cite{Adamo:2013tja}. To date, only the linearized conformal higher spin equations have been cast in twistor language \cite{Haehnel:2016mlb, Adamo:2016ple, Adamo:2018srx}, and a fully nonlinear formulation remains out of reach. 
\end{enumerate}
Also, there is another vantage point where partially massless fields make their appearance, namely, their embedding in Conformal Higher Spin theory. The fact that CHS's symmetries coincide with symmetries of massless HS theory in one higher dimension allows PM to be embedded inside conformal higher spin theory.  Indeed $(d+1)$-dimensional free conformal spin-$s$ field can be decomposed, around $AdS_{d+1}$, into the set of spin-$s$ PM fields with all depths \cite{Nutma:2014pua, Metsaev:2014lza, JoungMkrtchyan2012},
\[
\text{CHS}_{s}= \bigoplus\limits_{t=1}^{s+\frac{d-3}{2}} PM_{(s,t)}
\]
where $PM_{(s,t)}$ with $t>s$ are massive  fields. For a bottom-up approach to this, see \cite{Joung:2014aba}. CHS theory is non-unitary since its linearized spectrum is described by a higher derivative action. Also, this non-unitary nature is much more apparent if one sees that the CHS action can be decomposed as  (see e.g.,\cite{Joung:2015jza} and references therein),
\[
S_{s}^{CHS}=\sum_{t=1}^{s+\frac{d-3}{2}}(-1)^{(t+1)}S_{(s,t)}
\]
The alternating sign between even and odd depth in the action shows the non-unitary nature of CHS. This non-unitarity should not be discouraging, as unitarity is not mandatory for the physical
applications in condensed matter and statistical physics. 

In \cite{Basile:2022nou}, a fully non-linear, manifestly covariant action for conformal higher spin gravity in an arbitrary curved spacetime was constructed using Fedosov–type deformation quantization techniques \cite{Fedosov:1994zz,Fedosov:1996}. While the full consequences of this framework remain to be explored, it opens many intriguing directions—for instance, one may hope that its Fedosov formulation will guide the development of a complete twistor description of CHS fields. More broadly, it reveals a deep link between deformation quantization and higher spin geometry in general.

In the second part of the thesis, it is this formulation of CHS that we will explore and show how to couple matter to a higher spin background in a covariant manner.

\phantomsection
\chapter*{Summary and Structure of the Thesis}
\addcontentsline{toc}{chapter}{Summary and Structure}

In this thesis, I investigate the interactions of two classes of higher spin theories: partially massless fields and Conformal Higher Spin fields. The whole thesis is based on the following three papers:
\begin{itemize}
 \item Thomas Basile and Shailesh Dhasmana, 
    \emph{Partially massless higher spin algebras in four dimensions}, 
    JHEP \textbf{12} (2024) 152.
    \href{https://doi.org/10.1007/JHEP12%282024%29152}{DOI:10.1007/JHEP12(2024)152}.
  \item Thomas Basile, Shailesh Dhasmana and Evgeny Skvortsov,
    \emph{Chiral approach to partially massless fields},
    JHEP \textbf{05} (2023) 136.
    \href{https://doi.org/10.1007/JHEP05%282023%29136}{DOI:10.1007/JHEP05(2023)136}.
  \item Thomas Basile, Shailesh Dhasmana, and Evgeny Skvortsov, 
    \emph{Scalar field on a higher-spin background via Fedosov quantization}. JHEP, 149 (2025).
    \href{https://doi.org/10.1007/JHEP07%282025%29149},
    {DOI:10.1007/JHEP07(2025)149}.
\end{itemize}

\section*{Structure of the Thesis}
\begin{description}
  \item[Chapter 0: Introduction]  
    In this chapter, we introduce the history, motivation, overview of challenges, and the role of higher spin theories in various contexts.  
  \item[Chapter 1: Review of Free Higher Spin Theory]  
    An introduction to Frønsdal formalism, describing free massless higher spin fields, is given. We review both the metric-like and the frame-like formulation of higher spin fields.  
  \item[Chapter 2: Chiral Formulation of Partially Massless Fields]  
     Here we start by reviewing the chiral approach to gravity and generalize it to a new (chiral) description of partially massless higher spin fields in 4D. Then we will work out the interactions and construct complete examples of higher spin gravities with (partially)massless fields that feature Yang–Mills and current interactions. 
  \item[Chapter 3: (Partially) Massless Higher Spin Algebra]  
   Starting with a review on Higher spin algebra, we will discuss a new realisation of partially massless higher spin algebras in four dimensions in terms of bosonic and fermionic oscillators, using Howe duality between $\mathfrak{sp}(4,\mathbb{R})\simeq\mathfrak{so}(2,3)$ and $\mathfrak{osp}(1\vert2(\ell-1),\mathbb{R})$. We also discuss the possible existence of a deformation of this algebra, which would encode interactions for the type-A$_\ell$ theory.
  \item[Chapter 4: Conformal Higher Spin and Matter Coupling] In this chapter, we will first review Conformal Higher Spin and their formulation based on Tseytlin and Segal's approach. I will introduce Fedosov techniques used to formulate covariant action for Conformal Higher Spin fields, and we finally revisit the problem of a scalar field in a higher spin background and explain the proposed, manifestly covariant formulation.
\end{description}

\section*{Notation Used}
Even though in every chapter we explicitly stated the notation used here are some of the commonly used notations throughout the thesis. 
\begin{itemize}[leftmargin=!,labelwidth=\widthof{\bfseries $\mathbb{R}$}]
  \item $D$ is the dimension of spacetime. 
  \item $d=D-1$ is to denote the dimension of the boundary.
  \item $\mu, \nu..$ are spacetime indices $\mu=0,1,\dots ,D$ and $a,b\dots=0,1,\dots D-1=d$ are Lorentz indices
  \item Indices  $\AlgInd{A,B\dots}=0,1,\dots D$ are used to denote $\mathfrak{so}(2,d)$ tensor indices.
  \item Indices $A,B,..=0,1$ and $A',B'\dots=0',1'$ are used to denote $2$-component spinor indices mainly used in chapter $2$.
  \item $T_{\mu(s)}$ is used to denote symmetric indices with the convention\\ $T_{\mu(s)}=T_{(\mu_{1}\dots \mu_{s})}=\frac{1}{s!}\big(T_{\mu_{1}\dots \mu_{s}}+\text{permutation}\big)$.
  \item The trace of a tensor is denoted by prime as $\phi'_{\mu(s-2)}=\phi^{\nu}{}_{\nu\mu(s-2)}.$
  
\end{itemize}

\chapter{Review of Free Higher Spin  Fields}
In this chapter, we review free massless fields of arbitrary spin~$s$, considering both the metric and frame-like formulations. Our discussion is restricted to \textit{bosonic fields} and focuses solely on the \textit{free theory}, thus omitting several interesting features that are significant in their own right. For those aspects, we refer the interested reader to the original literature \cite{Fronsdal:1978rb,Fronsdal:1978vb, Francia:2002aa,Francia:2002pt,Francia:2005bu, Bekaert:2002dt} and to some recent reviews \cite{Didenko:2014dwa, Ponomarev2023, Pekar:2023Modave} (An expert can skip this chapter completely).

\section{Metric Formulation}

The \textit{metric-like formulation} is a specific approach to describe higher spin fields using totally symmetric rank-$s$ Lorentz tensors, denoted by $\varphi_{\mu(s)} \equiv \varphi_{\mu_1 \dots \mu_s}$, which represent integer spin-$s$ gauge fields. This is known as the \textit{Frønsdal formulation} of linear higher spin (HS) gauge theories. It naturally generalizes the familiar Maxwell and linearized Einstein equations. To provide context and highlight the natural progression to higher spins, we begin by briefly reviewing the spin-2 case below.

\subsection{Linearised Einstein Gravity and the Spin-2 Field}

In Einstein gravity, the fundamental dynamical object is the metric \( g_{\mu\nu} \). In the linearized theory, the dynamics is approximated by considering small fluctuations around a flat Minkowski background, i.e., we assume:
\begin{equation}
g_{\mu\nu} = \eta_{\mu\nu} + h_{\mu\nu}
\end{equation}
where $h_{\mu\nu}$ is some small perturbation around the flat background. The linearized Einstein equations describe massless spin-2 particles propagating on a flat background. At this level, dynamical gravity can be viewed as a classical field theory of gravitons in flat spacetime.

The Christoffel symbols are given by:
\begin{equation}\label{Crhristoffel symbol}
\Gamma_{\rho;\mu\nu} = \frac{1}{2}(\partial_{\mu}g_{\nu\rho} + \partial_{\nu}g_{\mu\rho} - \partial_{\rho}g_{\mu\nu}) = \frac{1}{2}(\partial_{\mu}h_{\nu\rho} + \partial_{\nu}h_{\mu\rho} - \partial_{\rho}h_{\mu\nu})
\end{equation}
Using this result, we find that the linearized Riemann tensor is:
\begin{equation}
2R_{\mu\nu\rho\sigma} = \partial_{\rho}\Gamma_{\mu;\sigma\nu} - \partial_{\sigma}\Gamma_{\mu;\rho\nu} + \Gamma_{\mu;\rho\lambda}\Gamma^{\lambda}_{\sigma\nu} - \Gamma_{\mu;\sigma\lambda}\Gamma^{\lambda}_{\rho\nu} = \partial_{\nu}\partial_{\rho}h_{\mu\sigma} + \partial_{\mu}\partial_{\sigma}h_{\nu\rho} - \partial_{\mu}\partial_{\rho}h_{\nu\sigma} - \partial_{\nu}\partial_{\sigma}h_{\mu\rho}
\end{equation}
The Ricci tensor is the contraction $R_{\mu\nu} = R_{\mu\rho\nu\sigma}g^{\rho\sigma}$. The linearised Ricci tensor is:
\begin{equation}
R_{\mu\nu} = \frac{1}{2}\left(\partial_{\nu}\partial_{\rho}h_{\mu}^{\ \rho} + \partial_{\mu}\partial_{\rho}h_{\nu}^{\ \rho} - \partial_{\mu}\partial_{\nu}h - \Box h_{\mu\nu}\right)
\end{equation}
The vacuum Einstein equations are:
\begin{equation}
G_{\mu\nu} = R_{\mu\nu} - \frac{1}{2}Rg_{\mu\nu} = 0
\end{equation}
Contracting with $g^{\mu\nu}$ gives $R = 0$, so the vacuum Einstein equations reduce to the Ricci flat condition $R_{\mu\nu} = 0$. Using this, we obtain a second-order equation for $h_{\mu\nu}$:
\begin{equation}\label{linearized eq.}
\Box h_{\mu\nu} - (\partial_{\nu}\partial^{\rho}h_{\mu\rho} + \partial_{\mu}\partial^{\rho}h_{\nu\rho}) + \partial_{\mu}\partial_{\nu}h = 0
\end{equation}
This is the equation that describes the spin-2 field classically, called the graviton field. It is this equation that is generalized to higher spin naturally. Note that this equation is gauge invariant under the linearized diffeomorphism (spin-2 gauge transformation):
\begin{equation}
\delta h_{\mu\nu} = \partial_\mu \epsilon_\nu + \partial_\nu \epsilon_\mu
\end{equation}
where $\epsilon_\mu(x)$ is an arbitrary vector field.
We will choose the gauge condition,
\begin{equation}\label{gauge condition}
\partial^{\mu}h_{\mu\nu} - \frac{1}{2}\partial_{\nu}h = 0
\end{equation}
This is sometimes called the De Donder gauge condition. Substituting this, we get the wave equation for a massless, symmetric tensor field as,
\begin{equation}
\Box h_{\mu\nu} = 0
\end{equation}
To confirm that the linearized Einstein equations indeed describe a \textit{massless spin-2 field}, (as per group representation theory, which is the dimension of the little group $SO(2)$ in four dimensions gives us 2 degrees of freedom). Achieving this requires further gauge fixing, which results in complete gauge fixing. 

Even after taking the De-Donder gauge, further gauge transformations are possible because the gauge variation of the left-hand side of equation (\ref{gauge condition})  takes the form:
\begin{align}
\partial^{\mu}\partial_{\mu}\epsilon_{\nu} + \partial^{\mu}\partial_{\nu}\epsilon_{\mu} - \partial_{\nu}\partial^{\mu}\epsilon_{\mu} &= \Box\epsilon_{\nu}
\end{align}
This implies that equation (\ref{gauge condition})  is invariant under gauge transformations whose gauge parameter \( \epsilon_\nu \) satisfies the wave equation \( \Box \epsilon_\nu = 0 \). Thus, there remains sufficient residual gauge freedom to impose additional constraints.

We now aim to use this freedom to \textit{gauge away the trace} of the metric perturbation \( h_{\mu\nu} \). Specifically, we want to find a gauge transformation such that the trace of the transformed field vanishes: $\tilde{h} = h + \delta h = 0$, where the trace is defined as $h = \eta^{\mu\nu}h_{\mu\nu}$. The gauge variation of the trace is given by $-\delta h = -\eta^{\mu\nu}\delta h_{\mu\nu} =-2\partial^{\rho}\epsilon_{\rho}=h$
To make this concrete, we consider plane wave solutions for both the field and the gauge parameter:
\[
h_{\mu\nu} = H_{\mu\nu}e^{ikx}, \quad \epsilon_{\mu} = C_{\mu}e^{ikx}
\]
where \( H_{\mu\nu} \) and \( C_{\mu} \) are constant tensors, and the wavevector \( k^\mu \) satisfies \( k^2 = 0 \), since we are dealing with massless excitations. We get,
\[
\delta h(x)
= 2\,\partial_\rho \varepsilon^\rho
= 2\,\bigl(i\,k_\rho\,C^\rho\bigr)\,e^{i k\cdot x}
= \bigl(2\,i\,k\!\cdot\!C\bigr)\,e^{i k\cdot x}
\]
Now demanding that the new trace vanish, i.e., we want \(\tilde{h}(x) \;=\; h(x) \;+\; \delta h(x)\)
to satisfy \(\tilde{h}(x)=0\). Equivalently,
\[
H\,e^{i k\cdot x} + \delta h(x) \;=\; 0
\quad\Longrightarrow\quad
H + 2\,i\,(k\cdot C) \;=\; 0 \quad\Longrightarrow\,k\cdot C =- \frac{H}{2\,i}.
\]
Choose light‐cone momentum \(k^\mu = (k^+,0,0,0)\).  Then gives \(k\!\cdot\!C \;=\; k^+\,C^+\) \(k^+\,C^+ \;=\;\frac{H}{2\,i}
\quad\Longrightarrow\quad
C^+ \;=\;-\,\frac{H}{2ik^+}\). Therefore, we can eliminate the trace of \( h_{\mu\nu} \) by an appropriate gauge choice. After removing the trace, the gauge condition (4.6) becomes:
\[
\partial^{\mu}h_{\mu\nu} = 0
\]
which is precisely the \textit{transversality condition}, ensuring that the field is divergence-free.

At this stage, residual gauge transformations are still possible, provided they do not alter the trace, i.e. $\delta \tilde{h} = -\partial^{\rho}\epsilon_{\rho} = 0$. This implies that the residual gauge parameter is divergenceless: \( \partial^\rho \epsilon_\rho = 0 \). Such a condition reduces the number of independent components in \( \epsilon_\mu \) from 4 to 3.

Let us now count degrees of freedom: The symmetric tensor \( h_{\mu\nu} \) initially has 10 components. The De Donder gauge condition \( \partial^\mu h_{\mu\nu} = 0 \) fixes 4 components. Gauging away the trace removes 1 more. The residual divergenceless gauge transformations remove 3 additional components. This leaves:
\[
10 - 4 - 1 - 3 = 2
\]
independent components, which correspond exactly to the two physical polarizations of a massless spin-2 particle.

Hence, we conclude that the \textit{transverse-traceless (TT) gauge} can be reached via gauge fixing, and that the linearized Einstein equations describe precisely the correct number of degrees of freedom. This validates the interpretation of linearized Einstein gravity as a classical field theory of massless spin-2 particles (gravitons) propagating in flat spacetime.
\subsection{Frønsdal Formulation}

Motivated by the spin-2 case discussed above, we seek to generalize to the case of arbitrary spin. We can easily construct higher spin generalizations of the linearised Christoffel symbols (\ref{Crhristoffel symbol}) and use these to derive a gauge-invariant second-order equation for higher spin fields, which serves as the analog of the vacuum equation (\ref{linearized eq.}) for the spin-2 field. This method was considered by de Wit and Freedman in \cite{deWit:1980lyi} and gives generalized Christoffel symbols for the spin-$s$ gauge field as\footnote{In fact, \emph{a priori} you could try many different linear combinations of first (or higher) derivatives of your rank-$s$ field $\varphi_{\mu_1\cdots\mu_s}$. De~Wit \& Freedman show that the only choice which carries the correct index symmetries, and transforms correctly under \(\delta_\xi\varphi_{\mu_1\cdots\mu_s}
=\partial_{(\mu_1}\xi_{\mu_2\cdots\mu_s)}
\)so that each successive Christoffel drops one derivative off the gauge parameter, and eventually becomes exactly gauge invariant, is unique as used above.}
,
\begin{equation}
\Gamma^{(1)}_{\nu;\mu_1...\mu_s} = \partial_\nu \phi_{\mu_1...\mu_s} - s \partial_{(\mu_1} \phi_{\mu_2...\mu_s)\nu}
\end{equation}
Clearly, for $s = 2$ we recover the usual formula up to a sign and numerical factors that we ignore,

The gauge-invariant second-order equation they obtained is precisely the one previously found by Frønsdal in \cite{Fronsdal:1978rb}. The Frønsdal equation reduces to the wave equation for massless fields upon gauge fixing, although the resulting free theory requires double-traceless fields and traceless gauge parameters.

In \cite{deWit:1980lyi}, they further recursively define the higher-rank Christoffel symbols,
\begin{equation}
\Gamma^{(m)}_{\nu_1...\nu_m,\mu_1...\mu_s} = \partial_{\nu_1} \Gamma^{(m-1)}_{\nu_2...\nu_m,\mu_1...\mu_s} - \frac{s}{m} \partial_{(\mu_1} \Gamma^{(m-1)}_{|\nu_2...\nu_m,\nu_1|\mu_2...\mu_s)}
\end{equation}
where $(a_1...|b_1...|c_1...)$ denotes symmetrisation over the indices $a_i$ and $c_i$ only and omission of $b_{i}$. We will not go into much detail and refer the reader to \cite{deWit:1980lyi} for details. For us, only the second order in recursion is enough, for which we get,
\begin{align}
\Gamma^{(2)}_{\nu_1\nu_2,\mu_1...\mu_s} &= \partial_{\nu_1} \Gamma^{(1)}_{\nu_2,\mu_1...\mu_s} - \frac{s}{2} \partial_{(\mu_1} \Gamma^{(1)}_{|\nu_2,\nu_1|\mu_2...\mu_s)} \\
&= \partial_{\nu_1} \partial_{\nu_2} \phi_{\mu_1...\mu_s} - s \partial_{\nu_1} \partial_{(\mu_1} \phi_{\mu_2...\mu_s)\nu_2} + \frac{1}{2} s(s-1) \partial_{(\mu_1} \partial_{\mu_2} \phi_{\mu_3...\mu_s)\nu_1\nu_2}
\end{align}
Now if we contract the first two indices to get $\Gamma^{(2)\rho}_{\ \ \ \rho \mu_1...\mu_s}$, the right-hand side becomes, what is known as, \emph{"Frønsdal tensor"}.
Let us call it $\mathcal{F}_{\mu_1...\mu_s} = \Gamma^{(2)\rho}_{\ \ \ \rho \mu_1...\mu_s}$, then  $\mathcal{F}_{\mu_1...\mu_s} = 0$ refers to Frønsdal equation of spin-$s$ field as:
\begin{equation}
\mathcal{F}_{\mu_1...\mu_s} = \Box \phi_{\mu_1...\mu_s} - s \partial_{(\mu_1} \partial^\rho \phi_{\mu_2...\mu_s)\rho} + \frac{1}{2} s(s-1) \partial_{(\mu_1} \partial_{\mu_2} \phi^\rho_{\ \ \mu_3...\mu_s)\rho} = 0
\end{equation}
This is exactly what was found by Frønsdal \cite{Fronsdal:1978rb} that describes massless bosons of arbitrary spin. 

Indeed, Frønsdal's equation generalises the case of Maxwell's and the linearised vacuum Einstein equations, as can be seen for the spin-1 field we get,
\begin{equation}
\Box \phi_\mu - \partial_\mu (\partial^\nu \phi_\nu) = \partial^\nu (\partial_\nu \phi_\mu - \partial_\mu \phi_\nu) = \partial^\nu F_{\mu\nu} = 0
\end{equation}
which are exactly Maxwell's equations with $F_{\mu\nu} = \partial_\nu \phi_\mu - \partial_\mu \phi_\nu$ as field-strength. Similarly, for the spin-2 case, the Frønsdal equation is reduced to,
\begin{equation}
\mathcal{F}_{\mu\nu} = \Box \phi_{\mu\nu} - (\partial_\mu \partial^\rho \phi_{\rho\nu} + \partial_\nu \partial^\rho \phi_{\rho\mu}) + \partial_\mu \partial_\nu \phi = 0
\end{equation}
which agrees exactly with (\ref{linearized eq.}). For convenience, let us use the prime over fields to indicate that a trace is being taken, i.e. $\varphi^{\prime}_{\mu_3...\mu_s)}\equiv\eta^{\mu_1\mu_2}\varphi_{\mu_1\mu_2...\mu_s}$. with this we get that in $D=d+1$-dimensional Minkowski spacetime the equation of motion for massless spin $s$ field is given by 
\begin{equation}\label{eq:fronsdal}
\mathcal{F}_{\mu(s)} \equiv \Box\phi_{\mu_1...\mu_s}-s\,\partial_{(\mu_1}\partial\cdot\phi_{\mu_2...\mu_s)}+\frac{s(s-1)}{2}\,\partial_{(\mu_1}\partial_{\mu_2}\phi^{\prime}_{\mu_3...\mu_s)} = 0 ,
\end{equation}
where the indices within parentheses are intended to be symmetrized. Note that the convention for symmetrization used is $T_{(\mu_{1}\mu_{2}}S_{\mu_{3}...\mu_{s})}=\frac{1}{s!}\big(T_{\mu_{1}\mu_{2}}S_{\mu_{3}...\mu_{s}}+\text{permutation}\big)$.

Let us check that (\ref{eq:fronsdal}) is gauge invariant under the spin-$s$ gauge transformation $\delta \phi_{\mu_1...\mu_s} = s \partial_{(\mu_1} \zeta_{\mu_2...\mu_s)}$ where $\zeta$ is totally symmetric.  We get under the gauge transformation that
\begin{align}
\delta (\partial^{\mu_1} \phi_{\mu_1...\mu_s})
&= \Box \zeta_{\mu_2...\mu_s} + (s-1) \partial_\rho \partial_{(\mu_2} \zeta^\rho_{\ \ \mu_3...\mu_s)}\nonumber\\
\delta (\eta^{\mu_1 \mu_2} \phi_{\mu_1 \mu_2 ... \mu_s})
&= 2 \partial^\rho \zeta_{\rho \mu_3 ... \mu_s} + (s-2) \partial_{(\mu_3} \zeta^\rho_{\ \ \mu_4 ... \mu_s)\rho}
\end{align}
Using these, we find that the variation of the Frønsdal tensor is:
\begin{align}
\delta \mathcal{F}_{\mu_{1}\dots\mu_{s}} &=\frac{1}{2} s(s-1)(s-2) \partial_{(\mu_1} \partial_{\mu_2} \partial_{\mu_3} \zeta'_{\ \ \mu_4 ... \mu_s)}
\end{align}
Clearly, this variation does not vanish for spin $> 2$ unless the trace of the gauge parameter vanishes, which is non-trivial only for spin $\geq 3$. Thus, to achieve invariance of Frønsdal operator $\mathcal{F}_{\mu_1...\mu_s}$, it is necessary to restrict the gauge freedom to \textit{traceless} gauge parameters,
\begin{equation}\label{eq:gauge_constraint}
\zeta^{\prime}_{\mu(s-3)}(x) = 0 .
\end{equation}
Recalling the notion of generalized Christoffel symbols from equation (\ref{Crhristoffel symbol}). One can further extend this analogy with the usual Christoffel symbol and construct a generalized curvature. Indeed, from \cite{Cnockaert:2006hp} we have that such a curvature is given in terms of derivatives of the field $\phi$ as,
\begin{equation}
    R_{\mu_1\nu_1\mu_2\nu_2\cdots\mu_s\nu_s} = -2\, \phi_{[\mu_1[\mu_2\cdots[\mu_s,\nu_s]\cdots\nu_2]\nu_1]}
\end{equation}
Here the symbol $,\nu$ indicate a partial derivative $\partial_{\nu}$. This curvature is gauge invariant even when the gauge parameter is not traceless. Importantly, this curvature is related to the antisymmetrisation of the Frønsdal tensor, given in \cite{Bekaert:2006ix, Bekaert:2003az ,Cnockaert:2006hp} as,
\begin{equation}
R_{\mu_1\nu_1...\mu_s\nu_s} \eta^{\nu_1\nu_2} = -\frac{1}{2} \mathcal{F}_{\mu_1\mu_2[\mu_3],...[\mu_s,\nu_s],...[\nu_3]}
\end{equation}
For clarity, one must check that for spin-2, it gives,
\begin{equation}
R_{\mu\rho\nu\sigma} \eta^{\rho\sigma} = R_{\mu\nu} = -\frac{1}{2} \mathcal{F}_{\mu\nu}
\end{equation}
Thus, Frønsdal equation for spin-2, $\mathcal{F} = 0$, gave us the Ricci flat condition.

Motivated by this, we can write a generalised Einstein tensor $\mathcal{G}$ for arbitrary spin. This is given in terms of the Frønsdal tensor as follows \cite{Fronsdal:1978rb}:
\begin{equation}
\mathcal{G}_{\mu_1...\mu_s} = \mathcal{F}_{\mu_1...\mu_s} - \frac{1}{4} s(s-1) \eta_{(\mu_1\mu_2} \mathcal{F}'_{\mu_3...\mu_s)}
\end{equation}
The equation $\mathcal{G}_{\mu_1...\mu_s} = 0$, then implies that the terms on the right above vanish. Also, contracting with $\eta^{\mu_1\mu_2}$ gives $\eta^{\mu_1\mu_2} \mathcal{G}_{\mu_1...\mu_s}\sim\mathcal{F}'_{\mu_3...\mu_s}$. So, considering the equation $\mathcal{G}=0$, the trace of the Frønsdal tensor has to vanish, which means that $\mathcal{G}_{\mu_1...\mu_s} =0$ implies the equation of motion $\mathcal{F}_{\mu_1...\mu_s}=0$.

\paragraph{Frønsdal Action:} Now, with this Einstein-like tensor for higher spin, one can write the action principle very easily as,
\begin{equation}\label{eq:fronsdal_action}
\begin{aligned}
S^{(s)}_2[\phi] &= -\frac{1}{2}\int d^Dx\left(\partial_\nu\phi_{\mu_1...\mu_s}\partial^\nu\phi^{\mu_1...\mu_s}\right. \\
&\left.-\frac{s(s-1)}{2}\,\partial_\nu\phi^{\lambda}{}_{\lambda\mu_3...\mu_s}\partial^\nu\phi_\rho^{\;\rho\mu_3...\mu_s}+s(s-1)\,\partial_\nu\phi^{\lambda}{}_{\lambda\mu_3...\mu_s}\partial_\rho\phi^{\nu\rho\mu_3...\mu_s}\right. \\
&\left.-s\,\partial_\nu\phi^{\nu}{}_{\mu_2...\mu_s}\partial_\rho\phi^{\rho\mu_2...\mu_s}-\frac{s(s-1)(s-2)}{4}\,\partial_\nu\phi^{\nu\rho}{}_{\rho\mu_2...\mu_s}\partial_\lambda\phi_\sigma^{\;\;\lambda\sigma\mu_2...\mu_s}\right).
\end{aligned}
\end{equation}
and in terms of $\mathcal{G}_{\mu_1...\mu_s}$ one can simplify the action as,
\begin{equation}\label{eq:fronsdal_action_rewritten}
\begin{aligned}
S^{(s)}_2[\phi] &\sim \int d^Dx\,\phi^{\mu_1...\mu_s}\mathcal{G}_{\mu_1...\mu_s} \\
&= \int d^Dx\,\,\phi^{\mu_1...\mu_s}\left(\mathcal{F}_{\mu_1...\mu_s}-\frac{s(s-1)}{4}\eta_{(\mu_1\mu_2}\mathcal{F}^{\prime}_{\mu_3...\mu_s)}\right) ,
\end{aligned}
\end{equation}
Under gauge transformation, one can easily see that up to boundary terms, $\delta S\sim \partial^{\mu} \mathcal{G}_{\mu....}$, where
\begin{equation}\label{eq:divergence_G}
\partial^\nu\mathcal{G}_{\nu\mu_1...\mu_{s-1}} = -\frac{(s-1)(s-2)(s-3)}{4}\,\partial_{(\mu_1}\partial_{\mu_2}\partial_{\mu_3}\varphi^{\prime\prime}_{\mu_4...\mu_{s-1})} ,
\end{equation}
Clearly, for $s>3$ the divergence does not vanish, and hence we have that for $s>3$ the so-called doubly-traceless condition, which is imposed by requiring a gauge invariant action principle for the Frønsdal field.

Now, similar to what we have seen in the $spin-2$ case, we would like to check that the Frønsdal equation indeed describes 2 degrees of freedom for any integer spin. We have to ensure that the transverse-traceless gauge can be reached. In the spin-2 case, choosing the De Donder gauge condition reduced the linearized equations to a wave equation. Clearly, by eliminating the last two terms on the right of (\ref{eq:fronsdal}) with some gauge choice reduces it to a wave equation. The correct choice is the generalized De Donder gauge condition:
\begin{equation}\label{gen DG}
D_{\mu_2...\mu_s} = \partial^\rho \phi_{\rho\mu_2...\mu_s} - \frac{1}{2} (s-1) \partial_{(\mu_2} \phi'_{\ \ \mu_3...\mu_s)} = 0
\end{equation}
Using this condition, the second term in (\ref{eq:fronsdal}) cancels the third term, and the wave equation becomes:
\begin{equation}
\Box \phi_{\mu_1...\mu_s} = 0
\end{equation}
Now the condition (\ref{gen DG}) is traceless i.e. $\eta^{\mu_2\mu_3} D_{\mu_2...\mu_s} \sim \partial \phi'' = 0$
, since the double-trace vanishes. Further gauge transformations are possible because 
\begin{align}
\delta D_{\mu_2...\mu_s} &= s \Box \zeta_{\mu_2...\mu_s}
\end{align}
where, as we recall, the trace of the gauge parameter vanishes. Therefore (\ref{gen DG}) holds for gauge transformations with parameter $\zeta$ satisfying the wave equation $\Box \zeta_{\mu_2...\mu_s} = 0$. 
In this gauge, the solution for the Frønsdal equation therefore takes the form
\begin{equation}
    \phi_{\mu_1 \dots \mu_s}(x) = \int d^Dk\, e^{ikx} e_{\mu_1 \dots \mu_s}(k)
\end{equation}
with $k^2 = 0$. The completely symmetric tensor $e_{\mu_1 \dots \mu_s}$ is double-traceless and therefore has
\begin{equation}
    \binom{D - 1 + s}{s} - \binom{D - 5 + s}{s - 4}
\end{equation}
independent components. Now, some components are related by the gauge condition (\ref{gen DG}), and we also know that the de Donder tensor is traceless; all of this imposes
\begin{equation}
    \binom{D - 2 + s}{s - 1} - \binom{D - 4 + s}{s - 3}
\end{equation}
conditions on $e_{\mu_1 \dots \mu_s}$. 
Now, similar to the wave equation, the residual gauge symmetry $\Box \zeta_{\mu_1 \dots \mu_{s-1}} = 0$ has a solution,
\begin{equation}
    \zeta_{\mu_1 \dots \mu_{s-1}}(x) = \int d^Dx\, e^{ikx} \tilde{\zeta}_{\mu_1 \dots \mu_{s-1}}(k)
\end{equation}
with $k^2 = 0$. Since $\tilde{\zeta}$ is traceless, again, this allows us to eliminate
\begin{equation}
    \binom{D - 2 + s}{s - 1} - \binom{D - 4 + s}{s - 3}
\end{equation}
components from $e_{\mu_1 \dots \mu_s}$. So overall, we get the total degree of freedom as,
\begin{equation}
    \binom{D - 1 + s}{s} - \binom{D - 5 + s}{s - 4} - 2\left\{\binom{D - 2 + s}{s - 1} - \binom{D - 4 + s}{s - 3}\right\}
\end{equation}
This further simplifies to 
\begin{equation}
    \binom{D - 3 + s}{s} - \binom{D - 5 + s}{s - 2}
\end{equation}
Note that this number coincides with the dimension of the space of completely symmetric, traceless, rank-$s$ tensors of the little group $\mathrm{SO}(D-2)$.  Indeed, a massless spin-$s$ particle in $D$ dimensions is classified by its transformation under the rotations that leave its null momentum invariant, namely $\mathrm{SO}(D-2)$.  The corresponding irreducible representation is precisely the space of symmetric, traceless, rank-$s$ tensors on $\mathbb{R}^{D-2}$.  Hence, the Frønsdal equation propagates exactly the correct number of physical degrees of freedom.

\subsection{Frønsdal Formulation in (A)dS}

\subsubsection{Maximally Symmetric Space-times}

 Now we wish to extend the flat spacetime construction to any maximally symmetric background, which is a space-time whose metric has the maximum number, $\frac{D(D+1)}{2}$, of isometries in $D$ dimensions, with nonvanishing cosmological constant. Notable examples are, together with $(A)dS_D$, their euclidean versions: the hyperbolic space $H_D$, obtained from $AdS_D$ through a "Wick rotation" of the time direction, and the sphere $S^D$, obtained from $dS_D$ through a "Wick rotation" of the time direction. In their turn, $AdS_D$ and $dS_D$ are connected by a change in the sign of the curvature (\textit{i.e.}, of the cosmological constant), and the same is true for $H_D$ and $S^D$. In other words, all such spaces admit a unified description characterized by two relevant parameters: the signature of their tangent-space metric $\eta_\AlgInd{AB}$ and the sign of the cosmological constant. The simplest one is given in terms of flat coordinates that describe the embedding of any $D$-dimensional maximally symmetric space-time in a flat, $(D+1)$-dimensional one via the condition
\begin{equation}\label{eq:embedding}
k\eta_{ab}x^a x^b + z^2 = L^2 \qquad a,b=0,1,...,D-1 ,
\end{equation}
where $L$ is a constant called radius of $(A)dS_{D}$, and for the moment we do not specify the signature of $\eta_{\mu\nu}$, with the flat embedding space metric
\begin{equation}\label{eq:embedding_metric}
ds^2 = \eta_{ab}dx^a dx^b + \frac{1}{k}dz^2 .
\end{equation}

Only the sign of $k$ will be of relevance, since any rescaling with a positive factor can be absorbed into the definition of the coordinates $x^\mu$. Solving $z$ from \eqref{eq:embedding}, differentiating and substituting $dz^2$ in \eqref{eq:embedding_metric} one gets
\begin{equation}\label{eq:induced_metric}
ds^2 = \eta_{ab}dx^a dx^b + k\frac{\eta_{ac}\eta_{bd}x^c x^d}{L^2 - k\eta_{ab}x^a x^b}dx^a dx^b ,
\end{equation}
From which it follows that the metric for a maximally symmetric space can be written as
\begin{equation}\label{eq:max_sym_metric}
g_{ab} = \eta_{ab} + k\frac{\eta_{ac}\eta_{bd}x^c x^d}{L^2 - k\eta_{ab}x^a x^b} ,
\end{equation}
that has the inverse
\begin{equation}\label{eq:inverse_metric}
g^{ab} = \eta^{ab} - k\frac{x^a x^b}{L^2} .
\end{equation}

It is a simple computation to show that, 
\begin{equation}\label{eq:riemann_components}
R_{cdab} = \frac{k}{L^2}(g_{ca}g_{db} - g_{cb}g_{da}) .
\end{equation}
The Ricci tensor is
\begin{equation}\label{eq:ricci_tensor}
R_{ab} \equiv R^c_{acb} = \frac{k}{L^2}(D-1)g_{ab} ,
\end{equation}
and the curvature scalar
\begin{equation}\label{eq:curvature_scalar}
R \equiv R^a_a = \frac{k}{L^2}D(D-1) .
\end{equation}

Therefore, the Riemann tensor for a constant curvature space-time is completely determined by the curvature scalar $R$, and
\begin{equation}\label{eq:riemann_in_terms_of_R}
R_{cdab} = \frac{1}{D(D-1)}R(g_{ca}g_{db} - g_{cb}g_{da}) ,
\end{equation}
Moreover, the scalar curvature is proportional to \(k\), whose sign distinguishes the corresponding space–time:
\begin{itemize}
  \item \(k=0\) gives a flat metric \(\eta_{\mu\nu}\) (of arbitrary signature).
  \item \(k=+1\) (resp.\ \(-1\)) yields a constant positive (resp.\ negative) curvature manifold.
\end{itemize}
If \(\eta_{\mu\nu}\) is Euclidean, then \(k=+1\) (resp.\ \(-1\)) defines the sphere \(S^D\) (resp.\ hyperbolic space \(H_D\)); if \(\eta_{\mu\nu}\) is Lorentzian, \(k=+1\) (resp.\ \(-1\)) defines de Sitter (\(dS\)) (resp.\ anti–de Sitter, \(AdS\)) space–time.

All of these are vacuum solutions of Einstein’s equations with cosmological constant \(\Lambda\).

\begin{equation}\label{eq:einstein_equations}
R_{ab} - \frac{1}{2}g_{ab}R = -\Lambda g_{ab} ,
\end{equation}
that are extrema of the Einstein-Hilbert action
\begin{equation}\label{eq:eh_action}
S = \frac{1}{16\pi G_D}\int d^D x\sqrt{-g}(R-2\Lambda) .
\end{equation}

From \eqref{eq:ricci_tensor} and \eqref{eq:curvature_scalar} it follows that
\begin{equation}\label{eq:einstein_tensor}
R_{ab} - \frac{1}{2}g_{ab}R = -k\frac{(D-1)(D-2)}{2L^2}g_{ab} ,
\end{equation}
and by comparison with \eqref{eq:einstein_equations} one has
\begin{equation}\label{eq:cosmological_constant}
\Lambda = k\frac{(D-1)(D-2)}{2L^2} ,
\end{equation}
from which one reads that the sign of $\Lambda$ is related to that of $k$, \textit{i.e.}, of the curvature, for any\footnote{$D=1,2$ are trivial cases, since in $D=1$ there is no curvature, and in $D=2$, although a curvature can be defined, the Einstein-Hilbert action, that encodes the dynamics of the gravitational field, is a topological invariant, the Euler characteristic.} $D>2$. Thus, $S^D$ and $dS_D$ space-times have a positive cosmological constant, and $H_D$ and $AdS_D$ have a negative one.

\subsubsection{Free Equations in (A)dS Space-time}

The interaction with the fixed gravitational background is introduced, as usual, by covariantizing derivatives with respect to the (A)dS Levi-Civita connection $\partial\to\nabla\equiv\partial+\Gamma$. Moreover,
\begin{equation}\label{eq:covariant_trace}
\varphi^{\prime}_{\mu_3...\mu_s}=g^{\mu_1\mu_2}\varphi_{\mu_1...\mu_s} ,
\end{equation}
where $g$ is the $(A)dS$ metric tensor, and we are assuming $\varphi''=0$ and $\epsilon'=0$. Now, because the commutator of two covariant derivatives does not vanish, it turns out that in $(A)dS$, these two conditions are no longer sufficient to ensure the invariance under the covariantized spin-$s$ gauge transformation
\begin{equation}\label{eq:cov_gauge_transformation}
\delta\varphi_{\mu_{1}\dots \mu_{s}} = s\nabla_{(\mu_1}\zeta_{\mu_2...\mu_s)} ,
\end{equation}
Indeed, we compute, say for $AdS$, that
\begin{equation}\label{eq:cov_commutator}
[\nabla_\mu,\nabla_\nu] \varphi_{\rho_1...\rho_s} = \frac{s}{L^2} \left(g_{\nu(\rho_1} \varphi_{\mu|\rho_2...\rho_s)} - g_{\mu(\rho_1|} \varphi_{\nu|\rho_2...\rho_s)} \right) ,
\end{equation}
Now using covariant derivatives, the covariantized Frønsdal kinetic operator,
\begin{equation}\label{eq:cov_fronsdal}
\mathcal{F}^{\text{cov}}_{\mu_1...\mu_s}(\varphi) = \Box\varphi_{\mu_1...\mu_s}-s\nabla_{(\mu_1}\nabla\cdot\varphi_{\mu_2...\mu_s)}+s(s-1) \nabla_{(\mu_1}\nabla_{\mu_2} \varphi^{\prime}_{\mu_3...\mu_s)}
\end{equation}
(where $\Box=g^{\mu\nu}\nabla_\mu\nabla_\nu$). Now, under gauge transformation \eqref{eq:cov_gauge_transformation}, the variation of the Frønsdal kinetic operator produces terms such as
\begin{equation}\label{eq:variation_terms}
s [\Box,\nabla_{(\mu_1}] \epsilon_{\mu_2...\mu_s)} + \frac{1}{L^2}s(s-1)(D+s-3) \nabla_{(\mu_1} \epsilon_{\mu_2...\mu_s)} .
\end{equation}

To eliminate these terms, it is necessary to modify the kinetic operator with appropriate terms of order $1/L^2$ that cancel the variation of \eqref{eq:cov_fronsdal} and vanish in the flat limit $L\to\infty$. By explicitly calculating the commutator in \eqref{eq:variation_terms}, one can check that the invariant Frønsdal equation in $AdS_D$ is
\begin{equation}\label{eq:ads_fronsdal}
\mathcal{F}^L_{\mu(s)} \equiv \mathcal{F}^{\text{cov}}_{\mu(s)} - \frac{1}{L^2} \left\{[(3-D-s)(2-s)-s]\varphi_{\mu(s)} + \frac{s(s-1)}{4} g_{(\mu_1\mu_2} \varphi^{\prime}_{\mu_3...\mu_s)}\right\} = 0 .
\end{equation}

Notice that, although we deal with massless fields, requiring the invariance of the Frønsdal 
equations in a space-time with a non-vanishing cosmological constant result in the appearance of a mass-like term, which originates from the coupling with the (constant) 
space-time curvature. One can repeat now for \eqref{eq:ads_fronsdal} the same considerations made above for the flat case. Again, the Frønsdal equations are non-Lagrangian, and one can
Define a generalized Einstein tensor
\begin{equation}\label{eq:generalized_einstein}
\mathcal{G}^L_{\mu(s)} = \mathcal{F}^L_{\mu(s)} - \frac{s(s-1)}{4} g_{(\mu_1\mu_2} \mathcal{F}^{\prime L}_{\mu_3...\mu_s)} ,
\end{equation}
In terms of which one can construct a Lagrangian from which \eqref{eq:ads_fronsdal} follows. 

Now it is important to note that, once the background deviates from (anti)de Sitter or Minkowski, the Frønsdal operator ceases to be gauge invariant. In verifying gauge invariance, one must commute covariant derivatives \(\nabla\). In Minkowski space, these simply commute, while in constant curvature (Anti-)de Sitter space, their commutator yields terms proportional to the background metric, producing harmless “mass-like” contributions. However, on a generic background, the commutator involves the full Riemann tensor, one finds schematically
\[
\delta\mathcal{F}
= R\cdot\nabla\,\xi + \nabla R\cdot\xi
\neq0,
\]
where \(R\) is the Riemann tensor. As a result, the Frønsdal operator loses its gauge invariance and propagates unwanted degrees of freedom on arbitrary metric backgrounds.

\section{Frame Formulation}

In gravitational theories, there's a well-established approach that uses a first-order formalism built on a frame field and a Lorentz connection. This framework highlights the role of gauging the isometries of the tangent space, such as the Poincaré or (A)dS groups, in a way that parallels the treatment of gauge symmetries in Yang-Mills theories. Inspired by this analogy, it becomes natural to explore whether massless higher spin theories can also be reformulated in terms of one-form fields that correspond to generators of a deeper symmetry structure. Doing so may shed light on the algebraic foundations needed for consistent higher spin interactions. If such a symmetry algebra exists, then the free field equations could be viewed as linear approximations of a more general interacting theory, where the fields take values in a non-abelian algebra whose generators align with the internal index structure of the connections themselves. Indeed, similar to spin-$2$ fields, spin-$s$ fields can also be described in a frame-like approach\cite{Vasiliev:1980as}.
This section is dedicated to presenting this frame-like formulation of higher spin theory. We begin by reviewing the Einstein–Cartan approach to gravity, which leads to an elegant action for gravity proposed by MacDowell and Mansouri, along with its refined version developed by Stelle and West, in which full covariance under all symmetries is made manifest. Finally, we will see how to extend such a formulation to higher spin.

\subsection{Frame Formulation of Gravity}

\paragraph{Einstein-Cartan action}
In the traditional formulation of gravity is given by the Einstein Hilbert action, which is expressed in terms of the spacetime metric and leads to second-order field equations for the metric tensor. In this formulation, vanishing of torsion is a fundamental assumption of the theory.  However, an alternative but equivalent formulation known as the Einstein–Cartan (EC) action, in which, instead of treating the metric as a fundamental field, one uses the vielbein to define the geometry, while the spin connection encodes how local Lorentz frames are patched together. A key feature of the Einstein–Cartan formalism is that the spin connection is not assumed a priori to be torsion-free; instead, its equation of motion determines the vanishing torsion. It is this formalism that allows a clearer geometric interpretation, especially in the context of gauge-theoretic formulations of gravity, making it closer to Yang-Mills theory in spirit. 
The conventional formulation of gravity in terms of metric tensor $g_{\mu\nu}$ is described by Einstein-Hilbert action
\begin{equation}
    S_{EH}=\alpha\int\,\,\sqrt{-g}R\,,
\end{equation}
where $R$ is the curvature scalar and $\alpha$ is a constant, later chosen conveniently. 

The basic dynamical fields in the Einstein-Cartan (EC) formulation are two Lorentz algebra $\mathfrak{so}(3,1)$ valued one-forms:
\begin{itemize}
    \item A 1-form field \( e^{a} = e^{a}_{\mu} dx^{\mu} \), called the \emph{frame field} (or \emph{vielbein}),
    \item A 1-form field \( \omega^{a}{}_{b} = \omega_{\mu}^{a}{}_{b} dx^{\mu} \), called the \emph{spin connection}.
\end{itemize}
Here, Latin indices denote Lorentz indices, which describe the local Lorentz frame on the tangent space \( TM \) of the spacetime manifold \( M \). In the language of principal fibre bundles, these are also referred to as \emph{fibre indices}; we avoid the language of Bundles here and refer interested readers to \cite{Krasnov:1970bpz} for further details on this aspect. 

The formalism in terms of differential form provides a powerful and geometrically natural framework for describing gravity, significantly simplifying both conceptual understanding and computational clarity. In this approach, the conventional metric tensor \( g_{\mu\nu} \) is reconstructed from the frame field through the relation:
\begin{equation}
    g_{\mu\nu}=\eta_{ab}e_{\mu}^ae_{\nu}^b.
\end{equation}
The first Cartan structure is given by, 
\begin{equation}
    T^{a}=de^{a} +\omega^{a}{}_{b}\wedge e^{b}.
\end{equation}
Here, the 2-form \( T^a \) represents the torsion. In conventional gravity theories, it is typically assumed that the torsion vanishes, and the metric satisfies the compatibility condition, which implies that the spin connection satisfies \( \omega_{ab} = -\omega_{ba} \). Furthermore, the vanishing of the torsion 2-form, expressed as  
\[  
de^{a} + \omega^{a}{}_{b} \wedge e^{b} = 0\,,  
\]  
ensures that \( \omega \) depends on the vielbein \( e \), rendering \( \omega \) an auxiliary field. The curvature 2-form \( R^{a}{}_{b} \) of the spin connection is given by (for a detailed discussion on the frame formulation, its relation to the standard formulation, and explicit computations, see \cite{Yepez2011EinsteinsVierbein}),
\begin{equation}
    R^{a}{}_{b}=d\omega^{a}{}_{b}+\omega^{a}{}_{c}\wedge \omega^{c}{}_{b}.
\end{equation}
These objects further satisfy the Bianchi identities, 
\begin{align}
    \nabla T^a:=dT^a+\omega^{a}{}_{b}\wedge T^b=R^{a}{}_{b}\wedge e^b\,,\nonumber\\
    \nabla R^{a}{}_{b}:= dR^{a}{}_{b}+\omega^{a}{}_{c}\wedge R^{c}{}_{b}-\omega^{c}{}_{b}\wedge R^{a}{}_{c}=0.
\end{align}
The familiar usual spacetime curvature tensor is given by 
\begin{align}
    R^{a}{}_{b}=\frac{1}{2}R^{a}{}_{bcd} e^{c}\wedge e^{d}\,,\nonumber\\
    R^{\mu}{}_{\nu\rho\sigma}=e^{\mu}_{a}e_{\nu}{}^{b}e_{\rho}{}^{c}e_{\sigma}{}^{d}R^{a}{}_{bcd}\,, 
\end{align}
where the first equation is the fact that $R^{a}{}_{b}$ is a $2$-form. For simplicity, we will work in four dimensions here.  
Now, one can start with the Einstein-Hilbert action and plug all these familiar objects into it, and we will get the action as (modulo some constant coefficient),
\begin{equation}\label{EH in form}
S[e] = \alpha \int d^4x \, |e| \, R(\omega(e))_{\mu\nu}^{\;\;\;ab} \, e^\mu_a \, e^\nu_b\,,
\end{equation}
where $e = \det(e^{a}_{\mu})$. While this action depend only on the frame field $e$ (since $R$ depends on $e$ through $\omega(e)$), Palatini observed \cite{Palatini1919} (and later work gave this a more geometrical interpretation using differential forms \cite{Cartan1922}) that promoting $\omega$ to an independent variable yields a new action:

\begin{equation}\label{Palatini}
    S[e,\omega] = \alpha\int d^4x \, |e| \, R(\omega)_{\mu\nu}^{\;\;\;ab} \, e^\mu_a \, e^\nu_b,
\end{equation}
which remains equivalent to \eqref{EH in form}. This equivalence arises because the equation of motion for $\omega$ enforces $T^{a} = 0$, thereby recovering the original relationship between $\omega$ and $e$.

Expressing everything in differential form notation - using the volume form $|e| \, d^4x = \frac{1}{4!} \epsilon_{abcd} \, e^a \wedge e^b \wedge e^c \wedge e^d$ and the curvature 2-form $R^{ab} = \frac{1}{2} R^{ab}_{\mu\nu} \, dx^\mu \wedge dx^\nu$  leads to the Einstein-Cartan action:

\begin{equation}
    S_{EC}[e, \omega] =  \int \epsilon_{abcd} \, e^a \wedge e^b \wedge R(\omega)^{cd},
\end{equation}
where $\alpha$ is chosen so that the overall factor is unity.

In the presence of cosmological constant $\Lambda$ this generalizes to,
\begin{equation}
    S_{EC}=\int_{M} \epsilon_{abcd}\,e^a\wedge e^b\wedge\big( R^{cd}[\omega]-\frac{\Lambda}{3} e^c\wedge e^d\big).
\end{equation}
The equations of motion obtained by varying with respect to $e$ and $\omega$ are:
\begin{align}
    \delta \omega &: \quad de^{a} + \omega^{a}{}_{b} \wedge e^{b} = 0 \quad \text{(Torsion-free condition)},\label{eq:torsion} \\
    \delta e &: \quad \epsilon_{abcd}\left(e^b \wedge R^{cd} - \frac{2\Lambda}{3}e^b \wedge e^c \wedge e^d\right) = 0. \label{eq:eom_e}
\end{align}
By solving the torsion-free condition \eqref{eq:torsion} to determine $\omega(e)$ and substituting into \eqref{eq:eom_e}, we recover the standard Einstein field equations with cosmological constant $\Lambda$. This demonstrates the equivalence between the Einstein-Cartan formulation and the conventional Einstein-Hilbert formulation of general relativity.

Now it is important to note that the EC formulation is a polynomial in fields even when $\Lambda\neq0$. This fact and the necessity of frame formulation in the presence of fermions make EC theory much more fundamental than metric formulation. The Einstein-Cartan action serves as the foundation for several important formulations of gravity, including the pure connection formulation and chiral formulation (see e.g., \cite{Krasnov:2021nsq,Krasnov:2011pp}). For comprehensive discussions of these developments based on the EC action, we refer to \cite{Krasnov:1970bpz}.

In what follows, we will focus on one particular line of development, the MacDowell-Mansouri formulation, which admits a natural frame-like extension to higher spin theories.

\paragraph{MacDowell-Mansouri Formulation} 
We begin with the fundamental observation that in Einstein-Cartan theory, the basic ingredients - the one-form fields $e^{a}$ and $\omega^{L\,ab}$ - can be naturally combined into a single object (Note that we have used the superscript $L$ to indicate the usual Lorentz connection we discussed above). This unification is motivated by the fact that in $D$-dimensional spacetime, the pair $(e,\omega)$ comprises $D + \frac{D(D-1)}{2} = \frac{D(D+1)}{2}$ independent one-forms, which precisely matches the dimension of the Poincaré group $ISO(D-1,1)$. 

The MacDowell-Mansouri approach elegantly combines $e^{a}$ and $\omega^{\,ab}$ into a single connection one-form valued in the Poincaré algebra:
\begin{equation}
    \Omega = e^a P_a + \frac{1}{2} \omega^{\,ab} M_{ab},
\end{equation}
where $P_a$ and $M_{ab}$ are the generators of translations and Lorentz transformations, respectively, in the Poincaré algebra $\mathfrak{iso}(D-1,1)$, 
\begin{align*}
[P_a, P_b] &= 0 \\
[M_{ab}, P_c] &=  \eta_{bc}P_a-\eta_{ac}P_b \\
[M_{ab}, M_{cd}] &=  \eta_{ad}M_{bc}-\eta_{ac}M_{bd} -\eta_{bd}M_{ac} + \eta_{bc}M_{ad}.
\end{align*}
The corresponding curvature two-form is given by:
\begin{equation}\label{eq:poincare_curvature}
    R^L = d\Omega + \Omega \wedge \Omega \equiv T^a P_a + \frac{1}{2} R^{L\,ab} M_{ab},
\end{equation}
where $R^L$ is the Lorentz algebra valued curvature 2-form, $T^a = de^a + \omega^{L\,a}{}_b \wedge e^b$ is the torsion.
 Now we know that the most symmetrical solutions are constant curvature solutions, for example, $AdS$ or $dS$, or Minkowski, corresponding to negative, positive, and zero cosmological constants, respectively. In the $D=4$ case, they are given by,
\begin{equation}
    R^{L\,ab}=\frac{2\Lambda}{3}e^{a}\wedge e^b.
\end{equation}
So, if we want to encode the most symmetrical solution by some flat connection, we need to define 
\begin{equation}
    R:=d\Omega+\frac{1}{2}[\Omega, \Omega]=T^a P_a + \frac{1}{2}\big( R^{L\,ab}-\frac{2\Lambda}{3}e^a\wedge e^b\big) M_{ab}.
\end{equation}
which amounts to modify the Poincare algebra to 
\begin{align*}
[P_a, P_b] &= -\frac{\Lambda}{3}M_{ab}, \\
[M_{ab}, P_c] &=  \eta_{bc}P_a-\eta_{ac}P_b , \\
[M_{ab}, M_{cd}] &=  \eta_{ad}M_{bc}-\eta_{ac}M_{bd} -\eta_{bd}M_{ac} + \eta_{bc}M_{ad}
\end{align*}
This is $\mathfrak{so}(3,2)$ or $\mathfrak{so}(4,1)$ depending on the sign of $\Lambda$. With this let us consider the the action 
\begin{equation}\label{MM action}
    S_{MM}=\frac{1}{\Lambda}\int_{M^D} R^{a_{1}a_{2}}\wedge R^{a_{3}a_{4}}\epsilon_{a_{1}\dots a_{4}}.
\end{equation}
We can see that 
\begin{equation}
   \epsilon_{abcd} R^{ab}\wedge R^{cd}=\epsilon_{abcd} R^{L\,ab}\wedge R^{L\,cd}-\frac{4\Lambda}{3}\epsilon_{abcd}\big(R^{L\,ab}-\frac{\Lambda}{3}e^a\wedge e^b\big)\wedge e^{c}\wedge e^d,
\end{equation}
Hence, the action becomes 
\begin{align}
S_{MM} = \frac{1}{\Lambda} \int_{M^D} R^{L\,a_{1}a_{2}} &\wedge R^{L\,a_{3}a_{4}} \, \epsilon_{a_{1} \dots a_{4}} \nonumber \\&
\quad - \frac{4}{3} \int_{M^D} \left(R^{L\,a_{1}a_{2}} - \frac{\Lambda}{3} e^{a_{1}} \wedge e^{a_{2}} \right) \wedge e^{a_{3}} \wedge e^{a_{4}} \, \epsilon_{a_{1} \dots a_{4}}.
\end{align}
The first term becomes singular in the limit \(\Lambda \to 0\), whereas the remaining terms remain well-defined.  In \(D = 4\), the first term is the Gauss-Bonnet term, which is a topological invariant and does not affect the equations of motion\footnote{It is important to note that in dimensions \(D > 4\), this singular term poses a problem because it is not topological and thus contributes nontrivially to the dynamics.}. This also explains the choice of the coefficient \(1/\Lambda\): it ensures that the second and third terms behave properly in the \(\Lambda \to 0\) limit. Therefore, modulo the topological term in four dimensions, the MacDowell--Mansouri action is equivalent to the Einstein--Cartan action. That is,
\[
S_{MM} = S_{EC} + \text{(Topological term)}.
\]
We will not explicitly verify it here, but it is straightforward to see that the symmetries of the MacDowell--Mansouri action~\eqref{MM action} are diffeomorphisms and local Lorentz transformations. Now the action $S_{MM}$ in the form (\ref{MM action}) can easily be generalized to $D$ dimensions by extending the epsilon tensor to $\epsilon_{a_{1}\dots a_{D}}$. Also one can make the full \(\mathfrak{so}(D{-}1,2)\) symmetry manifest by embedding the vielbein \(e^a\) and the spin connection \(\omega_\mu^{ab}\) into a single \(\mathfrak{so}(D{-}1,2)\)- or \(\mathfrak{so}(D,1)\)-valued connection 1-form:
\[
\Omega = \Omega_\mu^{\AlgInd{AB}} dx^\mu M_\AlgInd{AB},
\]
where the indices \(\AlgInd{A, B} = 0, 1, \dots, D\), and $\AlgInd{A}=(a,D)$and \(M_\AlgInd{AB}\) are the generators of \(\mathfrak{so}(D{-}1,2)\), satisfying the algebra:
\[
[M_\AlgInd{AB}, M_\AlgInd{CD}] = \eta_{\AlgInd{AD}} M_\AlgInd{BC} - \eta_{\AlgInd{AC}} M_\AlgInd{BD} - \eta_{\AlgInd{BD}} M_\AlgInd{AC} + \eta_{\AlgInd{BC}} M_\AlgInd{AD}.
\]
Here, \(\eta_\AlgInd{AB}\) denotes the invariant metric of either the \(\mathfrak{so}(D{-}1,2)\) or \(\mathfrak{so}(D,1)\) algebra.

The embedding is carried out by identifying the generators as \(M_{aD} = \sqrt{\frac{|\Lambda|}{3}} P_a\). Note that in the expression \(M_{aD}\), the symbol \(D\) refers to the numerical label of the extra dimension, not a tensor index. To avoid confusion, following conventions common in the literature, we will refer to the \(D^\text{th}\) direction as the “\(5^\text{th}\) component,” and accordingly write \(\AlgInd{A} = (a,5)\). In this notation, we have \(M_{a5} = \sqrt{\frac{|\Lambda|}{3}} P_a\). For further details and a more in-depth discussion of the underlying symmetry structure, see~\cite{Didenko:2014dwa}.
The curvature 2-form is given by:
\[
R^{\AlgInd{AB}} = d\Omega^{\AlgInd{AB}} + \Omega^{\AlgInd{A}}{}_{\AlgInd{C}} \wedge \Omega^{\AlgInd{CB}}.
\]
With this structure in place, we can now reformulate the action~\eqref{MM action} in a manifestly \(\mathfrak{so}(D{-}1,2)\)-invariant form.

Since we are now working in an extended algebra with an additional dimension, we must use the corresponding epsilon tensor \(\epsilon_{\AlgInd{A_{1} \dots A_{D+1}}}\). If we fix the last index to be \(5\), i.e., \(\epsilon_{\AlgInd{A_{1} \dots A_{D}} 5}\), then the remaining indices must be Lorentz indices, and we can write:
\[
\epsilon_{\AlgInd{A_{1} \dots A_{D}} 5} = \epsilon_{a_{1} \dots a_{D} 5} = \epsilon_{a_{1} \dots a_{D}}.
\]
Using this, we can extend the $ D$-dimensional MacDowell–Mansouri action to:
\begin{equation}
S_{MM} = \frac{1}{\Lambda} \int_{M^D} R^{a_{1} a_{2}} \wedge R^{a_{3} a_{4}} \wedge e^{a_{5}} \wedge \dots \wedge e^{a_{D}} \, \epsilon_{a_{1} \dots a_{D} 5}.
\end{equation}

Now, we define a new frame field \(E^\AlgInd{A}\), valued in the AdS algebra, such that:
\[
E^\AlgInd{A} = (e^a, \, E^5 = 0).
\]
With this, the action becomes:
\begin{equation}
S_{MM} = \frac{1}{\Lambda} \int R^{\AlgInd{A_{1} A_{2}}} \wedge R^{\AlgInd{A_{3} A_{4}}} \wedge E^{\AlgInd{A_{5}}} \wedge \dots \wedge E^{\AlgInd{A_{D}}} \, \epsilon_{\AlgInd{A_{1} \dots A_{D}} 5}.
\end{equation}
However, this expression is still not manifestly \(\mathfrak{so}(D{-}1,2)\)-invariant, as the appearance of the fixed index \(5\) in the epsilon tensor breaks manifest covariance. To restore full covariance, we introduce an additional field: a time-like vector \(V^{\AlgInd{A}}\), known as the \emph{compensator} field.

Using this compensator, we can now write the action in a manifestly \(\mathfrak{so}(D{-}1,2)\) or \(\mathfrak{so}(D,1)\)-invariant form as:
\begin{equation}\label{MMSW action}
S_{MM} = \frac{1}{\Lambda} \int R^{\AlgInd{A_{1} A_{2}}} \wedge R^{\AlgInd{A_{3} A_{4}}} \wedge E^{\AlgInd{A_{5}}} \wedge \dots \wedge E^{\AlgInd{A_{D}}} \wedge V^{\AlgInd{A_{D+1}}} \, \epsilon_{\AlgInd{A_{1} \dots A_{D+1}}}.
\end{equation}

This is called \textbf{MacDowell-Mansouri-Stelle-West} gravity. Clearly, choosing \(V^\AlgInd{A} = \delta^\AlgInd{A}_{5}\) recovers the previous expression. This choice is referred to as the \emph{standard gauge}. More generally, we impose that \(V^\AlgInd{A}\) has constant norm:
\[
V^\AlgInd{A} V^\AlgInd{B} \eta_{\AlgInd{AB}} = \rho,
\]
The reverse argument is, in fact, more natural: suppose we begin with a manifestly \(\mathfrak{so}(D{-}1,2)\)-invariant formulation. To recover a Lorentz-invariant theory, we must break the \(\mathfrak{so}(D{-}1,2)\) symmetry down to the Lorentz subalgebra \(\mathfrak{so}(D{-}1,1)\). This is achieved by introducing an AdS vector \(V^\AlgInd{A}\), which defines the Lorentz subalgebra as the stability subalgebra of \(V^\AlgInd{A}\), i.e., the set of generators that leave \(V^\AlgInd{A}\) invariant.

We know that the connection is \(\Omega = \Omega_\mu^{\AlgInd{AB}} dx^\mu M_{\AlgInd{AB}}\), and in the standard gauge, this implies \(\Omega^{a5} = e^a\). Since we also have \(E^a = e^a\), one might be tempted to write \(E^\AlgInd{A} = \Omega^{\AlgInd{AB}} V_\AlgInd{B}\). However, this expression does not transform covariantly under local \(\mathfrak{so}(D{-}1,2)\) transformations. To resolve this, we instead define the AdS frame field as:
\[
E^\AlgInd{A} = dV^\AlgInd{A} + \Omega^{\AlgInd{A}}{}_{\AlgInd{B}} V^\AlgInd{B}.
\]

With this definition, the symmetry transformations of the fields become transparent:
\begin{equation}
    \delta \Omega^{\AlgInd{AB}} = D_\Omega \epsilon^{\AlgInd{AB}}, \quad
    \delta V^\AlgInd{A} = -\epsilon^{\AlgInd{A}}{}_{\AlgInd{B}} V^\AlgInd{B}, \quad
    \text{and} \quad \delta E^\AlgInd{A} = -\epsilon^{\AlgInd{A}}{}_{\AlgInd{B}} E^\AlgInd{B}.
\end{equation}
These variations show that the action is invariant under local Lorentz transformations, local translations (via the compensator \(V^\AlgInd{A}\)), and diffeomorphisms, provided the variations above are assumed. The action thus possesses full \((A)dS\) gauge symmetry, since all \((A)dS\) indices are fully contracted.

Finally, note that the definition
\[
E^\AlgInd{A} = dV^\AlgInd{A} + \Omega^{\AlgInd{A}}{}_{\AlgInd{B}} V^\AlgInd{B}
\]
combined with the constraint \(V^\AlgInd{A} V^\AlgInd{B} \eta_\AlgInd{AB} = \rho\), implies
\[
E^\AlgInd{A} V_\AlgInd{A} = 0,
\]
as a consequence of the anti-symmetry of the connection \(\Omega^{\AlgInd{AB}} = -\Omega^{\AlgInd{BA}}\).

\subsection{Frame-like Formulation of Higher Spin Fields}
Similar to the spin-2 case reviewed above, spin-$s$ fields can also be described using the frame-like formalism~\cite{Aragone:1979hx,Aragone:1980rk,Vasiliev:1980as}. The core idea behind this approach to higher spin gauge fields is to extend the $(A)dS$ isometry algebra, \(\mathfrak{so}(D{-}1,2)\) or \(\mathfrak{so}(D,1)\), to a larger algebra known as the \emph{higher spin algebra}. In this framework, higher spin fields are described by one-forms valued in this extended algebra, in analogy with how gravity is formulated as a gauge theory in the MacDowell--Mansouri--Stelle--West construction for \(\mathfrak{so}(D{-}1,2)\) or \(\mathfrak{so}(D,1)\). Importantly, this higher spin algebra is infinite-dimensional, reflecting the fact that a consistent higher spin theory necessarily involves an infinite tower of gauge fields.
We extend the usual fields, $e^{a}$ and $\omega^{ab}$, of the spin-$2$ case to the generalized frame field $e^{a(s-1)}$, which is symmetric and traceless in its Lorentz indices. That is, it takes values in the irreducible representation of $\mathfrak{so}(D{-}1,1)$ labeled by a Young diagram with a single row of length $s{-}1$. The linearized gauge transformation of the frame field will be,
\[
\delta e^{a(s-1)} = \nabla\epsilon^{a(s-1)} + h^{b}\epsilon^{a(s-1)},{}_{b},
\]
where $h$ is the background frame field and the zero-form $\epsilon^{a(s-1)}$ is a gauge parameter. The second term is called shift-symmetry, with gauge parameter $\epsilon^{a(s-1),b}$ representing the generalized local Lorentz transformations. Now the gauge field associated with $\epsilon^{a(s-1),b}$ is a one-form is given by $\omega^{a(s-1),b}_{\mu}dx^{\mu}$, such that, the field strength of this $\omega^{a(s-1),b}$ is given as,

\[
R^{a(s-1)} = \nabla e^{a(s-1)} + h^{b} \wedge \omega^{a(s-1)},
\]
This is invariant not only under $\epsilon^{a(s-1)}$ and $\epsilon^{a(s-1),b}$ transformations but under certain algebraic transformations of $\omega^{a(s-1),b}_{\mu}dx^{\mu}$ so that the full gauge transformation,
\[
\delta\omega^{a(s-1),b} = \nabla\epsilon^{a(s-1),b} + h^{c}\epsilon^{a(s-1),b}{}_{c},
\]
Now again the parameter $\epsilon^{a(s-1),bb}$ suggests that there has to be another gauge field associated to it, say $\omega^{a(s-1),bb}$, and this process continues \cite{Lopatin:1987hz} till $\omega^{a(s-1),b(s-1)}$. In summary, we have a collection of $1$-forms as,
\[
\Yboxdim{11pt}
\begin{tabular}{@{}c@{\quad}c@{\quad}c@{\quad}c@{\quad}c@{}}
$e^{a(s-1)}$ &
$\omega^{a(s-1),b}$ &
$\omega^{a(s-1),bb}$ &
$\cdots$ &
$\omega^{a(s-1),b(s-1)}$
\\[1em]
\gyoung(_5<s-1>) &
\gyoung(_5<s-1>,;) &
\gyoung(_5<s-1>,;;) &
$\cdots$ &
\gyoung(_5<s-1>,_5<s-1>)
\end{tabular}
\]
Here, the Young diagrams indicate the irreducible \(\mathfrak{so}(D-1,1)\) representations carried by each field. The first field, \(e^{a(s-1)}\), is the generalized frame (or vielbein), and the second, \(\omega^{a(s-1),b}\), plays the role of the primary spin-\(s\) connection—exactly as \(e^a\) and \(\omega^{ab}\) do in the spin-2 (gravity) case. All subsequent fields, commonly referred to as “extra fields”, are algebraically expressed in terms of derivatives of the frame field. Although they do not introduce new propagating degrees of freedom at the free level, these extra fields are essential for constructing consistent cubic interactions (see \cite{Fradkin:1987ks}).

\begin{framed}
\textbf{Remark:} Just like spin-$2$ case, it is possible to connect to a metric-like Frønsdal formulation. Since the primary frame field $e^{a(s-1)}=e_{\mu}^{a(s-1)}dx^{\mu}$ is the fundamental variable, we take the background frame $h^{a}_{\mu}$ which is invertible and convert spacetime indices into fibre/Lorentz indices to get $e^{a(s-1)|b}$. This is a reducible Lorentz tensor that decomposes into irreducible parts,
\[
e^{a_1\cdots a_{s-1}\,\lvert\,b}
=
\underbrace{e^{(a_1\cdots a_{s-1}b)}\bigl|_{\mathrm{traceless}}}_{[\,s\,]}
\;\oplus\;
\underbrace{\bigl(e^{a_1\cdots a_{s-1}\,\lvert\,b}
- e^{(a_1\cdots a_{s-1}b)}\bigr)\bigl|_{\mathrm{traceless}}}_{[\,s-1,1\,]}
\;\oplus\;
\underbrace{g^{b(a_1}\,t^{a_2\cdots a_{s-1})}}_{[\,s-2\,]}\,.
\]
The last term represents the trace given by $
t^{a_2\cdots a_{s-1}}
:= 
\frac{1}{D+2s-4}\,
g_{b c}\,
e^{b\,a_2\cdots a_{s-1}\,\lvert\,c}\,.
$
In terms of young diagram this means $e^{a(s-1)|b}=\Yboxdim{11pt}\gyoung(_5<s>) \oplus
\,\gyoung(_5<s-2>)\oplus \gyoung(_5<s-1>,;)$.

Now we have shift symmetries $\delta e^{a(s-1)}= h_{b}\epsilon^{a(s-1),b}$ the parameter $\epsilon^{a(s-1),b}$ is of same symmetry type and hence we can gauge away the last irreducible part $\Yboxdim{11pt}\gyoung(_5<s-1>,;)$, which left us with 
\begin{equation}
   e^{a(s-1)|b}=\Yboxdim{11pt}\gyoung(_5<s>) \oplus
\,\gyoung(_5<s-2>) 
\end{equation}
These components combine into Frønsdal's field of rank-$s$. This follows from the following fact: Any symmetric, double-traceless rank-\(s\) tensor \(T_{\mu_1\cdots\mu_s}\) can be decomposed into a traceless rank-\(s\) part and a traceless rank-\((s-2)\) part.  Explicitly,
\[
T_{\mu_1\cdots\mu_s}
=\varphi_{\mu_1\cdots\mu_s}
+\eta_{(\mu_1\mu_2}\,U_{\mu_3\cdots\mu_s)}\,,
\]
where
\[
\eta^{\nu\rho}\,\varphi_{\nu\rho\mu_3\cdots\mu_s}=0
\quad\text{and}\quad
\eta^{\nu\rho}\,U_{\nu\rho\mu_5\cdots\mu_s}=0.
\]
Here \(\varphi_{\mu_1\cdots\mu_s}\) is the traceless spin-\(s\) part and \(U_{\mu_1\cdots\mu_{s-2}}\) is the traceless spin-\((s-2)\) part.  Double-tracelessness of \(T\) ensures no further trace components appear.  
\end{framed}

Furthermore, the pattern of Young diagrams—and the corresponding branching rules of representation theory—strongly suggests that one may combine all of these one-forms into a single \((A)dS\) connection. This master one-form takes values in the irreducible two-row Young diagram of length \((s-1)\), thereby unifying the entire spin-\(s\) field content in a manifestly \((A)dS\)–covariant way as,
\[
\Yboxdim{11pt}
\gyoung(_5<s-1>,_5<s-1>)|_{\mathfrak{so}(D-1,2)}=
\gyoung(_5<s-1>) \oplus
\gyoung(_5<s-1>,;)\,\oplus
\gyoung(_5<s-1>,;;) \oplus
\cdots \oplus
\gyoung(_5<s-1>,_5<s-1>)
\]
Let's call such a \(\mathfrak{so}(D-1,2)\) object,
\begin{equation}
    \Omega^{\AlgInd{A}(s-1),\AlgInd{B}(s-1)}=\Yboxdim{11pt}
\gyoung(_5<s-1>,_5<s-1>)|_{\mathfrak{so}(D-1,2)} \qquad\, \text{where now}\,\, \AlgInd{A,B}= 0,1,2\dots D
\end{equation}

\begin{framed}\label{PM remark}
\textbf{Remark.} In \((A)dS\) space, gauge fields include not only the usual massless higher spin fields but also the so-called \emph{partially massless} (PM) fields. A PM field of spin \(s\) and \emph{depth} \(t\) is described by a single one-form connection, say $W$, valued in an irreducible \((A)dS\) representation whose Young diagram has two rows of lengths \((s-1)\) and \((s-t)\). The integer \(t\) counts the number of derivatives in the gauge transformation,
\[
W^{\AlgInd{A}(s-1),\AlgInd{B}(s-t)}=\Yboxdim{11pt}
\gyoung(_6<s-1>,_3<s-t>)|_{\mathfrak{so}(D-1,2)} \qquad \text{with}\,\,\,\,
\delta\phi_{\mu(s)}
=
\underbrace{\nabla\!\cdots\!\nabla}_{t}\,\epsilon_{\,\mu(s-t)}\,.\qquad t\in 1,2\dots,s
\]
Here \(\phi_{\mu(s)}\) is the corresponding field in the metric-like approach (see chapter 2). The massless case is recovered for \(t=1\). We will discuss partially massless fields/algebra in four dimensions in detail in the next two chapters. For a very general discussion on possible gauge fields and their connection in $(A)dS$ see \cite{Skvortsov:2009zu} 
\end{framed}

Now, as in the case of spin-$2$, one also introduces a time-like compensator vector $V^\AlgInd{A}$ of constant norm $\rho$. Then the component of the connection $\Omega_{\mu}^{\AlgInd{A}(s-1), \AlgInd{B}(s-1)}$  that is mostly in the direction of $V^\AlgInd{A}$ is the frame-like field while contraction with less $V_{\AlgInd{A}}$- are the other connections 
\begin{align}\label{eq:frame_from_connection}
E^{\AlgInd{
A}(s-1)} = \Omega^{\AlgInd{A}(s-1), \AlgInd{B}_1...\AlgInd{B}_{s-1}} V_{\AlgInd{B}_1} ... V_{\AlgInd{B}_{s-1}}\\
\omega^{\AlgInd{A}(s-1),\AlgInd{B}(r)}=\Omega^{\AlgInd{A}(s-1),\AlgInd{B}(r)\AlgInd{B}_{r+1}\dots \AlgInd{B}_{s-1}}V_{\AlgInd{B}_{r+1}}\dots V_{\AlgInd{B}_{s-1}}
,\end{align}
Note that contraction with $V_{\AlgInd{A}}$ more than $s-1$ times will vanish due to the Young symmetry. Clearly in standard gauge $V^{\AlgInd{A}}=\delta^{\AlgInd{A}}_{5}$ we get, the Lorentz frame and connection field
\begin{align}
e^{a(s-1)} = \Omega^{a(s-1),5\dots 5}\nonumber \\
\omega^{a(s-1),B_{1}\dots b_{r}}=\Pi\big(\Omega^{\AlgInd{A}(s-1),\AlgInd{B}_{1}\dots \AlgInd{B}_{r}5\dots 5}\big)
\end{align}
where $\Pi$ is a projector to the Lorentz-traceless part of a Lorentz tensor, which is needed for $t>1$

The linearized curvature of the connection $\Omega^{\AlgInd{A}(s-1),\AlgInd{B}(s-1)}$ is defined as,
\begin{equation}\label{eq:linearized_curvature}
\begin{aligned}
R^{\AlgInd{A}(s-1),\AlgInd{B}(s-1)} &= d\Omega^{\AlgInd{A}(s-1),\AlgInd{B}(s-1)} + (s-1)\,\Omega_{0}^{\AlgInd{A}}{}_{\AlgInd{C}} \Omega^{\AlgInd{CA}_2...\AlgInd{A}_{s-1}, \AlgInd{B}_1...\AlgInd{B}_{s-1}} + \\
&+(s-1)\, \Omega_0^{\AlgInd{B}_1}{}_{\AlgInd{C}} \Omega^{\AlgInd{A}_1...\AlgInd{A}_{s-1}, \AlgInd{CB}_2...\AlgInd{B}_{s-1}},
\end{aligned}
\end{equation}
where $\Omega_{0}$ is the $\mathfrak{so}(d-1,2)$ connection associated with the $AdS$ space
solution defining the covariant derivative $D_{0}$ such that $D^2_{0}=0$.
\subsubsection{Action for HS gauge fields}

Just as in the MacDowell–Mansouri–Stelle–West formulation of gravity, one can construct the higher spin action quadatric in linearized curvature by using the compensator field \(V^A\)  the most general \(\mathfrak{so}(D-1,2)\)-invariant action is given in terms of background frame field $E_0^B=\nabla_0 V^B$ as,
\begin{equation}\label{eq:hs_action}
S^{(s)}[\Omega, V ] = \frac{1}{2} \sum_{p=0}^{s-2} a(s,p) S^{(s,p)}[ \Omega, V]
\end{equation}
where $a(s,p)$ is the \textit{a priori} arbitrary coefficient of the term
\begin{equation}\label{eq:hs_action_term}
\begin{aligned}
S^{(s,p)}[ \Omega,V ] &= \epsilon_{\AlgInd{A}_1...\AlgInd{A}_{D+1}} \int_{\mathcal{M}^D} E_0^{\AlgInd{A}_5} ... E_0^{\AlgInd{A}_D} V^{\AlgInd{A}_{D+1}} V_{\AlgInd{C}(2s-4-2p)} \times \\
&\times R^{\AlgInd{A}_1 \AlgInd{B}(s-2), \AlgInd{A}_2 \AlgInd{C}(s-2-p) \AlgInd{D}(p)} R^{\AlgInd{A}_3} {}_{\AlgInd{B}(s-2)}{}^{ \AlgInd{A}_4\,\AlgInd{C}(s-2-p)}{}_{\AlgInd{D}(p)} .
\end{aligned}
\end{equation}
Now, by imposing the higher‐spin analogue of the torsion constraint, one finds that each connection \(\omega^{a_1\cdots a_{s-1},\,b_1\cdots b_t}\) can be solved in terms of \(t\) derivatives of the frame‐like field. To achieve this, the coefficients \(a(s,p)\) in the action must be chosen so that the Euler–Lagrange equations are nontrivial only for the frame field and the first connection (\(t=1\)). All other connections, \(\omega^{a_1\cdots a_{s-1},\,b_1\cdots b_t}\) with \(t>1\), then enter the action only through total derivatives. This condition ensures the absence of higher‐derivative terms in the free theory and uniquely fixes the spin-\(s\) quadratic action—up to an overall normalization \(b(s)\). Concretely, one finds \cite{ref19}
\begin{equation}\label{eq:hs_coefficients}
a(s,p)
= b(s)\,(-\Lambda)^{-(s-p-1)}\,
\frac{(D-5+2(s-p-2))!!\,(s-p-1)}{(s-p-2)!}\,,
\end{equation}
where \(b(s)\) remains an arbitrary, spin‐dependent constant.

Thus, we get an action which is manifestly invariant under diffeomorphisms, local $\mathfrak{so}(D-1,2)$ transformations of the spin-$2$ sector,
\begin{equation}
    \delta \omega^{\AlgInd{AB}} = D_{0}\epsilon^{\AlgInd{AB}}, \quad
    \delta V^\AlgInd{A} = -\epsilon^{\AlgInd{A}}{}_{\AlgInd{B}} V^\AlgInd{B}, 
\end{equation}
and HS gauge transformations
\begin{equation}
    \delta \Omega^{\AlgInd{A}(s-1),\AlgInd{B}(s-1)}=D_{0} \epsilon ^{\AlgInd{A}(s-1),\AlgInd{B}(s-1)}.
\end{equation}
Finally we see that the equation of motion for $\omega^{a(s-1),b}$
 are 
 \begin{equation}
     R^{\AlgInd{A}(s-1),\AlgInd{B}(s-1)}V_{\AlgInd{B}}\dots V_{\AlgInd{B}}=0\,,
 \end{equation}
 which is nothing but the zero-torsion condition. Solving this, we can get the $\omega^{a(s-1),b}$ in terms of derivatives of the frame field, which leads to an action that depends on the frame field and its derivatives. Effectively, it depends on the Fronsdal field's component of the frame field.

\chapter{Chiral Approach to Partially Massless Fields}
\label{ch:chiral-pm}

\section*{Introduction}
General Relativity describes gravity as the geometry of spacetime. From a field-theoretic point of view, gravity is seen as mediated by a particle called the graviton, which is a self-interacting, massless spin-2 particle. A natural physical question is whether the graviton could have a small mass---that is, whether small mass corrections to General Relativity might be present. This question has attracted significant interest, particularly in the context of modifying gravity at large distances. Specifically, the observed acceleration of the expansion of the universe has led some physicists to propose that this phenomenon could be explained if the graviton were massive~\cite{Dvali:2000hr, Deffayet:2000uy, Deffayet:2001pu}.

Another motivation for considering a massive graviton arises when the background spacetime is not flat. In curved spacetimes, particularly those with constant curvature, such as de Sitter (dS) space, the spectrum of allowed gauge fields is broader than in flat spacetime. While flat spacetime admits only massless and massive particles, constant curvature spacetimes allow for additional possibilities. Notably, on the $dS$ space, there exists a mathematical possibility of a graviton that is neither fully massless nor fully massive. Such gravitons propagate more degrees of freedom than a massless graviton but fewer than a massive one. These are known as \emph{partially massless} (PM) gravitons, and they exhibit a scalar gauge symmetry that removes one of the degrees of freedom associated with a massive graviton.
\cite{Deser:2001wx,Deser:2012qg,Deser:2012euu,Higuchi:1986py}
.

This structure generalizes to higher spin fields (\( s \geq 2 \)), but with an important distinction: for spins greater than 2, there exists \emph{multiple} partially massless modes, each labeled by an integer \( t \), known as the \emph{depth}. The depth determines both the number of derivatives appearing in the gauge transformation and the number of propagating modes. We refer to such fields as \emph{partially massless higher spin fields}.

In this way, we see that partially massless fields, in general, constitute a novel class of gauge fields that emerge in the presence of a non-vanishing cosmological constant~\cite{Deser:1983mm,Higuchi:1986wu,Deser:2001us} (see also~\cite{Deser:1983tm,Higuchi:1986py,Higuchi:1989gz,Deser:2001wx,Zinoviev:2001dt,Deser:2001pe}). These fields correspond to special mass values at which we get an extra Noether identity of higher order, which further implies higher-derivative gauge symmetry. As a result, they propagate an intermediate number of degrees of freedom: more than a massless field (which has single-derivative gauge symmetry), but fewer than a massive field (which lacks any gauge symmetry).

Partially massless fields are unitary in de Sitter space
and may have phenomenological applications (see e.g.
\cite{Baumann:2017jvh, Goon:2018fyu} and references therein).
Despite being non-unitary around anti-de Sitter spacetime,
partially massless fields are nevertheless of interest, if only because they are dual to partially-conserved
currents, that is, currents which are annihilated after
taking several divergences \cite{Dolan:2001ih}. These kinds
of currents naturally appear in free conformal field theories
of higher-derivative scalar fields, i.e., scalar fields subject
to polywave equations of the type $\Box^\ell\phi=0$,
with $\ell>1$ \cite{Bekaert:2013zya}, which are known to describe special RG fixed points called `multi-critical isotropic Lifshitz points\footnote{A Lifshitz point is a special multicritical point in the phase diagram of a condensed matter system, where a disordered phase, a uniformly ordered phase, and a spatially modulated phase all meet. It occurs when competing interactions present in the system.}' \cite{Diehl:2002ri}. The holographic dual of this theory would be a theory of both massless and partially massless fields of arbitrary spin in anti-de Sitter
space, which has been studied in \cite{Bekaert:2013zya,Alkalaev:2014nsa, Brust:2016zns}
(see also \cite{Eastwood2008, Gover2009, Michel2014, Joung:2015jza}
for works on the corresponding higher spin algebras), but not worked out in full details yet. One reason is that holographic duals of vector models feature severe nonlocalities that invalidate the usual field theory methods to construct them \cite{Maldacena:2015iua, Bekaert:2015tva, Sleight:2017pcz, Ponomarev:2017qab}.

Nevertheless, cubic interactions for partially massless fields
of any spins have been studied
\cite{Joung:2012hz, Boulanger:2012dx, Joung:2012rv}, but complete interacting theories featuring partially massless fields in the spectrum are still lacking.
Particular attention has been given to the problem of finding
gravitational interactions and constructing what one might want
to call a theory of partially massless gravity, i.e., an interacting
theory of a massless and a partially massless spin-$2$ field.
Unfortunately, the search for such a non-linear theory led
to several no-go theorems, whether it is in relation to massive
and/or bimetric gravity
\cite{Hassan:2012gz,deRham:2013wv,Garcia-Saenz:2015mqi,Apolo:2016ort,Apolo:2016vkn},
with conformal gravity \cite{Deser:2012euu, Deser:2012qg},
or on general grounds \cite{Joung:2014aba, Joung:2019wwf}.
A notable exception is the recent work \cite{Boulanger:2019zic},
wherein an interacting theory of a multiplet of spin-$2$ partially massless fields has been found.

All of the aforementioned results were obtained by working
with symmetric rank-$s$ tensors to describe partially massless
fields of spin-$s$. In this chapter, we  introduce a new description
of partially massless fields in $4d$, inspired by twistor theory
and the description of massless fields given in \cite{Krasnov:2021nsq}\footnote{For more twistor literature on massless fields equation see \cite{Hughston:1979tq, Hitchin:1980hp, Eastwood:1981jy}.},
based on a pair of a $1$-form and a $0$-form which are also
$SL(2,\mathbb C)$ spin-tensors (see also \cite{Krasnov:2011pp,Krasnov:2016emc,Krasnov:1970bpz}
for a pure connection formulation of gravity, which is closely
related). In terms of these new field variables, the free action for partially massless fields takes a fairly simple form,
and more importantly, one can construct complete interacting theories
featuring partially massless fields. We will illustrate this last fact
by spelling out a partially massless higher spin extension of self-dual
Yang--Mills, which is a generalization of the higher spin
extension discussed in \cite{Krasnov:2021nsq}, and a theory featuring
current interactions between a couple of massless fields
with a partially massless one, which is complete at the cubic order.

The organization of this chapter is as follows: in section \ref{sec:PM graviton}, we will start by discussing the spin-$2$ case to motivate and illustrate the concept of partially masslessness. In section \ref{sec:free}, we briefly recall the metric- and frame-like description 
of free partially massless fields before introducing a new description based on twistor-inspired, two-component spin-tensors, in Section \ref{sec:twistor like}, describes the Plebanski formulation. This forms the basis for chiral higher spin formulation for both massless and partially massless fields, which is the content of section \ref{sec:chiral action}. Finally, we will discuss two simple examples of fully interacting theories featuring partially massless fields in section \ref{sec:interactions}, and we end up with a discussion on beyond maximal depth (i.e., $t>s$) in section \ref{sec: maximal depth} and a summary in section \ref{sec:disco}.

\section{Partially Massless Gravitons}
\label{sec:PM graviton}
As we have already seen in the previous chapter, how the massless spin-$2$ case provides the motivation for higher spin case, similarly, we will start by considering a massive spin-$2$ field in a 4-dimensional constant curvature background with background metric $\bar{g}_{\mu\nu}$
\begin{align*}
    g_{\mu\nu}=\bar{g}_{\mu\nu}+h_{\mu\nu}
\end{align*}
The Fierz--Pauli action for a massive spin-2 field \( h_{\mu\nu} \) on a 4-dimensional spacetime with constant curvature (i.e., \( R_{\mu\nu} = \Lambda g_{\mu\nu} \)) is:

\begin{equation*}
S = \int d^4 x \, \sqrt{-g} \left[ \mathcal{L}_{\text{kin}}(h_{\mu\nu}) - \frac{1}{2} m^2 \left( h_{\mu\nu} h^{\mu\nu} - h^2 \right) \right]
\end{equation*}
with the kinetic term \( \mathcal{L}_{\text{kin}}(h_{\mu\nu}) \) given by:

\begin{align*}
\mathcal{L}_{\text{kin}}(h_{\mu\nu}) =\;
& -\frac{1}{2} \nabla_\lambda h_{\mu\nu} \nabla^\lambda h^{\mu\nu}
+ \nabla_\mu h^{\mu\nu} \nabla^\lambda h_{\lambda\nu}
- \nabla_\mu h^{\mu\nu} \nabla_\nu h
+ \frac{1}{2} \nabla_\lambda h \nabla^\lambda h - \Lambda \left( h_{\mu\nu} h^{\mu\nu} - \frac{1}{2} h^2 \right)
\end{align*}
Here:
\begin{itemize}
  \item \( h = \bar{g}^{\mu\nu} h_{\mu\nu} \) is the trace of the field,
  \item \( \nabla_\mu \) is the covariant derivative compatible with the background metric \( \bar{g}_{\mu\nu} \),
  \item \( \Lambda \) is the cosmological constant.
\end{itemize}
The equation of motion is given by
\begin{equation}
\mathcal{K}_{\mu\nu}^{\phantom{\mu}\rho\sigma}h_{\rho\sigma}-\Lambda\left(h_{\mu\nu}-\frac{1}{2}\bar{g}_{\mu\nu}h\right)+\frac{m^{2}}{2}\left(h_{\mu\nu}-\bar{g}_{\mu\nu}h\right)=0\,,
\end{equation}
where 
\begin{equation}
\mathcal{K}_{\mu\nu}^{\phantom{\mu}\rho\sigma}h_{\rho\sigma}\equiv-\tfrac{1}{2}\Bigl{[}\delta_{\mu}^{\rho}\delta_{\nu}^{\sigma}\nabla^{2}+\bar{g}^{\rho\sigma}\nabla_{\mu}\nabla_{\nu}-\delta_{\mu}^{\rho}\nabla^{\sigma}\nabla_{\nu}-\delta_{\nu}^{\sigma}\nabla^{\sigma}\nabla_{\mu}-\bar{g}_{\mu\nu}\bar{g}^{\rho\sigma}\nabla^{2}+\bar{g}_{\mu\nu}\nabla^{\rho}\nabla^{\sigma}\Bigr{]}\,h_{\rho\sigma}\,.
\end{equation}

For convenience, let's use the notation $\mathcal{E}_{\mu\nu}$ for the L.H.S of the equation of motion as,
\begin{align}\label{equation symbol}                   \mathcal{E}_{\mu\nu}=\mathcal{K}_{\mu\nu}^{\phantom{\mu}\rho\sigma}h_{\rho\sigma}-\Lambda\left(h_{\mu\nu}-\frac{1}{2}\bar{g}_{\mu\nu}h\right)+\frac{m^{2}}{2}\left(h_{\mu\nu}-\bar{g}_{\mu\nu}h\right).
\end{align}
We now see how to count the degrees of freedom in this model in a covariant way. One first notices that, due to the Bianchi identities identically satisfied by the kinetic operator,
\begin{equation}
\nabla^{\mu}\left[\mathcal{K}_{\mu\nu}^{\phantom{\mu}\rho\sigma}h_{\rho\sigma}-\Lambda\left(h_{\mu\nu}-\frac{1}{2}\bar{g}_{\mu\nu}h\right)\right]=0\,,
\end{equation}
one has from the definition (\ref{equation symbol})
\begin{equation}
\nabla^{\mu}\mathcal{E}_{\mu\nu}=\frac{m^{2}}{2}\left(\nabla^{\mu}h_{\mu\nu}-\bar{g}^{\rho\sigma}\nabla_{\nu}h_{\rho\sigma}\right)\,.
\end{equation}
Thus on-shell we get the relation (assuming $m^{2}\neq 0$)
\begin{equation}\label{vector constraint}
\nabla^{\mu}h_{\mu\nu}-\nabla_{\nu}h\simeq 0.
\end{equation}
This is a vector constraint and thus provides four constraint equations. These eliminate four degrees of freedom from the original 10 components of the symmetric tensor $h_{\mu\nu}$, leaving six.

Now, a second covariant divergence of the field equations gives,
\begin{equation}
\nabla^{\mu}\nabla^{\nu}\mathcal{E}_{\mu\nu}=\frac{m^{2}}{2}\left(\nabla^{\mu}\nabla^{\nu}h_{\mu\nu}-\nabla^{2}h\right)\,.
\end{equation}
On the other hand, taking trace the field $\mathcal{E}_{\mu\nu}$ gives,
\begin{equation}
\bar{g}^{\mu\nu}\mathcal{E}_{\mu\nu}=\nabla^{2}h-\nabla^{\mu}\nabla^{\nu}h_{\mu\nu}+\left(\Lambda-\frac{3m^{2}}{2}\right)h\,.
\end{equation}
From the above two equations, we see that,
\begin{equation}
2\nabla^{\mu}\nabla^{\nu}\mathcal{E}_{\mu\nu}+m^{2}\bar{g}^{\mu\nu}\mathcal{E}_{\mu\nu}=\frac{m^{2}}{2}\left(2\Lambda-3m^{2}\right)h\,.
\end{equation}
Hence, on-shell, it constitutes a scalar constraint reading,
\begin{equation}\label{scalar constraint}
\left(2\Lambda-3m^{2}\right)h\simeq 0.
\end{equation}

\paragraph{Case : $m^{2}\neq 2\Lambda/3$} In this case this constraint implies $h\simeq 0$ which reduces the system to 5 degree of freedom. Together with vector constraint (\ref{vector constraint}) and scalar constraint (\ref{scalar constraint}) gives, $\nabla^{\mu}h_{\mu\nu}\simeq 0$, means that $h_{\mu\nu}$ is transverse-traceless in vacuum. By enforcing these constraints, the equations of motion are reduced to the following system,
\begin{equation}
\left(\nabla^{2}-m^{2}\right)h_{\mu\nu}+\frac{2\Lambda}{3}h_{\mu\nu}\simeq 0\,,\qquad\nabla^{\mu}h_{\mu\nu}\simeq 0\,,\qquad h\simeq 0\,.
\end{equation}
Hence, on a generic Einstein spacetime, the above theory describes a massive graviton with five degrees of freedom.

\paragraph{Case: $2\Lambda=3m^2$ (Higuchi bound \cite{Higuchi:1986py}) } In this case, we get an off-shell relation, signaling a Noether identity, as
\begin{equation}
2\nabla^{\mu}\nabla^{\nu}\mathcal{E}_{\mu\nu}+m^{2}\bar{g}^{\mu\nu}\mathcal{E}_{\mu\nu}=0\,.
\end{equation}
The existence of such a Noether identity indicates that there must be a two-derivative gauge symmetry available. Indeed the transformation 
\begin{equation}
    \delta{h_{\mu\nu}}=\left(\nabla_{\mu}\nabla_{\nu}+\frac{m^{2}}{2}\bar{g}_{\mu\nu}\right)\xi(x)=\left(\nabla_{\mu}\nabla_{\nu}+\frac{\Lambda}{3}\bar{g}_{\mu\nu}\right)\xi(x)\,,
\end{equation}
The scalar gauge symmetry, parametrized by the gauge parameter $\xi$, eliminates an additional degree of freedom, reducing the total number of propagating degrees of freedom to four. This places the partially massless graviton in an intermediate regime between the massless case (which has 2 degrees of freedom) and the fully massive case (which has 5 degrees of freedom)-hence the name \emph{partially massless}.

\newpage

\section{PM Higher Spin: Metric and Frame-Like descriptions}
\label{sec:free}
\subsection{Metric-like Approach}
Free fields are known to be in one-to-one correspondence with irreducible representations of the spacetime isometry group.
For de Sitter (dS) space in $(d+1)$-dimensions,
the isometry algebra is $\mathfrak{so}(1,d+1)$, whereas 
for anti-de Sitter (AdS) space in $(d+1)$-dimensions, it is
$\mathfrak{so}(2,d)$. We will hereafter denote these algebras
collectively by $\mathfrak{g}_\Lambda$.
One new feature of the representation theory
of (anti-)de Sitter algebras, as compared to that of
the Poincar\'e algebra,
is that they admit irreducible representations that are realized
as fields propagating
an intermediate number of degrees of freedom
between that of a massless field and that of a massive one, 
for a fixed value of the spin \cite{Deser:1983mm,Higuchi:1986wu,Deser:2001us}.
Consequently, these fields are called partially massless (PM).
A spin-$s$ partially massless field
of depth-$t$, with $1 \leq t \leq s$,
can be represented by a rank-$s$ symmetric tensor
$\Phi^{a_1 \dots a_s} \equiv \Phi^{a(s)}$ 
that is subject to\footnote{In trying to save letters we abbreviate
a group of symmetric indices $a_1 \dots a_s$ as $a(s)$ and, more generally,
denote all indices to be symmetrized by the same letter.}
\begin{align}\label{masslessdict}
     \delta_\xi\Phi^{a(s)}
    = \underbrace{\nabla^a \dots \nabla^a}_{t\ \text{times}} \xi^{a(s-t)}
    + \dots\,,
\end{align}
where the dots denote lower-order derivative terms.
In other words, the depth of a partially massless field 
is nothing but the number of derivatives in its gauge transformation,
and the massless case corresponds to $t=1$ in our convention.
Omitting the transversality and tracelessness constraints
for $\Phi$ and $\xi$, the equations of motion reduce to 
\begin{align}
    (\square -m^2)\,\Phi^{a(s)} & = 0\,,
    && m^2 = -\Lambda\,\big((d+s-t-1)(s-t-1)-s\big)\,.
\end{align}
where, as for the massless case, the mass-like term is
proportional to the cosmological constant and depends on
the spin-$s$, depth-$t$ and spacetime dimension $d+1$.
The mass-like term is fixed by the gauge symmetry. While equations
of motion are simple, the action requires an intricate pattern
of auxiliary fields\footnote{This is because partially massless fields are closer to the massive ones.
For a massive spin-$s$ field, one has to impose transversality
on top of the Klein--Gordon equation,
which starting from $s=2$ requires auxiliary fields \cite{Fierz:1939ix, Singh:1974qz}.} \cite{Zinoviev:2001dt}.

\subsection{Frame Formulation of PM Fields}

As noted in the previous chapter, in \((A)dS\) space the spectrum of gauge fields contains not only the familiar massless higher spin fields but also \emph{partially massless} (PM) fields. A PM field of spin \(s\) and \emph{depth} \(t\) is described by a single one-form connection taking values in the irreducible \((A)dS\) representation whose Young diagram has two rows of lengths \(s-1\) and \(s-t\). The integer \(t\) specifies the number of derivatives that appear in its gauge transformation as,
\[
 W^{\mathbb Y_{s,t}}\overset{\mathrm{def}}{=} W^{\AlgInd{A}(s-1),\AlgInd{B}(s-t)}=\Yboxdim{11pt}
\gyoung(_5<s-1>,_3<s-t>)|_{\mathfrak{so}(D-1,2)} \qquad \text{with}\,\,\,\,\,
t\in 1,2\dots,s.
\]
The massless case is recovered for \(t=1\). Similar to the massless case, PM fields also admit a frame formulation as follows.

The gauge transformation and curvature of the connection are given by,
\begin{align}
    \delta W^{\AlgInd{A}(s-1),\AlgInd{B}(s-t)}=D\xi^{\AlgInd{A}(s-1),\AlgInd{B}(s-t)},\nonumber\\
    R^{\AlgInd{A}(s-1),\AlgInd{B}(s-t)}=D W^{\AlgInd{A}(s-1),\AlgInd{B}(s-t)},
\end{align}
where $D$ is the covariant derivative with respect to the background connection $W_{0}$. The field strength is invariant because $D^2 = 0$, and it satisfies the usual Bianchi identity, $D R = 0$. 

\begin{framed}\label{Remark: decomp into Lorentz}
\paragraph{Remark:} For more clarity, it is instructive to see this in terms of the Lorentz tensor. Upon decomposing it with respect to the Lorentz algebra (using the branching rule), one gets many auxiliary fields,
\begin{equation}
    W^{\mathbb Y_{s,t}} =\{\omega^{a(s-k),b(s-m)}\}\,,
    \qquad\text{with}\qquad
    k \in \{1,2,\dots,t\}\,,
    \quad m \in \{t,t+1,\dots,s\}\,.
\end{equation}
The gauge-invariant curvature $R$ for $W$, then given by,
\begin{equation}
    R[W] = \nabla W + e^a\wedge\rho(P_a)\,W\,,
\end{equation}
where $\rho$ is the representation $\mathbb Y$ of the (anti-)de Sitter algebra\footnotemark.
This curvature is invariant under the gauge transformations generated by a 0-form \(\xi\) valued in the same representation \(\mathbb{Y}\)
\begin{equation}\label{eq:gauge_frame}
    \delta_\xi W = \nabla\xi + e^a\,\rho(P_a)\,\xi\,
\end{equation}
on an (anti-)de Sitter background, i.e. defined by a vielbein
$e^a$ and spin-connection $\varpi^{a,b}$ obeying
\begin{equation}
    \nabla e^a = 0\,,
    \qquad \qquad 
    R^{ab}-e^{[a} \wedge e^{b]} = 0\,,
\end{equation}
where $\nabla$ is the covariant derivative induced by $\varpi$
and $R^{a,b} = d\varpi^{a,b} + \varpi^a{}_c\wedge\varpi^{c,b}$
is its usual Lorentz curvature $2$-form. 
Note in particular that the second piece of this gauge transformation,
the one generated by the action of the transvection generators,
is algebraic (it is given by symmetrization and contraction 
of the background vielbein with the gauge parameters, and 
does not involve any derivatives).
\end{framed}

As in the massless case, the dynamical fields—those embedded in \(W^{A(s-1),B(s-t)}\) can be extracted using the compensator field \(V_{A}\) (\textbf{Note:} Dynamical fields are defined as fields that are neither pure gauge nor expressible in terms of derivatives of other fields via any constraints).
\footnotetext{This expression can be thought of as originating from the curvature $F[A]=dA+\frac{1}{2}[A,A]$ of a connection $A$ taking values in the algebra $\mathfrak{g}_\Lambda \inplus_\rho \mathbb{Y}$, which is the semi-direct sum of the (anti-)de Sitter algebra $\mathfrak{g}_\Lambda$ with the representation $\mathbb{Y}$, considered as an Abelian subalgebra. The component of this curvature taking values in $\mathfrak{g}_\Lambda$ is the usual curvature of the (A)dS algebra, and is assumed to vanish here, while the component in $\mathbb{Y}$ reproduces the above formula.}
Following \cite{Skvortsov:2006at}, the action in the frame formulation is given by,
\begin{align}
    S_{s,t}=\frac{1}{2}\sum_{k,m}a^{s,t}(k,m)\int  & \epsilon _{\AlgInd{A}_{1}\dots \AlgInd{A}_{D+1}}V_{\AlgInd{C}(2k+2m)}V^{\AlgInd{A}_{5}}E^{\AlgInd{A}_{6}}\dots E^{\AlgInd{A}_{D+1}}\nonumber \\& R^{\AlgInd{A}_{1}\AlgInd{B}(s-k-2)\AlgInd{C}(k),\AlgInd{A}_{2}\AlgInd{D}(s-t-m-1)\AlgInd{C}(m)}R^{\AlgInd{A}_{3}}{}_{\AlgInd{B}(s-k-2)}{}^{\AlgInd{C}(k),\AlgInd{A}_{4}}{}_{\AlgInd{D}(s-t-m-1)}{}^{\AlgInd{C}(m)},
\end{align}
where, 
\begin{align}
    a^{s,t}(k,m)=b^{s,t}(k,m)\theta(m)\theta(k)\theta(s-t-m-1)\theta(s-m-k-2)\nonumber\\
    \theta(n)=1\,\,\text{for}\,\,n\geq0 \qquad \text{and}\qquad\theta(n)=0\,\,\text{for}\,\,n<0 .
\end{align}
Clearly, for the massless case $t=1$, one can then convince oneself that, for \(k = 0\) and \((m = s - 2 - p)\), the structure of the massless action in \eqref{eq:hs_action_term} is recovered. Note that, in general, different choices of the coefficients ${b^{s,t}(k.m)}$ give rise to different dynamical systems; however, requiring gauge invariance, the correct number of degrees of freedom, and the proper flat-space limit fixes these coefficients up to an overall coefficient (for details see \cite{Skvortsov:2006at}), 
\[
{b^{s,t}(k.m)}=b^{s,t}\frac{(s-k-m-1)!(D-5+2(k+m))!!}{k!m!(s-k-2)!(s-m)!}.
\]
\paragraph{Dynamical Fields}
To isolate the dynamical fields, we decompose $W^{\AlgInd{A}(s-1),\AlgInd{B}(s-t)}$ into components transverse and longitudinal to the compensator $V^\AlgInd{A}$ (a fixed vector in $(A)dS$ space satisfying $V^\AlgInd{A} V_\AlgInd{A} = \text{const}$). This is done by contracting with $V^\AlgInd{A}$ or projecting orthogonally to it.

The compensator $V^\AlgInd{A}$ breaks $\mathfrak{so}(d,2)$ down to the Lorentz algebra $\mathfrak{so}(d,1)$, and the gauge field splits into Lorentz tensors. 
We define dynamical fields as
\paragraph{(1)}
$\phi^{a(s)} = \Pi \left[ W^{a(s-1),C_{1}\dots C_{s-t}|a} V_{C_1} \cdots V_{C_{s-t}} \right]$ \,,\,\,\,\\\\
 where $W^{...|a} = W^{...}_\mu e^{\mu|a}$ is the gauge field with one form index converted to a fiber index using the vielbein. The projection $\Pi$ enforces tracelessness, hence gives a symmetric, traceless rank-$s$ tensor $\phi^{a_1 \dots a_s}$. The contraction with $(s-t)$ compensators $V_C$ isolates the part of $W$ that transforms nontrivially under the gauge symmetry;
This is the "primary" dynamical field, analogous to the Frønsdal field in massless HS theories. Its gauge transformation involves $t$ derivatives of the gauge parameter $\xi^{a(t)}$, characteristic of partial masslessness:
\begin{equation}
\delta \phi^{a(s)} = \Pi \left[ \nabla^{(a_1} \cdots \nabla^{a_t} \xi^{a_{t+1} \dots a_s)} \right],
\end{equation}
where the parameter $\xi^{b(s-t)}$ is defined, via the gauge parameter $\xi^{A(s-1),B(s-t)}$ of $W$, as 
\begin{equation}
    \xi^{b(s-t)}= \xi^{\AlgInd{A}(s-1),b(s-t)}V_{\AlgInd{A}_{1}}\dots V_{\AlgInd{A}_{s-1}}.
\end{equation}
One can see that it is traceless by choosing the standard gauge $V^\AlgInd{A}=\delta^{\AlgInd{A}}_{5}$. Indeed by recalling that $\eta_{\AlgInd{BB}}\xi^{\AlgInd{A}(s-1),\AlgInd{B}(s-t)}=0$ we see that,
\begin{equation}\label{traceless parameter}
    \eta_{bb}\xi^{b(s-t)}\sim\,\eta_{\bullet\bullet}\xi^{\AlgInd{A}(s-1),\bullet\bullet\,b(s-t-2)}V_{A_{1}}\dots V_{\AlgInd{A}_{s-1}}\sim\,\eta_{\bullet\bullet}\xi^{\bullet\dots \bullet,\bullet\bullet\,b(s-t-2)}=0.
\end{equation}
We have used the Young condition in the last step.

\paragraph{(2)}\,$\phi^{a(s-t)} = \alpha W_{m}^{a(s-t)m\AlgInd{C}(t-2),\AlgInd{C}(s-t)} V_{\AlgInd{C}_1} \cdots V_{\AlgInd{C}_{s-2}} + \beta W_{m}^{a(s-t)\AlgInd{C}(t-1),\AlgInd{C}(s-t-1)|m} V_{\AlgInd{C}_1} \cdots V_{\AlgInd{C}_{s-2}}$.\\

 This field is a linear combination of two components of $W$. The first term involves a trace over the form index $m$ (after turned into fibre index via $e^{\mu}_{m}$) and one of the $a$ indices in the first group. The second term involves a contraction of the form index $m$ with one of the $b$ indices, and the coefficients $\alpha, \beta$ are chosen to ensure consistency with the gauge transformations \cite{Skvortsov:2009nv}. It is symmetric, and note that we didn't use the traceless projector $\Pi$ because these are automatically traceless; again, we can see this by using the standard gauge for the compensator, as in (\ref{traceless parameter}). 

\paragraph{(3)}\,$\phi^{a(s-t-1)} = W_{m}^{ma(s-t-1)\AlgInd{C}(t-1),\AlgInd{C}(s-t)} V_{\AlgInd{C}_1} \cdots V_{\AlgInd{C}_{s-1}}$.\\\\
  This is a lower-spin field (rank-$(s-t-1)$ tensor) and is absent for $t=s$ (called the maximal depth partially massless case). It ensures the consistency of the gauge algebra and the correct counting of degrees of freedom.

These three fields $\phi^{a(s)}$, $\phi^{a(t)}$, and $\phi^{a(t-1)}$ are sufficient because the gauge transformations of $W$ allow all other components to be either gauged away (Stueckelberg) or expressed in terms of derivatives (auxiliary). The Young symmetry of $W$ ensures that no additional independent tensors can be formed without violating tracelessness or symmetry properties. In the flat space limit, the theory must reduce to a sum of massless Frønsdal actions for spins $(s-t+1), \dots s$\cite{Deser:2001pe}. 

 \paragraph{Example: Spin-2 Case} The $s=2$ case is described by the connection $1$-form $W^{A,B(2-t)}$ which for $t=1$ becomes $W^{A,B}$ which is the adjoint representation $\parbox{10pt}{\gyoung(;,;)}$. This describes massless gravity for which the connection $W^{A,B}$ contains\footnote{Indices $\AlgInd{A},\AlgInd{B},...=0,...,d+1$ are of $\mathfrak{g}_\Lambda$ and we can decompose them as $A={a,\bullet}$, where indices $a,b,c,...$ are of the Lorentz algebra. } two one-forms valued
in finite-dimensional representations of the Lorentz subalgebra $\mathfrak{so}(1,d)$,
namely the vielbein $e^a=W^{a,\bullet}$ and the spin-connection
$\omega^{a,b}=W^{a,b}$. For $t=2$ we get a genuine and the only PM case for spin-$2$ called \text{partially massless graviton}. It is described in this language by a connection $W^{A}$, taking values in $\mathbb{Y}=\,\,\scriptscriptstyle\gyoung(;)$,
the fundamental (or vector) representation of the (anti-)de Sitter algebra $\mathfrak{g}_\Lambda$. The gauge transformation and curvature become, in this case,
 \begin{equation}
     \delta W^{\AlgInd{A}}=D\xi^{\AlgInd{A}},\qquad R^{\AlgInd{A}}=D W^{\AlgInd{A}}.
\end{equation}
Such a connection has components
$W^{\AlgInd{A}}=\{w^a,w\}$, i.e. it is composed of two $1$-forms,
valued in the vector and scalar representation of
the Lorentz algebra respectively. Their curvature simply read
\begin{equation}
    R^a = \nabla w^a + e^a \wedge w\,,
    \qquad 
    R = \nabla w - e^a \wedge w_a\,,
\end{equation}
while the gauge transformations are given by
\begin{equation}
    \delta_{\xi,\epsilon} w^a = \nabla\xi^a + e^a\,\epsilon\,,
    \qquad 
    \delta_{\xi,\epsilon} w = \nabla\epsilon - e^a\,\xi_a\,,
\end{equation}
where $\xi^a$ and $\epsilon$ are the two $0$-form gauge parameters.
Let us briefly review how one can recover the metric-like formulation
discussed previously \cite[Sec. 5.1]{Skvortsov:2006at}. First,
note that one can gauge-fix to zero the component $w$ upon using
its gauge symmetry generated by $\xi^a$. The residual gauge
transformations (i.e. which preserve the gauge choice $w=0$)
are those generated by $\epsilon$ and $\xi_a = -\nabla_a\epsilon$,
i.e.
\begin{equation}
    \delta_\epsilon w_{a|b} = -\nabla_a\nabla_b\,\epsilon
    + \eta_{ab}\,\epsilon\,,
\end{equation}
where $w_{b|a} = e_b^\mu\,w_\mu^c\,\eta_{ac}$. Imposing that
the curvature $R$ of $w$ vanishes in the gauge $w=0$ implies
that the antisymmetric part of $w_{a|b}$ vanishes,
\begin{equation}
    R\rvert_{w=0} = 0
    \qquad \Rightarrow \qquad 
    w_{[a|b]} = 0\,.
\end{equation}
This is a first sign that one can recover the symmetric
rank-$2$ tensor subject to a two-derivative gauge transformation,
which encodes the PM spin-$2$ field in the metric-like formulation,
as the symmetric part of the $1$-form $w^a$.
Inspecting the Bianchi identities for the curvature $R^a$,
one finds that its only possible non-trivial component 
is encoded by a hook, so that one can impose
\begin{equation}
    R^a = e_b\wedge e_c\,C^{ab,c}\,,
\end{equation}
where $C^{ab,c}$ is a $0$-form which takes values 
in the irrep $\gyoung(;;,;)$ of the Lorentz algebra.
The above example is representative of the frame-like
description of partially massless field: for a spin-$s$
and depth-$t$ field, one can impose the zero-curvature
equations
\begin{equation}
    R^{a(s-m),b(s-n)} = 0\,,
    \qquad 
    m\neq 1 \quad\text{and}\quad n \neq t\,,
\end{equation}
and 
\begin{equation}
    R^{a(s-1),b(s-t)} = C^{a(s-1)c,b(s-t)d}\,e_c\wedge e_d\,,
\end{equation}
where $C$ is a $0$-form, that can be thought of as
a partially massless version of the Weyl tensor. 
The metric-like partially massless field can be found
in the connection $e^{a(s-1)}$ valued in the totally symmetric
irrep of the Lorentz algebra, and the above zero-curvature equations
expresses the intermediate/auxiliary connections $\omega^{a(s-1),b(m)}$
with $m=1,\dots,s-t-1$ as $m$ derivatives of the PM field, 
while the last equation equates the $0$-form $C$ to a particular
traceless projection of $s-t+1$ derivatives of the PM field.

One can build a gauge-invariant action from the above curvature,
however, this action exhibits an intricate pattern involving
the `auxiliary connections' \cite{Skvortsov:2006at}.
Let us specialize this construction to $4d$, where it is advantageous
to use the two-component spinor language (explained below).

\section{ Chiral Approach}
\label{sec:twistor like}

The chiral approach to gravity, and its extension to higher spin, provides an alternative formulation in which the degrees of freedom are encoded using variables that naturally split into self-dual and anti-self-dual components. In four dimensions, this framework leverages the fact that at the level of complexified lie algebra the Lorentz algebra $\mathfrak{so}(3,1)$ is locally isomorphic to $\mathfrak{sl}(2,\mathbb{C})$, allowing the spin connection and curvature to be decomposed into left-handed (self-dual) and right-handed (anti-self-dual) parts. This decomposition simplifies the structure of the theory, particularly in the self-dual (chiral) sector, where the theory becomes more tractable and admits elegant reformulations. Chiral formulation has been instrumental in developing twistor methods\cite{Penrose:1968me, Penrose:1976js}, Ashtekar variables for canonical quantum gravity\cite{Ashtekar:1986yd}, and more recently, in the study of chiral higher spin theories\cite{Skvortsov:2022syz, Sharapov:2022faa,Sharapov:2022wpz,Sharapov:2022nps,Sharapov:2022awp,Sharapov:2023erv}. In these contexts, the chiral sector offers a useful laboratory for exploring quantum properties of gravity and possible UV-completions. With this backdrop, it is natural to explore a chiral-type formulation for partially massless fields. We begin by reviewing the chiral formulation of gravity introduced by Plebanski~\cite{Plebanski:1977zz}, and then show how its linearized version can be naturally extended to a chiral formulation of partially massless higher spin fields.

\subsection{Plebanski Formulation}
Let $M$ be an oriented 4-dimensional manifold. As we have seen before, in the frame formulation of gravity, the basic objects are $e_{\mu}{}^{a}$ such that the metric is given by 
\begin{equation}
    g_{\mu\nu}=\eta_{ab}e_{\mu}^ae_{\nu}^b
\end{equation}
For convenience, let us collect all the formulas here. The first Cartan structure is given by, 
\begin{equation}
    T^{a}=de^{a} +\omega^{a}{}_{b}\wedge e^{b}
\end{equation}
The curvature 2-form $R^{a}{}_{b}$ 
\begin{equation}
    R^{a}{}_{b}=d\omega^{a}{}_{b}+\omega^{a}{}_{c}\wedge \omega^{c}{}_{b}
\end{equation}
Using these objects, one describes what is called Einstein-Cartan theory in the presence of cosmological constant $\Lambda$ as,
\begin{equation}
    S_{EC}=\int_{M} \epsilon_{abcd}\big(e^a\wedge e^b\wedge R^{cd}[\omega]-e^a\wedge e^b\wedge e^c\wedge e^d\big)
\end{equation}
Now, since we are working in 4 dimensions, we must use for simplicity in computation, the isomorphism $\mathfrak{sl}(2,\mathbb C) \cong \mathfrak{so}(1,3)$.
The latter relates a Lorentz vector $V^a$ to 
a $\mathfrak{sl}(2,\mathbb C)$-bi-spinor $V^{AA'}$, via an object (constant matrices) $\sigma_{a}^{AA'}$ such that $V^{AA'}=\sigma_{a}^{AA'} V^a$ 
where both $A=1,2$ and $A'=1,2$ are two-component spinor indices.
These two-component spinor techniques are very handy when it comes to explicit computation in 4-dimension. 
More generally, finite-dimensional irreducible
representations of $\mathfrak{so}(1,3)$, which are
mixed-symmetric traceless tensor $T^{a(m),b(n)}$
correspond to a spin-tensor carrying two groups of $m+n$ and $m-n$ totally symmetrized (un)primed indices,
\begin{equation}
    T^{a(m),b(n)}
    \qquad\longleftrightarrow\qquad
    \big(T^{A(m+n),A'(m-n)},\,\, T^{A(m-n),A'(m+n)}\big)\,.
\end{equation}
As usual, in the Lorentzian signature, the two spin-tensors
are complex conjugates of each other. In the Euclidean
or split signature, they are independent real spin-tensors.  

For example vierbein $e^{a}$, the spin connection $\omega ^{ab}$, curvature 2-form $R_{ab}$, epsilon tensor and $\epsilon_{abcd}$ in two-component spinor notation becomes
\begin{align}
    \omega_{ab}\leftrightarrow\omega_{AB}\epsilon_{A'B'}+\omega_{A'B'}\epsilon_{AB}\\
    R_{ab}\leftrightarrow\
    R_{AB}\epsilon_{A'B'}+R_{A'B'}\epsilon_{AB}\\
    \epsilon_{abcd} \leftrightarrow\big(\epsilon_{AC}\epsilon_{BD}\epsilon_{A'D'}\epsilon_{B'C'}-\epsilon_{AD}\epsilon_{BC}\epsilon_{A'C'}\epsilon_{B'D'}\big),
\end{align}
where $R_{AB}$, $\omega_{AB}$ are symmetric and $\epsilon_{AB}$ and $\epsilon_{A'B'}$ are the component of the symplectic form defined on (un)primed spaces such that $g_{ab}\propto\epsilon_{AB}\epsilon_{A'B'}$. One important thing to note is that, since $\epsilon_{AB}$ is antisymmetric, raising and lowering using them requires a choice of convention. We choose that unprimed spinor indices are raised and lowered
with the invariant tensor $\epsilon_{AB}$ and its inverse
$\epsilon^{AB}$, in the sense that 
$\epsilon^{AC}\,\epsilon_{BC} = \delta^A_B$, via
\begin{equation}
    \xi^A = \epsilon^{AB}\,\xi_B\,,
    \qquad\qquad
    \xi_B = \xi^A\,\epsilon_{AB}\,,
\end{equation}
and similarly for primed indices. In this two-component
spinor language, we call the primed part $\omega^{A'A'}$ and $R_{A'B'}$ are called anti-self-dual part of the spin connection and curvature. Similarly, the unprimed parts are called self-dual.
Now will all of this technology we can write our action in terms of two-component spinor language, we get,
\begin{equation}
    S_{EC}=-2 \int_{M}e^{AA'}\wedge e^{BB'}\wedge \big(R_{AB}\epsilon_{A'B'}-R_{A'B'}\epsilon_{AB}\big)-\frac{\Lambda}{3}e^{AA'}\wedge e^{B}{}_{A'}\wedge e^{C'}_{A}\wedge e_{BC'}
\end{equation}
Note that the self-dual and anti-self-dual part of the curvature is given by 
\begin{align}
    R_{AB}=d\omega_{AB}+\omega_{A}{}^{C}\wedge \omega_{CB}
\end{align}
and similarly for $R_{A'B'}$. 
\paragraph{Plebanski's formulation} 
In the above EC action, the fundamental dynamical variables are 1-forms $e^{AA'}$ and $\omega^{AB}$. Plebanski's formulation has a triple of 2-forms $H_{AB}$ as fundamental variables replacing $e^{AA'}$. The key observation is that the vielbein field in the above action appears only in the combination 
\begin{equation}
    H^{AB}=e^{A}{}_{C'}\wedge e^{BC'} \qquad \,\, H^{A'B'}=e^{A'}{}_{C}\wedge e^{B'C},
\end{equation}
which appears in the decomposition, 
\begin{equation}
    e^{AA'}\wedge e^{BB'}=\frac{1}{2}\big(H^{AB}\epsilon^{A'B'}+H^{A'B'}\epsilon^{AB}\big)
\end{equation}
Note that $H^{AB},\,\, H^{A'B'}$ are symmetric by construction. Also if $d^4x=dx^{0}\wedge dx^1\wedge dx^2\wedge dx^3\wedge dx^4$, one finds that, 
\begin{equation}
    H^{AB}\wedge H_{CD}=4\delta_{C}^{(A}\delta_{D}^{B)}\sqrt{|g|}\,\,d^4x
\end{equation}
There is also
the $3$-form basis, defined as
\begin{equation}\label{eq:basis_3-forms}
    \hat e_{AA'} := H_{AB} \wedge e\fud{B}{A'}\,.
\end{equation}
In particular, the $2$-forms $H_{AA}$ and $H_{A'A'}$
verify
\begin{equation}
    H_{AB} \wedge H_{A'B'} = 0\,,
\end{equation}
and the identities
\begin{equation}\label{eq:sym_prop_H}
    H_{AA} \wedge e_{AB'} = 0
    \qquad\Rightarrow\qquad 
    H_{AA} \wedge H_{AB} = 0\,,
\end{equation}
which will be useful later on (for more details,
see e.g. \cite{Krasnov:2020lku}). 
Now, in terms of these two 2-forms, the action becomes,
\begin{equation}
    S_{EC}=2\int H^{AB}\wedge R_{AB}-H^{A'B'}\wedge R_{A'B'}+\frac{\Lambda}{3}H^{AB}\wedge H_{AB},
\end{equation}
\noindent where it is understood that \(R\) depends on \(\omega\) and \(H\) depends on \(e\).  Plebanski’s insight was to promote the two-form \(H^{AB}\) and the connection one-form \(\omega^{AB}\) (and their complex conjugates) to fundamental variables, rather than \(e\) and \(\omega\).  However, the resulting action—with \(H^{AB}\) and \(\omega^{AB}\) treated independently—fails to reproduce Einstein–Cartan gravity: since \(H^{AB}\) no longer depends on \(e^{AA'}\), the theory is not equivalent to the usual formulation.

To overcome this, Plebanski introduced the simplicity constraint \(H^{(AB}\wedge H^{CD)}=0\), which, by the classical result that a two-form \(\tau\) in four dimensions is decomposable (\(\tau=\sigma\wedge\rho\)) if and only if \(\tau\wedge\tau=0\), guarantees that \(H^{AB}=e^A{}_{C'}\wedge e^{BC'}\) (and similarly in the primed sector).  This dynamically recovers the tetrad \(e^{AA'}\) from \(H^{AB}\).  Imposing this constraint at the level of the action leads to the Plebanski action for general relativity  
\begin{align}
    S_{P}[H,\omega]=2\int_{M} \Big[H^{AB}\wedge R_{AB}-H^{A'B'}\wedge R_{A'B'}-\frac{\Lambda}{6}H^{AB}\wedge H_{AB}+\frac{\Lambda}{6}H^{A'B'}\wedge H_{A'B'}\nonumber\\ -\frac{1}{2}\Psi_{ABCD}H^{AB}\wedge H^{CD}+\frac{1}{2}\Psi_{A'B'C'D'}H^{A'B'}\wedge H^{C'D'}\Big],
\end{align}
where $\Psi_{ABCD},\,\,\Psi_{A'B'C'D'}$ are completely symmetric traceless objects called Lagrange multipliers for the aforementioned constraints. Now, the equations of motion coming from the above action are 
\begin{align}
    dH^{AB}-2\omega^{(A}{}_{C}\wedge H^{B)C}=0\\
    R_{AB}-\Psi_{ABCD}H^{CD}-\frac{\Lambda}{3}H_{AB}=0\\
    H^{(AB}\wedge H^{CD)}=0
\end{align}
and a similar equation for the primed sector. All these equations together constitute the Einstein equation as before and hence are equivalent to Einstein gravity in the presence of a cosmological constant. Note that the equation $H^{(AB}\wedge H^{CD)}=0$ clearly has a solution given by $H^{AB}=e^{A}_{C'}\wedge e^{BC'}$ because $e^{(A}_{A'}\wedge e^{B}_{B'}\wedge e^{C)}_{C'}=0$ since the expression is totally antisymmetric in $A'B'C'$ and hence vanishes because the spin space is 2-dimensional. This is sometimes in the literature called the Fierz identity. The advantage of Plebanski's action is that it has scope to give different possible formulations for gravity by integrating out different fields from the action. For example, integrating out $H^{AB}$, one gets the pure connection formulation. We will not go into this interesting line of thought. But for the detail see \cite{Krasnov:2011pp,Krasnov:2012pd} and references therein.
The other advantage is that the action gets nicely separated between the primed and unprimed sectors (self-dual and anti-self-dual, respectively). Now, the idea behind the chiral formulation is that it is sufficient to have access to only one of the curvatures, out of SD and ASD parts, because it then becomes possible to
impose the Einstein condition, working with only one of the chiral parts of the
spin connection, for details and proofs, see \cite{Krasnov:1970bpz}. Hence, from here on, we will only consider the unprimed sector,
\begin{equation}
    S_{P}[H,\omega]=2\int_{M} \Big[H^{AB}\wedge R_{AB}-\frac{1}{2}\big(\Psi_{ABCD}+\frac{\Lambda}{3}\epsilon_{CA}\epsilon_{DB}\big)H^{AB}\wedge H^{CD}\Big].
\end{equation}
Now the equation of motion is $ R_{AB}-\Psi_{ABCD}H^{CD}-\frac{\Lambda}{3}H_{AB}=0$ or $R^{AB}=(\Psi^{AB}{}_{CD}+\frac{\Lambda}{3}\delta_{C}^{(A}\delta_{D}^{B)})H^{CD}$.
If we assume that $(\Psi^{AB}{}_{CD}+\frac{\Lambda}{3}\delta_{C}^{(A}\delta_{D}^{B)})$ is not singular, then we can invert this relation and get 
\begin{equation}
    H^{CD}=\big[\big(\Psi+\frac{\Lambda}{3}\big)^{-1}\big]^{CD}{}{}_{AB}\,\,R^{AB}.
\end{equation}
We get Plebanski Chiral action as,
\begin{equation}
     S_{P}[H,\omega]=\int_{M} \big[\big(\Psi+\frac{\Lambda}{3}\big)^{-1}\big]^{CDAB}\,\,R_{AB}\wedge R_{CD}.
\end{equation}
Now, on expanding, the first factor we get is a topological term $\int R\wedge R$, which we ignore, and a linear term in $\Psi$ as
\begin{equation}
   S_{Chiral}=\int_{M}\Psi^{ABCD}R_{AB}\wedge R_{CD}
\end{equation}
and we ignore higher order terms $\mathcal{O}(\Psi^2)$.
Note that now $\Psi$ is no longer a Lagrange multiplier; rather, it becomes a dynamical field. This action is an action for self-dual GR and the reason it is called self-dual because the equation of motion in the inverted form $H^{CD}=\big[\big(\Psi+\frac{\Lambda}{3}\big)^{-1}\big]^{CD}{}{}_{AB}\,\,R^{AB}$ becomes in the leading order as $H^{AB}=R^{AB}$, that is the chiral 2-form becomes the curvature itself. For such self-dual curvature, the metric is famously given by 
\begin{equation}
    g_{ab} \propto \, \epsilon^{cdef} \, (R^{A}_{\;B})_{ac} \, (R^{B}_{\;C})_{bd} \, (R^{C}_{\;A})_{ef}.
\end{equation}
This metric now describes a self-dual solution of GR with cosmological constant, with vanishing left-handed Weyl curvature. The field $\Psi_{ABCD}$ then describes a linearized left-handed Weyl tensor, not derived from $g_{ab}$, but propagating on top of it.
Finally, we will expand this chiral action around some background (note that the background value of $\Psi$ is zero) and see that the free action is given by, 
\begin{equation}\label{chiral action}
    S_{Chiral}=\int_{M} \Psi^{ABCD}H_{AB}\wedge \nabla\omega_{CD},
\end{equation}
where $H_{AB}$ is the background self-dual 2-form basis. It is this linearization that we are interested in because this simple-looking linearized action can be generalized to higher spin massless theory, "Chiral Higher Spin theory". Such a massless higher spin case was first given \cite{Krasnov:2021nsq} and has seen great progress in recent times, see, for example, \cite{Skvortsov:2022syz}. Hence, in the next section, we will explore its further generalization to Partially massless (PM) field and study in detail its chiral action for arbitrary spin and depth in 4-dimensions.

\subsection{Chiral Action for PM Higher Spin Fields}
\label{sec:chiral action}
In this section we will seek a generalization of the chiral action in (\ref{chiral action}) to an action for  Partially massless higher spin fields. It was shown in \cite{Krasnov:2021nsq} that
for massless fields, we can take the self-dual parts
of the very `last' spin-connection (by which we mean
the component of the $\mathfrak{g}_\Lambda$-connection
valued in the `biggest' Lorentz Young diagram, that is,
the Young diagram with the same shape as the one labelling
the $\mathfrak{g}_\Lambda$-irrep) and of the Weyl tensor
as our dynamical variables. Indeed, we will see that, similar to massless case \cite{Krasnov:2021nsq}, the PM generalization of (\ref{chiral action}) leads to
a simple action. In tensor language, the last spin-connection for a spin-$s$ and depth-$t$ partially massless field is a one-form $\omega^{a(s-1),b(s-t)}$ and the Weyl tensor is of the form $C^{a(s),b(s-t+1)}$, where the indices merely indicate the symmetry type of a tensor. In the spinorial language, the self-dual components of these two fields are thus 
\begin{align}
    \omega^{A(2s-t-1),A'(t-1)}
    &&& \Psi^{A(2s-t+1),A'(t-1)}\,,
\end{align}
and their anti-self-dual cousins can be obtained via $t \to 2s-t$
for $\omega$ and $t \to 2s-t+2$ for $\Psi$.
The chiral approach deals with one pair of such fields
and ignores the duals thereof (See figure below), 
\begin{figure}[!ht]
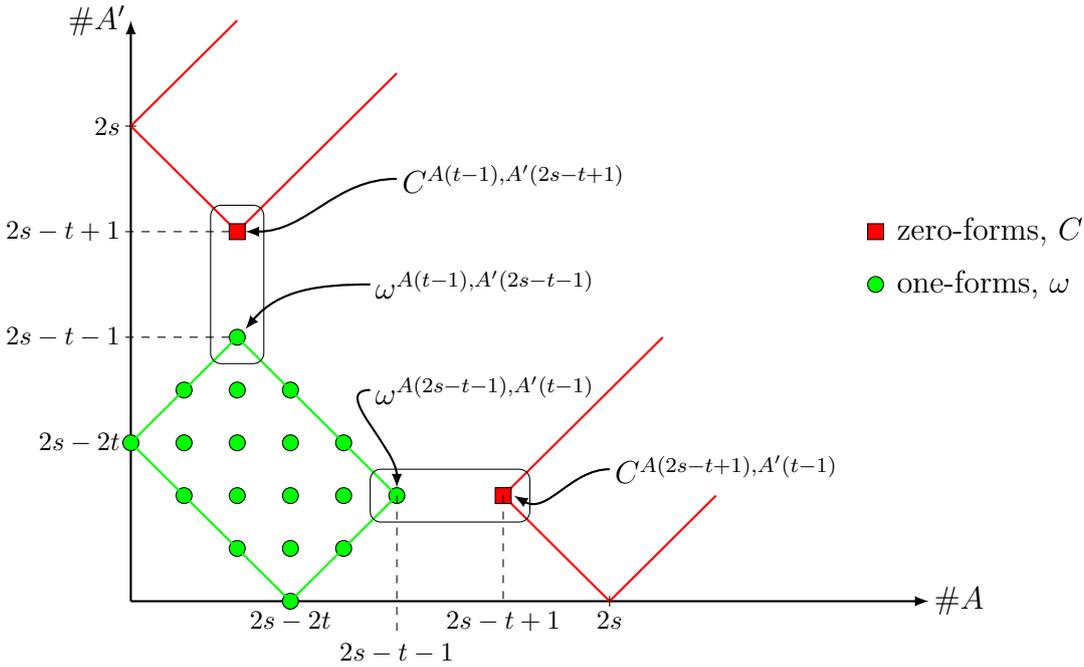

    \FIELDS
    \caption{A diagram to show fields/coordinates involved
    into the description of partially massless higher spin
    fields. Along the horizontal/vertical axe, we have
    the number of unprimed/primed indices on a spin-tensor. Components of the $1$-form connection are represented
    by green circles, while the $0$-forms (the Weyl tensor
    and its descendants) are represented by red rectangles. By descendants we mean the on-shell nontrivial derivatives of the Weyl tensor, which are associated with the coordinates on the on-shell jet space \cite{Penrose:1986ca}.}
    \label{fig:fields}
\end{figure}
\paragraph{{Action}}
Using the fields description discussed above we generalize the Chiral action of (\ref{chiral action}) to PM higher spin field of spin-$s$ and depth-$t$ as, 
\begin{align}
   {S_{s,t}[\omega,\Psi]
   =\int \Psi^{A(2s-t+1),A'(t-1)}\,H_{AA}
   \wedge \nabla \omega_{A(2s-t-1),A'(t-1)}\,.}
   \label{eq:free_PM_action}
\end{align}
In particular, for the spin-$2$ field with depth $t=1$, we recover the standard massless action shown in~(\ref{chiral action}). For the spin-$2$ field with depth $t=2$, we obtain the chiral description of the so-called partially massless graviton, expressed in terms of $\omega^{A,A'}$ and $\Psi^{A(3),A'}$.

This action is invariant under the gauge transformation of $\omega$ reads as,
\begin{equation}\label{eq:gauge_transformations}
    \begin{aligned}
        \delta_{\xi,\eta}\omega^{A(2s-t-1),A'(t-1)}
        & = \nabla \xi^{A(2s-t-1),A'(t-1)} \\
        & \qquad\qquad + e^{AA'}\eta^{A(2s-t-2),A'(t-2)}
        + e\fud{A}{B'}\eta^{A(2s-t-2),A'(t-1)B'}\,.
    \end{aligned}
\end{equation}
To show the invariance, we have used the well-known Fierz identities
shown in (\ref{eq:sym_prop_H}). Note that the transformation contains both a differential part (the first term) and an algebraic part (the second and third terms). The latter, hereafter, is referred to as a {\it shift symmetry}. This is nothing but a direct translation of (\ref{eq:gauge_frame}) into 2-component spinor notation. Now, in general, the field $\omega$ decomposes into irreducible spin-tensors as,
\begin{equation}
    \begin{aligned}\label{eq:irreducible_decompo}
        \omega^{A(2s-t-1),A'(t-1)} &
        = e\fud{A}{B'}\Phi^{A(2s-t-2),A'(t-1)B'}
        + e\fdu{B}{A'}\Phi^{A(2s-t-1)B,A'(t-2)} \\
        & \qquad +e_{BB'} \Phi^{A(2s-t-1)B,A'(t-1)B'}
        + e^{AA'}\Phi^{A(2s-t-2),A'(t-2)}\,,
    \end{aligned}
\end{equation}
where $\Phi$ are $0$-forms. Two of these components are unphysical as they can be gauged away. Indeed {\it shift symmetry}, can be used to gauge away the first and fourth terms in the irreducible decomposition \eqref{eq:irreducible_decompo}.
After this gauge fixing, the connection $\omega$ is given by
\begin{align}
    \omega^{A(2s-t-1),A'(t-1)}&= e\fdu{B}{A'}\Phi^{A(2s-t-1)B,A'(t-2)}
    +e_{BB'} \Phi^{A(2s-t-1)B,A'(t-1)B'}\,,
\end{align}
and is subject to the residual gauge symmetry
\begin{subequations}
    \begin{align}
        \delta \Phi^{A(2s-t),A'(t-2)}
        & = \nabla\fud{A}{B'} \xi^{A(2s-t-1),A'(t-2)B'}\,,\\
        \delta \Phi^{A(2s-t),A'(t)} & = \nabla^{AA'} \xi^{A(2s-t-1),A'(t-1)}\,,
    \end{align}
\end{subequations}
expressed in terms of its two irreducible components. 
Note that the gauge symmetry \eqref{eq:gauge_transformations}
is nothing but the two-component spinor translation of
the gauge symmetry \eqref{eq:gauge_frame} in the frame-like
approach, and in particular, the shift symmetry here is
simply the algebraic part of the gauge symmetry of the `last
connection'.

 Notice also that this action is of presymplectic AKSZ-type \cite{Alkalaev:2013hta},
which is not that surprising considering that the frame-like action for Gravity \cite{Grigoriev:2020xec} and Conformal/Weyl Gravity \cite{Dneprov:2022jyn} are also of this type, and that the relevance
of this approach for higher spin theories is established
\cite{Sharapov:2016qne, Sharapov:2021drr}. 

Also, this new description of PM higher spin fields (which includes the massless case) in terms of $(\omega\, ,\Psi)$ is very "twistor friendly" in the sense that such fields occur very naturally in Twistor theory via Penrose transform. Twistor theory leads to a new geometrical understanding of massless fields in $4d$ in terms of holomorphic structures on a $3d$ complex manifold that is twistor space
\cite{Eastwood:1981jy, Atiyah:1979iu, Hitchin:1980hp} (see also
the textbooks \cite{Huggett:1986fs, Penrose:1986ca, Ward:1990vs, Mason:1991rf} and, for instance, the recent review
\cite{Adamo:2017qyl}). Although we will not use twistor theory directly in our description of partially massless fields, it is very much inspired by it, and is a straightforward
extension of the approach proposed for massless fields
in \cite{Hitchin:1980hp, Krasnov:2021nsq}. The current formulation strongly suggests that a completely twistor-based chiral description of PM fields exists, and it will be interesting to explore this line of research, similarly to the massless case.

Another noteworthy feature of the above action is that
it is not manifestly real in the Lorentzian signature,
as is the well-known cases of (self-dual) Yang--Mills theory \cite{Chalmers:1996rq} and gravity \cite{Krasnov:2011pp,Krasnov:2016emc,Krasnov:1970bpz} that can be formulated in terms of chiral field variables. Nevertheless, it is worth mentioning that the use of chiral field variables does not imply that the theory is actually chiral (parity-violating) or non-unitary. This is always true for free theories that have the same degrees of freedom as their non-chiral relatives. The free action of \cite{Chalmers:1996rq} corresponds to $s=1$, $t=1$ of \eqref{eq:free_PM_action}.

Now the equations of motion obtained from \eqref{eq:free_PM_action} are 
\begin{align}\label{eq:free_EOM}
   H_{AA} \wedge \nabla \omega_{A(2s-t-1),A'(t-1)}=0\nonumber \qquad
  H_{AA} \wedge \nabla\Psi^{A(2s-t+1),A'(t-1)} & = 0\,.
\end{align}
There are two noteworthy cases: $t=1$ which corresponds to
massless fields, and in which case the above action reproduces
the one proposed in \cite{Krasnov:2021nsq}, and $t=s$,
which corresponds to maximal depth partially massless fields,
and for which the spin-connection is balanced (meaning it has the same number of primed and unprimed indices, as opposed
to the massless case where it is completely unbalanced).

\begin{figure}[!ht]
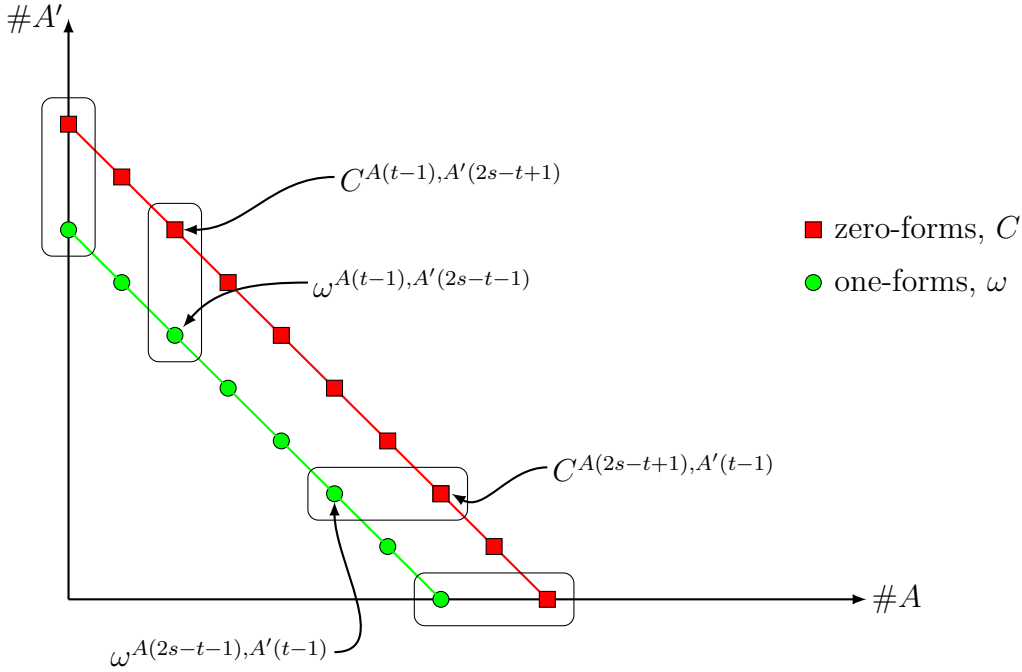

    \FIELDSDescr
    \caption{For a given spin-$s$, the fields grouped horizontally/vertically
    correspond to chiral/anti-chiral description of depth-$t$
    partially massless fields. There are two descriptions for each admissible $s$, and $t$. The group on each of the axes describes
    massless fields in terms of (anti-)chiral variables. It is clear that extrapolation of one description
    beyond $t>s$ does give the other one.}
    \label{fig:scale}
\end{figure}

These equations can be taken as a starting point
to build a free differential algebra (FDA) formulation of
partially massless fields, see  \cite{Skvortsov:2006at,Skvortsov:2009zu,Skvortsov:2009nv,Ponomarev:2010st,Alkalaev:2011zv,Khabarov:2019dvi}. Indeed, they can be read as expressing 
the fact that the first derivatives of $\omega$ and $\Psi$
are in the kernel of an operator determined by the background
self-dual $2$-form $H_{AA}$ (symmetrization for $\omega$,
contraction for $\Psi$). These operators are nothing but
components of the presymplectic
form used to build the action \eqref{eq:free_PM_action}.
The FDA is obtained by parametrizing $\nabla\omega$
and $\nabla\Psi$ as the most general elements in the kernel
of this presymplectic form, i.e.
\begin{subequations}\label{firststepsFDA}
    \begin{align}
        \nabla\omega_{A(2s-t-1),A'(t-1)}
        & = e\fdu{A}{B'}\,\omega_{A(2s-t-2),A'(t-1)B'}
        + e_{AA'}\,\omega_{A(2s-t-2),A'(t-2)}\,, \\
        \nabla\Psi_{A(2s-t+1),A'(t-1)}
        & = e\fud{B}{A'}\,\Psi_{A(2s-t+1)B,A'(t-2)}
        + e^{BB'}\,\Psi_{A(2s-t+1)B,A'(t-1)B'}\,,
    \end{align}
\end{subequations}
and imposing that the resulting equations
are integrable. Typically, this condition leads to constraints
on the first derivatives of the components of the elements
in the kernel of the symplectic form, and one should repeat
the procedure (i.e. find the most general form of the first
derivatives of these new fields compatible with integrability,
thereby introducing new fields, and imposing once more
the integrability of this equation, etc \dots). See e.g.
\cite{Alkalaev:2013hta} or \cite[Sec. 4]{Sharapov:2021drr}
for a review. The outcome of this procedure is to build two modules
of the (A)dS algebra $\mathfrak{g}_\Lambda$:
\begin{itemize}
\item A finite-dimensional one, which is spanned by the $1$-forms
$\omega^{A(2s-m-n),A'(n-m)}$ and their complex conjugate,
with $1 \leq m \leq t$ and $t \leq n \leq s$. This corresponds
to the $\mathfrak{g}_\Lambda$-module $\scriptscriptstyle \gyoung(_5{s-1},_4{s-t})$
used in the frame-like formulation;
\item An infinite-dimensional one, spanned by the $0$-forms
$\Psi^{A(2s-t+m+n),A'(t-m+n)}$ with $n\geq0$ and $1 \leq m \leq t$,
which corresponds to the derivatives of the self-dual Weyl tensors
unconstrained by equations of motion or Bianchi identities.
\end{itemize}
The pattern of connections, and descendants of the Weyl tensor, for a fixed spin-$s$ and depth-$t$ is illustrated in Figure \ref{fig:fields} and was already detailed in \cite{Skvortsov:2006at} (see also \cite{Boulanger:2008up, Boulanger:2008kw, Skvortsov:2009zu, Skvortsov:2009nv,Alkalaev:2009vm, Alkalaev:2011zv, Ponomarev:2010st}), while the pattern of pairs made of a connection one-form and a Weyl tensor zero-form, for a fixed spin-$s$ and different values of the depth-$t$ is displayed in Figure \ref{fig:scale}.

Let us dwell a little on the maximal depth case $t=s$.
In vector language, the last connection decomposes as
\begin{equation}
    \Yboxdim{10pt}
    \omega^{a(s-1)} \simeq {\gyoung(_5{s})
    \oplus \gyoung(_5{s-1},;) \oplus \gyoung(_4{s-2})}
\end{equation}
under the Lorentz group, and is subject to the algebraic symmetry
\begin{equation}
    \delta_\epsilon \omega^{a(s-1)} = e^{\{a}\,\epsilon^{a(s-2)\}}\,,
\end{equation}
where $\{\dots\}$ denotes the traceless projection
of symmetrized indices. This algebraic symmetry removes
the trace part ${\scriptscriptstyle\gyoung(_4{s-2})}$ in the irreducible
decomposition of $\omega^{a(s-1)}$. It may, however, be surprising at first glance that in the two-component spinor language, one has two parameters for the algebraic symmetry of $\omega$, namely $\eta^{A(s-2),A'(s-2)}$ and $\eta^{A(s-2),A'(s)}$. The first one simply corresponds to $\epsilon$, converted in spinor language, but the second one appears to have no counterpart in the vector language. This is not accidental: in fact, this additional parameter has the same symmetry as the anti-self-dual part of the hook component of $\omega$, and its role is simply to remove it. This is consistent with the fact that, in spinor language, $\omega$ has two irreducible components, corresponding respectively to a symmetric rank-$s$ tensor and the self-dual part of a hook tensor, and is also in accordance with the counting of degrees of freedom detailed below. Such additional symmetry is also present in the FDA form \cite{Ponomarev:2010st,Khabarov:2019dvi} of Zinoviev's description of partially massless fields \cite{Zinoviev:2001dt,Zinoviev:2008ze}. 

Massless spinning fields, described as in \cite{Hitchin:1980hp,Krasnov:2021nsq},
can propagate on self-dual backgrounds. This is due to the fact that
the fields $\Psi^{A(2s)}$ and $\omega^{A(2s-2)}$
do not have any primed indices, hence, $\nabla^2 \xi^{A(2s-2)}\equiv0$
on a self-dual background, which ensures the gauge invariance
of the action. However, partially massless fields are always
described by mixed spin-tensors, i.e. have both primed and
unprimed indices. 
The action \eqref{eq:free_PM_action} as well as the equations
of motion \eqref{eq:free_EOM} remain consistent in Minkowski space,
the difference being that the corresponding solution space
is not an irreducible representation of the Poincar\'e group
(see e.g. \cite{Brink:2000ag, Boulanger:2008up, Boulanger:2008kw, Alkalaev:2009vm, Alkalaev:2011zv}). 

\paragraph{Degrees of freedom.}
Let us justify the main claim of the previous paragraphs,
which is that the action \eqref{eq:free_PM_action} does describe
a partially massless spin-$s$ and depth-$t$ field in $4d$.
To do so, we will show that the solutions of the resulting
equations of motion propagate the correct number of degrees of freedom,
namely $2t$ (irrespectively of the spin).
In our case, the equations of motion are first order differential
equations for the fields $\Psi$ and $\omega$.
The number of physical degrees of freedom 
propagated by an arbitrary field, which is a solution of
an involutive system of equations,
is given by the formula \cite{Kaparulin:2012px}
\begin{equation}
    N_{dof} = \tfrac12\,\sum ^{\infty }_{k=0}
    k\left(e_k-i_k-g_k\right)\,,
\end{equation}
where $e_k$ is the number of equations of order $k$ in the system,
$i_k$ number of (gauge) identities of $k$-th order, and $g_k$ is
the number of gauge symmetry generators of order $k$ (here,
the order is the number of derivatives).
Let us recall that an involutive system of order $n$ is defined
in \cite{Kaparulin:2012px} as a system of equations such that
any differential consequence of these equations, of order $n$
or less, is already a part of the system.
In our case, the equation of motion for the field $\Psi$ is given by,
\begin{align}
    H_{AA} \wedge \nabla\Psi^{A(2s-t+1),A'(t-1)}
    \propto \widehat{e}_{AB'}\nabla\fdu{A}{B'}\Psi^{A(2s-t+1),A'(t-1)}=0\,,
\end{align}
where $\hat e_{AB'}$ are the basis $3$-forms introduced
in \eqref{eq:basis_3-forms} above. 
Using it, we can write down the set of independent equations
of motion as
\begin{equation}
    E^{A(2s-t),A'(t-1)|B'}
    = \nabla\fdu{B}{B'}\Psi^{BA(2s-t),A'(t-1)}=0\,,
\end{equation}
and easily count that these are $e_1=2t(2s-t+1)$ equations
of first order.
The field $\Psi$ does not have any gauge symmetry, 
hence $g_k=0$ for all $k$. Now since the field $\omega$ has a first
order gauge symmetry, the $\Psi$-field, after integrating by parts in the action, satisfies the Bianchi identity of second order. 
Explicitly, this identity is given by,
\begin{align}
    \nabla_{FF'} E^{FA(2s-t-1),F'A'(t-1)} = 0\,,
\end{align}
which consists of $i_2=t\,(2s-t)$ identities of the second order. 
Thus, the number of  physical degrees of freedom described by
the field $\Psi$ is
\begin{equation}
    N_{dof}(\Psi)=\tfrac12\,\big[2t(2s-t+1)-2t(2s-t)\big]=t\,.
\end{equation}
Similarly, the equations of motion for the field $\omega$
read
\begin{align}
    H_{AA}\wedge \nabla\omega_{A(2s-t-1),A'(t-1)}=H_{AA}\wedge e_{DD'}\nabla^{DD'}\omega_{A(2s-t-1),A'(t-1)}=0\,,
\end{align}
and, upon using the decomposition of $\omega$ into
its irreducible components,
\begin{equation}
    \begin{aligned}
        \omega^{A(2s-t-1),A'(t-1)} 
        & = e\fud{A}{B'}\Phi^{A(2s-t-2),A'(t-1)B'}
        +e\fdu{B}{A'}\Phi^{A(2s-t-1)B,A'(t-2)} \\
        & \qquad + e_{BB'} \Phi^{A(2s-t-1)B,A'(t-1)B'}
        +e^{AA'}\Phi^{A(2s-t-2),A'(t-2)}\,,
    \end{aligned}
\end{equation}
takes the form
\begin{align}
    \nabla\fdu{A}{F'}\Phi_{A(2s-t),F'A'(t-1)}
    + \nabla_{AA'}\Phi_{A(2s-t),A'(t-2)} = 0\,.
\end{align}
These are $e_1=t\,(2s-t+2)$ equations of first order.
The gauge transformations
are of first order, and generated by $g_1=t\,(2s-t)$ parameters. 
Since there are no additional identities, the number 
of degrees of freedom propagated by $\omega$ is
\begin{equation}
    N_{dof}(\omega) = \tfrac12\,\big[(2s-t+2)t-(2s-t)t\big]=t\,,
\end{equation}
and hence $\Psi$ and $\omega$ contain, in total, $2t$ physical
degrees of freedom. 
In particular, for massless field ($t=1$), we recover
$2$ degrees of freedom, as expected, while for
the partially massless graviton ($t=2$), we find $4$ degrees
of freedom, in conformity with expectations.\footnote{The same counting of degrees of freedom is suggested by the first step \eqref{firststepsFDA} towards the FDA form of the equations.}

Note that the counting of degrees of freedom presented
here applies for any values of $t$. In particular, when $t>s$,
we see that the number of degrees of freedom keeps increasing
and is larger than the one expected for a spin-$s$ field
of any depth. This is another indication that, despite the fact
that the pairs of fields $(\omega,\Psi)$ can still be considered
for $t>s$, and the action \eqref{eq:free_PM_action} still
makes sense, their interpretation remains elusive and should not
be related to PM fields (see the discussion in the section \ref{sec: maximal depth}).

\section{Interactions}
\label{sec:interactions}
Since we have a well-defined free action, the next task
is to look for interacting theories. In this section,
we will consider two simple types of possible interactions
using the new description discussed above. 

\subsection{Yang--Mills Interactions}
\label{sec:YM}
First, we will consider Yang--Mills interactions
for partially massless fields, which are straightforward
generalization of the higher spin extension of self-dual
Yang--Mills theory introduced in \cite{Krasnov:2021nsq},
and recently revisited in \cite{Tran:2021ukl,Herfray:2022prf, Adamo:2022lah,Tran:2022tft}, see also \cite{Steinacker:2022jjv}.
This type of interaction is obtained by first extending
the spin-connection $\omega^{A(2s-t-1),A'(t-1)}$
and the Weyl tensor $\Psi^{A(2s-t+1),A'(t-1)}$ of
a partially massless spin-$s$ and depth-$t$ field to take
values in a Lie algebra $\mathfrak{g}$ equipped with an ad-invariant
bilinear form\footnote{Recall that a bilinear form is called ad-invariant
if it verifies $\pmb([x,y],z\pmb)=\pmb(x,[y,z]\pmb)$ for any elements
$x,y,z \in \mathfrak{g}$.} that we will denote by $\pmb(-,-\pmb)$. Next,
we can pack up together the spin-connections for partially massless
fields of all spin and depth into a single $1$-form, 
\begin{align}
    \omega = \sum_{s=1}^\infty\sum_{t=1}^s\,\omega_{s,t}(x|y)\,,
    \qquad
    \omega_{s,t}(x|y)
    := \frac{\omega_{A(2s-t-1)A'(t-1)}}{(2s-t-1)!(t-1)!}\,
    y^A \dots y^A\,\bar{y}^{A'} \dots \bar{y}^{A'}\,,
\end{align}
whose curvature is defined by the usual formula
\begin{equation}
    F = \nabla \omega + \tfrac12\,[\omega,\omega]\,,
\end{equation}
where the bracket above should be understood as
the $\mathbb{C}[y,\bar y]$-linear extension of the Lie bracket
of the Yang--Mills algebra $\mathfrak{g}$. 
More concretely, the Lie algebraof $\omega$ with itself 
is given by
\begin{equation}
    [\omega,\omega]_{s,t}
    = \sum_{\substack{s_1+s_2=s+1\\ t_1+t_2=t+1}}\,
    [\omega_{s_1,t_1},\omega_{s_2,t_2}]\,,
\end{equation}
where the subscript $(s,t)$ denotes the component
of degree $2s-t-1$ in $y$ and $t-1$ in $\bar y$.
Packing up in a similar way the differential gauge parameters
associated with each spin-connection into a $0$-form $\xi$,
we can define an extension of the free gauge symmetry
\eqref{eq:gauge_transformations} via
\begin{equation}\label{eq:YM_gauge_transfo}
    \delta_\xi\omega = \nabla\xi+[\omega,\xi]\,,
    \qquad\qquad
    \delta_\xi\Psi = [\Psi,\xi]\,,
\end{equation}
under which the curvature transforms according to
\begin{equation}
    \delta_\xi F = \nabla^2\xi + [F,\xi]\,,
\end{equation}
where the first term can be re-written as
\begin{equation}\label{eq:curvature_gauge}
    \nabla^2\xi = (H\fdu{A}{B}\,y^A\,\partial_B
    + H\fdu{A'}{B'}\,\bar y^{A'}\,\partial_{B'})\,\xi\,.
\end{equation}
Similarly, we can pack up the shift symmetry parameters
into a single $0$-form $\eta$, and write it as
\begin{equation}
    \delta_\eta\omega = e_{AA'}\,y^A\,(\bar y^{A'}
    +\partial^{A'})\,\eta\,,
\end{equation}
so that the curvature transforms as
\begin{equation}\label{eq:shift_curvature}
    \delta_\eta F = -e_{AA'}\,y^A\,(\bar y^{A'} + \partial^{A'})\,
    (\nabla\eta+[\omega,\eta])\,,
\end{equation}
since the vierbein is torsionless and does not take values 
in the Lie algebra $\mathfrak{g}$. We will consider the action
\begin{equation}
    \begin{aligned}
        S_{PMYM}[\omega,\Psi] & = \langle\Psi \mid \tfrac12\,H_{AA}\,y^A y^A
        \wedge F \rangle\\
        & := \sum_{1 \leq t \leq s}\,
        \tfrac1{(2s-t-1)!(t-1)!}\,\int \pmb(\Psi^{A(2s-t+1),A'(t-1)},
        H_{AA} \wedge F_{A(2s-t-1),A'(t-1)}\pmb)\,,
    \end{aligned}
\end{equation}
which defines a complete interacting theory for partially massless fields. The interactions are of Yang--Mills type. 
This action is invariant under shift symmetry
since its variation under this transformation will produce
a term $H_{AA} \wedge e_{AB'} = 0$, as can be seen
from \eqref{eq:shift_curvature}. Its variation under the gauge
transformations \eqref{eq:YM_gauge_transfo} is given by
\begin{equation}
    \delta_\xi S_{PMYM} = \langle[\Psi,\xi] \mid
    \tfrac12\,H_{AA}\,y^A y^A \wedge F \rangle
    + \langle\Psi \mid \tfrac12\,H_{AA}\,y^A y^A \wedge
    (\nabla^2\xi+[F,\xi])\rangle = 0\,,
\end{equation}
and vanishes due to the fact that the term $\nabla^2\xi$
produces $H_{AA} \wedge H_{AB}=0=H_{AA} \wedge H_{A'B'}$
according to \eqref{eq:curvature_gauge}, and the two remaining terms cancel one another due to the ad-invariance of the bilinear 
form on $\mathfrak{g}$.\footnote{Very recently, in \cite{Serrani:2025owx}, a full solution to the holomorphic quartic constraint was obtained, leading to a complete classification of chiral higher spin theories involving one- and two-derivative vertices. This suggests that consistent truncations in the partially massless one-derivative actions exists \cite{Basile:2022mif}, thereby enabling finite higher spin models involving partially massless fields and the standard self-dual Yang-Mills vertex.}

\subsection{Current Interactions}
\label{sec:current}
Consider the functional
\begin{equation}
    S_{int}[\omega,\Psi] = \int T^{A(2s-t),A'(t)}(\Psi)\,
    \omega_{A(2s-t-1),A'(t-1)}\,\hat e_{AA'}
\end{equation}
where $\hat e_{AA'} := H_{AB} \wedge e\fud{B}{A'}$ and the spin-tensor $T^{A(2s-t+1),A'(t)}(\Psi)$ is
a $0$-form built out of Weyl tensors of some (partially-)massless
fields (of possibly different spins and depths), which verifies
\begin{equation}
    \nabla_{BB'} T^{A(2s-t)B,A'(t-1)B'}(\Psi) \approx 0\,,
\end{equation}
where the symbol $\approx$ signifies that the spin-tensor
$T(\Psi)$ is divergenceless only on-shell. This term is invariant
under the shift symmetry, as a consequence of the fact that
\begin{equation}
    e_{AA'}\wedge\hat e_{BB'} = -\tfrac14\,
    \epsilon_{AB}\,\epsilon_{A'B'}\, {\rm vol}
    \qquad\Rightarrow\qquad 
    e_{AA'}\wedge\hat e_{AB'} = 0\,,
\end{equation}
where `${\rm vol}$' denotes a volume form on the background,
and the fact that $\Psi$ is assumed to be inert under this symmetry.
Under the differential gauge symmetry, the variation of this term
reads
\begin{subequations}
    \begin{align}
        \delta_\xi S_{int}[\omega,\Psi]
        & = \int T^{A(2s-t),A'(t)}(\Psi)\,
        \nabla\xi_{A(2s-t-1),A'(t-1)}\,\hat e_{AA'} \\
        & = -\int \nabla^{BB'}\,T^{A(2s-t),A'(t)}(\Psi)\,
        \xi_{A(2s-t-1),A'(t-1)}\,e_{BB'}\,\hat e_{AA'} \\
        & = \tfrac14\,\int \nabla_{BB'} T^{A(2s-t-1)B,A'(t-1)B'}(\Psi)\,
        \xi_{A(2s-t-1),A'(t-1)}\,{\rm vol} \approx 0\,,
    \end{align}
\end{subequations}
and vanishes on-shell. It therefore provides a good starting point
to construct interactions for partially massless fields.

Indeed, divergenceless spin-tensors are fairly easy to construct
out of the Weyl tensors of a pair of massless fields.
Consider for instance the Bell--Robinson tensor
\begin{equation}
    T_{abcd} = \tfrac14\,\left(C_{a}{}^{p}{}_{b}{}^{q}C_{cpdq}
    + *C_{a}{}^{p}{}_{b}{}^{q}*C_{cpdq}\right)\,,
\end{equation}  
where $C_{abcd}$ is the gravitational Weyl tensor
and $*$ is the Hodge dual operator, i.e. 
$*C_{abcd}=\epsilon_{ab}{}^{pq}C_{pqcd}$. This tensor is divergenceless
as a consequence of Einstein's equation in vacuum. In spinor
notations, this tensor takes an especially simple form,
namely it is given by the product of the self-dual and anti-self-dual
Weyl tensor,
\begin{equation}
    T_{A(4),A'(4)} = \Psi_{A(4)}\, \Psi_{A'(4)}\,,
\end{equation}
and suggests the generalization (see \cite{Gelfond:2006be} for a complete set of currents)
\begin{equation}\label{simplecurrents}
    T_{A(2s_1),A'(2s_2)} = \Psi_{A(2s_1)}\,\Psi_{A'(2s_2)}\,,
\end{equation}
given by the product of the Weyl tensors of two massless fields
of spin $s_1$ and $s_2$. This spin-tensor will be divergence-free
as a consequence of the equation of motion
\begin{equation}
    \nabla\fud{B}{B'}\,\Psi_{A(2s_1-1)B} \approx 0\,,
    \qquad \qquad 
    \nabla\fdu{B}{B'}\,\Psi_{A'(2s_2-1)B'} \approx 0\,,
\end{equation}
for these Weyl tensors.

We will consider the one-parameter family of actions
\begin{equation}\label{eq:current_interaction}
    S[\omega,\Psi] = S_{free}[\omega,\Psi]
    + \alpha\,S_{int}[\omega,\Psi]\,,
    \qquad 
    \alpha \in \mathbb C\,,
\end{equation}
whose first piece,
\begin{equation}
    \begin{aligned}
        S_{free}[\omega,\Psi]
        & = \int \Psi^{A(2s-t)}\,H_{AA}\wedge\nabla\omega_{A(2s-t-2)}
        + \Psi^{A'(t)}\,H_{A'A'}\wedge\nabla\omega_{A'(t-2)} \\
        & \qquad\qquad + \Psi^{A(2s-t+1),A'(t-1)}\,H_{AA}\wedge
        \nabla\omega_{A(2s-t-1),A'(t-1)}\,,
    \end{aligned}
\end{equation}
is the sum of the free actions for the massless fields
of spin $s-\tfrac{t}2$ and $\tfrac{t}2$ as well as for
the partially massless field of spin-$s$ and depth-$t$,
and the second piece is the current interaction
\begin{equation}
    S_{int}[\omega,\Psi] = \int \Psi^{A(2s-t)}\,\Psi^{A'(t)}\,
    \hat e_{AA'}\wedge\omega_{A(2s-t-1),A'(t-1)}
\end{equation}
made out of the current associated with the previous pair
of massless fields and the partially massless field. Note that
we will restrict ourselves to bosonic fields, and hence
will assume that $t$ is even.
As already argued before, all of these pieces are invariant under
shift symmetry. Moreover, the free action is invariant under 
the differential gauge symmetry
\begin{equation}
    \delta_\epsilon\omega_{A(2s-t-2)} = \nabla\epsilon_{A(2s-t-2)}\,,
    \qquad 
    \delta_\epsilon\omega_{A'(t-2)} = \nabla\epsilon_{A'(t-2)}\,,
\end{equation}
for the massless fields, and 
\begin{equation}
    \delta_\xi\omega_{A(2s-t-1),A'(t-1)} = \nabla\xi_{A(2s-t-1),A'(t-1)}\,,
\end{equation}
for the partially massless field. Under this last gauge transformation,
the variation of the current interaction term reads
\begin{equation}
    \delta_\xi S_{int}[\omega,\Psi] = \int \nabla(\Psi^{A(2s-t)}\,
    \Psi^{A'(t)})\,\hat e_{AA'}\,\xi_{A(2s-t-1),A'(t-1)}\,,
\end{equation}
and vanishes only on-shell as explained before. It can be compensated off-shell by deforming the gauge symmetry of the pair of massless fields as follows,
\begin{subequations}\label{currentsymm}
    \begin{align}
        \delta_\xi\omega_{A(2s-t-2)} &= +\tfrac32\,\alpha\,\Psi^{A'(t)}\,e\fud{B}{A'}\,
        \xi_{A(2s-t-2)B,A'(t-1)}\,,\\
        \delta_\xi\omega_{A'(t-2)} & = -\tfrac32\,\alpha\,\Psi^{A(2s-t)}\,e\fdu{A}{B'}\,
        \xi_{A(2s-t-1),A'(t-2)B'}\,,
    \end{align}
\end{subequations}
i.e. with terms depending on the gauge parameter of
the partially massless field. The variation of the free
actions for the massless fields under this modification
of their gauge symmetry then reads
\begin{equation}
    \begin{aligned}
        \delta_\xi S_{free}[\omega,\Psi]
        & = -\tfrac32\,\alpha\,\int \nabla\Psi^{A(2s-t)}\,
        \Psi^{A'(t)}\,H_{AA}\,e\fud{B}{A'}\,
        \xi_{A(2s-t-2)B,A'(t-1)}\\
        & \quad {+} \tfrac32\,\alpha\,
        \int \Psi^{A(2s-t)}\,\nabla\Psi^{A'(t)}\,H_{A'A'}\,
        e\fdu{A}{B'}\,\xi_{A(2s-t-1),A'(t-2)B'}\,,
    \end{aligned}
\end{equation}
which, upon using
\begin{equation}
    H_{AA}\,e\fud{B}{A'} = +\tfrac23\,\hat e_{AA'}\,\delta^B_A\,,
    \qquad 
    H_{A'A'}\,e\fdu{A}{B'} = -\tfrac23\,
    \hat e_{AA'}\,\delta^{B'}_{A'\,,}
\end{equation}
can be brought to the form
\begin{equation}
    \delta_\xi S_{free}[\omega,\Psi]
    = -{\alpha}\int \nabla\,(\Psi^{A(2s-t)}\,\Psi^{A'(t)})\,
    \hat e_{AA}\,\xi_{A(2s-t-1),A'(t-1)}\,,
\end{equation}
so that the full action \eqref{eq:current_interaction}
is gauge invariant. Note that the deformations \eqref{currentsymm} of the gauge symmetries are Abelian, which is not the case for the current interactions in the non-chiral formulation. A straightforward generalization of these current interactions is to take advantage of other conserved currents that involve derivatives, see e.g. \cite{Gelfond:2006be}. Schematically they read $J_{2s_1+k,2s_2+k}\sim\Psi_{2s_1} \nabla^{k}\bar \Psi_{2s_2}$. In all these cases, except for $s_1=s_2=0$, the action does not require any higher order corrections. 

Note also that this type of interaction is simply a Noether coupling, which is similar
to the one explored in \cite{Boulanger:2019zic}. The spectrum
of the two resulting theories are however different: 
here, we find interactions between a partially massless
field of spin-$s$ and even depth-$t$, and two massless
fields of spin $s-\tfrac t2$ and $\tfrac t2$, whereas
the interacting theory constructed in \cite{Boulanger:2019zic}
involves only partially massless spin-$2$ fields.

\section{Beyond Maximal Depth}
\label{sec: maximal depth}
As is clear from the discussion in Section \ref{sec:free},
the action \eqref{eq:free_PM_action} and equations of motion
\eqref{eq:free_EOM} are formally well-defined beyond the maximal
depth $t=s$. Moreover, the number of physical degrees of freedom
still follows the $2t$-track. While it is beyond the scope of the present thesis to analyze the $t>s$ case in detail, let us make a few remarks. 

For $t=s+1$, we are presented with the puzzle
that the $0$-form $\Psi^{A(s),A'(s)}$ is balanced, and hence
in vector language corresponds to a symmetric tensor.
It therefore cannot be related to any Weyl tensor, since
the latter are always valued in two-row diagrams.
For $t=s+2,\dots,2s-1$, let us define $t=2s-\tau$,
with $\tau=1,\dots,s-2$, so that the pairs of fields 
in these cases take the forms
$(\omega^{A(\tau-1),A'(2s-\tau-1)},\Psi^{A(\tau+1),A'(2s-\tau-1)})$.
In this parametrization, the $1$-form $\omega$ seems like
the anti-self-dual part of the last connection for a spin-$s$ field
of depth-$\tau$, but the $0$-form does not have the required
symmetry to be considered as the corresponding Weyl tensor.
This can be traced back to the fact that we used the self-dual
basis $2$-forms $H_{AA}$ in the action to contract the $0$-form
$\Psi$. Consequently, the number of unprimed indices in $\omega$
and $\Psi$ differs by $2$, but when crossing the boundary
$t=s+1$, this difference is now the source of the mismatch 
between the pairs of indices for them to be identified with
the anti-self-dual part of the last connection and Weyl tensor
for a partially massless field.

More importantly, the equations of motion obtained in these cases
do not describe the propagation of a partially massless field:
one can check that the first few descendants of the Weyl tensor
which are not constrained by Bianchi identities do not generate
the usual module of a PM anti-self-dual Weyl tensor. Indeed,
consider a $0$-form $\Psi^{A(t-1),A'(2s-t+1)}$ where the parametrization
of its indices suggests that it corresponds to the anti-self-dual
part of the Weyl tensor of a spin-$s$ and depth-$t$ PM field,
subject to the equation of motion
\begin{equation}
    H^{BB}\,\nabla\Psi_{A(t-3)BB,A'(2s-t-1)} \approx 0\,.
\end{equation}
Then, one finds 
\begin{equation}\label{eq:wrong}
    \nabla\Psi_{A(t-1),A'(2s-t+1)}
    = e\fud{B}{A'}\,\Psi_{A(t-1)B,A'(2s-t)}
    + e^{BB'}\,\Psi_{A(t-1)B,A'(2s-t+1)B'}\,,
\end{equation}
instead of
\begin{equation}\label{eq:right}
    \nabla\Psi_{A(t-1),A'(2s-t+1)}
    = e\fdu{A}{B'}\,\Psi_{A(t-2),A'(2s-t+1)B'}
    + e^{BB'}\,\Psi_{A(t-1)B,A'(2s-t+1)B'}\,,
\end{equation}
as would be expected for the anti-self-dual part
of a spin-$s$ and depth-$t$ Weyl tensor. One can notice
that, though the second term on the right hand side of
these two expressions are identical, the first one is not.
In vector language, the expected spectrum of $0$-forms
is given by Young diagrams of the Lorentz group of the form
\begin{equation}
    \Yboxdim{12pt}\scriptstyle
    \gyoung(_9{s}_4{n},_6{s-t+1}_3{m})\,,
\end{equation}
with $n\geq0$ and $m=0,\dots,t-1$. This simply corresponds
to the fact that the derivatives of the Weyl tensor that
are unconstrained by equations of motion and Bianchi identities
are those projected in the first two rows of the Weyl tensor
Young diagram (in arbitrary number in the first row, or only up to $t-1$ in the second row). The equation \eqref{eq:wrong}
is not compatible with this because the two $0$-forms
appearing on the right hand side correspond to the diagrams
\begin{equation}
    \Yboxdim{12pt}\scriptstyle
    \gyoung(_9{s},_5{s-t};{\times})
    \qquad \qquad
    \gyoung(_9{s};,_5{s-t+1})
\end{equation}
so that in particular, the first diagram is unexpected
(see \cite{Skvortsov:2006at,Ponomarev:2010st}), due to the fact
that a box has been removed in the second row (crossed
hereabove) instead of being added. Due to this early
departure in the descendants of $\Psi$, the whole module
generated by the infinite tower of $0$-form required
to build an FDA will not correspond to that of a PM Weyl 
tensor. Once again, this can be traced back to the fact
that the expected equations \eqref{eq:right} is
the parametrization of a generic element in the kernel
of the symplectic form determined by $H_{A'A'}$, i.e.
it is a solution of
$H^{B'B'}\Psi_{A(t-1),A'(2s-t-1)B'B'} \approx 0$.

\begin{figure}[!ht]
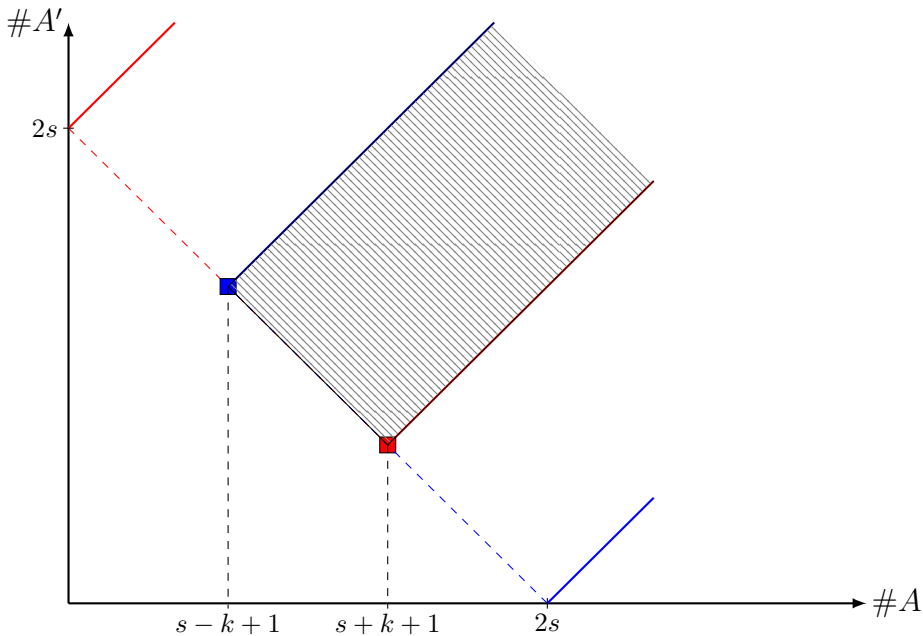

    \FIELDSBeyond
    \label{fig:beyond}
    \caption{In blue, the region covered by descendants
    of $\Psi^{A(s-k+1),A'(s+k-1)}$, in red the descendants
    of $\Psi^{A(s+k-1),A'(s-k+1)}$ and in gray the overlap
    between these two regions.}
\end{figure}

A possible scenario would be that this system, for $t=s+k$
and $k=1,\dots,s-1$, describes a reducible representation
of $\mathfrak{g}_\Lambda$, composed of two massive fields
of spin-$s$ and $k-1$. A trivial, but necessary, check
is that the counting of degrees of freedom is consistent,
since $2t = 2s+1 + 2(k-1)+1$. A more significant hint,
which motivates our conjecture, is that the spectrum
of $0$-forms in this case, represented in Fig. \ref{fig:beyond},
agrees with this proposal. Indeed, when the depth $t$
goes beyond $s$, the two strips of $0$-forms start
overlapping. The whole
region covered by these strips corresponds to the spectrum
of $0$-forms of a massive spin-$s$ field \cite{Ponomarev:2010st},
when each $0$-form appears with multiplicity $1$. 
The overlapping region could similarly be interpreted as
the collection of $0$-forms describing a massive spin-$(k-1)$
field, due to the width of this strip, but that would be represented
by spin-tensors of higher ranks than expected. In other words,
this massive spin-$(k-1)$ field could appear in our system
as a spin-tensor, which, due to some equation of motion,
should be expressed as derivative of a lower rank spin-tensor,
the latter being the genuine massive spin-$(k-1)$ field.
Note that this is to be taken, for the time being, only as
a proposal since proving rigorously the above statement
would go beyond the scope of this chapter, and is left for
potential future work.

\section{Discussion and Summary}
\label{sec:disco}
We have studied the simplest types of interactions: Yang--Mills
and current ones. It would be interesting to classify all possible
interactions within the new approach to partially massless fields
advocated in the present chapter. For example, there should exist
partially massless theories featuring gravitational interactions.
Another important omission is to have genuine non-Abelian higher spin higher derivative interactions. 
Such interactions, as different from, say, the Yang--Mills ones,
introduce nontrivial constraints that fix the spectrum of a theory
together with all the couplings.  

The elephant in the room is twistor theory, which played an important,
but silent, r\^ole in the chapter. Indeed, the twistor approach directly
leads to field variables $\Psi^{A(2s)}$ and $\omega^{A(2s-2)}$
for massless fields \cite{Hitchin:1980hp}. This was the starting point
of our generalization to partially massless fields. However,
the original twistor formulation of partially massless fields
seems to be missing at the moment. It would be interesting
to bridge this gap. 

At least for the purely massless case there exists a complete,
local higher spin gravity --- Chiral Theory 
\cite{Metsaev:1991mt,Metsaev:1991nb,Ponomarev:2016lrm,Skvortsov:2018jea},
which in addition to Yang--Mills and gravitational interactions
incorporates genuine higher spin interactions. The theory admits
any value of the cosmological constant, including zero. As was shown
in \cite{Ponomarev:2017nrr}, Chiral Theory has two contractions
where the scalar field can be dropped while either Yang--Mills
or gravitational interactions are kept (no genuine higher spin
interactions are present). These two contractions have simple
covariant actions \cite{Krasnov:2021nsq} and twistor origin
\cite{Tran:2021ukl,Herfray:2022prf,Adamo:2022lah}. Within AdS/CFT duality, Chiral Theory
should be dual to a subsector of Chern--Simons matter theories
\cite{Sharapov:2022awp}. 

In view of the facts collected here-above, it looks plausible
that there exist (Chiral) higher spin gravities with partially massless
fields in the spectrum \cite{Sharapov:2022awp}. These theories should admit
contractions that feature either Yang--Mills or gravitational
interactions, the former of which are considered in the present chapter.
Within AdS/CFT duality, such theories should be dual to a subsector
of isotropic (Chern--Simons) Lifshitz CFT's \cite{Bekaert:2013zya},
i.e. of vector models with higher-derivative kinetic terms.\footnote{Chern--Simons extension of these models have not been explored so far. It also remains unclear if the $3d$ bosonization duality can be extended to these models.} 

Lastly, it would be interesting to explore a family of deformations
of the actions proposed in the chapter via the $\Psi^2$-terms.
Such deformation mimics the well-known result on how Yang-Mills theory
can be represented as a deformation of the self-dual Yang-Mills theory
\cite{Chalmers:1996rq}: $\Psi F(\omega)$-type actions need to be completed
with $\Psi^2$-terms. This idea can be interesting already for free fields,
resulting in a new second-order action for partially massless fields,
which is still simpler than its cousins in terms of non-chiral field
variables. For massless fields, the $\Psi^2$-deformation was also shown to give higher spin theories with nontrivial scattering already in flat space case.
\cite{Adamo:2022lah}.

\chapter{(Partially massless) Higher Spin algebras}\label{sec:HSA}

\section*{Introduction}

In the preceding chapters, we have seen that free higher spin (HS) gravity can be formulated via an action principle in either the metric-like or the frame-like language. We also reviewed how chiral formulations of both massless and partially massless higher spin theories emerge naturally within the frame-like approach in four dimensions. In each of these formulations—just as in conventional lower-spin ($s<3$) gauge theories—a gauge algebra, known as the higher spin algebra (HSA), underlies the theory.

Conversely, one may begin with this algebra as a global symmetry and then construct the corresponding field theory by gauging it: for instance, Maxwell’s theory and the Cartan (frame) formulation of gravity arise in precisely this way. Accordingly, a detailed understanding of the global higher spin algebra is essential, since gauging it yields the associated higher spin gravity theory.

From a field-theoretic perspective, the higher spin algebra is simply the Lie algebra of the theory’s global symmetries, encompassing its entire higher spin spectrum. In four dimensions, Fradkin and Vasiliev demonstrated its relevance for cubic interactions \cite{Fradkin:1987ks, Fradkin:1986qy}; extensions to higher dimensions have likewise been developed \cite{Eastwood:2002su, Vasiliev:2003ev}. Moreover, from the standpoint of holography, a theory of massless higher spin fields in $\mathrm{AdS}_{D}$ should be dual to a free conformal field theory (CFT) in $(D-1)$ dimensions. Consequently, the higher spin algebra in $\mathrm{AdS}_{D}$ is isomorphic to the conformal higher spin symmetry algebra of a free massless scalar in $d$ dimensions. This is sometimes even taken as the definition of higher spin algebras in many contexts. A detailed review of massless and partially massless higher spin algebra can be found in \cite{Joung:2014qya, Joung:2015jza} and references therein.

Although the explicit form of the gauge symmetries depends on whether one employs a frame-like or metric-like formalism, the underlying global higher spin algebra remains the same. Consequently, there exists a formulation of the higher spin algebra that is independent of any particular field-theoretic description: namely, as a quotient of the universal enveloping algebra (often called the coset construction) of the spacetime symmetry algebra \(\mathfrak{g}\). Every higher spin algebra arising in field theory admits such a coset realisation. Concretely, the choice of higher spin algebra is determined by two ingredients: the background symmetry algebra \(\mathfrak{g}\), and the two-sided ideal by which one quotients of \(U(\mathfrak{g})\). Different ideals correspond to different spectra of higher spin fields.

Even though the coset construction provides a conceptually elegant definition of the higher spin algebra, implementing the quotient by various two-sided ideals in \(U(\mathfrak{g})\) can be technically challenging.  To circumvent this difficulty, one often employs alternative realisations of \(\mathfrak{g}\) that either simplify or automatically enforce the desired quotient.  In particular, oscillator realisations have proven to be invaluable computational tools in higher spin theory.  In four dimensions, these oscillators are naturally represented by two-component spinors, whereas in higher dimensions, one works with vector oscillators.

Another important tool for identifying higher spin algebras is the notion of a "Reductive (Howe) dual pair" (See \cite{Joung:2015jza} for a review in the HS context).  A reductive dual pair \((G_{1},G_{2})\) in a group \(G\) consists of two subgroups whose Lie algebras \(\mathfrak{g}_{1}\) and \(\mathfrak{g}_{2}\) are maximal mutual centralizers in \(\mathfrak{g}\),
\[
[\mathfrak{g}_{1}, \mathfrak{g}_{2}] \;=\; 0.
\]
Whenever a reductive dual pair \((G_1,G_2)\subset G\) acts on a \(G\) module \(M\), a lack of multiplicity freeness i.e.\ when some irreducible representation of \(G_1\) appears more than once prevents a clean one-to-one correspondence between the irreducibles of \(G_1\) and those of \(G_2\) in \(M\).  By contrast, when the action is multiplicity-free, there is a natural bijection between the two sets of irreducible representations.  

In practical applications, one often uses such a dual pair to realize the higher spin algebra associated to, say, \(\mathfrak{g}_1\) as the centralizer of \(\mathfrak{g}_2\).  In many examples, this construction automatically enforces the necessary quotient of the universal enveloping algebra, without the need to impose the ideal explicitly.  

Combined with an oscillator realisation, Howe duality thus provides a powerful and computationally efficient framework for constructing higher spin algebras. 

In this chapter, we will motivate the definition of higher spin algebra in terms of the quotient universal enveloping algebra, then present oscillator realisations and the role of Howe dual pairs in implementing the ideal quotient. Finally, we will specialize to four dimensions and define the partially massless higher spin algebra via bosonic and fermionic oscillators and explore the possibility of deformations of these algebras to incorporate interactions.

\section{Motivation}

From the previous chapters, we see that higher spin gravity is a theory that contains gauge fields with spin taking any integer value. We know that the gauge symmetry for the lower spins is given by,
\begin{equation}
  \begin{aligned}
    \delta A_{\mu} &= \partial_{\mu}\,\xi + \mathcal{O}(A),
      \quad \text{spin-1}\\
    \delta h_{\mu\nu} &= \nabla_{(\mu}\xi_{\nu)} + \mathcal{O}(h),
      \quad \text{spin-2}
  \end{aligned}
\end{equation}
As we have mentioned in the introduction that a gauge symmetry must come from gauging a global symmetry. Such global symmetry can be seen from above mentioned gauge transformation. Indeed the first term in the above equation defines the global symmetry if the gauge parameter satisfies the Killing equation as,
\begin{equation}
  \begin{aligned}
    & \partial_{\mu}\,\xi=0\quad\,\implies\xi=\text{constant}  \quad \text{spin-1},\\
   &\nabla_{(\mu}\xi_{\nu)}=0,\quad\,\implies\,\xi_{\mu}=P_{\mu}+M_{\mu\nu}x^\nu
      \quad \text{spin-2},
  \end{aligned}
\end{equation}
where the $P$ and $M$ are translation and Lorentz generators. Similarly, for higher spin fields in arbitrary dimension we have\footnote{We consider symmetric fields, but in higher dimension we can also have fields with more complicated symmetry structure, which we do not consider here.}
\begin{equation}
    \delta \phi_{(\mu_{1}\mu_{2}...\mu_{s})}=\nabla_{(\mu_{1}}\xi_{\mu_{2}...\mu_{s})}+\mathcal{(\phi)}
\end{equation}
with $\xi_{\mu_{2}...\mu_{s}}$ are traceless as we have seen in chapter one. The Killing equation is given by 
\begin{equation}
    \nabla_{(\mu_{1}}\xi_{\mu_{2}...\mu_{s})}=0.
\end{equation}
This implies 
\begin{equation} 
   \xi_{\mu_{1}...\mu_{r}}=M_{\mu_{1}...\mu_{r}}+M_{\mu_{1}...\mu_{r},\,\nu}x^{\nu}+...\frac{1}{n!}M_{\mu_{1}...\mu_{r}\,\nu_{1}...\nu_{n}}x^{\nu_{1}}...x^{\nu_{n}}.
\end{equation}
Here $M_{\mu_{1}...\mu_{r}\,\nu_{1}...\nu_{n}}$ the translation an Lorentz generator to higher spin. It is easy to see that the Killing condition $ \nabla_{(\mu_{1}}\xi_{\mu_{2}...\mu_{s})}=0$ implies that 
\begin{equation}
    M_{(\mu_{1}...\mu_{r}\,\nu_{1})...\nu_{n}}=0.
\end{equation}
This condition is nothing but the Young condition, which means such tensor objects have the symmetry of the  Young diagram of the type 
\begin{equation}
    \Yboxdim{11pt}
    \gyoung(_6<r>,_4<n>)
    =M_{\mu_{1}...\mu_{r}\,\nu_{1}...\nu_{n}}\,, \quad n\leq r\,.
\end{equation}
Thus, in this way we see that a global symmetry associated with a massless spin-s field is a collection of finite-dimensional Lorentz irreducible representations labeled by Young diagrams of the form, 
\begin{equation}
    \Yboxdim{11pt}
    \gyoung(_6<s-1>),\,\, \,\,\Yboxdim{11pt}
    \gyoung(_6<s-1>,;),\,\dots \quad\dots\Yboxdim{11pt}
    \gyoung(_6<s-1>,_6<s-1>)\,.
\end{equation}
In other words, if we begin with an internal symmetry algebra and the usual Lorentz (spin) symmetry—and then seek a massless higher spin extension—we are forced to enlarge the gauge symmetry so that the resulting spectrum matches the desired higher spin multiplets. Concretely, this amounts to extending the Poincaré algebra into a higher spin algebra.

A further subtlety arises from general covariance (diffeomorphism invariance): any Killing tensor defining a higher spin gauge parameter must itself furnish a representation of the background’s isometry group. This requirement is neatly met in anti-de Sitter space, where the nonzero cosmological constant enhances the isometry algebra to $\mathfrak{so}(2,d)$. In fact, one can show that all of the higher spin generators—those descending from the “internal” and spin symmetries—fit together precisely as the components of a single representation of $\mathfrak{so}(2,d)$, once one performs the appropriate branching.
\begin{equation}
   \Yboxdim{11pt}
    \gyoung(_6<s-1>,_6<s-1>)|_{\mathfrak{so}(2,d)}= \Yboxdim{11pt}
    \gyoung(_6<s-1>)|_{\mathfrak{so}(1,d-1)},\,\, \,\,\Yboxdim{11pt}
    \gyoung(_6<s-1>,;)_{\mathfrak{so}(1,d-1)},\,\dots\quad\dots\Yboxdim{11pt}
    \gyoung(_6<s-1>,_6<s-1>)|_{\mathfrak{so}(1,d-1)}\,.
\end{equation}
Thus if a theory contains gauge fields ${\phi_{\mu_{1}...\mu_{s}}}$ then there should exist a global symmetry (Lie algebra)
\begin{equation}
    \mathfrak{hs}=\textbf{Span}_{\mathbb{R}}\bigg\{\Yboxdim{11pt}
    \gyoung(_6<s-1>,_6<s-1>)|_{\mathfrak{so}(2,d)}\bigg\}
\end{equation}
Note that these are traceless tensors. The resulting structure—first identified by Fradkin and Vasiliev—is known as the higher–spin algebra. This observation implies that no consistent higher–spin theory can exist without such an algebra. Accordingly, one naturally begins by constructing the higher–spin algebra, which encodes the global symmetry, and then gauges this algebra to obtain the full higher–spin theory.
Let us set some notation first: We will denote by $D=d+1$ the spacetime dimension and denote by $M_\AlgInd{AB}$ the generator of the Lie algebra $\mathfrak{so}(2,d)$, with indices
$\AlgInd{A,B},\dots$ taking $d+2$ values, and by $\eta_\AlgInd{AB}$
the (components of the diagonal) metric of signature $(-,-,+,\dots,+)$. Then, $\mathfrak{so}(2,d)$ traceless generators are given as 
\begin{equation}
    T_{\AlgInd{A_{1}...A_{s-1}\,B_{1}...B_{s-1}}}=\Yboxdim{11pt}
    \gyoung(_6<s-1>,_6<s-1>)|_{\mathfrak{so}(2,d)}\,.
\end{equation}
It turns out that such a higher spin generator can be formed using the $\mathfrak{so}(2,d)$ generator $M_{AB}$ as,
\begin{equation}
    T_{\AlgInd{A_{1}...A_{s-1}\,B_{1}...B_{s-1}}}=T_{\AlgInd{A(s-1)B(s-1)}}=\text{traceless part of }\{M_{\AlgInd{A_{1}B_{1}}}M_{\AlgInd{A_{2}B_{2}}}...M_{\AlgInd{A_{s-1}B_{s-1}}}\},
\end{equation}
where it is understood that all \(\AlgInd{A}\)–indices and \(\AlgInd{B}\)–indices are symmetrized separately. This construction shows that the higher–spin generators \(T_{\AlgInd{A(r)B(r)}}\) lie naturally in the universal enveloping algebra \(\mathcal{U}(\mathfrak{so}(2,d))\).  Indeed, motivated by holography, one finds that the higher–spin symmetry in \(\mathrm{AdS}_{d+1}\) is isomorphic to the conformal higher–spin symmetry of the free scalar “singleton” satisfying \(\Box\phi=0\) in \(d\) dimensions.  

However, \(\mathcal{U}(\mathfrak{so}(2,d))\) is much larger than the subspace spanned by the traceless, two–row tensors \(T_{\AlgInd{A(r)B(r)}}\).  To extract the genuine higher–spin algebra, one therefore quotients \(\mathcal{U}(\mathfrak{so}(2,d))\) by the two–sided ideal generated by all elements that are not of the Killing–tensor type \(T_{\AlgInd{A(r)B(r)}}\).  This quotienting procedure simultaneously fixes the values of all Casimir operators and isolates the correct spectrum of representations. Eastwood in \cite{Eastwood:2002su} first established that the space of all the symmetries of \(\Box\phi=0\) is isomorphic to \(\mathcal{U}(\mathfrak{so}(2,d))\) modulo a certain ideal (more details in the next section). 

We can further enlarge the massless higher spin algebra $\mathfrak{hs}$ that we discussed above in constant-curvature spacetime, say $AdS_{d+1}$, by generalizing the gauge transformation to a higher derivative gauge transformation as 
\begin{equation}
    \delta\phi^{{t}}_{\mu_{1}...\mu_{s}}=\nabla_{(\mu_{1}}...\nabla_{\mu_{t}}\xi^{t}_{\mu_{t+1}...\mu_{s})} +\mathcal{O}(\Lambda).
\end{equation}
 Here \(\Lambda\) is the cosmological constant, \(t=1,2,\dots,s\) is the so-called “depth,” and \(\nabla\) denotes the $AdS$ covariant derivative.  For \(t=1\), one recovers the standard massless spin–\(s\) gauge transformation, while each \(t>1\) corresponds to a gauge parameter \(\xi\) of lower rank than in the massless case.  Consequently, a depth-\(t\) field propagates more degrees of freedom than a massless spin–\(s\) field but fewer than a fully massive one.  Such fields are therefore called \emph{partially massless} (PM) fields. Similarly to what we did above, we can write the corresponding  Killing equation for PM symmetries from the above gauge transformation as,
 \begin{equation}
     \nabla_{(\mu_{1}}...\nabla_{\mu_{t}}\xi^{(t)}_{\mu_{t+1}...\mu_{s})}=0.
 \end{equation}
The solution is again given by a two-row $\mathfrak{so}(2,d)$ young diagram, but this time the length of the second row depends on depth $t$ as $s-t$, i.e,
\begin{equation}
     \Yboxdim{11pt}
    \gyoung(_6<s-1>,_4<s-t>)
\end{equation}
Remarkably, the algebra generated by these generators for $t=1,3...(2\ell-1)$ is isomorphic to the higher symmetries of the so–called “higher–order singleton” (satisfying \(\Box^{\ell}\phi=0\)). This algebra is called the partially massless higher–spin algebra. In this way, the well–known correspondence between higher–spin algebras and singleton symmetries generalizes directly to a bijection between PM higher–spin algebras and the symmetries of higher–order singletons.  We will examine this in detail in the next section.

\subsection{Universal Enveloping Algebra: Quotient Construction}
Let \(\mathfrak{g}\) be an arbitrary Lie algebra. We define the Universal Enveloping algebra (UEA) as the tensor algebra \(\mathcal{T(\mathfrak{g})}\) quotiented by the two-sided ideal \(\mathcal{I}=\langle x\otimes y-y\otimes x-[x,y]\rangle\) for any $x,y\in \mathfrak{g}\subset \mathcal{T(\mathfrak{g})}$, i.e,
\begin{equation}
    \mathcal{U(\mathfrak{g})}=\mathcal{T(\mathfrak{g})}/\mathcal{I}.
\end{equation}
Now, if we consider $\mathfrak{g}$ to be the isometry algebra of the spacetime, then we define an abstract higher spin algebra as follows.
\begin{definition}
    The higher spin algebra associated with a Lie  algebra $\mathfrak{g}$ is defined as the quotient of the universal enveloping algebra ${\Enveloping\big(\mathfrak{g}\big)}$ by its Joseph ideal $\Ideal$, i.e.
    \begin{equation}
    \hs = {\Enveloping\big(\mathfrak{so}(2,d)\big)}
    \big/{\Ideal}\,,
    \end{equation}    
\end{definition}
where the \emph{Joseph ideal} \(\Ideal\subset U(\mathfrak{g})\) is the (unique) ideal in the universal enveloping algebra of a simple Lie algebra \(\mathfrak{g}\) that annihilates its minimal nontrivial representation\cite{Joseph:1974hr}.

Different Lie groups and ideals chosen will give different higher spin algebras\cite{Joung:2014qya, Joung:2015jza}. As an illustration, let us briefly consider the construction of higher spin algebra in the three-dimensional case. We know that the $3d$-gravity is topological and is equivalent to Chern-Simons theory. More precisely $3d$-gravity on $AdS_{3}$ is based on Lie algebra $\mathfrak{sl}(2,\mathbb{R}))$. A higher spin extension of the above scenario will be to have HS-gravity on $AdS_{3}$ with $\mathfrak{sl}(2,\mathbb{R}))$ replaced with its higher spin extension, usually denoted by $\mathfrak{\mathfrak{hs}[\lambda]}$.  In other words we need to consider the UEA  $\mathcal{U}(\mathfrak{sl}(2,\mathbb{R}))$ and define $3d$ HS algebra as 
\begin{equation}
    B[\lambda]=\frac{\mathcal{U}(\mathfrak{sl}(2,\mathbb{R}))}{\langle \big(\Casimir_2-\frac{(\lambda^2-1)}{4}\big)\rangle}.
\end{equation}
Here $\Casimir_2$ is the quadratic Casimir\footnote{The notation 
$\big\langle (\cdots) \big\rangle$ means that
the ideal is generated by the elements inside
the bracket.}. Now we know that the $\mathfrak{sl}(2,\mathbb{R})$ generators are given by $J_{0},\,J_{+},\,J_{-}$ and satisfy,
\begin{equation}\label{sl algebra}
    [J_{+},\,J_{-}]=2J_{0}\,\,\qquad\,\,[J_{+},\,J_{0}]=J_{0}\,\,\quad\,\,[J_{-},\,J_{0}]=-J_{0}
\end{equation}
\begin{equation}
    C_{2}=J_{0}^2-\frac{1}{2}(J_{+}J_{-}+J_{-}J_{+})
\end{equation}

An easy way to remember what is the universal enveloping algebra, of any finite dimensional Lie algebra, is to assume the generators of the Lie algebra as letters of an alphabet, then the UEA will be the collection of all the words you can write (using these generators) provided we assume the commutation relation of the Lie algebra. For $\mathfrak{sl}(2,\mathbb{R})$ this means that $\mathcal{U}(\mathfrak{sl}(2,\mathbb{R}))$ contains all the words formed by $J_{0},\,J_{+},\,J_{-}$ with commutation relation (\ref{sl algebra}) imposed on them. So for example, if the word is $J_{+}J_{+}J_{0}J_{-}$ then using commutation relation it can be always be rearranged in a given order say, $J_{0}J_{+}J_{+}J_{-}$ plus lower order terms. We can always do this to any string of $J's$ and put them in the form $J_{0}J_{0}J_{0}....J_{+}J_{+}J_{+}....J_{-}J_{-}J_{-}....$ plus lower order terms which themselves are in the same order. Hence a typical element of $\mathcal{U}(\mathfrak{sl}(2,\mathbb{R}))$ can be written as a sum of elements (monomials), each of which is written as $J_{0}J_{0}....J_{+}J_{+}....J_{-}J_{-}....$ with specific number of $J_{0}$, $J_{+}$ and $J_{-}$ (This in the literature is called Poincare-Birkhoff-Witt (PBW) theorem). Hence the set $\{J^{m}_{0}J^n_{+}J^r_{-}\}$ forms a basis of $\mathcal{U}(\mathfrak{sl}(2,\mathbb{R}))$.  Also, if you have two such words, you can multiply them, and using the commutation relation, rearrange them again in the desired order. Now to get the HS-algebra $\mathcal{U}(\mathfrak{sl}(2,\mathbb{R}))$ we need to quotient $\mathcal{U}(\mathfrak{sl}(2,\mathbb{R}))$ by the ideal generated by $\Casimir_2-\frac{(\lambda^2-1)}{4}$, which means that we need to impose the relation
\begin{equation}
    J_{0}^2-\frac{1}{2}(J_{+}J_{-}+J_{-}J_{+})=\frac{(\lambda^2-1)}{4}
\end{equation}
Now, we have learned that the basis set for $B[\lambda]$ is given by $\{J^{m}_{0}J^n_{+}J^r_{-}\}$, we can look at some lower order basis (lower values of $m,n,r$) to get more clarity\footnote{This further elaboration is just for the sake of completeness but can be skipped without any loss.}. We start with zero letter (generator) words which means no $J's$, clearly we only have identity in $\mathcal{U}(\mathfrak{sl}(2,\mathbb{R}))$ as a vector space, one letter words are $J_{0},\,J_{+},\,J_{-}$, two letter words are $J_{+}^{2}, J_{+}J_{0}, J_{+}J_{-}, J_{0}^{2}, J_{0}J_{-}$ and $J_{-}^{2}$. These are six words but since we have a non-trivial relation due to ideal relation $ J_{0}^2-\frac{1}{2}(J_{+}J_{-}+J_{-}J_{+})=\frac{(\lambda^2-1)}{4}$ we can eliminate one of the word say $J_{+}J_{-}$, hence it gives us only 5 independent two letter words. Similarly we can go on up to arbitrary numbers.

\begin{table}[ht]
  \centering
  \begin{tabular}{c|l|c}
    \textbf{\# of generators} & \textbf{Words (basis)}                          & \textbf{Independent \# of words} \\ \hline\hline
    $0$                       & $1$                                            & $1$                               \\ \hline
    $1$                       & $J_{0},\,J_{+},\,J_{-}$                        & $3$                               \\ \hline
    $2$                       & $J_{+}^{2},\,J_{+}J_{0},\,J_{0}^{2},\,J_{0}J_{-},\,J_{-}^{2}$ 
                              & $5$                               \\ \hline
    $3$                       & $J_{+}^{3},\;\dots$                            & $7$                               \\ \hline
    $4$                       & $J_{+}^4,\;\dots$                                        & $9$                               \\
  \end{tabular}
  \caption{Number of independent words in the universal‐enveloping algebra up to given word‐length (generator count).}
  \label{tab:uea-words}
\end{table}
We can generate the basis of any order as follows: We start from $J_{+}^{s-1}$ and successively commute it with $J_{-}$ and use the quadratic ideal relation. We will see that the basis elements are generated by,
\[
\begin{aligned}
  V^{s}_{s-1} &= J_{+}^{\,s-1},\\[6pt]
  V^{s}_{n}   &= (-1)^{\,s-n-1}\,\frac{(n+s-1)!}{(2s-2)!}\,
                 \underbrace{[\,J_{-},[\,J_{-},\dots,[\,J_{-}}_{(s-n-1)\text{ commutators}},
                 V^{s}_{s-1}]\,]\dots]\,.\,\qquad\,|n|\leq (s-1)
\end{aligned}
\]
In this way, we generate all the basis of the vector space $\mathcal{U}(\mathfrak{sl}(2,\mathbb{R}))$ and the associative multiplication between two such basis elements is done by concatenation or juxtaposition\footnote{Concatenation means joining together, for example, concatenating two basis elements $J_{0}J_{+}$ and $J_{0}J_{-}$ is just writing them together as $J_{0}J_{+}J_{0}J_{-}$ and one can use commutation rule to bring them into desired order.} of the basis elements. Now, given an associative algebra with a multiplication, we can convert it into a Lie algebra by simply defining the Lie algebra's commutator as $[A,B]=AB-BA$. In our case $A,B$ are of the form $A=J_{0}^{m_{1}}J_{+}^{n_{1}}J_{-}^{r_{1}}$ and $B=J_{0}^{m_{2}}J_{+}^{n_{2}}J_{-}^{r_{2}}$. Hence, we get an infinite-dimensional Lie algebra, which we again call $B[\lambda]$. Now, clearly, if we take two arbitrary basis elements from table 3.1, except for the first row, their commutator will never give us the identity. It means that the first row itself constitutes a Lie algebra. Since its basis is just the identity, it is nothing but the abelian Lie algebra $\mathbb{C}$. Hence, we can write  $B[\lambda]=\mathbb{C}\oplus \mathfrak{hs}[\lambda]$. Note that in the literature it is this infinite-dimensional, one-parameter family of algebras, $\mathfrak{hs}[\lambda]$, that is called $3d$ higher spin algebra. One important observation worth noting is that if one takes $[V^{s}_{s-1},\,V^{s}_{n}]$ for $s=3$, then $n=-2,-1,0,1,2$ and five commutators will give you five quadratic basis elements which we showed in the third row of the table. In particular for $n=-1$ we get,
\begin{equation}
    [V^{3}_{2},V^{3}_{-1}]=6V^4_{1}+\frac{1}{5}(\lambda^2-4)V^2_{1}
\end{equation}
This shows that for an arbitrary parameter $\lambda$, the commutator of two quadratic basis elements gives you one cubic basis element and one linear basis element. From the table's point of view, it means that the commutator of two elements of the $3^{rd}$-row lies in $2^{nd}$ and $4^{th}$ rows. But if we take an integer value for $\lambda=2$ then $[V^{3}_{2},V^{3}_{-1}]=6V^4_{1}$,i.e, the commutator lies below $2^{nd}$-row. It is true for any two basis elements from the second row or below that the commutator lies below the second row. This means that the basis generators below the second row form an infinite-dimensional ideal and can be mod out and leave behind a finite-dimensional Lie algebra with basis from the first and second row, which is nothing but $B[\lambda=2]=\mathbb{C}\oplus \mathfrak{sl}(2)$. Similarly, it can be checked by explicit computation that when $\lambda=3,4,..$ we get an infinite-dimensional ideal below the third row, fourth row, and so on, and by modding them out we are left with a finite-dimensional Lie algebra. Thus, in conclusion, we have a one-parameter family of infinite-dimensional higher spin extensions of $3d$ gravity, and when the parameter takes an integer value, it develops an infinite-dimensional ideal and gives us a finite-dimensional higher spin extension of $3d$ gravity. It is important to remark that this $3d$ example is one of the simplest examples and also a unique instance, where there exists a finite-dimensional higher spin algebra, see also for recent developments \cite{Serrani:2025owx}. This is all possible because the ideal, which needs to be taken out, simply fixes the quadratic Casimir in terms of the parameter $\lambda$.

In higher dimensions, the ideal will be much more difficult to handle, and we need different techniques to implement the quotient of the UEA. This is what we will discuss next.

\subsection{Quotient Construction: In Arbitrary dimension $>$ 3}
Higher spin algebras in dimensions greater than three have a complicated ideal. As we have already mentioned that HS-algebra requires two things to be identified: one is the Lie algebra of which we will construct the UEA, and the second is the ideal $\mathcal{I}$. We will consider the isometry algebra of the $AdS$ i.e, $\mathfrak{so}(2,d)$, to build $\mathcal{U}(\mathfrak{so}(2,d))$.

First, we recall some notation again: we will denote by $M_\AlgInd{AB}$
the generator of the Lie algebra $\mathfrak{so}(2,d)$, with indices
$\AlgInd{A,B},\dots$ taking $d+2$ values, and by $\eta_\AlgInd{AB}$
the (components of the diagonal) metric of signature $(-,-,+,\dots,+)$.
The Lie bracket of these $\mathfrak{so}(2,d)$ generators reads
\begin{equation}
    [M_\AlgInd{AB},M_\AlgInd{CD}]
    = \eta_\AlgInd{BC}\,M_\AlgInd{AD}
    -\eta_\AlgInd{AC}\,M_\AlgInd{BD}
    -\eta_\AlgInd{BD}\,M_\AlgInd{AC}
    +\eta_\AlgInd{AD}\,M_\AlgInd{BC}\,.
\end{equation}
And, as we have seen above, we will simply denote the associative product in the universal enveloping algebra $\Enveloping\big(\mathfrak{so}(2,d)\big)$ by juxtaposition, for instance, we will write
\begin{equation}
    \Casimir_2 = -\tfrac12\,M_\AlgInd{AB}\,M^\AlgInd{AB}\,,
\end{equation}
for the quadratic Casimir operator of $\mathfrak{so}(2,d)$,
where the indices have been raised with the inverse metric
$\eta^\AlgInd{AB}$. The higher spin algebra of type-A$_\ell$
is the quotient \cite{Gover2009, Michel2014}
\begin{equation}
    \hs_\ell = {\Enveloping\big(\mathfrak{so}(2,d)\big)}
    \big/{\Ideal_\ell}\,,
\end{equation}
of the universal enveloping algebra of $\mathfrak{so}(2,d)$
by the (two-sided) ideal
\begin{equation}
    \Ideal_\ell = \Big\langle V_\AlgInd{ABCD}
    \oplus \big(\Casimir_2 + \tfrac{(d-2\ell)(d+2\ell)}4\,\1\big)
    \oplus \Joseph_{\AlgInd{A}(2\ell)}\Big\rangle\,,
\end{equation}
where 
\begin{equation}
    V_\AlgInd{ABCD} := M_\AlgInd{[AB}\,M_\AlgInd{CD]}\,,
    \qquad 
    \Joseph_{\AlgInd{A}(2\ell)} := M_\AlgInd{A}{}^\AlgInd{B_1}\,M_\AlgInd{AB_1} \dots M_\AlgInd{A}{}^\AlgInd{B_\ell}\,M_\AlgInd{AB_\ell} - \text{traces}\,,
\end{equation}
and where we used the convention
(standard in the higher spin literature)
that symmetrised indices are denoted by
the same letter, with their number being indicated
in parenthesis when necessary,
e.g. $A(l)=(A_1...A_l)$.

Recall that the universal enveloping algebra
of a Lie algebra $\mathfrak{g}$ is isomorphic,
as vector space%
\footnote{Actually as a $\mathfrak{g}$-module,
and as a (co-commutative) coalgebra.},
to the symmetric algebra $S(\mathfrak{g})$. This space
is, by definition, the symmetrised tensor product of
the adjoint representation $\gyoung(;,;)$ of $\mathfrak{so}(2,d)$,
and can be decomposed into a direct sum of finite-dimensional
irreducible representations that we will denote by
the corresponding Young diagram. In such terms,
the subspace of elements quadratic in the Lie algebra
generators reads
\begin{equation}
    \gyoung(;,;)^{\odot2} \cong \gyoung(;;,;;)
    \oplus \gyoung(;,;,;,;) \oplus \gyoung(;;)
    \oplus \bullet\,,
\end{equation}
where in particular
\begin{equation}
    \gyoung(;,;,;,;)
    \quad\longleftrightarrow\quad 
    V_\AlgInd{ABCD}
    \qquad\text{and}\qquad
    \bullet \quad\longleftrightarrow\quad \Casimir_2\,.
\end{equation}
When modding out the ideal $\Ideal_\ell$, the totally antisymmetric
diagram is removed, whereas the quadratic Casimir operator
is related to a multiple of the identity. More precisely,
the quadratic Casimir operator is set to take the value $-\tfrac14(d-2\ell)(d+2\ell)$,
which is the same value it takes when acting on the order-$\ell$
singleton module. Next we can look at the subspace of the universal
enveloping algebra spanned by elements cubic in the Lie algebra,
\begin{equation}
    \gyoung(;,;)^{\odot3} \cong \gyoung(;;;,;;;)
    \oplus \gyoung(;;,;;,;,;) \oplus \gyoung(;,;,;,;)
    \oplus \gyoung(;;,;,;) \oplus \gyoung(;;;,;)
    \oplus \gyoung(;,;)\,,
\end{equation}
and make the following observations.
\begin{enumerate}[label=$(\roman*)$]
\item First, the three diagrams
with more than two rows are contained in the product
of the ideal generators $V_\AlgInd{ABCD}$ and the Lie algebra
generators $M_\AlgInd{AB}$, and hence belong to the ideal
$\Ideal_\ell$,
\begin{equation}
    \gyoung(;,;,;,;) \oplus \gyoung(;;,;;,;,;)
    \oplus \gyoung(;;,;,;) \subset \Ideal_\ell\,,
\end{equation}
so that they are removed once $\Ideal_\ell$ is modded out 
from the universal enveloping algebra of $\mathfrak{so}(2,d)$. This is,
in fact, a general pattern: Young diagrams with more than
two rows appearing in the decomposition of
$\Enveloping\big(\mathfrak{so}(2,d)\big)$ all belong to the ideal
$\Ideal_\ell$, and more specifically, to the ideal generated
by $V_\AlgInd{ABCD}$. As a consequence, the higher spin algebra 
$\hs_\ell$ contains only Young diagrams with one or two rows.
\item Second, the diagram $\gyoung(;,;)$ is obtained
as the product of the quadratic Casimir operator $\Casimir_2$
with the Lie algebra generators $M_\AlgInd{AB}$.
Since, after modding out the ideal $\Ideal_\ell$,
the value of $\Casimir_2$ is fixed, the adjoint representation
only appears with multiplicity one in $\hs_\ell$.
\end{enumerate}
One can immediately extract from the previous item
the following lesson: in order
for the quotient algebra $\hs_\ell$ to admit
a \emph{multiplicity-free} decomposition under $\mathfrak{so}(2,d)$,
i.e. that all irreducible representations (irreps)
appear only once in the decomposition of $\hs_\ell$
under the adjoint action of $\mathfrak{so}(2,d)$,
the center of the universal enveloping algebra
has to be fixed. In other words, modding out the ideal
$\Ideal_\ell$ should fix the values of all Casimir operators
(quadratic and higher), as the latter form a basis
of the center of $\Enveloping\big(\mathfrak{so}(2,d)\big)$.

Finally, since $\mathfrak{so}(2,d)$-module appearing
in the decomposition of the universal enveloping algebra
$\Enveloping\big(\mathfrak{so}(2,d)\big)$ are contained,
by definition, in tensor product of its adjoint
representation, all these irreps are characterised by
Young diagrams with an \emph{even} number of boxes.
In view of the previous discussion, this means that
 they are necessarily of the form $\gyoung(_5,_3)$,
where difference between the number of boxes
in the first and in the second row is \emph{even}.
This difference equals $t-1$ where, as before,
$t$ is the depth of the partially massless field.
Modding out by the symmetric diagram 
$\gyoung(_5{\scriptstyle 2\ell})$
effectively removes all diagrams with $t>2\ell-1$,
as they would belong to the product of the former
with another diagram in the spectrum.

\paragraph{Howe duality in the Weyl algebra.}
Now consider the Weyl algebra $\Weyl_{2(d+2)}$ generated by
$\{Y^\AlgInd{A}_i\}$ where $i=\pm$, and with the Moyal--Weyl
star-product
\begin{equation}
    f \star g = f\,\exp
    \big(\tfrac{\overleftarrow{\partial}}{\partial Y^\AlgInd{A}_i}\,
    \eta^\AlgInd{AB} \epsilon_{ij}\,
    \tfrac{\overrightarrow{\partial}}{\partial Y^\AlgInd{B}_j}\big)\,g\,,
\end{equation}
where $\epsilon_{ij}$ are the components of the canonical
$2 \times 2$ symplectic matrix, as associative product.
Quadratic monomials in $Y^\AlgInd{A}_i$, i.e. linear
combinations of the generators
\begin{equation}
    K^\AlgInd{AB}_{ij}
    := \tfrac12\,Y^\AlgInd{A}_i Y^\AlgInd{B}_j\,,
\end{equation}
span a Lie subalgebra isomorphic to $sp\big(2(d+2),\R\big)$,
\begin{equation}
    [K^\AlgInd{AB}_{ij}, K^\AlgInd{CD}_{kl}]_\star
    = \eta^\AlgInd{BC}\epsilon_{jk}\,K^\AlgInd{AD}_{il}
    + \eta^\AlgInd{AC}\epsilon_{ik}\,K^\AlgInd{BD}_{jl}
    + \eta^\AlgInd{BD}\epsilon_{jl}\,K^\AlgInd{AC}_{ik}
    + \eta^\AlgInd{AD}\epsilon_{il}\,K^\AlgInd{BC}_{jk}\,,
\end{equation}
where $[-,-]_\star$ denotes the commutator with respect to 
the star-product. The index structure on display here allows
one to easily identify two mutually commuting Lie subalgebras,
\begin{equation}
    \mathfrak{so}(2,d) \oplus \mathfrak{sp}(2,\R) \subset sp\big(2(d+2),\R\big)\,,
\end{equation}
respectively generated by
\begin{equation}
    M_\AlgInd{AB} := \tfrac12\,\epsilon^{ij}\,
    Y^\AlgInd{A}_i Y^\AlgInd{B}_j\,,
    \qquad \text{and} \qquad 
    \tau_{ij} := \tfrac12\,\eta_\AlgInd{AB}\,
    Y^\AlgInd{A}_i Y^\AlgInd{B}_j\,.
\end{equation}
Such pairs of algebras are usually called reductive dual pairs,
or Howe dual pairs \cite{Howe1989i, Howe1989ii, Goodman2000} (this will be discussed in more generality in the next subsection),
and can be used to construct a realisation of the type-A$_\ell$ higher spin algebra in the Weyl algebra.

To construct such a realisation, we will first need to identify the centraliser 
$\Centraliser_{\Weyl_{2(d+2)}}\big(\mathfrak{sp}(2,\R)\big)$
of $\mathfrak{sp}(2,\R)$ in the Weyl algebra $\Weyl_{2(d+2)}$,
which is the space of elements annihilated by
\begin{equation}
    [\tau_{ij},-]_\star = Y^\AlgInd{A}_{(i}\,\epsilon_{j)k}\,
    \tfrac{\partial}{\partial Y^\AlgInd{A}_k}\,,
\end{equation}
or equivalently by the three operators
\begin{equation}
    Y^\AlgInd{A}_+\,
    \tfrac{\partial}{\partial Y^\AlgInd{A}_+}
    - Y^\AlgInd{A}_-\,
    \tfrac{\partial}{\partial Y^\AlgInd{A}_-}\,,
    \qquad\qquad 
    Y^\AlgInd{A}_+\,
    \tfrac{\partial}{\partial Y^\AlgInd{A}_-}\,,
    \qquad\qquad 
    Y^\AlgInd{A}_-\,
    \tfrac{\partial}{\partial Y^\AlgInd{A}_+}\,.
\end{equation}
The first operator imposes that elements in the centraliser 
of $\mathfrak{sp}(2,\R)$ be of the same degree in $Y^\AlgInd{A}_+$
and $Y^\AlgInd{A}_-$, while the other two operators
both impose that the coefficients of monomials
in $Y_\pm^\AlgInd{A}$ have the symmetry of a rectangular
Young diagram in the $\mathfrak{so}(2,d)$ indices. In other words,
\begin{equation}
    f(Y) \in \Centraliser_{\Weyl_{2(d+2)}}\big(\mathfrak{sp}(2,\R)\big)
    \qquad \Longleftrightarrow \qquad 
    f(Y) = \sum_{s=1}^\infty f_{\AlgInd{A}(s-1),\AlgInd{B}(s-1)}\,
    Y^{\AlgInd{A}(s-1)}_+\,Y^{\AlgInd{B}(s-1)}_-\,,
\end{equation}
with $f_{\AlgInd{A}(s-1),\AlgInd{AB}(s-2)} = 0$.
Note, however, that these tensors are still traceful,
and hence are \emph{reducible} representations of $\mathfrak{so}(2,d)$.
Decomposing them into irreducible representations,
one would find all possible finite-dimensional irreps
of $\mathfrak{so}(2,d)$ labelled by Young diagrams of the form
\begin{equation}
    \Yboxdim{11pt}
    \gyoung(_7<s-1>,_4<s-t>)
    \qquad\text{with}\qquad
    s \geq 1\,, \quad t \in 2\,\mathbb N+1\,,
\end{equation}
i.e., all Young diagrams with two rows whose lengths differ
by an even number of boxes. Compared to the universal
enveloping algebra construction reviewed previously, we found ourselves with the same content as we do
after modding out $\Enveloping\big(\mathfrak{so}(2,d)\big)$ by
the ideal generated by $V_\AlgInd{ABCD}$. We also
face the same multiplicity problem: recall that we need
to fix the center of the universal enveloping algebra
in order to obtain a multiplicity-free spectrum. Here, the source of multiplicities is not only the center
of the universal enveloping algebra of $\mathfrak{so}(2,d)$,
but also that of $\mathfrak{sp}(2,\R)$ which is, by definition,
also contained in the centraliser of $\mathfrak{sp}(2,\R)$
in $\Weyl_{2(d+2)}$. Fortunately, both problems
can be solved at once, thanks to the fact that the quadratic
Casimir operators of $\mathfrak{so}(2,d)$ and $\mathfrak{sp}(2,\R)$ are related
via
\begin{equation}
    \Casimir_2[\mathfrak{so}(2,d)] + \Casimir_2[\mathfrak{sp}(2,\R)]
    = -\tfrac14\,(d-2)(d+2)\,,
\end{equation}
and similarly for higher order Casimir operators
(see e.g. \cite{Klink1988, Leung1993, Leung1994, Itoh2003} and \cite[Sec. 9]{Basile:2020gqi} for more details).
Since $\mathfrak{sp}(2,\R)$ only has one independent Casimir operator, it is sufficient to fix its value to also fix the values
of all Casimir operators of $\mathfrak{so}(2,d)$. In particular,
imposing
\begin{equation}
    \Casimir_2[\mathfrak{sp}(2,\R)] = -(\ell-1)(\ell+1)\,,
\end{equation}
sets the quadratic Casimir operator of $\mathfrak{so}(2,d)$ to
\begin{equation}
    \Casimir_2[\mathfrak{so}(2,d)] = -\tfrac14\,(d-2\ell)(d+2\ell)\,,
\end{equation}
as it should in the type-A$_\ell$ algebra $\hs_\ell$.
Finally, notice that the diagrams of shape $(s-1,s-t)$
with $t=2k+1$ and $k \geq 0$ appear as the $k$th trace
of rectangular diagrams, and that these traces are proportional
to $k$ times the $\mathfrak{sp}(2,\R)$ generators. As a consequence, one can recover the partially massless higher spin algebra
as the quotient\footnote{Note that
$\tau_{(i_1j_1} \dots \tau_{i_\ell j_\ell)}$ generate
the annihilator of the finite-dimensional
$\mathfrak{sp}(2,\R)$-irrep of highest weight $\ell-1$,
which is a reflection of the fact that the order-$\ell$
singleton is Howe dual to this $\ell$-dimensional
irrep of $\mathfrak{sp}(2,\R)$ \cite{Alkalaev:2014nsa, Joung:2015jza}.}
\begin{equation}
    \hs_\ell \cong \Centraliser_{\Weyl_{2(d+2)}}\big(\mathfrak{sp}(2,\R)\big)\Big/
    \big\langle \tau_{(i_1 j_1} \dots \tau_{i_\ell j_\ell)} \oplus
    \Casimir_2[\mathfrak{sp}(2,\R)] - \tfrac12\,(\ell-1)(\ell+1)\1\big\rangle\,,
\end{equation}
as modding out the ideal generated by the elements 
$\tau_{(i_1j_1} \dots \tau_{i_\ell j_\ell)}$ guarantees
that only $\mathfrak{so}(2,d)$ Young diagrams corresponding to
partially massless fields of depth $t=1,3,\dots,2\ell-1$
remain. See e.g. \cite{Vasiliev:2003ev, Bekaert:2004qos, Bekaert:2008sa, Didenko:2014dwa, Joung:2015jza, Sharapov:2018kjz}
for more details on the construction of higher spin algebras
from the perspective of Howe duality.

\subsection{Four-dimensional Specificities}
\paragraph{Dual pairs and the Weyl--Clifford algebra.}
Consider a set of bosonic $(Y^A)$ and fermionic $(\phi^A_i)$
oscillators, where the capital indices $A,B,\dots$ take $2n$ values
and the lower case indices $i,j, \dots$ take $2p$ values.
These oscillators are subject to the commutation
and anticommutation relations
\begin{subequations}
    \begin{align}
        \big[\hat Y^A, \hat Y^B\big]
        & = 2\,C^{AB}\,\1\,, \label{eq:Y} \\
        \big\{\hat \phi^A_i, \hat \phi^B_j\big\}
        & = 2\,C^{AB}\epsilon_{ij}\1\,,
    \end{align}
\end{subequations}
where $C^{AB}=-C^{BA}$ and $\epsilon_{ij}=-\epsilon_{ji}$ are 
two antisymmetric, non-degenerate matrices, with inverses given by
\begin{equation}
    C^{AC}\,C_{BC} = \delta^A_B\,,
    \qquad
    \epsilon^{ik}\,\epsilon_{jk} = \delta^i_j\,.
\end{equation}
We can therefore use these matrices to raise and lower indices,
which we will do using the convention 
\begin{equation}
    C^{AB}\,X_B = X^A\,,
    \qquad
    X^A\,C_{AB} = X_B\,,
\end{equation}
and similar convention for $\epsilon_{ij}$. The associative
algebra generated by these oscillators modulo the above
anti/commutation relations forms the Weyl--Clifford algebra
$\WeylClifford_{2n|4np}$, which is simply the tensor product
of the Weyl algebra generated by the bosonic oscillators,
and the Clifford algebra generated by the fermionic ones.

The elements quadratic in these oscillators
(modulo the previous anti/commutation relations),
\begin{equation}
    K^{AB} := \tfrac14\big\{\hat Y^A, \hat Y^B\big\}\,,
    \qquad 
    M^{AB}_{ij} := \tfrac14\,\big[\hat \phi^A_i, \hat \phi^B_j\big]\,,
    \qquad 
    Q^{A|B}_i := \tfrac12\,\hat Y^A\,\hat \phi^B_i\,,
\end{equation}
form a subalgebra isomorphic to $\mathfrak{osp}(4np|2n,\mathbb R)$,
whose bosonic subalgebra $\mathfrak{o}(4np) \oplus \mathfrak{sp}(2n,\mathbb R)$
is generated by $M^{AB}_{ij}$ and $K^{AB}$,
and the odd/fermionic generators --- the supercharges
--- correspond to $Q^{A|B}_i$. Their anti/commutation
relations read
\begin{subequations}
    \begin{align}
        \big[K^{AB}, K^{CD}\big]
        & = C^{BC}\,K^{AD} + C^{AC}\,K^{BD}
        + C^{BD}\,K^{AC} + C^{AD}\,K^{BC}\,,\\
        \big[M^{AB}_{ij}, M^{CD}_{kl}\big]
        & = C^{BC}\epsilon_{jk}\,M^{AD}_{il}
        - C^{AC}\epsilon_{ik}\,M^{BD}_{jl}
        - C^{BD}\epsilon_{jl}\,M^{AC}_{ik}
        + C^{AD}\epsilon_{il}\,M^{BC}_{jk}\,,\\
        \big[K^{AB}, Q^{C|D}_i\big]
        & = C^{BC}\,Q^{A|D}_i + C^{AC}\,Q^{B|D}_i\,,\\
        \big[M^{AB}_{ij}, Q^{C|D}_k\big]
        & = C^{BD}\epsilon_{jk}\,Q^{C|A}_i
        -C^{AD}\epsilon_{ik}\,Q^{C|B}_j\,,\\
        \big\{Q^{A|C}_i, Q^{B|D_j}\big\}
        & = C^{AB}\,M^{CD}_{ij}
        + C^{CD}\epsilon_{ij}\,K^{AB}\,.
    \end{align}
\end{subequations}
Note that the orthogonal algebra is presented
in a slightly unconventional basis here: one should
think of the pair of indices $(A,i)$ on the generators
$M^{AB}_{ij}$ and $Q^{B|A}_i$ as a single index
for the fundamental representation of $\mathfrak{o}(4np)$.
This is in accordance with the fact that only the first 
capital index of the fermionic generators $Q^{A|B}_i$
(the index `$A$' here) is rotated by the $\mathfrak{sp}(2n,\mathbb R)$
generators $K^{AB}$, whereas the second capital index
is rotated, along with the lower case index (the indices
`$B$' and `$i$' here) are rotated together by the $\mathfrak{o}(4np)$
generators $M^{AB}_{ij}$.

This unusual structure of indices for the $\mathfrak{o}(4np)$
generators, which stems from the choice of indices
carried out by the fermionic oscillators $\phi^A_i$,
is motivated by the fact that we are interested in
singling out the pair of subalgebras
\begin{equation}
    \mathfrak{sp}(2n, \mathbb R) \oplus \mathfrak{sp}(2p, \mathbb R) \subset \mathfrak{o}(4np)\,,
\end{equation}
generated by
\begin{equation}
    J^{AB} := \epsilon^{ij}\,M^{AB}_{ij}\,,
    \qquad \text{and} \qquad 
    \tau_{ij} := C_{AB}\,M^{AB}_{ij}\,,
\end{equation}
i.e. the generators obtained by contracting
those of $\mathfrak{o}(4np)$ with the invariant tensors
$\epsilon^{ij}$ of $\mathfrak{sp}(2p,\mathbb R)$, and
$C_{AB}$ of $\mathfrak{sp}(2n,\mathbb R)$, respectively.
As is clear from the index structure of these generators,
these two subalgebras commute with one another, 
i.e. they are contained in each other's centraliser
in $\mathfrak{o}(4np)$, and in fact they are exactly
their respective centralisers.

\paragraph{An interlude on Howe duality.}
Such a pair of subalgebras are usually called `dual pairs'
and have particularly interesting applications in representation
theory and physics. The most famous examples come from dual
pairs in a symplectic group $Sp(2N,\mathbb R)$,
which are the central objects of study of Howe duality 
\cite{Howe1989i, Howe1989ii}. In this case, one can show
that the oscillator representation of $Sp(2N,\mathbb R)$,
i.e. the Fock space generated by $N$ pairs of bosonic
creation-annihilation operators admit a decomposition
into the direct sum of the tensor product of a representation
of each group of the dual pair.

Another variation on the same theme consists of
considering dual pairs in an \emph{orthogonal} group,
say $O(2N)$.
This is precisely the case we are presented
with above, with the pair $\big(\mathfrak{sp}(2n,\R), \mathfrak{sp}(2p, \R)\big)
\subset \mathfrak{o}(4np)$. For such dual pairs, the natural
representation of the orthogonal group is the Fock space
generated by \emph{fermionic} pairs of creation-annihilation
operators. Indeed, bilinears in these operators define
a representation of the orthogonal group (or the double
cover thereof) on the fermionic Fock space, which can then
be decomposed into irreducible representations of
the dual pair of interest.
See e.g. \cite{Rowe:2011zz, Rowe:2012ym} for more details 
on this `skew-Howe' duality.

Since we have both bosonic and fermionic oscillators
at hand, we can consider dual pairs in the orthosymplectic
group \cite{Cheng2000, Cheng2010, Cheng2012}. In our case,
the relevant pair is composed of $\mathfrak{sp}(2n,\R)$, generated by
\begin{align}
   T^{AB} := K^{AB} - \epsilon^{ij}\,M^{AB}_{ij}
   = \tfrac14\,\big\{\hat Y^A, \hat Y^B\big\}
   - \tfrac14\,\big[\hat \phi^A_i, \hat \phi^{Bi}\big]\,,
\end{align}
and satisfying the commutation relations
\begin{subequations}
    \begin{align}
        [T^{AB}, \hat Y^C] &= C^{AC}\,\hat Y^B + C^{BC}\,\hat Y^A\,,\\
        [T^{AB}, \hat \phi^C_i] &= C^{AC}\,\hat \phi^B_i
        + C^{BC}\,\hat \phi^A_i\,,\\
        [T^{AB}, T^{CD}] &= C^{AC}\,T^{BD} + C^{AD}\,T^{BC}
        +C^{BD}\,T^{AC} + C^{BC}\,T^{AD}\,,
    \end{align}
\end{subequations}
and $\mathfrak{osp}(1|2p,\R)$, generated by
\begin{equation}
    Q_i = \tfrac12\,C_{AB}\,\hat Y^A\,\hat \phi^B_i\,,
    \qquad \text{and} \qquad 
    \tau_{ij} \equiv \big\{Q_i, Q_j\big\}
    = \tfrac14\,C_{AB}\,[\hat \phi^A_i, \hat \phi^B_j]\,,
\end{equation}
obeying,
\begin{subequations}
    \begin{align}
        [\tau_{ij}, Q_k] &= \epsilon_{kj}\,Q_i+\epsilon_{ki}\,Q_j\,,\\
        [\tau_{ij}, \tau_{kl}] &= \epsilon_{ki}\tau_{jl}
        + \epsilon_{kj}\tau_{il} + \epsilon_{li}\tau_{jk}
        + \epsilon_{lj}\tau_{ik}\,.
    \end{align}
\end{subequations}

\paragraph{Casimir operators.}
The quadratic Casimir operators for $\mathfrak{sp}(2n,\mathbb R)$
and $\mathfrak{osp}(1|2p,\mathbb R)$ are respectively given by,
\begin{align}
    \Casimir_2\big[\mathfrak{sp}(2n,\mathbb R)\big]
    = -\tfrac14\,T_{AB}\,T^{AB}\,,
    \qquad 
    \Casimir_2\big[\mathfrak{osp}(1|2p,\mathbb R)\big]
    = -\tfrac12\,Q_i Q^i - \tfrac14\,\tau_{ij} \tau^{ij}\,,
\end{align}
and a direct computation shows that, in the previously
described oscillator realisation, these Casimir operators
are related to one another via
\begin{equation}
    \Casimir_2\big[\mathfrak{sp}(2n,\mathbb R)\big]
    = \tfrac{n}8\,(2p-1)(2p+2n+1)
    -\Casimir_2\big[\mathfrak{osp}(1|2p,\mathbb R)\big]\,.
\end{equation}
In particular, for $n=2$ and $p=\ell-1$,
one finds
\begin{equation}
    \Casimir_2\big[\mathfrak{sp}(4,\mathbb R)\big]
    + \Casimir_2\big[\mathfrak{osp}(1|2(\ell-1),\mathbb R)\big]
    = -\tfrac14\,(3-2\ell)(3+2\ell)\,,
\end{equation}
this last number being the value of the quadratic
Casimir operator of $\mathfrak{so}(2,3) \cong \mathfrak{sp}(4, \mathbb R)$
on the module of the order-$\ell$ scalar singleton.
This is a first hint that one may recover the type-A$_\ell$
higher spin algebra as the centraliser of $\mathfrak{osp}(1|2(\ell-1),\R)$
in the Weyl--Clifford algebra, modulo $\mathfrak{osp}(1|2(\ell-1),\R)$
generators, as we shall prove in the next paragraphs.

\paragraph{Partially massless higher spin algebra.}
In order to identify the type-A$_\ell$ higher spin algebra, 
let us first give an equivalent presentation of the Weyl--Clifford algebra in terms of symbols of the previous oscillators,
that we will denote by $Y^A$ and $\phi^A_i$ and which are
commuting and anticommuting, respectively. Their product
is the graded version of the previously discussed
Moyal--Weyl product\footnote{Note that, for a homogeneous
element $f \in \WeylClifford_{2n|4np}$ of degree $|f|$,
the left and right derivatives with respect to $Y^A$
and $\phi^A_i$ are related by
$f\,\tfrac{\overleftarrow{\partial}}{\partial Y^A}
= \tfrac{\partial}{\partial Y^A} f$ and 
$f\,\tfrac{\overleftarrow{\partial}}{\partial \phi^A_i}
= -(-1)^{|f|}\,\tfrac{\partial}{\partial \phi^A_i} f$.}
\begin{equation}
    f \star g = f\,\exp
    \big(\tfrac{\overleftarrow{\partial}}{\partial Y^A}\,
    C^{AB}\,\tfrac{\overrightarrow{\partial}}{\partial Y^B}
    +\tfrac{\overleftarrow{\partial}}{\partial \phi^A_i}\,
    C^{AB}\,\epsilon_{ij}\,
    \tfrac{\overrightarrow{\partial}}{\partial \phi^B_j}\big)\,g\,,
\end{equation}
where $f$ and $g$ are arbitrary polynomials in $Y^A$ and $\phi^A_i$.
The symbols of the $\mathfrak{sp}(2n,\R)$ and $\mathfrak{osp}(1|2p,\R)$ generators
are simply
\begin{equation}
    T^{AB} = \tfrac12\,Y^A Y^B
    - \tfrac12\,\epsilon^{ij}\,\phi^A_i \phi^B_j\,,
    \qquad 
    Q_i = \tfrac12\,C_{AB}\,Y^A \phi^B_i\,,
    \qquad 
    \tau_{ij} = \frac12\,C_{AB}\,\phi^A_i \phi^B_j\,,
\end{equation}
respectively.

Now let us characterise the centraliser of $\mathfrak{osp}(1|2p,\R)$
in $\WeylClifford_{2n|4np}$, that is the space of elements
annihilated by
\begin{equation}
    [Q_i,-]_\star = \phi^A_i\,\tfrac{\partial}{\partial Y^A}
    - \epsilon_{ij}\,Y^A\,\tfrac{\partial}{\partial \phi^A_j}\,,
\end{equation}
where $[-,-]_\star$ should be understood as the graded commutator
(i.e. $[f,g]_\star=f \star g - (-1)^{|f||g|}\,g \star f$
for homogeneous elements of the Weyl--Clifford algebra $f$ and $g$).
This condition is solved by considering any function
of the symbol of the $\mathfrak{sp}(2n,\R)$ generators $T^{AB}$,
\begin{equation}
    f(Y^A,\phi^B_i) \in \Centraliser_{\WeylClifford_{2n|4np}}\big(\mathfrak{osp}(1|2p,\R)\big)
    \qquad \Leftrightarrow \qquad 
    f(Y^A,\phi^B_i) = f(T^{AB})\,,
\end{equation}
since the symbols $T^{AB}$ are characteristics of
the first order partial differential equations
$[Q_i,f]_\star=0$. Due to the fact that the $\mathfrak{sp}(2n,\R)$
algebra contain a piece quadratic in the anticommuting variables
$\phi^A_i$, the only possible diagram that can appear
when decomposing the centraliser of $\mathfrak{osp}(1|2p,\R)$
are those whose second row (and by extension, all rows 
except the first one) are of length smaller than $2p$.
Indeed, upon splitting the $\mathfrak{sp}(2p,\R)$ indices as
$i=(+\alpha,-\alpha)$ with $\alpha=1,\dots,p$,
we have
\begin{equation}
    \tfrac12\,\epsilon^{ij}\,\phi^A_i\,\phi^B_j
    = \sum_{\alpha=1}^p \varphi^{AB}_\alpha\,,
    \qquad\quad
    \varphi^{AB}_\alpha
    := \phi^{(A}_{+\alpha}\,\phi^{B)}_{-\alpha}\,,
\end{equation}
where
\begin{equation}
    \varphi_\alpha^{(AB}\,\varphi^{CD)}_\alpha = 0\,,
\end{equation}
by virtue of the fact that $\phi^A_i$ are anticommuting.
Note also that since the `building blocks' of the centraliser
of $\mathfrak{osp}(1|2p,\R)$ are rank-$2$ symmetric tensors of $\mathfrak{sp}(2n,\R)$, all diagrams appearing will have an even number of boxes, and in particular, each row will be of even length. This means
that, for $n=2$, diagrams appearing in the centraliser
of $\mathfrak{osp}(1|2p,\R)$ will be of the form
\begin{equation}\label{eq:spectrum}
    \Yboxdim{11pt}
    \gyoung(_7<2s-t-1>,_4<t-1>)\,,
\end{equation}
with $s\geq1$ and $t=1,3,\dots,2p+1$, so that upon setting
$p=\ell-1$ as before, we recover exactly the spectrum
of diagrams expected to appear in the higher spin algebra
$\hs_\ell$ (in the $\mathfrak{sp}(4,\R)$ basis). However, these diagrams
are not traceless in $\mathfrak{sp}(4,\R)$ sense a priori: consider
for instance the product of two $\mathfrak{sp}(4,\R)$ generators,
which can be projected onto a totally symmetric part,
\begin{equation}
    T^{ABCD} := T^{(AB}\,T^{CD)}
    \qquad \longleftrightarrow \qquad 
    \gyoung(;;;;)\,,
\end{equation}
and a piece with the symmetry of a `window-shaped' diagram,
\begin{equation}
    T^{AB,CD} := T^{AB}\,T^{CD} - T^{A(C}\,T^{D)B}
    \qquad \longleftrightarrow \qquad 
    \gyoung(;;,;;)\,.
\end{equation}
While the first one, the totally symmetric part,
is trivially traceless, the second one is not since
\begin{equation}
    C_{BC}\,T^{AB,CD}\
    \propto\ 2\,\epsilon^{ij}\,\phi^{[A}_i Y^{D]}\,Q_j
    - \epsilon^{ij}\epsilon^{kl}\,\phi^A_i \phi^D_k\,\tau_{jl}\,,
\end{equation}
does not vanish identically, but is proportional to 
the $\mathfrak{osp}(1|2(\ell-1),\R)$ generators. This is in fact a general
feature, namely all traces are proportional to these generators.
Indeed, taking a trace in the $\mathfrak{sp}(4,\R)$ sense
means contracting the capital latin indices $A,B,\dots$
with the invariant tensor $C_{AB}$, which thereby produces
the generators $Q_i$ and $\tau_{ij}$. Consequently, we can remove
traces by modding out the centraliser of $\mathfrak{osp}(1|2(\ell-1),\R)$
by the ideal generated by $Q_i$ and $\tau_{ij}$, and thereby
obtain the type-A$_\ell$ higher spin algebra in four dimensions
as the quotient\footnote{Note that this definition
also works for $\ell=1$, even though this case may seem degenerate
at first glance. Indeed, in this case the Howe dual algebra
becomes trivial, which is simply a consequence of the fact
the relevant Howe dual \emph{group} is the finite group
$\mathbb{Z}_2$. This group acts on the Weyl algebra 
by reflections $Y^A \to - Y^A$, so that its centraliser
is nothing but the \emph{even} subalgebra, the subalgebra
of polynomials in an even number of $Y^A$'s.}
\begin{equation}
    \hs_\ell \cong \Centraliser_{\WeylClifford_{4|8(\ell-1)}}
    \big(\mathfrak{osp}(1|2(\ell-1),\R)\big)\Big/
    \big\langle Q_i \oplus \tau_{ij}\big\rangle\,.
\end{equation}
The main difference compared to the realisation reviewed
in the previous section is that here, the `order' $\ell$
of the theory is no longer controlled by choosing different
ideal to mod out from the centraliser of the Howe dual algebra,
but by the choice of the Howe dual algebra itself. This allows
us to slightly simplify the identification of the type-A$_\ell$
higher spin algebra in four dimensions with respect to
the arbitrary dimension construction. 

Given that $\hs_\ell$ is the symmetry algebra
of the order-$\ell$ scalar singleton, and having found
a realisation of it within the Weyl–Clifford algebra,
it is natural to seek a realisation of the order-$\ell$ singleton
in the Fock space generated by the bosonic and fermionic oscillators 
used above, which we will do in the next section.

\section{Higher order singleton module}
\label{sec:singleton}
The Weyl--Clifford algebra generated by the oscillators
$\hat Y^A$ and $\hat \phi^A_i$ introduced
in Section \ref{sec:HSA} naturally acts on the Fock space
generated by $n$ pairs of bosonic creation-annihilation
operators,
\begin{equation}
    [\bosOsc_a, \bar\bosOsc^b] = \delta^b_a\,\1\,,
    \qquad 
    \bar\bosOsc^a := (\bosOsc_a)^\dagger\,,
    \qquad
    a,b,\dots=1,\dots,n\,,
\end{equation}
and $2np$ fermionic ones,
\begin{equation}
    \{\ferOsc^i_a, \bar\ferOsc_j^b\}
    = \delta^i_j\,\delta^b_a\,\1\,,
    \qquad
    \bar \ferOsc^a_i := (\ferOsc_a^i)^\dagger\,,
    \qquad 
    i,j,\dots=1,\dots,2p\,.
\end{equation}
In fact, the Weyl--Clifford algebra is the algebra
of endomorphisms of this Fock space.
Bilinears in these creation-annihilation operators
form a Lie subalgebra isomorphic to $\mathfrak{osp}(2n|4np,\R)$,
which contain the dual pair $\mathfrak{sp}(2n,\R) \oplus \mathfrak{osp}(1|2p,\R)$
discussed previously. Introducing the notation,
\begin{equation}
    v \cdot w := \epsilon^{ij}\,v_i\,w_j
    = \epsilon_{ij}\,v^i\,w^j\,,
\end{equation}
for the contraction of the $\mathfrak{sp}(2p,\R)$ indices,
the generators of $\mathfrak{sp}(2n,\R)$ are given by
\begin{subequations}
    \begin{equation}
        T^{ab} := \bar\bosOsc^a\,\bar\bosOsc^b
        - \bar\ferOsc^a \cdot \bar\ferOsc^b\,,
        \qquad
        \qquad 
        T_{ab} := \bosOsc_a\,\bosOsc_b
        - \ferOsc_a \cdot \ferOsc_b\,,
    \end{equation}
    \begin{equation}
        T^a{}_b := \bar\bosOsc^a\,\bosOsc_b
        + \bar\ferOsc^a\cdot\ferOsc_b
        + \tfrac{1-2p}2\,\delta^a_b\,\1\,,
    \end{equation}
\end{subequations}
while the generators of $\mathfrak{osp}(1|2p,\R)$ read
\begin{equation}
    Q_i := \tfrac12\,(\bar\ferOsc^a_i\,\bosOsc_a
    - \bar\bosOsc^a\,\epsilon_{ij}\,\ferOsc_a^j)\,,
    \qquad 
    \tau_{ij} := \epsilon_{k(i}\,\bar\ferOsc_{j)}^a\,\ferOsc^k_a\,.
\end{equation}

Now let us isolate the $\mathfrak{sp}(2n,\R)$ representation
dual to the trivial irrep of $\mathfrak{osp}(1|2p,\R)$. Doing so
amounts to finding states in the Fock space
which are annihilated by the action of the $\mathfrak{osp}(1|2p,\R)$
supercharges, i.e. 
\begin{equation}
    Q_i\,f(\bar\bosOsc,\bar\ferOsc) \lvert 0 \rangle = 0\,,
\end{equation}
which is solved by 
\begin{equation}
    f(\bar\bosOsc,\bar\ferOsc) = f(T^{ab})\,,
\end{equation}
that is, any function of the $\mathfrak{sp}(2n,\R)$ raising operators
$T^{ab}$. Since the vacuum $\lvert 0 \rangle$ of the Fock space
is $\mathfrak{osp}(1|2p,\R)$-invariant, it defines a lowest weight vector
for the dual $\mathfrak{sp}(2n,\R)$-module, with weight 
\begin{equation}
    (\underbrace{\tfrac{1-2p}2,\dots,
    \tfrac{1-2p}2}_{n\,\text{times}})\,,
\end{equation}
with respect to the Cartan subalgebra spanned by
the generators $T^a{}_a$ (no summation implied).
The subspace of homogeneous polynomials of degree $k$
in $T^{ab}$ is preserved by the action of the $\mathfrak{u}(n)$ subalgebra
generated by $T^a{}_b$ (since the latter preserve
the number of creation/annihilation operators). The decomposition
of these subspaces into irreducible representations of $\mathfrak{u}(n)$
consists of all Young diagram with $2k$ boxes, whose rows
are all of even length and such that the second row is of length
at most $2p$. In particular, for $n=2$, the lowest weight 
$\mathfrak{sp}(4,\R)$-module dual to the trivial $\mathfrak{osp}(1|2p,\R)$-representation
admits the decomposition
\begin{equation}
    \mathcal{D}_{\mathfrak{sp}(4,\R)}\big(\tfrac{1-2p}2,\tfrac{1-2p}2\big)
    \cong \bigoplus_{s=0}^\infty \bigoplus_{k=0}^p
    \big[\tfrac{1-2p}2+2s+2k,\tfrac{1-2p}2+2k\big]_{\mathfrak{u}(2)}\,,
\end{equation}
under the maximal compact subalgebra $\mathfrak{u}(2) \subset \mathfrak{sp}(4,\R)$.
Taking into account the isomorphism
\begin{equation}
    [\lambda_1,\lambda_2]_{\mathfrak{u}(2)}
    \cong \big[\tfrac{\lambda_1+\lambda_2}2,
    \tfrac{\lambda_1-\lambda_2}2\big]_{\mathfrak{so}(2) \oplus \mathfrak{so}(3)}\,,
\end{equation}
between finite-dimensional irreps of $\mathfrak{u}(2)$
and $\mathfrak{so}(2) \oplus \mathfrak{so}(3)$ and setting $p=\ell-1$,
this decomposition matches the one of the order-$\ell$
singleton module
\begin{equation}
    \mathcal{D}_{\mathfrak{so}(2,3)}\big(\tfrac{3-2\ell}2,0\big)
    \cong \bigoplus_{s=0}^\infty
    \bigoplus_{t=1,3,\dots}^{2\ell-1}
    \big[\tfrac{3-2\ell}2+t-1+s,s\big]_{\mathfrak{so}(2) \oplus \mathfrak{so}(3)}
\end{equation}
in three dimensions (see e.g. \cite[Sec. 3.4.2]{Iazeolla:2008ix}).

\paragraph{A word about non-unitarity.}
Let us conclude this section by commenting
on the non-unitarity of these modules. Recall that 
higher-order singletons are lowest weight irreps
of $\mathfrak{so}(2,d)$, which can therefore be described 
by its lowest weight vector $\ket\phi$, obeying
\begin{equation}
    (D-\Delta_\phi)\ket{\phi} = 0\,,
    \qquad J_{ab}\ket\phi = 0\,,
    \qquad K_a\ket{\phi} = 0\,,
\end{equation}
where $D$, $J_{ab}$ and $K_a$ are the dilation,
Lorentz, and special conformal transformation generators.
All states of the modules are obtained
by repeated application of the translation generators
$P_a$ on $\ket\phi$. Using the relations
\begin{equation}
    [K_a, P_b] = \eta_{ab}\,D - M_{ab}\,,
    \qquad 
    [D,P_a] = P_a\,,
    \qquad 
    [M_{ab},P_c] = 2\,\eta_{c[b}\,P_{a]}\,,
\end{equation}
one finds
\begin{equation}
    K_a\,P^2\,\ket{\phi}
    = 2\,\big(\Delta_\phi-\tfrac{d-2}2\big)\,
    P_a\ket{\phi}\,,
\end{equation}
and with $P_a^\dagger = K_a$, this implies
\begin{equation}
    \|P^2\,\ket{\phi}\|^2 = 2d\,\Delta_\phi\,
    \big(\Delta_\phi-\tfrac{d-2}2\big)\,
    \langle\phi\!\mid\!\phi\rangle\,.
\end{equation}
The above identities tells us that $P^2\,\ket\phi$
is singular and null for $\Delta_\phi=\tfrac{d-2}2$,
while for $\Delta_\phi<\tfrac{d-2}2$ it is not singular
but acquires a negative norm.
For the order-$\ell$ singleton,
$\Delta_\phi=\tfrac{d-2\ell}2$ and hence the presence 
of $P^2\ket\phi$ is one of the first indications
that the module is non-unitary.

Now coming back to our construction,
it may be surprising that such a non-unitary module 
can be realised in a Fock space, which is usually 
itself a unitary module (for the Heisenberg algebra,
or its supersymmetric version relevant here).
A first consistency check is that this negative
norm state $P^2\ket{\phi}$ is indeed present,
since we recover the correct $\mathfrak{u}(2)$ decomposition.
More importantly, the Hermitian conjugation
\emph{does not preserve} the $\mathfrak{osp}(1|2p,\R)$
generators in this realisation, which is why
we have a non-unitary module in a Fock space.

\paragraph{Higher order spinor singleton?}
Note that one could look for other representation
of $\mathfrak{osp}(1|2p,\R)$ than the trivial one. For instance,
the `next-to-simplest' representation is of dimension
$2p+1 \equiv 2\ell-1$ and splits into a direct sum
of $\mathfrak{sp}(2p,\R)$ irreps, the trivial and the vector
(or fundamental) one. It can be realised
in the Fock space considered here as the subspace
with basis
\begin{equation}
    \bar\bosOsc^a\ket0 
    \qquad\text{and}\qquad
    \bar\ferOsc^a_i\ket0\,,
\end{equation}
which are indeed, for $\mathfrak{osp}(1|2p,\R)$, 
a scalar and a vector respectively.
This subspace is preserved by the action
of $\mathfrak{osp}(1|2p,\R)$ since
\begin{equation}
    Q_i\,\bar\bosOsc^a\ket0
    = \tfrac12\,\bar\ferOsc^a_i\ket0\,,
    \qquad 
    Q_i\,\bar\ferOsc_j^a\ket0
    = -\tfrac12\,\epsilon_{ij}\,
    \bar\bosOsc^a\ket0\,,
\end{equation}
while the $\mathfrak{sp}(2p,\R)$ generators $\tau_{ij}$
merely rotate this states, as expected.
As usual in the context of Howe duality, 
this representation appears with a \emph{multiplicity},
as indicated by the fact that the above basis vectors
also carry an $\mathfrak{sp}(2n,\R)$ index. In fact, 
as in the case of the trivial representation,
any state obtained from the above basis vectors
by the action of $\mathfrak{osp}(1|2p,\R)$-invariant operators,
which are generated by the Howe dual algebra $\mathfrak{sp}(2n,\R)$,
will not change the representation. In other words, 
this finite-dimensional representation of $\mathfrak{osp}(1|2p,\R)$
appears with \emph{infinite multiplicity} in the Fock space,
but this feature is merely the reflection of the fact
that it is Howe dual to a lowest weight module
of $\mathfrak{sp}(2n,\R)$, which is infinite-dimensional.

The lowest weight $\mathfrak{sp}(2n,\R)$-module in question
is induced by the lowest $u(n)$-irrep spanned by
the state $\bar\bosOsc^a\ket0$ and $\bar\ferOsc^a_i\ket0$,
and generated by the action of the raising operators $T^{ab}$.
The lowest weight reads
\begin{equation}
    \big(\tfrac{3-2p}2, \tfrac{1-2p}2, \dots, \tfrac{1-2p}2\big)\,,
\end{equation}
which, in the case of $\mathfrak{sp}(4,\R) \cong \mathfrak{so}(2,3)$,
corresponds to the lowest weight 
\begin{equation}
    \big[\tfrac{3-2p}2, \tfrac{1-2p}2\big]_{\mathfrak{u}(2)}
    = \big[2-\ell,\tfrac12\big]_{\mathfrak{so}(2) \oplus \mathfrak{so}(3)}\,,
\end{equation}
whose components are respectively the conformal weight
and spin of the spinor singleton of order $\ell$
(i.e. a free spinor $\psi$ subject to the higher order
Dirac equation ${\partial}^{2\ell-1}\psi\approx0$
as recalled in the Introduction). In other words,
we find that the higher order spinor singleton
\begin{equation}
    \mathcal{D}_{\mathfrak{sp}(4,\R)}(\tfrac{3-2p}2,\tfrac{1-2p}2\big)
    \cong \mathcal{D}_{\mathfrak{so}(2,3)}\big(2-\ell,\tfrac12\big)\,,
\end{equation}
is Howe dual to the finite-dimensional $\mathfrak{osp}(1|2p,\R)$
representation made out of the trivial and vector
$\mathfrak{sp}(2p,\R)$-irreps.

This therefore begs the question: can we find the type-B$_\ell$ 
higher spin algebra in our construction? To do so, 
one would need to quotient the centraliser of $\mathfrak{osp}(1|2p,\R)$
in the Weyl--Clifford algebra by a different ideal 
than the one generated by $\mathfrak{osp}(1|2p,\R)$. Indeed, 
we saw previously that the scalar singleton is Howe dual
to the \emph{trivial} representation of $\mathfrak{osp}(1|2p,\R)$,
and hence the full algebra is the annihilator
of this representation. The quotient by $\mathfrak{osp}(1|2p,\R)$
should be understood as the quotient by the annihilator
of this trivial representation---as recalled above
when we discussed the definition of the type-A$_\ell$
algebra in arbitrary dimensions. Having this framework
in mind, we should quotient the centraliser of $\mathfrak{osp}(1|2p,\R)$
by the annihilator of its $(2\ell-1)$-dimensional irrep
in order to obtain the type-B$_\ell$ higher spin algebra.
Schematically, this means modding out by higher powers
of the $\mathfrak{osp}(1|2p,\R)$ generators, which in turn amounts
to keeping some of the traces in the diagrams \eqref{eq:spectrum},
as may be expected to reproduce the spectrum
of the type-B$_\ell$ algebra. Such an analysis is however
beyond the scope of this chapter, and we leave for potential
future work.

\section{Formal partially massless higher spin gravity}
\label{sec:formalHiSGRA}
Having built an oscillator realisation of the higher spin
algebra $\hs_\ell$ in four dimensions, we will now use it
to try and construct an interacting theory
of partially massless higher spin fields.

The most common way of constructing formal higher spin gravities
is to consider a gauge connection $\omega$ of the relevant
higher spin algebra $\hs$, together with a zero-form $C$
taking value in a module of this algebra
(see e.g. \cite{Vasiliev:1988sa, Vasiliev:1990vu, Vasiliev:2003ev, Bekaert:2004qos, Boulanger:2008up, Boulanger:2008kw, Alkalaev:2014nsa, Didenko:2014dwa, Grigoriev:2018wrx, Sharapov:2019vyd}).
This data is associated with the coordinates
on a $Q$-manifold, which we denote by the same symbols,
and whose (co)homological vector field $Q$ encodes
the interactions. More precisely, one is then charged
with constructing equations of motion  
\begin{align}\label{eq:formal_EOM}
    d\omega & = \mathcal{V}(\omega,\omega)
    + \mathcal{V}(\omega,\omega,C) + \dots\,, \\
    dC & = \mathcal{U}(\omega,C) + \mathcal{U}(\omega,C,C)
    + \dots\,,
\end{align}
where $\mathcal V$ and $\mathcal U$ are the component of $Q$,
and the initial data for the deformation problem reads 
\begin{align}
    \mathcal{V}(a,b) & = a \star b\,,
    & \mathcal{U}(a,u) & = a \star u - u \star \pi(a)\,,
\end{align}
where $\pi$ is an anti-involution of the higher spin algebra.
At this point it is convenient to define
${}^{\mathbb Z_2}\hs=\hs \rtimes \mathbb{Z}_2$,
where $\mathbb{Z}_2=\{1, \pi\}$. In practice,
one adds an element $k$ such that $k^2=1$
and $k \star a \star k = \pi(a)$. 

Under some fairly general assumptions, one can show that
the problem of constructing the $A_\infty$-algebra
underlying the $Q$-manifold reduces to a much simpler problem
of deforming ${}^{\mathbb Z_2}\hs$ as an associative algebra \cite{Sharapov:2018hnl, Sharapov:2018ioy, Sharapov:2018kjz, Sharapov:2019vyd}.
Moreover, often times it is easy to see that
${}^{\mathbb Z_2}\hs$ can be deformed and even construct
such a deformation, which we call $\hsdeformed$, explicitly.
Once $\hsdeformed$ is available, there is an explicit
procedure to construct all vertices. For example, 
\begin{align}\label{eq:V3}
    \mathcal{V}(a,b,u) & = \phi_1(a,b) \star \pi(u)\,,
\end{align}
where $\phi_1$ is a (Hochschild) $2$-cocycle that determines
the first order deformation of ${}^{\mathbb Z_2}\hs$
to $\hsdeformed$:
\begin{align}
    a \circ b & = a \star b + u\,\phi_1(a,b) k
    + \mathcal{O}(u^2)\,.
\end{align}
    It has to be noted that the above form of the vertices
    is non-minimal: the equations, in general,
    `mix' different spins even at the free level.
    One therefore needs to find a suitable field redefinition
    to bring the vertex in its `minimal' form 
    wherein such mixing are absent.

The deformed algebra $\hsdeformed$ is defined from $\hs$,
the latter being usually obtained via either one
of the following constructions:
\begin{enumerate}[label=$(\alph*)$]
\item A quotient of the universal enveloping algebra
$\Enveloping\big(\mathfrak{so}(d,2)\big)$ by a two-sided ideal $\Ideal$
(in most cases called the Joseph ideal), which corresponds
to the annihilator of a given irreducible $\mathfrak{so}(2,d)$-module,
e.g. \cite{Eastwood:2002su, Vasiliev:2003ev, Joung:2015jza, Fernando:2015tiu, Gunaydin:2016bqx, Campoleoni:2021blr};
\item Using an oscillator realisation, wherein
one embeds $\mathfrak{so}(2,d)$ and its enveloping algebra
in a Weyl(--Clifford) algebra and typically obtain $\hs$
as the quotient of the centraliser of a Howe dual
algebra, as discussed above for the type-A$_\ell$ algebra,
as well as in \cite{Vasiliev:2003ev, Bekaert:2004qos,
Govil:2013uta, Didenko:2014dwa, Govil:2014uwa, Gunaydin:2016bqx}
and references therein;
\item Via the \emph{quasi-conformal} realisation,
which consists in explicitly solving the defining relations
of the (Joseph) ideal mentioned previously, see e.g. 
\cite{Gunaydin:2007vc, Fernando:2009fq, Gunaydin:2016bqx,
Sharapov:2019pdu}. 
\end{enumerate}

The first order deformation defined by the $2$-cocycle
$\phi_1$ makes its presence felt already at the free level.
Indeed, linearizing the above equations around an (A)dS$_{d+1}$
background,
\begin{equation}
    \omega_0 = e^a\,P_a + \tfrac12\,\varpi^{a,b}\,L_{ab}\,,
    \qquad 
    C_0 = 0\,,
    \qquad 
    d\omega_0 + \tfrac12\,[\omega_0,\omega_0] = 0\,,
\end{equation}
their first order in the field fluctuations
should reproduce the free field equations
for partially massless fields in the frame-like formalism
\cite{Skvortsov:2006at}, whose schematic form reads
\begin{equation}
    R[\omega_1]^{a(s-1),b(s-t)}
    = e_c \wedge e_d\, C_1^{a(s-1)c,b(s-t)d}\,,
    \qquad 
    R[\omega_1]^{a(s-1-m),b(s-t-n)} = 0\,,
\end{equation}
where $R[\omega_1]^{a(s-1-m),b(s-t-n)}
= \nabla \omega_1^{a(s-1-m),b(s-t-n)} + \dots$,
with $\omega_1$ the first order fluctuations
of a $1$-form valued in $\hs_\ell$\,.
More specifically, the components of $1$-form
$\omega_1$ take values in the finite-dimensional
representations of $\mathfrak{so}(2,d)$ labelled by the two-row
Young diagrams of the form 
\begin{equation}
    \Yboxdim{11pt}
    \gyoung(_6<s-1>,_4<s-t>)
    \qquad 
    t=1,3,\dots,2\ell-1\,, \quad s=t, t+1, \dots,
\end{equation}
which corresponds to generators of the form
\begin{equation}
    M_{\AlgInd{A}(s-1),\AlgInd{B}(s-t)}
    = \underbrace{M_\AlgInd{AB}
        \cdots M_\AlgInd{AB}}_{s-t}\,
    \underbrace{M_\AlgInd{A}{}^\AlgInd{C}\,
    M_\AlgInd{AC} \cdots M_\AlgInd{A}{}^\AlgInd{C}\,
    M_\AlgInd{AC}}_{\frac{t-1}2} + \dots\,,
\end{equation}
where the dots denote terms ensuring
that the right hand side has the symmetry
of the above Young diagram, and is traceless.
The first order fluctuation of the zero-form
takes values in a representation of $\hs_\ell$,
usually called the `twisted-adjoint representation'.%
\footnote{Although it may be more relevant
to think of it as a coadjoint module
\cite[App. B]{Skvortsov:2022syz}.}
This module of the type-A$_\ell$ algebra is defined
on the same vector space as $\hs_\ell$,
but where the latter acts via a `twisted commutator'
\begin{equation}
    \mathcal{U}(\omega_0,C_1)
    = \omega_0 \star C_1 - C_1 \star \pi(\omega_0)\,,
\end{equation}
hence the name of this representation. The zero-forms
can therefore be expanded in a basis of generators
of the (partially massless) higher spin algebra
\cite{Vasiliev:2003ev, Iazeolla:2008ix}. Typically, the Weyl tensor
$C^{a(s),b(s-t+1)}$, for the spin-$s$
and depth-$t$ partially massless field
is the component of $C_1$ along the generator
of $\hs_\ell$ which schematically reads,
\begin{equation}
    \underbrace{L_{ab} \dots L_{ab}}_{s-t+1}
    \underbrace{P_a \dots P_a}_{t-1} P^{2\ell-t-1}
    + \dots\,,
\end{equation}
where we separated generators of $\mathfrak{so}(2,d)$
into those of the Lorentz subalgebra $\mathfrak{so}(1,d)$,
denoted by $L_{ab}$, and the transvection
(or AdS-translation) generators denoted by $P_a$.

The low spin ($s=1$ and $2$) components
of these fluctuations are given by
\begin{equation}
    \omega_1 = A \cdot \1 + h^a\,P_a
    + \tfrac12\,\omega^{ab}\,L_{ab} + \dots\,,
    \qquad 
    C_1 = \tfrac12\,F^{a,b}\,\Maxwell_{a,b} + \dots\,,
\end{equation}
where, to keep this discussion fairly general,
we denoted by $\Maxwell_{a,b}$ the generator of $\hs$
along which one finds the Maxwell tensor,
independently of the higher spin algebra of interest.
In the type-A case, it would simply be $\Maxwell_{a,b}=L_{ab}$,
while in the type-A$_\ell$ case, it would be of the form
$L_{ab}P^{2(\ell-1)} + (\dots)$ instead. 
When comparing \eqref{eq:formal_EOM}
to the previous free equations of motion,
e.g. in the spin-$1$ sector,
\begin{equation}
    dA = e_a \wedge e_b\, F^{a,b} + \dots\,,
\end{equation}
we can deduce that $\mathcal{V}$ yields
\begin{equation}
    \mathcal{V}(P_a, P_b; \Maxwell_{c,d})
    = 2\,\eta_{a[c}\,\eta_{d]b}\,\1 + \dots\,,
\end{equation}
when evaluated on $P_a \otimes P_b \otimes \Maxwell_{c,d}$.
From \eqref{eq:V3}, we know that the dots
in the previous equation originate from the expression
\begin{equation}
    \phi_1(P_a,P_b) \star \Maxwell_{a,b}
    = 2\,\eta_{a[c}\,\eta_{d]b}\,\1 + \dots\,,
\end{equation}
 modulo the field-redefinition
bringing the vertex in its `minimal' form.
In other words, the product of $\phi_1(P_a, P_b)$
and the generator that corresponds to the Maxwell tensor
must contain the unit of $\hs$. Let us note
that $\phi_1(P_a, P_b)$ is also the simplest term
to probe the deformation since $\phi_1(\1,\bullet)=0$
and $\phi_1(L_{ab}, \bullet)=0$. The first condition
means that the unit is not deformed and the second one
protects Lorentz symmetry. 

Recalling that $\hs$ has an invariant trace $\tr$
(defined as the projection onto the unit),
the above condition can also be rewritten as
\begin{align}
    \tr\big(\phi_1(P_a, P_b) \star \Maxwell_{a,b}\big) \neq 0\,.
\end{align}
Since the basis of any higher spin algebra $\hs$
can be decomposed into finite-dimensional
$\mathfrak{so}(d,2)$-modules, and that the trace respects $\mathfrak{so}(d,2)$, 
different generators are orthogonal to each other.
As a result, we have to have 
\begin{align}
    \phi_1(P_a, P_b)\, & \propto\, \eta_{ab}
    + T_{ab} + \dots
    && \tr\big(T_{ab} \star \Maxwell_{a,b})
    \neq 0\,,
\end{align}
where $T_{ab}$ is a generator of $\hsdeformed$
that \emph{deforms} the commutator $[P_a, P_b]$.
In other words, the generator $T_{ab}$ of this deformation 
is a multiple of the \emph{dual} of the Maxwell tensor 
generator $\Maxwell_{a,b}$. In order to define
$\hsdeformed$, one needs to define
$[P_a,P_b]=L_{ab}+ u k\,T_{ab}$
and deform the Joseph ideal accordingly.

The Maxwell tensor (and the whole decomposition)
can be found by decomposing the twisted-adjoint action
$\{P_a,\bullet\}$ of translations on $C$.
The adjoint of the Lorentz algebra
may appear with multiplicity greater than $1$
(this happens for instance in the type-B or Type-A$_\ell$,
$\ell>1$, cases). The Maxwell equations
should have the form
\begin{align}
    \nabla F^{a,b} &= h_c F^{ac,b} \\
    \nabla F^{ab,c} & \propto h^{(a} F^{b),c}
    -\tfrac1d\,h_\times\,\big(\eta^{ab}F^{c,\times}
    - \eta^{c(a} F^{b),\times}\big) + \dots\,,
\end{align}
where the first line comes from the anticommutator
$\{P_m, \Maxwell_{ac,b}\}$, where $\Maxwell_{ab,c}$
is a traceless and hook-symmetric generator of the form
$\Maxwell_{a,b}P_c + (\dots)$,  and in the second line
from $\{P_m, \Maxwell_{a,b}\}$. Most importantly, $F^{a,b}$
must not contribute anywhere else. The second equation
means that, at the algebra level, one finds
\begin{align}
    \{P_{(a}, \Maxwell_{b),c}\} & = \Maxwell_{ab,c}\,.
\end{align}
which implies that $\{P^a ,\Maxwell_{a,b}\} = 0$, 
and hence this anticommutator must be a part
of the two-sided ideal defining $\hs$.
This is indeed the case for the Type-A algebra,
whose Joseph ideal contains $\{P^m ,L_{mb}\}=0$.
For the type-A$_\ell$ case, $\Maxwell_{a,b}$ must be
in the adjoint representation that sits inside
the subspace of monomials of order $\ell-1$
in the $\mathfrak{so}(2,d)$ generator (i.e. one degree less
than the generator $\Joseph_{\AlgInd{A}(2\ell)}$
of the Joseph ideal).

\paragraph{Probing deformation through cycles.}
Cocycles are more complicated to derive than cycles
since cocycles are defined on the whole algebra
(must be assigned some value for all possible arguments),
while cycles involve few specific elements of the algebra.
Nontrivial cocycles can be evaluated on nontrivial cycles,
the result being nonzero. We start by recalling the definition of the Hochschild chain complex,
\begin{definition}
    Let \( A \) be an associative algebra over a field, and let \( M \) be an \( A \)-bimodule. The \textbf{Hochschild chain complex} \( C_n(A, M) \) is defined by $C_n(A, M) := M \otimes A^{\otimes n}$, with boundary map \( \partial: C_n(A, M) \to C_{n-1}(A, M) \) given by
\[
\begin{aligned}
\partial(m \otimes a_1 \otimes \cdots \otimes a_n) =\ & m \lhd a_1 \otimes a_2 \otimes \cdots \otimes a_n + \sum_{i=1}^{n-1} (-1)^i m \otimes a_1 \otimes \cdots \otimes a_i a_{i+1} \otimes \cdots \otimes a_n \\
&+ (-1)^n a_n\rhd m \otimes a_1 \otimes \cdots \otimes a_{n-1}.
\end{aligned}
\]
for $a_{1}\dots a_{n} \in A$ and $m\in M$ and $\lhd,\,\,\rhd$ are the left and right action. An element \( c \in C_n(A, M) \) is called a \textbf{Hochschild cycle} if it satisfies $\partial(c) = 0$.\footnote{Now in our case we take $A=\mathfrak{hs}_{2}$ and $M=\mathfrak{hs}_{2}^{\pi}$ called twisted adjoint module. This means that the $m\lhd a=ma$ while the right multiplication is \textbf{twisted} as $a\rhd m=\pi(a)m$. In our case, $\pi$ is implemented as $\pi(P)=-P$ and $\pi(L_{a,b})=L_{a,b}$.}
\end{definition}

Now, in the type-A case, the Maxwell equation
probes the cycle $c_{(1)}\in M\otimes A^{\otimes2}$ \cite[App. B]{Sharapov:2018kjz}
\begin{equation}
    c_{(1)} = L_{ab} \otimes P^a \otimes P^b
    +\tfrac14\,(\1 \otimes L_{ab} \otimes L^{ab})
    - \tfrac14\,C_L(\1 \otimes \1 \otimes \1)\,,
    \qquad
    C_L = -\tfrac{d(d-2)}4\,,
\end{equation}
which is closed by virtue of the fact that
$\{L_{ab}, P^b\} \sim 0$
and $-\tfrac12\,L_{ab} L^{ab} \sim C_L\,\1$
due to the quotient by the Joseph ideal. Note that we have taken the module $M$ to be the algebra $A=\mathfrak{hs}$ itself.
As it turns out, one can find a counterpart of this cycle
in $\hs_\ell$, namely
\begin{equation}
    c_{(\ell)} = \Maxwell_{a,b} \otimes P^a \otimes P^b
    + \tfrac12\,\big(\1 \otimes \Maxwell_{a,b} \otimes L^{ab}
    - P^a \otimes P^b \otimes \Maxwell_{a,b}
    - P^a \otimes \Maxwell_{a,b} \otimes P^b\big)
    + \dots\,,
\end{equation}
which is closed as a consequence of the fact that
$\{\Maxwell_{a,b}, P^b\} \sim 0$, \emph{up to a term}
$\1 \otimes \Maxwell_{a,b} L^{ab}$ (hence the dots).
To verify that this is indeed a cycle,
first note that
\begin{equation}
    \partial(\Maxwell_{a,b} \otimes P^a \otimes P^b)
    = \{\Maxwell_{a,b},P^a\} \otimes P^b
    - \tfrac12 \Maxwell_{a,b} \otimes L^{ab}
    \sim - \tfrac12 \Maxwell_{a,b} \otimes L^{ab}\,,
\end{equation}
as we have previously argued that $\{\Maxwell_{a,b},P^a\}$
belongs to the defining ideal of $\hs_\ell$
(in fact, any higher spin algebra
containing a massless spin-$1$ fields
in its spectrum), and hence is modded out.
The remaining term is compensated thanks to
\begin{equation}
    \partial(\1 \otimes \Maxwell_{a,b} \otimes L^{ab})
    = \Maxwell_{a,b} \otimes L^{ab}
    - \1 \otimes \Maxwell_{a,b}\,L^{ab}
    + L^{ab} \otimes \Maxwell_{a,b}\,,
\end{equation}
which however brings in two other terms. The last one
can be cancelled using
\begin{equation}
    \partial(P^a \otimes P^b \otimes \Maxwell_{a,b}
    + P^a \otimes \Maxwell_{a,b} \otimes P^b)
    \sim L^{ab} \otimes \Maxwell_{a,b}\,,
\end{equation}
where we made use of $\{\Maxwell_{a,b},P^b\}\sim0$ again,
which leaves us with the final task of eliminating
the term proportional to $\1 \otimes \Maxwell_{a,b}\,L^{ab}$.
In the type-A example, we could take advantage
of the fact that the term $L_{ab}L^{ab}$ is proportional
to the identity in $\hs$. This is a simple consequence
of quotienting $\Enveloping\big(\mathfrak{so}(2,d)\big)$
by the Joseph ideal.\footnote{More precisely,
the scalar component of the generator $\Joseph_{AB}$,
when decomposed under the Lorentz algebra, relates $P^2$
to the quadratic Casimir operator of $\mathfrak{so}(2,d)$
which is itself proportional to the identity.
Since $\Casimir_2 = -\tfrac12\,L_{ab}L^{ab} + P^2$,
one therefore concludes that $-\tfrac12\,L_{ab}L^{ab}$
is also proportional to the identity.}
We can expect that a similar property also holds
for type-A$_\ell$ algebras, by inspecting its spectrum:
since $\Maxwell_{a,b} = L_{ab}P^{2(\ell-1)} + (\dots)$
belongs to the $\mathfrak{so}(2,d)$-irrep $(2\ell-1,1)$,
the contraction $\Maxwell_{a,b}\,L^{ab}$
belongs to $(2\ell)$,
\begin{equation}
    \Yboxdim{11pt}
    \Maxwell_{a,b} \in \gyoung(_7{2\ell-1},;)
    \qquad\Longrightarrow\qquad
    \Maxwell_{a,b}\,L^{ab} \in \gyoung(_8{2\ell})
    \subset \Ideal_\ell\,,
\end{equation}
since the latter is the only $\mathfrak{so}(2,d)$ diagram
susceptible to contain a Lorentz scalar.
In other words, $\Maxwell_{a,b}\,L^{ab}$
is related to the scalar part of the generator
$\Joseph_\AlgInd{A(2\ell)}$, whose structure
is discussed in Appendix \ref{app:A2}.
We can expect that $\Maxwell_{a,b}\,L^{ab}$
is proportional to $P^{2(\ell-1)}$, or a polynomial
in $P^2$ of degree $\ell-1$ more generally.

This is indeed the case for the type-A$_2$ algebra,
where $\Maxwell_{a,b}\,L^{ab} \sim \#\,P^2$,
as we show in Appendix \ref{app:A2}. We can therefore
use this identity and compensate the term
$\1 \otimes \Maxwell_{a,b}\,L^{ab}$
by adding
\begin{equation}
    \1 \otimes P_a \otimes P^a
    \qquad\Longrightarrow\qquad
    \partial\big(\1 \otimes P_a \otimes P^a\big)
    = -\1 \otimes P^2\,,
\end{equation}
which, when added with the proper coefficient
to $c_{(2)}$ above, defines a cycle of $\hs_2$.

\paragraph{Oscillator realisation for type-A$_2$.}
Let us compute the Maxwell generator
in our oscillator realisation. To do so,
first recall that the Lorentz generators 
are embedded in $\mathfrak{sp}(4,\R)$ as
\begin{equation}
    L^{\alpha\beta}
    = \tfrac14\,\{\hat y^\alpha, \hat y^\beta\}
    - \tfrac14\,\epsilon^{ij}\,
    [\hat \phi^\alpha_i, \hat\phi^\beta_j]\,,
    \qquad 
    L^{\alpha'\beta'}
    = \tfrac14\,\{\hat y^{\alpha'}, \hat y^{\beta'}\}
    - \tfrac14\,\epsilon^{ij}\,
    [\hat \phi^{\alpha'}_i, \hat\phi^{\beta'}_j]\,,
\end{equation}
where we split the oscillators \eqref{eq:Y}
as $\hat Y^A=(\hat y^\alpha, \hat y^{\alpha'})$
with $\alpha,\alpha' \in \{1,2\}$ indices
for two-components spinors. Similarly,
the transvection generators read
\begin{equation}
    P^{\alpha\alpha'}
    = \tfrac14\,\{\hat y^{\alpha}, \hat y^{\alpha'}\}
    - \tfrac14\,\epsilon^{ij}\,
    [\hat \phi^{\alpha}_i, \hat\phi^{\alpha'}_j]\,.
\end{equation}
Let us also introduce the notation
\begin{equation}
    q_i = \tfrac12\,\hat y_\alpha\,\hat \phi^\alpha_i\,,
    \qquad 
    \bar q_i = \tfrac12\,\hat y_{\alpha'}\,
    \hat \phi^{\alpha'}_i\,,
    \qquad
    t_{ij} = \tfrac14\,\epsilon_{\alpha\beta}\,
    \big[\hat\phi_i^\alpha, \hat\phi_j^\beta\big]\,,
    \qquad 
    \bar t_{ij} = \tfrac14\,\epsilon_{\alpha'\beta'}\,
    \big[\hat\phi_i^{\alpha'}, \hat\phi_j^{\beta'}\big]\,,
\end{equation}
in terms of which the $\mathfrak{osp}(1|2p,\R)$ generators read
\begin{equation}
    Q_i = q_i + \bar q_i\,,
    \qquad 
    \tau_{ij} = t_{ij} + \bar t_{ij}\,.
\end{equation}
Note that $q_i$ and $t_{ij}$ form an $\mathfrak{osp}(1|2p,\R)$
algebra, and $\bar q_i$ and $\bar t_{ij}$ as well.
The square of the translation generators
can be written as
\begin{equation}\label{eq:P2}
    P^2 = -\tfrac12\,P_{\alpha\alpha'}\,P^{\alpha\alpha'}
    = \tfrac{2p-1}2 + q_i\,\bar q^i
    + \tfrac12\,t_{ij}\,\bar t^{ij}\,, 
\end{equation}
where the factor $-\tfrac12$ comes from
the $\gamma$-matrices used to convert vector indices
into spinor ones.\footnote{This can also be check by
comparing $[L_{ab}, P^b]$ and $[P_a, P^2]$
with their spinor counterparts, which shows that
one should use
$L_a \to \tfrac12\,\big(\epsilon_{\alpha\beta}
L_{\alpha'\beta'}+\epsilon_{\alpha'\beta'}
L_{\alpha\beta}\big)$ and
$P_a \to \tfrac{i}{\sqrt2}\,P_{\alpha\alpha'}$.}

The Maxwell generator for $p=1 \Leftrightarrow \ell=2$ 
therefore becomes
\begin{equation}
    \Maxwell_{\alpha\beta} = L_{\alpha\beta}\,
    \big(q_i\,\bar q^i + \tfrac12\,\bar t_{ij}\,t^{ij}
    +\tfrac12\big)\,,
\end{equation}
and similarly for $\Maxwell_{\alpha'\beta'}$,
upon exchanging $L_{\alpha\beta}$ with $L_{\alpha'\beta'}$.
A direct computation leads to
\begin{equation}
    \big\{\Maxwell_{\alpha\beta},P^\beta{}_{\alpha'}\big\}
    \sim 0\,,
\end{equation}
upon using the identities
\begin{equation}\label{eq:useful}
    \hat\phi^\alpha_i\,t^2
    = -2\,\hat\phi^\alpha_j\,t_i{}^j\,,
    \qquad
    t_{ik}\,t_j{}^k = -2\,t_{ij}
    +\tfrac12\,\epsilon_{ij}\,t^2\,,
    \qquad\text{and}\qquad
    (t_{ij}+2\epsilon_{ij})\,t^2 = 6\,t_{ij}\,,
\end{equation}
with $t^2 \equiv t_{ij}\,t^{ij}$,
which can be proved thanks to Fierz identities.

Let us conclude this section by pointing a subtlety
in the computation of the ideal generators
in our oscillator realisation. Introducing $\hbar$
in the canonical anti/commutation relations as
\begin{equation}
    [\hat Y^A, \hat Y^B] = 2\hbar\,C^{AB}\,,
    \qquad 
    \{\hat\phi^A_i, \hat\phi^B_j\}
    = 2\hbar\,C^{AB}\,\epsilon_{ij}\,,
\end{equation}
the $\mathfrak{osp}(1|2p,\R)$ anti/commutation relations read
\begin{equation}
    \{Q_i, Q_j\} = \hbar\,\tau_{ij}\,,
    \qquad 
    [\tau_{ij}, Q_k] = 2\hbar\,\epsilon_{k(i}\,Q_{j)}\,,
    \qquad 
    [\tau_{ij}, \tau_{kl}]
    = \hbar\,\big(\epsilon_{kj}\,\tau_{il} + \dots\big)\,,
\end{equation}
i.e. the right hand side of any anti/commutator
is proportional to $\hbar$. Contracting the second
relation with $\epsilon^{jk}$ yields
\begin{equation}
    \hbar\,q_i = -\tfrac1{2p+1}\,[t_{ij}, Q^j]
    \qquad\Longrightarrow\qquad
    \hbar^2\,t_{ij} = -\tfrac1{2p+1}\,
    \big\{[t_{ij}, Q^j], q_j\big\}\,,
\end{equation}
which could, in the absence of $\hbar$, lead one
to conclude that $q_i$ and $t_{ij}$ can be set to zero
(and similarly for $\bar q_i$ and $\bar t_{ij}$),
when taking the quotient by $\mathfrak{osp}(1|2p,\R)$. This would
however be incorrect since it would amount to quotienting
by $\mathfrak{osp}(1|2p,\R) \oplus \mathfrak{osp}(1|2p,\R)$, one copy
generated by $q_i$ and $t_{ij}$, and another copy by
$\bar q_i$ and $\bar t_{ij}$. This direct sum
is Howe dual to $\mathfrak{sp}(2,\R) \oplus \mathfrak{sp}(2,\R)$,
and not to $\mathfrak{sp}(4,\R)$, as each copy of $\mathfrak{osp}(1|2p,\R)$
does not commute with the transvection generators
$P_{\alpha\alpha'}$. Consequently, it would be inconsistent
to mod out $q_i$ and $\bar q_i$ separately (and similarly
for $t_{ij}$ and $\bar t_{ij}$) in the centraliser
of $\mathfrak{sp}(4,\R)$ --- in the sense that the resulting algebra
would not be related to the type-A$_\ell$ higher spin
algebra. 

The introduction of $\hbar$ in computation also proves
useful when it comes to checking that the scalar generator
of the ideal also vanishes: the expression \eqref{eq:P2}
of $P^2$ can be re-written as
\begin{align}
    -\tfrac12\,P_{\alpha\alpha'}\,P^{\alpha\alpha'}
    = \hbar^2\,\tfrac{2p-1}2 + q_i\,\bar q^i
    + \tfrac12\,t_{ij}\,\bar t^{ij}
    = \hbar^2\,\tfrac{2p-1}2 + q_i\,(Q^i - q^i)
    + \tfrac12\,t_{ij}\,(\tau^{ij} - t^{ij})\,,
\end{align}
Evaluating $\Joseph_\bullet^{(2)}$, which involves
the previous equation for $p=1$ and modulo $Q_i$
and $\tau_{ij}$, yields
\begin{equation}
    \Joseph_\bullet^{(2)} = \big(P^2-\tfrac{\hbar^2}2\big)\,
    \big(P^2-\tfrac{5\hbar^2}2\big)
    \sim \big(q_i\,q^i + \tfrac12\,t_{ij}\,t^{ij}\big)\,
    \big(q_k\,q^k + \tfrac12\,t_{kl}\,t^{kl}
    +2\,\hbar^2\big)
    \sim \big(q_i\,q^i + \tfrac12\,t_{ij}\,t^{ij}\big)^2\,,
\end{equation}
upon using $\hbar\,q_i \sim 0 \sim \hbar^2\,t_{ij}$.
Using again Fierz identity and \eqref{eq:useful},
one can show that 
\begin{equation}
    \big(q_i\,q^i + \tfrac12\,t_{ij}\,t^{ij}\big)^2 \sim 0
    \qquad\text{i.e.}\qquad
    \Joseph_\bullet^{(2)} \sim 0\,,
\end{equation}
modulo $Q_i$ and $\tau_{ij}$, as required.

\section{Discussion}
\label{sec:discu}
In this chapter, we proposed a new realisation
of the type-A$_\ell$ higher spin algebra
in four dimensions, based on extending 
the Weyl algebra with a Clifford algebra.
This allows for an arguably simpler realisation
of $\hs_\ell$, wherein the limit of the range
of values of the depth of the partially massless fields
is constrained by the dimension of the Clifford algebra.
We also exhibited a Hochschild $3$-cycle
of $\hs_\ell$, which suggests that there should exist
non-trivial deformations of the partially massless
higher spin algebras.

Unfortunately, the usual technique used 
to construct deformation of higher spin algebra
that consists in using deformed oscillators \cite{Vasiliev:1992av},
i.e. trading $\hat Y^A$ for $\hat q^A$
which satisfy
\begin{equation}
    [\hat q^A, \hat q^B] = 2\,C^{AB}\,(\1+k\,\nu\big)\,,
\end{equation}
where $k$ is the generator of the $\mathbb Z_2$
action on the Weyl algebra,
discussed in the previous section, does \emph{not}
seem to work: we were unable to use this deformation
and preserve a realisation of the $\mathfrak{osp}(1|2p,\R)$ algebra
\emph{undeformed}. %
Indeed, note first that,
assuming that the deformed oscillators $\hat q^A$
still \emph{commute} with the fermionic oscillators
implies
\begin{equation}
    0 = \big[\phi^A_i, [\hat q^B, \hat q^C]\big] 
    = 2\nu\,C^{BC}\,[\phi^A_i, k]
    \qquad\Longrightarrow\qquad 
    [\phi^A_i, k]=0\,,
\end{equation}
i.e. the fermionic oscillators should also \emph{commute}
with the generator of the $\mathbb Z_2$ action%
---also called the Klein operator.
A direct computation yields
\begin{equation}
    \{Q_i,Q_j\} = (\1+k\,\nu)\,\tau_{ij}\,,
    \qquad \text{with} \qquad 
    Q_i \equiv \tfrac12\,C_{AB}\,\hat q^A\,\hat \phi^B_i\,,
    \qquad
    \tau_{ij} = \tfrac14\,C_{AB}\,
    [\hat\phi^A_i, \hat\phi^B_j]\,,
\end{equation}
which deforms the $\mathfrak{osp}(1|2p,\R)$ algebra.
One could think of modifying the odd generators as
\begin{equation}
    Q_i^{\rm new} = f(k,\nu)\,Q_i\,,
    \qquad 
    f(k,\nu) = a(\nu)\,\1 + b(\nu)\,k\,,
    \qquad 
    a(0)=1\,, \quad b(0)=0\,,
\end{equation}
however, this leads to
\begin{equation}
    \{Q_i^{\rm new},Q_j^{\rm new}\}
    = f(k,\nu)f(-k,\nu)\,(\1+k\,\nu)\,\tau_{ij}\,,
\end{equation}
which does not allow us to remove the factor
$\1+k\nu$ by suitably choosing $f(k,\nu)$,
since the combination that appears,
$f(k,\nu)f(-k,\nu) = \big(a(\nu)^2-b(\nu)^2\big)\1$,
as it is only proportional to the identity.
If one could require that the $\mathbb Z_2$ generator
anticommute with the fermionic oscillators,
$\{k,\phi^A_i\} = 0$, the right hand side
of the above equation could be fixed to be
the (undeformed) generators of the $\mathfrak{sp}(2p,\R)$
subalgebra $\tau_{ij}$ by suitably choosing $f(k,\nu)$,
thereby providing us with a realisation of $\mathfrak{osp}(1|2p,\R)$
in the deformed oscillator algebra. Unfortunately,
we saw that requiring the Klein operator and the fermionic
oscillators to anticommute is inconsistent. %

This situation seems surprising
since in the case of the type-B algebra, whose realisation
is also based on a quotient of the Weyl--Clifford algebra
\cite{Vasiliev:2004cm}, and are known to admit deformations
of this type \cite{Grigoriev:2018wrx, Sharapov:2019vyd}.
The deformed oscillator algebra, which first appeared
in a paper of Wigner \cite{Wigner1950},
is one of the simplest example
of a \emph{symplectic reflection algebra}
(originally introduced by Etingof and Ginzburg
\cite{Etingof2000}, see also 
\cite{Gordon2007, Bellamy2012, Chlouveraki2013}
for more recent reviews). The algebras are deformations
of the smash product of the Weyl algebra
with a finite group (acting on it by automorphisms).
The latter naturally contains reductive dual pairs
$(\mathfrak{g},\mathfrak{g}')$ of bosonic type,
which can --- at least in some cases
\cite{Feigin:2014yha, DeBie2009, Ciubotaru2018} --- 
be deformed by finding
a realisation of one of the algebra of the pair,
say $\mathfrak{g}$, in a symplectic reflection algebra.
Typically, the other algebra $\mathfrak{g}'$
is deformed to an \emph{associative} (not Lie) algebra.
In any case, both algebras are mutual centralisers
of one another, and hence one again finds a bijection
between their representations (appearing in the appropriate
Fock space). Recently, some examples of dual pairs
of Lie \emph{superalgebras} have been deformed
\cite{Ciubotaru2020, Calvert2022} using symplectic
reflection algebras. The difference with respect to 
the pair $\big(\mathfrak{sp}(2n,\R), \mathfrak{osp}(1|2p,\R)\big)$ of interest
for us is that the superalgebra $\mathfrak{osp}(1|2p,\R)$
that we would like to preserve when using the deformed
oscillator has its bosonic subalgebra $\mathfrak{sp}(2p,\R)$
realised using only fermionic oscillators
which are not deformed (since they generate
a Clifford algebra which is finite-dimensional,
it does not admit a non-trivial deformation).
This seems to be one of the reasons why preserving
$\mathfrak{osp}(1|2p,\R)$ appears impossible, at least
if we simply replace the bosonic oscillators $\hat Y^A$
by deformed ones $\hat q^A$ in our realisation.

\newpage

\begin{appendices}
\section{More on the Type-\texorpdfstring{A$_2$}{} Algebra}
\label{app:A2}
Any higher spin algebra whose spectrum consists of
totally symmetric fields \emph{only}, and defined as
a quotient of $\Enveloping\big(\mathfrak{so}(2,d)\big)$
by an ideal $\Ideal$, will necessarily contain
the antisymmetric generator\footnote{The factor
$4$ in $V_\AlgInd{ABCD}$ has been added for simplicity.}
\begin{equation}
    V_\AlgInd{ABCD} = 4\,M_\AlgInd{[AB}\,M_\AlgInd{CD]}
    = M_\AlgInd{[AB}\,M_\AlgInd{C]D}
    -M_\AlgInd{[AB}\,\eta_\AlgInd{C]D}\,,
\end{equation}
in its defining ideal $\Ideal$.
We will therefore start this appendix
by reviewing how factoring out $V_\AlgInd{ABDC}$
relates all Casimir operators to the quadratic one
(see also \cite[Sec. 2.1]{Iazeolla:2008ix}
and \cite{Boulanger:2011se}). Let us illustrate
this mechanism in the case of the quartic
Casimir operator, defined as\footnote{%
More generally, we follow the convention
that the Casimir operator of $\mathfrak{so}(2,d)$
of order $2n$ is given by
$\Casimir_{2n} := \tfrac12\,M_\AlgInd{A_1}{}^\AlgInd{A_2}\,
M_\AlgInd{A_2}{}^\AlgInd{A_3}\,\dots\,
M_\AlgInd{A_{2n}}{}^\AlgInd{A_1}$.}
\begin{equation}
    \Casimir_4 = \tfrac12\,M_\AlgInd{A}{}^\AlgInd{B}\,
    M_\AlgInd{B}{}^\AlgInd{C}\,
    M_\AlgInd{C}{}^\AlgInd{D}\,M_\AlgInd{D}{}^\AlgInd{A}\,,
    \qquad 
    \Casimir_2= -\tfrac12\,M_\AlgInd{AB}\,M^\AlgInd{AB}\,.
\end{equation}
A direct computation yields\footnote{%
For all computations in this appendix,
one needs to use a few identities
that are specific to orthogonal algebra,
which we will list here. For $so_N$,
with generators $R_{IJ}=-R_{JI}$ obeying
$[R_{IJ}, R_{KL}] = \eta_{JK}\,R_{IL} + (\dots)$
with $\eta$ of \emph{arbitrary signature}, one has
$$[R_I{}^\bullet,R_{J\bullet}]=-(N-2)\,R_{IJ}\,,
\qquad [V^\bullet, R_{I\bullet}]=-(N-1)\,V_I\,,
\qquad R_I{}^J\,R_J{}^K\,R_K{}^I
=-\tfrac{N-2}2\,R_{IJ}\,R^{IJ}\,,$$
where $V_I$ is any vector of $so_N$.}
\begin{equation}
    V_\AlgInd{ABC}{}^\bullet\,M_\AlgInd{D\bullet}
    = M_\AlgInd{AB}\,M_\AlgInd{C}{}^\bullet\,M_\AlgInd{D\bullet}
    + 2\,M_\AlgInd{C[A}\,M_\AlgInd{D}{}^\bullet\,
    M_\AlgInd{B]\bullet} + M_\AlgInd{AB}\,M_\AlgInd{CD}
    -2\,(d-1)\,M_\AlgInd{C[A}\,M_\AlgInd{B]D}\,,
\end{equation}
which, upon taking a trace in $\AlgInd{CD}$
and contracting with $M_\AlgInd{AB}$, gives
\begin{equation}\label{eq:C4-C2}
    V_\AlgInd{ABCD} \sim 0 
    \qquad\Longrightarrow\qquad
    \Casimir_4 \sim \Casimir_2\,\big(\Casimir_2
    + \tfrac{d(d-1)}2\big)\,,
\end{equation}
in agreement with \cite[Sec. 2.1]{Iazeolla:2008ix}
in the special case of the singleton,
and with \cite{Dolan:2011dv} in general.
Similarly, taking the Lorentz components $V_{abcd}$,
and contracting them with $L^{ab}$ (on the left)
and $L^{cd}$ (on the right), one finds
\begin{equation}\label{eq:C4_Lor}
    V_{abcd} \sim 0
    \qquad\Longrightarrow\qquad
    \Casimir_4^L \sim \big(\Casimir_2-P^2\big)
    \Big(\Casimir_2-P^2 + \tfrac12\,(d-1)(d-2)\Big)\,,
\end{equation}
where $\Casimir_4^L = \tfrac12\,L_a{}^b\,L_b{}^c\,
L_c{}^d\,L_d{}^a$ is the quartic Casimir operator
of the Lorentz subalgebra. This is the same type
of relation as \eqref{eq:C4-C2} with $d \to d-1$,
upon using the fact the quadratic Casimir operator
$\Casimir_2^L$ of $\mathfrak{so}(1,d)$
is given in terms of that of $\mathfrak{so}(2,d)$ by
$\Casimir_2^L = \Casimir_2 - P^2$.
Contracting $V_\AlgInd{ABCD}$ with more generators
produces similar identities,
relating Casimir operators of order $2n$
to lower order ones, and ultimately to $\Casimir_2$.

When decomposing the generator $V_\AlgInd{ABCD}$
under $\mathfrak{so}(1,d)$, one finds an additional antisymmetric
generator of rank $3$, namely $V_{abc0'}$.
Contracting it with $L^{ab}$ (on the left)
and $P^c$ (on the right) yields the identity
\begin{equation}\label{eq:LLPP}
    V_{abc0'} \sim 0
    \qquad\Longrightarrow\qquad
    L_a{}^\bullet\,L_{b\bullet}\,\{P^a,P^b\}
    \sim -2\,\big(\Casimir_2-P^2\big)\,
    \Big(P^2 + \tfrac{d-1}2\big)\,,
\end{equation}
which will be useful for us later on.

\paragraph{Type-A$_2$.}
Let us define 
\begin{equation}
    W_\AlgInd{AB}
    := M_\AlgInd{(A}{}^\AlgInd{C}\,M_\AlgInd{B)C}\,,
\end{equation}
and consider the symmetric generator for the ideal
defining the partially massless higher spin
algebra $A_2$, which is the traceless part
of $W^\AlgInd{(AB}\,W^\AlgInd{CD)}$, given by,
\begin{equation}
    \begin{aligned}
        \Joseph_\AlgInd{ABCD}
        & := W_\AlgInd{(AB}\,W_\AlgInd{CD)}
        -\tfrac4{d+6}\,\eta_\AlgInd{(AB}\,
        \big(W_\AlgInd{C}{}^\AlgInd{M}\,W_\AlgInd{D)M}
        - \Casimir_2\,W_\AlgInd{CD)}\big)\\
        & \hspace{120pt} + \tfrac{4}{(d+4)(d+6)}\,
        \eta_\AlgInd{(AB}\,\eta_\AlgInd{CD)}\,
        \big(\Casimir_4 + \Casimir_2\,
        [\Casimir_2 - (\tfrac{d}2)^2]\big)\,,
    \end{aligned}
\end{equation}
where we used the relation
\begin{equation}
    \tfrac12\,W_\AlgInd{AB}\,W^\AlgInd{AB}
    = \Casimir_4 - (\tfrac{d}2)^2\,\Casimir_2\,,
\end{equation}
relating the contraction of the generator $W_{AB}$
with itself and the quadratic and quartic Casimir
operators. Note that we can also express
this generator of the ideal $\Ideal_2$ as
\begin{equation}
    \Joseph_\AlgInd{ABCD}
    = \Joseph_\AlgInd{(AB}\,\Joseph_\AlgInd{CD)}
    - \tfrac4{d+6}\,\eta_\AlgInd{(AB}\,
    \Joseph_\AlgInd{C}{}^\bullet\,\Joseph_\AlgInd{D)\bullet}
    + \tfrac4{(d+4)(d+6)}\,\eta_\AlgInd{(AB}\eta_\AlgInd{CD)}\,
    \big(\Casimir_4 - \tfrac2{d+2}\,\Casimir_2^2\,
    -(\tfrac{d}2)^2\,\Casimir_2\big)\,,
\end{equation}
where
\begin{equation}
    \Joseph_\AlgInd{AB}
    := M_\AlgInd{(A}{}^\AlgInd{C}\,M_\AlgInd{B)C}
    +\tfrac2{d+2}\,\eta_\AlgInd{AB}\,\Casimir_2\,,
\end{equation}
is the traceless part of $W_\AlgInd{AB}$,
which is also one of the generator of the Joseph ideal
of the type-A algebra, and where we used
\begin{equation}
    \tfrac12\,\Joseph_\AlgInd{AB}\,\Joseph^\AlgInd{AB}
    = \Casimir_4 - \tfrac2{d+2}\,\Casimir_2^2\,
    -(\tfrac{d}2)^2\,\Casimir_2\,.
\end{equation}
This generator can be decomposed
under the Lorentz subalgebra, and in particular
contains a scalar piece,
\begin{equation}
    \Joseph^{(2)}_\bullet = \tfrac{d+2}{d+6}\,P^4
    + \tfrac4{(d+6)}\,
    \big(\tfrac14\,\{L_{ab},P^b\}\,\{L^{ac},P_c\}
    - \Casimir_2\,P^2\big) + \tfrac4{(d+4)(d+6)}\,
    \big(\Casimir_4
    + \Casimir_2[\Casimir_2-(\tfrac{d}2)^2]\big)\,,
\end{equation}
which can be re-written in terms of $\Casimir_4$,
$\Casimir_2$, $P^4$ and $P^2$, using some previously
discussed results. To do so, notice first that
\begin{equation}
    \tfrac14\,\{L_{ab},P^b\}\{L^{ac},P_c\}
    = \tfrac12\,L_a{}^\bullet\,L_{b\bullet}\,\{P^a,P^b\}
    + \tfrac{d+1}2\,(\Casimir_2-P^2)
    - \tfrac{d^2}4\,P^2\,,
\end{equation}
The first term on the right hand side can be eliminated
using \eqref{eq:LLPP},
and using the relation \eqref{eq:C4-C2}
between $\Casimir_4$ and $\Casimir_2$,
as well as imposing $\Casimir_2 \sim -\tfrac14(d-4)(d+4)$,
we end up with
\begin{equation}
    \Joseph_\bullet^{(2)}
    = \big(P^2 + \tfrac{d-4}2\big)
    \big(P^2 + \tfrac{d-8}2\big)\,.
\end{equation}
The symmetric generator $\Joseph_{\AlgInd{A}(2\ell)}$
of the defining ideal for the type-A$_\ell$
higher algebra verifies
\begin{equation}
    [M_\AlgInd{AB}, \Joseph_{\AlgInd{C}(2\ell)}]
    = 4\,\ell\,\eta_\AlgInd{C[B}\,
    \Joseph_{\AlgInd{A]C}(2\ell-1)}\,,
\end{equation}
by definition. Decomposing this identity
under the Lorentz subalgebra yields
\begin{equation}
    [P_a, \Joseph^{(\ell)}_{b(2\ell-k)}]
    = (2\ell-k)\,\eta_{ab}\,\Joseph^{(\ell)}_{b(2\ell-k-1)}
    + k\,\Joseph^{(\ell)}_{ab(2\ell-k)}\,,
    \label{eq:rec_V}
\end{equation}
for $k=0,\dots,2\ell$. We can use the above equation
to express the various Lorentz generators, obtained
by decomposing $\Joseph_{\AlgInd{A}(2\ell)}$,
in terms of the scalar one
\begin{equation}\label{eq:J0}
    \Joseph^{(\ell)}_\bullet
    = \sum_{k=0}^\ell \nu_{2k}(\Casimir_{2n})\,P^{2k}\,,
\end{equation}
where $\nu_k$ are polynomials in the Casimir
operators of $\mathfrak{so}(2,d)$, and $\nu_{2\ell}=1$.
Indeed, for $k=2\ell$ the equality \eqref{eq:rec_V}
yields
\begin{equation}
    \Joseph^{(\ell)}_a
    = \tfrac1{2\ell}\,[P_a, \Joseph_\bullet^{(\ell)}]\,,
\end{equation}
while for $k=2\ell-1$ it gives,
\begin{equation}
    \Joseph^{(\ell)}_{ab} = \tfrac1{2\ell\,(2\ell-1)}\,
    \big[P_a, [P_b, \Joseph_\bullet^{(\ell)}]\big]
    + \tfrac1{(2\ell-1)}\,\eta_{ab}\,\Joseph_\bullet^{(\ell)}\,,
\end{equation}
which can then be used to obtain, recursively,
expressions for all generators $\Joseph^{(\ell)}_{a(2\ell-k)}$
given by various linear combinations of nested
commutators of $P_a$ and $\Joseph^{(\ell)}$. Schematically,
\begin{equation}
    V_{a(k)}^{(\ell)} = \sum_{j=0}^{[k/2]} \#\,
    \underbrace{\eta_{aa} \dots \eta_{aa}}_{j\,\text{times}}\,
    [\underbrace{P_a, \dots, [P_a}_{k-2j\,\text{times}},
    \Joseph^{(\ell)}_\bullet] \dots]\,,
\end{equation}
where $\#$ generically denotes combinatorial coefficients
that can be obtained by recursion. For instance,
for the $\ell=1$ case, i.e. the usual type-A
higher spin algebra, the symmetric generator
\begin{equation}
    \Joseph_\AlgInd{AB}
    = M_\AlgInd{(A}{}^\AlgInd{C}\,M_\AlgInd{B)C}
    + \tfrac2{d+2}\,\eta_\AlgInd{AB}\,\Casimir_2\,,
\end{equation}
decomposes into three generators,
\begin{equation}
    \Joseph^{(1)}_{ab} = L_{(a}{}^c\,L_{b)c} - P_{(a}\,P_{b)}
    + \tfrac2{d+2}\,\eta_{ab}\,\Casimir_2\,,
    \quad 
    \Joseph^{(1)}_a = \tfrac12\,\{L_{ab}, P^b\}\,,
    \quad 
    \Joseph^{(1)}_\bullet = P^2 -\tfrac2{d+2}\,\Casimir_2\,,
\end{equation}
and one can check that the rank-$2$ symmetric
and the vector generators can be re-written as
\begin{equation}
    \Joseph^{(1)}_{ab} = \tfrac12\,\big[P_a, [P_b, P^2]\big]
    + \eta_{ab}\,(P^2 - \tfrac2{d+2}\,\Casimir_2)\,,
    \qquad 
    \Joseph^{(1)}_a = \tfrac12\,[P_a, P^2]\,.
\end{equation}
For $\ell=2$, one finds
\begin{equation}
    \Joseph^{(2)}_a = \tfrac12\,\Big\{L_{ab}\,
    \big(P^2 + d-3\big), P^b\Big\}\,,
\end{equation}
which is similar to the $\ell=1$ case, 
in that it is given by the anticommutator of $P^b$
with a monomial of order $2\ell-1=3$ in generators,
which is an antisymmetric Lorentz tensor.
In light of the discussion
in Section \ref{sec:formalHiSGRA}, 
the generator in $\tfrac12\,(L_{ab}P^2 + d-3)$
can be identified as the Maxwell generator
in type-A$_2$. In fact, this pattern holds
for arbitrary values of $\ell$:
a simple recursion leads to 
\begin{equation}
    [P_a,P^{2k}] = \sum_{j=1}^k (2-\delta_{j,k})\,
    d^{j-1}\,\big\{L_{ab}\,P^{2(k-j)},P^b\big\}\,,
    \qquad k\geq1\,,
\end{equation}
which yields
\begin{equation}
    \Joseph_a^{(\ell)}
    = \tfrac1{2\ell}\,\sum_{k=0}^{\ell-1}
    a_{2k}\,\big\{L_{ab}\,P^{2k}, P^b\big\}\,,
    \qquad\text{with}\qquad
    a_{2k} = (2-\delta_{k,0})
    \sum_{j=k+1}^\ell d^{j-k-1}\,\nu_{2j}\,,
\end{equation}
where $\nu_{2j}$ denote the coefficients
in the expression of $\Joseph_\bullet^{(\ell)}$
as a polynomial in $P^2$ \eqref{eq:J0}.

\end{appendices}

\newpage

\chapter{Conformal Higher Spin and Matter Coupling}
\section{Introduction}

Conformal higher spin (CHS) gravity extends both conformal gravity and higher spin gauge theory, describing an infinite tower of symmetric, traceless tensor fields $\{\Phi_{\mu_1\cdots\mu_s}\}_{s=0}^\infty$ in any dimension $d>2$.  Just as Weyl gravity is formulated in terms of a conformal equivalence class of metrics, subject to the transformation,
\[
\delta_{\xi,\sigma}g_{\mu\nu}=\mathcal{L}_\xi g_{\mu\nu}+2\sigma\,g_{\mu\nu},
\]
with $\xi^\mu(x)\partial_\mu$ generating diffeomorphisms and $\sigma(x)$ parameterizing local Weyl rescalings $g'_{\mu\nu}=\Omega^2 g_{\mu\nu}$ with $\Omega=e^{\sigma(x)}$, the CHS gauge algebra enlarges these transformations to act on each field $\Phi_{\mu_{1}\dots\mu_{s}}$ by
\begin{equation}\label{CHS-transformation}
\delta_{\xi,\sigma}\Phi_{\mu_1\dots\mu_s}
=\mathcal{L}_\xi\Phi_{\mu_1\dots\mu_s}
+ w_s\,\sigma\,\Phi_{\mu_1\dots\mu_s}
+\nabla_{(\mu_1}\xi_{\mu_2\dots\mu_s)}
+g_{(\mu_1\mu_2}\,\sigma_{\mu_3\dots\mu_s)}
+\cdots.
\end{equation}
This infinite-dimensional algebra underlies the consistent coupling of all spins $s\ge0$ in a conformally invariant background.

Two complementary frameworks establish the existence, locality, and gauge invariance of CHS gravity.  In Tseytlin’s induced-action approach, one couples a conformal scalar to higher spin currents and identifies the logarithmically divergent part of its one-loop effective action with the CHS action \cite{Tseytlin:2002gz,Bekaert:2010ky}.  In Segal’s worldline formulation\footnote{Strictly speaking, Segal gives formal arguments, very convincing, but only formal, but perturbative computations agree.}, a first-quantized particle propagating in a higher spin background reproduces the same action through Fedosov–type deformation quantization of the cotangent bundle \cite{Segal:2002gd}.  Although these two perspectives coincide via a symbol map, neither is manifestly coordinate-independent.

This shortcoming was resolved in \cite{Basile:2022nou}, where Segal’s off-shell system was embedded into a Fedosov parent formalism.  There, the higher spin gauge symmetries are encoded in the flatness of an extended connection on the cotangent bundle, while the action arises from an invariant trace functional furnished by the Feigin–Felder–Shoikhet cocycle—an explicit realisation of Shoikhet–Tsygan–Kontsevich formality \cite{FFS,Shoikhet:2000gw,Tsygan:1999}.  The result is a fully nonlinear, manifestly covariant CHS action valid on any curved spacetime.

In this chapter, we focus on coupling a massless scalar field to an off-shell background of conformal higher spin fields within that covariant Fedosov framework.  While Segal’s original construction effectively fixed the CHS action by gauge invariance, the true challenge lies in covariantizing the scalar matter coupling.  After reviewing Tseytlin's and Segal's approach in section (\ref{review of CHS}), we will discuss Fedosov deformation quantization and the covariant action for CHS called parent formulation in Section \ref{sec:Fedosov}. In Section \ref{sec:Wigner}, we introduce a Wigner-function–inspired formalism to build the scalar action.  Section \ref{sec:HS} illustrates the conformally coupled scalar case, demonstrating how Weyl symmetry and higher spin gauge invariance are realized.  We conclude in Section \ref{sec:discu} with possible future directions.  Technical details on Weyl calculus, the FFS cocycle, and the curvature expansion of the Fedosov connection are collected in Appendices \ref{app:Weyl_calculus}–\ref{app:curvature}.

\section{Review of Conformal Higher Spin Fields}\label{review of CHS}

In the following sub-sections, we review two complementary formulations of conformal higher spin gravity. First, we summarize Tseytlin’s induced-action approach, and then we turn to Segal’s worldline framework, concluding with a discussion of their equivalence.

\subsection{Induced Action  Approach}

The route between the free scalar field and conformal higher spin fields is as follows. Being a free theory, the massless scalar on flat space
possesses an infinite tower of on-shell conserved currents
for all integer spin $s\geq0$,
which comes from the invariance of the d'Alembert equation
under the action of conformal Killing tensors
\cite{Eastwood:2002su}. The existence of these conserved currents
opens the possibility of introducing interactions
between the scalar field $\phi$, and gauge fields
of arbitrary spin, starting with the Noether coupling
and completing it to all orders. For instance,
the currents of spin $1$ and $2$,
\begin{equation}
    J_\mu := \tfrac{i}2\,\big(\phi^*\,\partial_\mu\phi
    - \phi\,\partial_\mu\phi^*\big)\,,
    \qquad 
    T_{\mu\nu} := \phi^*\,\partial_\mu\partial_\nu\phi
    -\tfrac{2D}{D-1}\,\partial_{(\mu}\phi^*\partial_{\nu)}\phi
    + \phi\,\partial_\mu\partial_\nu\phi^* - \text{(traces)}\,,
\end{equation}
can be used to introduce gauge fields, say $A_\mu$
and $h_{\mu\nu}$ respectively, to the free scalar action, via
\begin{equation}
    S[\phi,A] = \int_{\R^D} \dR^Dx\,\tfrac12\,\phi^*\Box\phi
    + e\,A_\mu\,J^\mu\,,
    \qquad\text{and}\qquad 
    S[\phi,h] = \int_{\R^D} \dR^Dx\,\tfrac12\,\phi^*\Box\phi
    + \kappa\,T^{\mu\nu}\,h_{\mu\nu}\,,
\end{equation}
where $e$ and $\kappa$ are coupling constants.
Since both currents are divergenceless on-shell (meaning 
modulo the scalar field equation of motion $\Box\phi\approx0$),
and the spin $2$ one is also traceless,
the gauge transformations
\begin{equation}\label{eq:linear}
    \delta_\varepsilon A_\mu = \partial_\mu\varepsilon\,,
    \qquad 
    \delta_\xi h_{\mu\nu}
        = \partial_{(\mu} \xi_{\nu)} + \eta_{\mu\nu}\,\sigma\,,
\end{equation}
together with the transformations of the scalar field
\begin{equation}
    \delta_\varepsilon\phi
        = -ie\,\varepsilon\,\phi\,,
    \qquad\text{and}\qquad
    \delta_\xi \phi = \xi^\mu\partial_\mu\phi\,,
\end{equation}
leave the respective actions invariant up to second order
in the coupling constants. The spin $1$ case can be completed
to a gauge-invariant action to all orders by adding
a quadratic term in the gauge field $A_\mu$, which amounts
to reconstruction of scalar electrodynamics,
\begin{equation}
    S[\phi,A] = \tfrac12\,\int_{\R^D} \dR^Dx\,
    \phi^*\Box_A\phi\,,
    \qquad\text{where}\qquad
    \Box_A = (\partial_\mu + ie\,A_\mu)
    (\partial^\mu + ie\,A^\mu)\,.
\end{equation}
The spin $2$ case is technically more involved,
though similar in spirit. It requires infinitely many 
correction terms, which can be summed up into
the action for the conformally-coupled scalar field,
\begin{equation}\label{eq:action_conformal_sc}
    S[\phi,g] = \tfrac1{2\kappa}\,\int_M \dR^Dx\,\sqrt{-g}\
    \phi^*\,\big(\nabla^2 - \tfrac{D-2}{4(D-1)}\,R\big)\phi\,,
    \qquad 
    g_{\mu\nu} := \eta_{\mu\nu} + h_{\mu\nu}\,,
    \qquad 
    \nabla^2 := g^{\mu\nu}\,\nabla_\mu\nabla_\nu\,,
\end{equation}
expanded around flat spacetime. In both of these low-spin cases,
The Noether coupling is completed by higher-order terms
in the gauge fields and suitable deformations
of their gauge symmetries. The output of this procedure
is an action, quadratic in the scalar field,
and non-linear in the gauge fields. The all order coupling
of the former to the latter is encoded in a covariant
differential operator---the square of the covariant derivative
in the spin $1$ case and the conformal Laplacian
in the spin $2$ case. From this point of view,
these gauge fields are \emph{background fields}
for the scalar field $\phi$.

One can then integrate out the scalar field to derive an action for the background fields. To be more precise, the effective action for the scalar field $\phi$
can be interpreted as an action for the background fields,
a point of view already advocated by Sakharov
\cite{Sakharov:1967pk} in his approach to gravity
as an `induced theory'.
This procedure generalizes to the higher spin currents,
thereby producing a coupling of the original complex scalar
to a background of higher spin gauge fields,
via a differential operator, covariant under the associated
higher spin symmetries, which define a non-linear completion
of the linear gauge transformations
\begin{equation}\label{linearized transformation}
    \delta_{\xi,\sigma} h_{\mu_1 \dots \mu_s}
    = \partial_{(\mu_1} \xi_{\mu_2 \dots \mu_s)}
    + \eta_{(\mu_1 \mu_2}\,\sigma_{\mu_3 \dots \mu_s)}\,,
\end{equation}
for all integers $s \geq 1$. These were identified
as the linear symmetries of \emph{conformal higher spin
gravity} (CHSGra), a higher spin generalization
of conformal (super)gravity proposed by Fradkin and Tseytlin 
\cite{Fradkin:1985am} at the free level, and studied further
at the cubic level \cite{Fradkin:1990ps}

More concretely, the idea of Tseytlin \cite{Tseytlin:2002gz}, which was fully developed in detail in \cite{Bekaert:2010ky}, was to consider a massless complex scalar field in the flat spacetime of even dimension $D$. The free action is given by,
\begin{equation}
S_0[\phi] = \int d^D x \, \partial^\mu \phi^* \partial_\mu \phi \,.
\end{equation}
Such a complex scalar field gives rise to an infinite tower of conserved currents of the form,
\[
 J_{\mu(s)} = \phi^* \partial_{\mu_1} \ldots \partial_{\mu_s} \phi + \ldots,
\]
for an arbitrary integer spin $s$. These currents are conserved ($\partial^\nu J_{\nu \mu(s-1)} \approx 0$) and traceless ($J^\alpha_{\ \ \alpha \mu(s-2)} \approx 0$)  on-shell $\Box \phi \approx 0$. Using these currents, one introduces another field, called conformal higher spin fields (CHS), as Noether interactions
\begin{equation}
S_{\text{int}}[\phi,h] = \sum_{s=0}^\infty \frac{(i)^s}{s!} \int d^D x \, J^{\mu(s)} h_{\mu(s)}\,.
\end{equation}
Let's introduce generating functions as,
\begin{equation}
J(x,u) := \sum_{s=0}^\infty \frac{1}{s!} J_{\mu_1 \ldots \mu_s}(x) u^{\mu_1} \ldots u^{\mu_s} = \sum_{s=0}^\infty J_s(x,u).
\end{equation}
With this, the conservation law becomes $\partial_u \cdot \partial_x J(x,u) \approx 0$ and the tracelessness becomes $\partial_u^2 J(x,u) \approx 0$. One can write 
\begin{equation}
J(x,u) = \Pi_D \mathcal{J}(x,u)\,, \quad \mathcal{J}(x,u) := \phi^*(x + u/2) \phi(x - u/2)\,,
\end{equation}
where \(\mathcal{J}(x,u)\) generates conserved (though not traceless) currents, and the operator \(\Pi_{D}\) projects these onto their traceless sector, see e.g \cite{Bekaert:2010ky}.
\begin{equation}
\Pi_D := \sum_{n=0}^\infty \frac{1}{n! (-\hat{N} - \frac{D-5}{2})_n} \left[ \frac{\partial^2 - g}{16} \Box \right]^n\,,
\end{equation}
where $(a)_n := \frac{\Gamma(a+n)}{\Gamma(a)}$, and $\hat{N} := u \cdot \partial_u$, $\partial := u \cdot \partial_x$, $g := u^2$. Let's also write the higher spin field in terms of a generating function as,
\begin{equation}
h(x,u) := \sum_{s=0}^\infty \frac{1}{s!} h_{\mu_1 \ldots \mu_s}(x) u^{\mu_1} \ldots u^{\mu_s} = \sum_{s=0}^\infty h_s(x,u)\,,
\end{equation}
Now, with this and along with the conserved current, the Noether interaction can be written as
\begin{align}
S_{\text{int}}[\phi,h] &= \int d^D x \, \mathcal{J}(x, i \partial_u) \mathcal{H}(x,u)|_{u=0} \\
&= \int d^D x \, e^{i \partial_u \cdot \partial_v} \mathcal{J}(x,v) \mathcal{H}(x,u)|_{u,v=0}\,,
\end{align}
where we have used the integration by parts, the derivatives in $\Pi$ such that,
\begin{equation}
\mathcal{H}(x,u) = \mathcal{P}_D h(x,u)\,, \quad \mathcal{P}_D := \sum_{n=0}^\infty \frac{1}{n! (\hat{N} + n + \frac{D-3}{2})_n} \left[ \frac{\partial^{*2} - \text{Tr} \Box}{16} \right]^n\,,
\end{equation}
This operator $\mathcal{P}_{D}$ admits an inverse, given by,
\begin{equation}
h(x,u) = \mathcal{P}_D^{-1} \mathcal{H}(x,u)\,, \quad \mathcal{P}_D^{-1} := \sum_{n=0}^\infty \frac{(-1)^n}{n! (\hat{N} + \frac{D-1}{2})_n} \left[ \frac{(\partial_u \cdot \partial_x)^{2} - \text{Tr} \Box}{16} \right]^n\,,
\end{equation}
where we defined trace $\text{Tr} := \partial_u^2$ operators. Despite the infinite series, each spin-$s$ component of the conformal fields $h_s$ produces a finite tail of traces and divergences, as can be seen by rewriting
\begin{equation}
\mathcal{H}(x,u) = \sum_{s=0}^\infty \sum_{n=0}^{[s/2]} \frac{1}{n! (s - n + \frac{D-3}{2})_n} \left[ \frac{\partial^{*2} - \text{Tr} \Box}{16} \right]^n h_s(x,u)\,.
\end{equation}
By introducing the Fourier transform of $\mathcal{J}(x,v)$ in $v$-space:
\begin{equation}
\mathcal{J}(x,v) = \int \frac{d^D p}{(2\pi)^D} e^{-i v \cdot p} \rho(x,p)\,,
\end{equation}
the interaction term becomes,
\begin{equation}
S_{\text{int}}[\phi,h] = \int \frac{d^D x d^D p}{(2\pi)^D} \rho(x,p) \mathcal{H}(x,p)\,.
\end{equation}
This form of the interaction is very suggestive of the fact that $\rho$ can be interpreted as 'density matrix'. But we know from quantum mechanics that the density matrix is nothing but a "Weyl-symbol" of the projector operator $|\phi\rangle \langle \phi|$. In this language, the field $\phi(x)$ can be written as the wave function $\langle x | \phi \rangle$. This allows us to employ the tools of Weyl-quantization (more details in the Appendix), to get the action as,
\begin{equation}
S_{\text{int}}[\phi,h] = \text{Tr} \left[ |\phi\rangle \langle \phi| \hat{H} \right] = \langle \phi | \hat{H} | \phi \rangle\,,
\end{equation}
where $\hat{H}(\hat{X}, \hat{P})$ is the operator with Weyl symbol given by $\mathcal{H}(x,p)$, i.e.,
\begin{equation}
\hat{H}(\hat{X}, \hat{P}) = \int \frac{d^D x d^D p}{(2\pi)^D} \mathcal{H}(x,p) \int \frac{d^D y d^D k}{(2\pi)^D} e^{i k \cdot (x - \hat{X}) - i y \cdot (p - \hat{P})}\,.
\end{equation}

The total action entering the path integral (12) can thus be written as
\begin{equation}
S[\phi,h] = \langle \phi | \Box + \hat{H} | \phi \rangle\,=\langle \phi | \hat{F}| \phi \rangle\,
\end{equation}
for $\hat{F}=\Box+\hat{H}$. 
\begin{framed}
In this form, it is now much easier to see the full symmetries of the action as it is invariant under infinitesimal transformation,
\begin{equation}
    \delta_u \hat{F}
    = \hat{u}^{\,\dagger} \circ \hat{F}
    + \hat{F} \circ \hat{u}\,,
    \qquad 
    \delta_u \ket\phi
        = -\hat{u}\ket\phi\,,
\end{equation}
If we define $\hat{u}=\hbar^{-1}\hat{\xi}+\hat{w}$, we get 
\begin{align}
\delta_{\xi,w}\hat{F} =\frac{1}{\hbar}\,[\hat{F},\hat{\xi}]+\{\hat{F},\hat{w}\}\,,\\
\delta_{\hat{\xi},\hat{w}}\phi =-\left(\tfrac{1}{\hbar}\,\hat{\xi}+\hat{w}\right)\phi\,,
\end{align}
where the gauge parameters $\hat{\xi}$ and $\hat{w}$ are Hermitian operators: $\hat{\xi}^{\dagger}=\hat{\xi}$, $\hat{w}^{\dagger}=\hat{w}$. We will see all of this in much more detail in later sections, but for the moment, we quote a result here that these transformations give 
\[ \delta_{\xi,\sigma} h^{a_1 \dots a_s}= 2\nabla^{(a_1} \xi^{a_2 \dots a_s)}
    +2\,\eta^{(a_1 a_2} \sigma^{a_3 \dots a_s)}
    +(s-2)\,\sigma\,h^{a_1 \dots a_s} + \dots
\] 
Here, the parameter $\sigma$ is related to $w$. It is because of this transformation structure, the parameter $\hat{\xi}$  is said to correspond to gauge symmetries, while the generator $\hat{w}$ corresponds to the generalized Weyl symmetry.
\end{framed}

\paragraph{Effective Action}

The effective action is given by
\begin{equation}
W[h] = \text{Tr} \log \left( \Box + \hat{H} \right) = -\int_{\frac{1}{\Lambda^2}}^\infty \frac{dt}{t} \text{Tr} \left[ e^{-t (\Box + \hat{H})} \right],
\end{equation}
where we have introduced an ultraviolet cutoff \(\Lambda\) in the small-\(t\) regime and $K_{t}[h]\;:=\;\text{Tr}\!\bigl[e^{-\,t\bigl(\Box+\hat{H}\bigr)}\bigr]\,$, is called the heat‐kernel trace, which, upon treating the higher‐spin terms in \(\hat{H}\) as a perturbation of the flat‐space operator \(\hat{P}^{2}\), admits an expansion in the parameter \(t\). The regularized effective action can then be organized according to its divergent contributions as,
\begin{equation}
W_\Lambda[h] := -\int_{\frac{1}{\Lambda^2}}^\infty \frac{dt}{t} K_{t}[h] = \sum_{n=1}^\infty \Lambda^{2n} W_n[h] + W_{\frac{D}{2}}[h]\log \Lambda  + W_{fin}[h] + \mathcal{O}(\Lambda^{-2})\,.
\end{equation}
Some important remarks to note,
\begin{enumerate}
    \item The coefficient $W_{\frac{D}{2}}$ is non-vanishing only in even dimensions and is gauge invariant. 
    \item The finite part $W_{fin}$ contains the contribution from all the coefficient $W_{n}$ for $n\geq \frac{D}{2}$. 
    \item  Coefficients $W_{n}$ as well as the finite part of the effective action $W_{fin}$ are invariant under gauge transformation with parameter $\xi$. This is because the heat kernel $K_{t}[h]$ is gauge invariant under this part of the transformation. 
    \item Under Weyl symmetry: In odd dimensions, no coefficient is Weyl invariant. In even dimension only the coefficient $W_{D/2}$ is Weyl-invariant 
\end{enumerate}
The last point is the most crucial because it suggests that in even dimensions, the only coefficient that is invariant under both parameters, $\xi$ and $w$, is the coefficient in the logarithmically divergent term. Thus, it is the perfect candidate for the definition of the conformal higher spin action. So we define,
\begin{equation}
    S_{CHS}:=W_{\frac{D}{2}}.
\end{equation}
In words, the CHS gravity action in Tseytlin’s approach is defined as the logarithmically divergent piece of the effective action of a scalar field conformally coupled to a background of CHS fields.  It is also well-known that for the low-spin background fields \(A_\mu\) and \(g_{\mu\nu}\) in, four dimensions, one has \cite{BeccariaTseytlin2017} the following conformal anomaly
\[
W_{2}=
\frac{1}{(4\pi)^{2}}
\int d^{4}x\,\sqrt{g}\,\big(-\frac{1}{12}F_{\mu\nu}^2+\frac{1}{120}C_{\mu\nu,\rho\sigma}C^{\mu\nu,\rho\sigma}\big),
\]
where \(C_{\mu\nu,\lambda\rho}\) is the Weyl tensor, and the topological Euler term has been omitted.  Therefore, it should not be so surprising that the $log$-divergence coefficient acquires well-defined, local higher spin corrections once the higher spin background fields are turned on. Even though conceptually this definition of CHS is clean and simple, in practice, we have to compute this action perturbatively, and it becomes cumbersome as we go higher in spin. For higher spin, the linearized action turned out to be 
\begin{equation}
    S_{s}[h]=\int d^Dx\,h^{a(s)}P^{b(s)}_{a(s)}(\partial)\Box^{s+\frac{D-4}{2}}h_{b(s)},
\end{equation}
where $P$ is the traceless and transverse projector, ensuring the invariance under linear transformation,
\begin{equation}
    \delta_{\xi,\sigma} h_{\mu_1 \dots \mu_s}
    = \partial_{(\mu_1} \xi_{\mu_2 \dots \mu_s)}
    + \eta_{(\mu_1 \mu_2}\,\sigma_{\mu_3 \dots \mu_s)}\,.
\end{equation}
The action can be put into a more suggestive form by integrating by parts as,
\begin{equation}
    S_{s}[h]=\int\,d^D x\,C^{a(s),b(s)}\Box^{\frac{D-4}{2}}C_{a(s),b(s)},
\end{equation}
where $C^{a(s),b(s)}$ is the linearized Weyl-tensor. This mimics the structure of the familiar spin-$2$ conformal gravity action in four dimensions.

\subsection{Segal’s Approach}
As we have seen that Tseytlin proposed to define conformal higher spin gravity as the coefficient of the logarithmically divergent piece of the effective action of a scalar field
in a higher spin background \cite{Tseytlin:2002gz}.
However,
working out \emph{perturbatively} the exact expression
of the relevant differential operator encoding this coupling 
for all spins $s > 2$ seems unrealistic.
This is not to say that with a perturbative approach
to this problem, it is impossible to get/recover
manifestly (higher spin) covariant objects. It allows one
to compute the conformal higher spin gravity action
at the lowest orders, and confirm that the quadratic piece
is the expected one \cite{Fradkin:1985am}, as argued
in \cite{Tseytlin:2002gz} and worked out in details
in \cite{Bekaert:2010ky}. But as we go to higher orders, the procedure quickly becomes impractically difficult. 

A. Segal proposed an elegant solution to the problem of coupling
a (complex) scalar field to a background of higher spin fields 
and computing its effective action, by resorting
to symbol calculus, and more generally, 
to deformation quantization \cite{Segal:2002gd}.
That is to work with Moyal-Weyl algebra. The idea is to translate
action (as we already did above in \ref{act}) and its gauge symmetries, which formally read
\begin{equation}\label{eq:Segal}
    S[\phi, h_s]
    = \tfrac12\,\langle\phi|\hat{H}[h_s]\ket\phi\,,
    \qquad\qquad 
    \delta_u \hat{H}
    = \hat{u}^{\,\dagger}\hat{H}
    + \hat{H}\hat{u}\,,
    \qquad 
    \delta_\varepsilon \ket\phi
        = -\hat{u}\ket\phi\,,
\end{equation}
where $\hat{H}$ and $\hat{\varepsilon}$
are differential operators, respectively encoding
the coupling to background fields $h_s$ and gauge parameters
(which appear as coefficients of these operators), 
into the language of symbols,
i.e., functions on the cotangent bundle $T^*\R^D$.
This approach has some computational advantages,
and in particular, the cubic part of the action
for CHSGra was derived \cite{Segal:2002gd} in this framework.

Concretely, in phase‐space coordinates, the star product is
\[
(f\star g)(x,p)
= f\,\exp\!\Bigl[\tfrac{\hbar}{2}\bigl(\overleftarrow\partial_{x^\mu}\,\overrightarrow\partial_{p_\mu}
-\overleftarrow\partial_{p_\mu}\,\overrightarrow\partial_{x^\mu}\bigr)\Bigr]\,g,
\]
so that $[x^\mu,p_\nu]_\star=\hbar\,\delta^\mu_\nu$.  Hermitian conjugation acts as\footnote{Note that in our notation, we have absorbed the factor of $i$, which is present in the conventional formula for Moyal-Weyl product, in the constant $\hbar$ and hence we have that $\hbar^{\dagger}=-\hbar$. }
\[
x^\dagger=x,\quad p^\dagger=p,\quad \hbar^\dagger=-\hbar,\quad (a\star b)^\dagger
=b^\dagger\star a^\dagger.
\]
The quantization map sends any symbol $f(x,p)$ to the symmetrically‐ordered operator $\widehat f(x,\partial_x)$, with the inverse given by the symbol map.  The adjoint $\dagger$ then matches the formal adjoint under the standard inner product
\[
\langle\phi,\psi\rangle
=\int d^D x\,\phi^*(x)\,\psi(x),
\]
provided $\phi,\psi$ are half‐densities. 
A generic Hermitian symbol $F(x,p)=\sum_s F^{\mu_1\cdots\mu_s}(x)\,p_{\mu_1}\cdots p_{\mu_s}$ encodes background fields.  The scalar action reads
\begin{equation}\label{act}
S[\phi,F]=\langle\phi,\widehat F\,\phi\rangle,
\end{equation}
with $\widehat F=\Box+\widehat H$.  Its full gauge symmetry is
\begin{align}\label{segal system}
\delta_{\xi,w}F &= \tfrac1\hbar\,[F,\xi]_\star + \{F,w\}_\star,
&
\delta_{\xi,w}\phi &= -\bigl(\tfrac1\hbar\,\widehat\xi + \widehat w\bigr)\,\phi,
\end{align}
where $\xi(x,p)$ and $w(x,p)$ are Hermitian symbols.  Defining $u=\hbar^{-1}\xi+w$, one writes
\[
\delta_u F = u^\dagger\star F + F\star u,
\quad
\delta_u\phi = -\widehat u\,\phi.
\]

Note that in terms of symbols of operators, $\widehat u\,\phi$ should be understood as an operator $\widehat u$ acting on $\phi$ via quantization map given by,
\begin{equation}
    \big(\widehat u\,\phi\big)(x)=\int \tfrac{\dR^Dv\,\dR^Dp}{(2\pi\hbar)^D}\,
    u(\tfrac{x+v}2,p)\,e^{\frac{i}\hbar\,p \cdot (x-v)}\,
    \varphi(v)\,,
\end{equation}
For more details on the Quantization map, see the appendix (\ref{app:Weyl_calculus}).

Now the system with field $F$ subject to above mention gauge symmetries gives us what is known as {\it off‐shell Segal system}.  In other words, the off-shell Segal system gives completion of the CHS transformations (\ref{CHS-transformation}) and also solves the problem of how to couple matter fields to a higher spin background by including the $\phi$-part of the transformation.
Now restricting to
\[
\xi=\xi^\mu(x)\,p_\mu,\quad w=\sigma(x),
\]
one finds that the commutator two such parameter  $u_{i}=\frac{1}{\hbar}\,\xi_{i}^{\mu}(x)p_{\mu}+\sigma_{i}(x)$ for $i=1,2$, reproduces 
\[[u_{1},u_{2}]_{\star}=\tfrac{1}{\hbar}\,(\xi_{2}^{\mu}\partial_{\mu}\xi_{1}^{\nu}-\xi_{1}^{\mu}\partial_{\mu}\xi_{2}^{\nu})p_{\nu}+\xi_{2}^{\mu}\partial_{\mu}\sigma_{1}-\xi_{1}^{\mu}\partial_{\mu}\sigma_{2}\,,\]
and hence, this sub-algebra is isomorphic to the semidirect product of the algebra of diffeomorphisms and Weyl rescalings. Indeed one can verify that for $F=\frac{1}{2}g^{\mu\nu}p_{\mu}p_{\nu}$ we get,
\begin{equation}
    \delta_{\xi,\omega}g^{\mu\nu}=\mathcal{L}_{\xi}g^{\mu\nu}+2\sigma g^{\mu\nu},
\end{equation}
which clearly shows that $g^{\mu\nu}$ transforms as conformal metric.
To see linearized CHS transformation we expand around the vacuum $F^{(0)}=\tfrac12\,p^2$, the transformation becomes,
\[
\delta_{\xi,w}f
=\tfrac1{2\hbar}\,[p^2,\xi]_\star + \tfrac12\,\{p^2,w\}_\star,
\]
We see that upon using the Taylor expansion, 
\begin{align}
    &f=\dots h^{\nu_{1}\dots\nu_{s}}p_{\nu_{1}}\dots p_{\nu_{s}}+\dots \nonumber\\&
    \xi=\dots+\xi^{\nu_{1}\dots\nu_{s-1}}p_{\nu_{1}}\dots p_{\nu_{s-1}}+\dots \nonumber\\&
    \sigma=\dots+\sigma^{\nu_{1}\dots\nu_{s-2}}p_{\nu_{1}}\dots p_{\nu_{s-2}}+\dots,
\end{align}
the transformation becomes,
\begin{equation}
    \delta_{\xi,\omega}h^{\nu_{1}\dots\nu_{s}}=\partial^{(\nu_{1}}\xi^{\nu_{2}\dots \nu_{s})}+\eta^{(\nu_{1}\nu_{2}}\sigma^{\nu_{3}\dots\nu_{s})},
\end{equation}
which is the standard higher spin diffeomorphisms and Weyl transformations to the leading order in $\hbar$.

Also, the vacuum symmetry algebra solves
\[
u^\dagger\star p^2 + p^2\star u = 0,
\]
modulo trivial generators $u=i\,v\star p^2$.  This reproduces the higher‐symmetry algebra of the Laplacian, which is the deformation quantization of the coadjoint orbit for the free conformal scalar $\Box\phi=0$, equivalent to a quotient of $\mathcal U(\mathfrak{so}(n,2))$ by the Joseph ideal.
Finally, Segal’s action for the background fields arises as the trace of a phase‐space Lagrangian,
\[
S[F]=\text{tr} \mathcal L_{x,p}(F):=\int d^D x\, d^D p\,\mathcal L_{x,p}(F),
\]
where $\text{tr}$ is the usual trace in Weyl quantization. Now under the gauge transformation $\delta_{\xi,w}F = \tfrac1\hbar\,[F,\xi]_\star + \{F,w\}_\star$, the variation $\delta_{\xi,w}S[F]$ involves commutator and anti-commutator. Because of the cyclic property of trace operation ($\text{tr}[f,g]_\star=0$), the commutator term vanishes. The anti-commutator term gives us  
\begin{equation}
    \delta_{\xi,w}S[F]=2\text{tr}\Big[\mathcal L'_{x,p}(F)\star F\star w\Big]
\end{equation}
Thus, to ensure invariance, we get that 
\[
\mathcal L'_{x,p}(F)\star F = 0,
\]
Clearly, the solution is $\mathcal L_{x,p}(F)=\Theta_\star(F)$, the star‐product Heaviside function, which ensures the invariance under Weyl‐type gauge transformations. This defines the conformal higher spin gravity action in Segal’s framework as,
\begin{equation}\label{SA}
    S[F]=\int d^D x\, d^D p\,\Theta_\star(F).
\end{equation}
Note that the star Heaviside function is defined in the integral form as,
\[
\Theta_\star(F)=\text{lim}_{\epsilon\rightarrow 0^+}\frac{1}{2\pi i}\int\frac{d\tau}{\tau-i\epsilon}e_{\star}^{i\tau F},
\]
where $e_{\star}^{i\tau F}$ is the usual exponential function with the usual product replaced by $\star$-product.

\paragraph{Tseytlin/Segal Dictionary}

At first glance, Tseytlin’s and Segal’s constructions appear unrelated: Tseytlin extracts the logarithmic divergence from one-loop matter diagrams with higher spin external legs, while Segal expands a star-product Heaviside function.  However, one can show \cite{Segal:2002gd} that they coincide.

Starting from Tseytlin’s definition
\[
S[\widehat F] \;=\; a_{\frac\alpha2}[\widehat F]\,,
\]
and using the heat-kernel expansion
\[
\Tr\bigl(e^{-t\widehat F}\bigr)
= t^{-\frac\alpha2}\sum_{k=0}^\infty t^k\,a_k[\widehat F]\,,
\]
we write
\[
S[\widehat F]
= \frac{1}{2\pi i}\oint\frac{dt}{t}\,\Tr\bigl(e^{-t\widehat F}\bigr)\,.
\]
Noting that the 'symbol' of $e^{-t\widehat F}$ is the star-exponential
\[
e_\star^{-tF}
:= \sum_{k=0}^\infty \frac{(-t)^k}{k!}\,\underbrace{F\star\cdots\star F}_{k}
\]
and that trace on operators $\Tr(\widehat D)$ is replaced by trace on its Weyl 'symbol' $\tr(D)$, we obtain CHS gravity action as 
\[
S[\widehat F]
= \frac{1}{2\pi i}\oint\frac{dt}{t}\,\tr\bigl(e_\star^{-tF}\bigr)\,.
\]
Now, consider a circle $|t|=\delta>0$ around $t=0$ as contour we get,
\[
S[\widehat F]
= \frac{1}{2\pi i}
  \oint_{|t|=\delta}
  \frac{dt}{t}\,\tr\bigl(e_\star^{-tF}\bigr)\,.
\]
Under the assumptions
\begin{itemize}
  \item $\tr(e_\star^{-tF})$ is analytic for $\Re\,t>0$,  
  \item $\tr(e_\star^{-tF})\to0$ as $|t|\to\infty$ in the half‐plane $\Re\,t>0$,
\end{itemize}
we deform the small circle into the boundary of the half‐plane $\Re\,t>-\varepsilon$, consisting of
\[
\bigl\{\,t = -\varepsilon + i\tau : \tau\in(-\infty,\infty)\bigr\}
\quad\text{and}\quad
\text{the large semicircle at }|t|\to\infty.
\]
On the large arc, $\Re\,t\to+\infty$ implies $e_\star^{-tF}\to0$, so its contribution vanishes.  Hence, only the integral along the vertical line remains:
\[
S[\widehat F]
= \lim_{\varepsilon\to0^+}
  \frac{1}{2\pi i}
  \int_{-\infty}^{\infty}
    \frac{dt}{t}\,\tr\bigl(e_\star^{-tF}\bigr)
  \bigg|_{\,t=-\varepsilon + i\tau}\,.
\]
Parameterizing \(t = -\varepsilon + i\tau\) gives $\frac{dt}{t}= \frac{d\tau}{\tau - i\varepsilon}\,$ and substituting into the integral yields,
\[
S[\widehat F]
= \lim_{\varepsilon\to0^+}
  \frac{1}{2\pi i}
  \int_{-\infty}^{\infty}
    \frac{d\tau}{\tau - i\varepsilon}\,
    \tr\bigl(e_\star^{\,i\tau F}\bigr)\,,
\]
Exchanging trace with integral and noting that $\Theta_\star(a)
:= \lim_{\epsilon\to0^+}
\frac{1}{2\pi i}
\int_{-\infty}^{+\infty}\frac{d\tau}{\tau - i\epsilon}\,
e_\star^{\,i\tau a}$, will give us,
\[
S[\widehat F]
= \tr\bigl(\Theta_\star(F)\bigr),
\]
which is precisely Segal’s action (\ref{SA}) in terms of star-Heaviside function. Thus, we get the formal equivalence of the Segal and Tseytlin approaches. 

One of the drawbacks of both approaches outlined above, however, is that they are defined around flat spacetime.
Working out the expression of conserved currents
for a free scalar field on a more general background
can be rather challenging, although the case of Weyl-flat
space (and $\mathcal{N}=1$ supersymmetrization thereof)
has been successfully worked out \cite{Kuzenko:2022hdv}.
More generally, formulating CHSGra around an arbitrary 
background or in a manifestly covariant manner, 
has been the subject of several works \cite{Grigoriev:2016bzl, Beccaria:2017nco, Kuzenko:2019ill, Kuzenko:2019eni, Kuzenko:2022hdv} (see also \cite{Kuzenko:2017ujh, Kuzenko:2021pqm, Kuzenko:2022qeq, Kuzenko:2024vms,Buchbinder:2024pjm} for supersymmetric extensions, and \cite{Joung:2021bhf}
for an approach to conformal gravity using `unfolding').

In the next section, we describe the tools of Fedosov quantization and related concepts, which leads us to a way to write a covariant action for Conformal higher spin theory in arbitrary spacetime as in \cite{Basile:2022nou} followed by a section describing matter coupling in a covariant way.

\section{Elements of Fedosov quantization}
\label{sec:Fedosov}
Before spelling out our action for a complex scalar
coupled to an arbitrary higher spin background, 
we shall briefly review some constructions proposed
by Fedosov in his seminal paper \cite{Fedosov:1994zz}
on the deformation quantization of symplectic manifolds
(see also his textbook \cite{Fedosov:1996} for more details).
Readers familiar with these ideas may safely skip 
this section, while unfamiliar readers interested
in complementary references may consult
\cite[App. A]{Grigoriev:2016bzl} which we closely follow,
as well as \cite{Basile:2022nou} where these techniques
have been used in the context of conformal higher spin gravity.

\paragraph{Building the Fedosov connection.}
The ingredient we need is a flat connection on the Weyl bundle,
\begin{equation}
    \WeylBundle_\Base := S(T\Base) \otimes \hat S(T^*\Base)
    \twoheadrightarrow \Base\,,
\end{equation}
where $\Base$ denotes our $D$-dimensional spacetime manifold,
and $\hat S(\dots)$ the completion of the symmetric algebra.
To be concrete, a typical section of this bundle
locally takes the form
\begin{equation}
    \Gamma(\WeylBundle_\Base)\ \ni\ {\tt a}(x;y,p)
    = \sum_{k,l} {\tt a}_{a_1 \dots a_k}^{b_1 \dots b_l}(x)\,
    y^{a_1} \dots y^{a_k}\,p_{b_1} \dots p_{b_l}\,,
\end{equation}
where $\{y^a\}$ and $\{p_b\}$, for $a,b=1,\dots,D:=\dim\Base$,
respectively define a basis of its cotangent
and tangent space over the point $x \in \Base$.
The above section is \emph{polynomial} in $p$, but is allowed
to be a \emph{formal power series} in $y$, in accordance 
with the fact that the Weyl bundle is the tensor product
of the symmetric algebra of $T\Base$, and the \emph{completion}
of the symmetric algebra of $T^*\Base$. 

The fiber at each point is isomorphic, upon extending it
over $\R\llbracket\hbar\rrbracket$, to that of the Weyl algebra 
$\WeylAlg_{2D}$ generated by the $2D$ variables $y$ and $p$,
whose associative (but non-commutative) product $\ast$
is given by
\begin{equation}
    \big(f \ast g\big)(y,p) = f(y,p)\,
    \exp\Big(\tfrac\hbar2\,
    \big[\tfrac{\overleftarrow{\partial}}{\partial y}
    \cdot \tfrac{\overrightarrow{\partial}}{\partial p}
    -\tfrac{\overleftarrow{\partial}}{\partial p}
    \cdot \tfrac{\overrightarrow{\partial}}{\partial y}\big]\Big)\,g(y,p)\,,
\end{equation}
where we denoted the contraction of Latin indices
by a dot, i.e. $y \cdot p = y^a\,p_a$.
This product is called the Moyal--Weyl product,
see Appendix \ref{app:Weyl_calculus} for a review
of its derivation from the perspective of symbol calculus.
Note that the operation\footnote{One can think
of it as essentially complex conjugation,
upon considering $\hbar$ as a \emph{purely imaginary}
formal parameter. When deriving the Moyal--Weyl product
from  the point of view of symbol calculus, as recalled
in Appendix \ref{app:Weyl_calculus}, the $\hbar$ factor
in its definition appears multiplied by the imaginary unit,
which we chose to absorb in $\hbar$ itself here
to simplify computations.}
\begin{equation}
    \hbar^\dagger = -\hbar\,,
    \qquad 
    (y^a)^\dagger = y^a\,,
    \qquad 
    (p_a)^\dagger = p_a\,,
\end{equation}
which also acts by complex conjugation on coefficients,
defines an anti-involution of the Weyl algebra, that is 
\begin{equation}
    (f \ast g)^\dagger = g^\dagger \ast f^\dagger\,,
\end{equation}
for any pair of elements $f$ and $g$.
The sections of the Weyl bundle can therefore be multiplied,
using the Moyal--Weyl product fiberwise, and thereby making 
$\WeylBundle_\Base$ into a bundle of associative algebras.
The Weyl algebra can be endowed with a grading, namely
\begin{equation}\label{eq:deg}
    \deg(y^a) = 1 = \deg(\hbar)\,,
    \qquad \deg(p_a) = 0\,,
\end{equation}
with respect to which the Moyal--Weyl product is of degree $0$.

Having recalled the definition of the Weyl bundle,
we can come back to our initial goal which is to construct
a flat connection on it. As it turns out, this is relatively
simple, as one can show that any $1$-form connection of the form
$A_0 = \dR x^\mu\,e_\mu^a\,p_a + \dots$,
where $e_\mu^a$ are the components of an invertible
frame field on $\Base$ and the dots denote higher order terms
in $y$ and $p$, can be extended into a flat connection
on $\WeylBundle_\Base$,
\begin{equation}
    \dR A + \tfrac1{2\hbar}\,[A,A]_\ast = 0\,,
    \qquad\text{with}\qquad 
    A = A_0 + \text{(corrections)}\,.
\end{equation}
A simple way of constructing such a flat connection
is to start from
\begin{equation}
    A_0 = \dR x^\mu\,\big(e_\mu^a\,p_a
    + \omega_\mu^{a,b}\,p_a\,y_b\big)\,,
\end{equation}
where $\omega^{a,b} := \dR x^\mu\,\omega_\mu^{a,b}$
are the components of the torsionless spin-connection
with respect to the vielbein $e^a_\mu$, which preserves
the fiber metric $\eta^{ab}$, used to raise and lower
the fiber (i.e. Latin) indices. Let us introduce
\begin{equation}
    \delta := -\tfrac1\hbar\,[\dR x^\mu\,e^a_\mu\,p_a,-]_\ast,
    \quad 
    \nabla := \dR + \tfrac1\hbar\,
    [\dR x^\mu\,\omega_\mu^{a,b}\,p_a\,y_b, -]_\ast,
    \quad
    R^\nabla := \big(\dR\omega^{a,b}
    + \omega^{a,}{}_c\,\omega^{c,b}\big)\,p_a\,y_b,
\end{equation}
so that the curvature of $\nabla$ is simply given by
\begin{equation}
    \nabla^2 = \tfrac1\hbar\,[R^\nabla,-]_\ast\,,
\end{equation}
and one can easily check that
\begin{equation}
    \delta\nabla + \nabla\delta = 0\,,
\end{equation}
as a consequence of the torsionlessness of $\nabla$.
Note that $\delta$ and $\nabla$ are respectively
of degree $-1$ and $0$ with respect to the previously 
introduced grading \eqref{eq:deg}.
One can then show that there exists a unique $1$-form
$\Completion \in \Omega^1(\Base, \WeylBundle_\Base)$
of degree $\geq 2$ such that
\begin{equation}\label{eq:Fedosov}
    A = A_0 + \Completion\,,
\end{equation}
defines a \emph{flat} connection on the Weyl bundle,
with $\Completion$ linear in $p$ and obeying $h\Completion=0$,
and where
\begin{equation}\label{eq:defh&N}
    h := \tfrac1N\,y^a\,e_a^\mu\,\tfrac{\partial}{\partial (\dR x^\mu)}\,,
    \qquad\qquad
    N := y^a\,\tfrac{\partial}{\partial y^a}
        + \dR x^\mu\,\tfrac{\partial}{\partial (\dR x^\mu)}\,,
\end{equation}
with $N$ the number operator returning the sum
of the form degree and $y$-degree of its argument.
Equivalently, the associated covariant derivative
\begin{equation}
    \Fedosov := \dR + \tfrac1\hbar\,[A,-]_\ast
    \equiv -\delta + \nabla + \tfrac1\hbar[\Completion,-]_\ast\,,
\end{equation}
defines a \emph{differential}, i.e. squares to zero,
on the Weyl bundle. The $1$-form $\Completion$
can be computed order by order in $y$ via the recursive formulae
\begin{equation}\label{eq:recA}
    \Completion_{(2)} = h(R^\nabla)
    \qquad\text{and}\qquad
    \Completion_{(k+1)} = h\Big(\nabla \Completion_{(k)}
    + \tfrac1{2\hbar}\,\sum_{l=2}^{k-1}
    [\Completion_{(l)}, \Completion_{(k+1-l)}]_\ast\Big)
    \quad\text{for}\quad k\geq2\,,
\end{equation}
which yield
\begin{equation}
    \begin{aligned}
        \Completion
        & = - \tfrac13\,\dR x^\mu\,R_{\mu a}{}^c{}_b\,y^a y^b p_c
        - \tfrac1{12}\,\dR x^\mu\,\nabla_a R_{\mu b}{}^d{}_c\,y^a y^b y^c p_d \\
        & \hspace{50pt}
        -\dR x^\mu\,\big[\tfrac1{60}\,\nabla_a \nabla_b R_{\mu c}{}^e{}_d
        + \tfrac2{45}\,R_{\times a}{}^e{}_b\,
        R_{\mu c}{}^\times{}_d\big]\,y^a y^b y^c y^d p_e + (\cdots)
    \end{aligned}
\end{equation}
where the dots denote terms of higher order in $y$.%
\footnote{Remark that the grading \eqref{eq:deg}
with respect to which the defining recursion relation
for $\Completion$ is given, reduces to the degree
of homogeneity in $y$. This is a consequence of the fact
that the first correction $\Completion_{(2)}$ is linear in $p$
so that, not only all higher order correction stay linear in $p$, 
but also the star-commutator in \eqref{eq:recA} reduces
to the Poisson bracket piece, i.e. to its piece
of order $\hbar^1$. Consequently, no $\hbar$ correction
appear in $\Completion$ in the case of interest here.}
Introducing the notation
\begin{equation}
    \Curvature := hR^\nabla\,,
    \qquad\text{and}\qquad 
    \SymDer := h\nabla\,,
\end{equation}
we can re-sum the defining relations of $\Completion$ as
\begin{equation}\label{eq:rec_A}
    \Completion = \Curvature + \SymDer \Completion
    + \tfrac1{2\hbar}\,h[\Completion,\Completion]_\ast\,,
\end{equation}
so that the first few orders of $\Completion$ in $y$
can be re-written as
\begin{equation}
    \Completion_{(2)} = \Curvature\,,
    \quad \Completion_{(3)} = \SymDer\Curvature\,,
    \quad \Completion_{(4)} = \SymDer^2\Curvature
    + \tfrac1{2\hbar}\,h\big[\Curvature,\Curvature\big]_\ast\,.
\end{equation}

As mentioned above, any $1$-form connection
valued in the Weyl algebra whose component along $p_a$
is an invertible vielbein can be extended to a flat connection
by the same mechanism as above: the vielbein piece
gives rise to the differential $\delta$, and the components
of the $1$-form valued in the Weyl bundle needed to flatten
the original connection can be computed recursively
using its contracting homotopy $h$. In particular,
one may start from a connection containing higher spin
components which appear as terms of higher order in $p$
(and $y$) in the initial data, e.g.
\begin{equation}\label{hsconnec}
    A_0 = e^a\,p_a + \omega^{a,b}\,p_a\,y_b
    + e^{ab}\,p_a p_b + \omega^{ab,c}\,p_a p_b\,y_c + \dots\,,
\end{equation}
and find $\Completion$ so that $A=A_0+\Completion$
is flat, though the $1$-form $\Completion$ will also involve
the curvature of these higher spin components.%
\footnote{Note that in the case where the initial data
$A_0$ contain higher spin components (higher orders in $p$),
one should use a slightly different degree, namely
one should assign degree $1$ to both $y$ and $p$
and degree $2$ to $\hbar$, so that the Moyal--Weyl product
remains of degree $0$ with respect to this new grading.
This is actually the gradation used originally by Fedosov
\cite{Fedosov:1994zz, Fedosov:1996}, for more details
see also, e.g., \cite[App. E]{Basile:2022nou}.} The higher components of \eqref{hsconnec} correspond to vielbeins and spin-connections of conformal higher spin fields within the frame-like formulation, which was developed in \cite{Fradkin:1989md,Vasiliev:2009ck}.

\paragraph{Lift of symbols and invariant trace.}
Once the Fedosov connection $\Fedosov$ is constructed,
we can define the lift of the symbol
of a differential operator on $\Base$, that is a function
on the cotangent bundle $T^*\Base$, say
$f(x,p) \in \Functions_{pol.}(T^*\Base) \cong \Gamma(ST\Base)$,
as the (unique) section
$F(x;y,p) \in \Gamma(\Base,\WeylBundle_\Base)$ verifying
\begin{equation}
    \Fedosov F = 0\,, 
    \qquad 
    F\rvert_{y=0} = f\,,
\end{equation}
i.e. the (unique) covariantly constant section
of the Weyl bundle whose order $0$ in $y$ is $f$.
In other words, starting from a function
only of $x^\mu$ and $p_a$, one reconstruct
a flat section of the Weyl bundle, which is a function
of $x^\mu$, $p_a$ and $y^a$, whose dependency on $y$
is completely determined by the covariant constancy condition,
and the coefficients of these terms proportional to $y$
are obtained from the original function of $x$ and $p$.
To do so, one simply needs to solve the covariant
constancy condition, which can be done iteratively via
\begin{equation}
    \label{eq:rec_F}
    F_{(0)} = f
    \qquad\text{and}\qquad
    F_{(k+1)} = h\Big(\nabla F_{(k)} + \tfrac1\hbar\,
    \sum_{l=2}^{k+1} [\Completion_{(l)}, F_{(k+1-l)}]_\ast\Big)
    \quad\text{for}\quad k \geq 0\,,
\end{equation}
where $F_{(n)}$ denotes the component of the lift $F$ 
homogeneous of degree $n$ with respect to grading \eqref{eq:deg},
i.e. it corresponds to the homogeneity both in $y$ and $\hbar$.
This leads to
\begin{equation}\label{eq:lift_y2}
    F(x;y,p) = f + y^a\,\nabla_a f + \tfrac12\,y^a y^b\,
    \big(\nabla_a \nabla_b + \tfrac13\,R_{da}{}^c{}_b\,
    p_c\,\tfrac{\partial}{\partial p_d}\big)\,f
    + (\dots)\,,
\end{equation}
at the first few orders. This lift of (fiberwise polynomial) 
functions establishes a bijection between the latter
and covariantly constant sections of the Weyl bundle, 
\begin{equation}
    \begin{aligned}
        \tau: \Functions(T^*\Base)\
        & \overset{\sim}{\longrightarrow}\
        {\rm Ker}(\Fedosov) \subset \Gamma(\WeylBundle_\Base) \\
        f(x,p)\,\ & \longmapsto\ F(x;y,p)
                                \equiv \tau(f)(x;y,p)\,,
    \end{aligned}
\end{equation}
and allows us to define a \emph{star-product},
i.e. an associative but non-commutative deformation
of the pointwise product, via the simple formula%
\footnote{Remark that, by construction, the evaluation
of a covariantly constant section at $y=0$ yields
the function on the cotangent bundle that it is the lift of.
In other words, this simple operation is the inverse
of the lift $\tau$, i.e. $\tau^{-1}(-) = (-)\rvert_{y=0}$.
The star-product on $T^*\Base$ can therefore be written as
$f \star g = \tau^{-1}\big(\tau(f) \ast \tau(f)\big)$
which makes it clear that the lift $\tau$ is a morphism
of algebras between $\big({\rm Ker}(\Fedosov), \ast\big)$
and $\big(\Functions(T^*\Base), \star\big)$,
the star-product on the latter being `pulled-back'
from the Moyal--Weyl one defined fiberwise.}
\begin{equation}\label{eq:star-product}
    f \star g = (F \ast G)\big|_{y=0}\,,
    \qquad 
    f, g \in \Functions_{pol}(T^*\Base)\,,
\end{equation}
where $F, G \in \Gamma(\WeylBundle_\Base)$ are the lifts
of $f$ and $g$ respectively. Associativity simply follows
from the fact that the Moyal--Weyl product in the fiber
is itself associative. To summarize, we are able to define
a star-product on the cotangent bundle of our spacetime 
$T^*\Base$ thanks to the fact that any function can be lifted 
to a flat section of the Weyl bundle, wherein we can use
the Moyal--Weyl star-product to multiply the flat sections 
corresponding to two functions on $\Base$, and evaluate
the result at $y=0$ thereby producing another function
on $T^*\Base$. 

\paragraph{Invariant Trace:} There exists a trace (essentially unique) on the space
of covariantly constant sections of the Weyl bundle,
which takes the form \cite{Basile:2022nou}
\begin{equation}\label{FFS trace}
    \Tr_A(F) = \int_{x \in \Base} \int_{T^*_x\Base} \dR^Dp\ 
    \mu(F|\underbrace{A,\cdots,A}_{D\,\text{times}})\,,
\end{equation}
where $\mu: \WeylAlg_{2D}^{\wedge D} \otimes \WeylAlg_{2D}
\longrightarrow \R[p_a]$ is a multilinear map
valued in polynomials in $p$,
obtained from the Feigin--Felder--Shoikhet cocycle
\cite{Feigin:2005}. The fact that $\mu$ is obtained
from a Hochschild cocycle for the Weyl algebra ensures
two important properties of this trace: it is invariant
under the gauge transformations
\begin{equation}
    \delta_\xi A = \dR\xi + \tfrac1\hbar\,[A,\xi]_\ast\,,
    \qquad 
    \xi \in \Gamma(\WeylBundle_\Base)\,,
\end{equation}
of the flat connection $A$ up to boundary terms, i.e.
\begin{equation}\label{var of trace}
    \delta_\xi \Tr_A(F) = \int_\Base \int \dR^Dp\
    \big[\dR (\dots) + \tfrac{\partial}{\partial p_a} (\dots)_a\big]\,,
\end{equation}
and it is cyclic, also up to boundary terms, 
\begin{equation}
    \Tr_A([F, G]_\ast) = \int_\Base \int \dR^Dp\
    \big[\dR (\dots) + \tfrac{\partial}{\partial p_a} (\dots)_a\big]\,,
\end{equation}
for any covariantly constant sections $F$ and $G$.

The detailed expression for $\mu$
is given in Appendix \ref{app:FFS}, for the moment
it is enough for our purpose to know that,
for a flat connection $A$ which is linear in $p$
as the example reviewed previously,
the associated trace of any lifted symbol $F$
boils down to
\begin{equation}\label{eq:trace_linear_case}
    \Tr_A(F) = \int_\Base \dR^Dx\,|e|\,
    \int_{T^*_x\Base} \dR^Dp\ \sum_{k\geq0} 
    \mu^{\scriptscriptstyle\nabla}_{a_1 \dots a_k}(x)\,
    \frac{\partial^k}{\partial p_{a_1} \dots \partial p_{a_k}}
    F\big|_{y=0}\,,
\end{equation}
where $\mu^{\scriptscriptstyle\nabla}_{a_1 \dots a_k}(x)$
are polynomials in the curvature of $\nabla$
and covariant derivatives thereof. 

What has been reviewed above is just the Fedosov approach
to deformation quantization for the particular case
of the symplectic manifold being the cotangent bundle
(of the spacetime), which was also studied by Fedosov himself 
\cite{Fedosov2001}. A development since \cite{Fedosov2001}
is the construction of the invariant trace
by Feigin, Felder and Shoikhet \cite{Feigin:2005}.
Let us briefly explain now, see \cite{Basile:2022nou}
for more details, how this is related to conformal
higher spin fields. To begin with, an off-shell description
of conformal higher spin fields requires
the Fedosov connection $A$ and a covariantly constant section
of the Weyl bundle $F$. Different types of scalar matter,
i.e. whether we start out with $\mathcal{L}\sim\phi \square \phi$ 
or $\mathcal{L}\sim\phi \square^k \phi$, $k>1$,
lead to different spectra of (higher spin) currents and,
hence, to different spectra of sources/background
(higher spin) fields. An immediate consequence
is that it is necessary to fix the background value for $F$
to land on a specific theory. We will consider $F$ of the form
\begin{align}
    F & = p^2 + \sum_{s>2} h^{a_1 \dots a_s}(x)\,
    p_{a_1} \dots p_{a_s} + \dots\,,
\end{align}
where the presence of $p^2$ here implies that we are coupling
the usual free scalar field $\mathcal{L}\sim\phi \square \phi$
to (higher spin) background fields $h^{a_1 \dots a_s}$.
The formalism is flexible enough to allow one
to realize conformal higher spin fields both in the frame-like, 
cf. \eqref{hsconnec}, and in the metric-like ways, as below.
In this chapter, we prefer to keep $A$ purely gravitational,
i.e. it is completely expressed in terms of a vielbein $e^a$. 
With the help of the $\xi$ gauge symmetry one can move
between the frame-like and metric-like formulations.

\paragraph{Fock space bundle.}
Having constructed a bundle of Weyl algebra, let us now 
proceed with the definition of a vector bundle associated
with the Fock representation. As a vector space, the latter
can be identified with the subspace of 
$\WeylAlg_{2D} \cong \R[y^a, p_b]$ consisting of 
polynomials (or even formal power series) in $y$,
that we shall denote by $\Fock_D \equiv \R[y^a]$.
The representation is given by the \emph{quantization map},
\begin{equation}
    \big(\rho(f) \varphi\big)(y)
    = f(y,p)\,\exp\Big(\!-\hbar\,
    \tfrac{\overleftarrow{\partial}}{\partial p} \cdot 
    \big[\tfrac12\,\tfrac{\overleftarrow{\partial}}{\partial y}
    +\tfrac{\overrightarrow{\partial}}{\partial y}\big]\Big)\,
    \varphi(y)\big|_{p=0}\,,
\end{equation}
for any element $f(y,p) \in \WeylAlg_{2D}$ of the Weyl algebra 
and $\varphi(y) \in \Fock_n$ of the Fock space. That it defines
a representation of the Weyl algebra means that it verifies
\begin{equation}
    \rho(f) \circ \rho(g) = \rho(f \ast g)\,,
    \qquad 
    f, g \in \WeylAlg_{2D}\,.
\end{equation}
The name `quantization map' comes from the fact
that it allows one to associate, to any (polynomial)
function of $\R^{2D} \cong T^*\R^D$, which are nothing but
elements of the Weyl algebra, a differential operator 
acting on the space of `wave functions',
i.e. smooth functions on $\R^D$, which we consider
as elements of the Fock space (via for instance
their Taylor series). Put differently, the pair
$(\WeylBundle_{2D}, \Fock_D)$ can be thought of
as a \emph{flat model} for the quantization
of a cotangent bundle $T^*\Base$ with $\dim\Base=n$,
wherein the Weyl algebra models the algebra of functions
on $T^*\Base$, while the Fock space models smooth functions
on the base manifold $\Base$, on which functions
on the cotangent bundle act as a differential operators.

Given now an arbitrary smooth manifold $\Base$,
one can consider the `bundle of Fock spaces'
defined as
\begin{equation}
    \Fock_\Base := S(T^*\Base)
                    \twoheadrightarrow \Base\,,
\end{equation}
whose sections are
\begin{equation}
    \Gamma(\Fock_\Base)\, \ni\, \Phi(x;y)
    = \sum_{k\geq0} \tfrac1{n!}\,\Phi_{a_1 \dots a_k}(x)\,
    y^{a_1} \dots y^{a_k}\,,
\end{equation}
that we shall extend as formal power series in $\hbar$.
A Fedosov connection $A$ defines a flat covariant derivative
on this Fock bundle, whose local expression is
\begin{equation}
    \Fedosov = \dR + \tfrac1\hbar\,\rho(A)\,,
\end{equation}
with $\rho$ is the quantization map above.
A simple computation leads to
\begin{equation}
    \rho(p_a) = -\hbar\,\tfrac{\partial}{\partial y^a}\,,
    \qquad 
    \rho(y^a p_b) = -\tfrac\hbar2\,
    \big(y^a\,\tfrac{\partial}{\partial y^b} + \tfrac{\partial}{\partial y^b}\,y^a\big)
    = -\hbar\,\big(y^a\tfrac{\partial}{\partial y^b}
        + \tfrac12\,\delta^a_b\big)\,,
\end{equation}
and more generally, 
\begin{equation}
    -\tfrac1\hbar\,\rho(y^{a_1} \dots y^{a_n} p_b)
    = y^{a_1} \dots y^{a_n}\,\tfrac{\partial}{\partial y^b}
    + \tfrac{n}2\,\delta_b^{(a_1} y^{a_2} \dots y^{a_n)}\,,
\end{equation}
so that, upon choosing $\nabla$ to be a metric connection,
and $A$ the flat connection \eqref{eq:Fedosov} built from it as explained above, one finds
\begin{equation}
    \Fedosov\Phi = \Big(\!-\delta + \nabla
                    + \rho(\Completion)\Big)\Phi\,,
    \qquad\text{with}\qquad 
    \delta = e^a\,\tfrac{\partial}{\partial y^a}\,,
\end{equation}
which in particular, contains the \emph{same acyclic piece}
$\delta$ as in the Fedosov connection \eqref{eq:Fedosov},
which is an operator of degree $n-1$ in $y$. As a consequence,
we can solve for covariantly constant sections
of the Fock bundle in a similar manner as we did 
in the Weyl bundle: expanding the condition
\begin{equation}
    \Fedosov\Phi = 0\,,
    \qquad\text{with}\qquad 
    \Phi\big|_{y=0} = \phi\,,
\end{equation}
order by order in $y$ yields
\begin{equation}
    \delta\Phi_{(n+1)} = \nabla\Phi_{(n)}
    + \tfrac1\hbar\,\sum_{k=2}^{n+1} \rho(\Completion_{(k)})
    \Phi_{(n+1-k)}\,,
\end{equation}
which gives us a definition of the order $n+1$ term
in the $y$ expansion of $\Phi$ thanks to the contracting
homotopy \eqref{eq:defh&N} introduced before, i.e.
\begin{equation}\label{eq:rec_phi}
    \Phi_{(n+1)} = h\Big(\nabla\Phi_{(n)}
    + \tfrac1\hbar\,\sum_{k=2}^{n+1} \rho(\Completion_{(k)})
    \Phi_{(n+1-k)}\Big)\,.
\end{equation}
The whole covariant section $\ket\Phi$ only depends
on its value at $y=0$, which is a function on $\Base$,
thereby establishing a bijection
\begin{equation}
    \begin{aligned}
        \tau: \Functions(\Base)\
        & \overset{\sim}{\longrightarrow}\
        {\rm Ker}(\Fedosov) \subset \Gamma(\Fock_\Base) \\
        \phi(x)\,\ & \longmapsto\ \Phi(x;y)
                                \equiv \tau(\phi)(x;y)\,,
    \end{aligned}
\end{equation}
between $\Functions(\Base)$ and covariantly constant sections 
of the Fock bundle (that we denoted by the same symbol $\tau$
as the isomorphism between functions on the cotangent bundle
and flat sections of Weyl bundle,
in a slight abuse of notation).
The first few order of the covariantly constant section
associated with $\phi(x)$ read
\begin{equation}\label{eq:lift_phi}
    \Phi(x;y) = \phi + y^a\nabla_a\phi
    + \tfrac12\,y^a y^b\,\big(\nabla_a\nabla_b
    - \tfrac16\,R_{ab}\big)\phi + \dots
\end{equation}
where $R_{ab}$ denotes the Ricci tensor of $\nabla$,
and the dots denote terms of order $3$ or higher in $y$.

We can now define a quantization map in this curved setting,
that is to say, a way to associate to any symbol
$f \in \Functions_{pol}(T^*\Base)$, that is any fiberwise 
polynomial function on the cotangent bundle of $\Base$,
a differential operator $\widehat f$
which acts on `wave functions', i.e. functions
$\phi\in\Functions(\Base)$ on the base, defined as follows%
\footnote{Let us note that, as for the star-product,
writing the quantization map in terms of the lift $\tau$,
namely $\widehat{f}\phi
= \tau^{-1}\big[\rho\big(\tau(f)\big)\tau(\phi)\big]$,
makes it apparent that the latter defines a morphism
of pairs algebra-module between
$\big(\Functions(T^*\Base), \Functions(\Base)\big)$
and flat sections of the Weyl and Fock bundles.
This also shows that the quantization map
on $\Functions(\Base)$ is `pulled-back' from that 
on flat sections of the Fock bundle, in complete parallel
with the definition of the star-product on $T^*\Base$.}
\begin{equation}
    \big(\widehat f\phi\big)(x)
        := \rho(F) \Phi \big|_{y=0}\,,
\end{equation}
where $F \in \Gamma(\WeylBundle_\Base)$
and $\Phi \in \Gamma(\Fock_\Base)$ are the lifts
of $f$ and $\phi$ respectively as a covariantly constant
section of the Weyl and Fock bundles. This defines
a representation of the star-product algebra
$\big(\Functions_{pol}(T^*\Base), \star\big)$
on the space of `wave functions' $\Functions(\Base)$, i.e.
\begin{equation}
    \widehat f \circ \widehat g = \widehat{f \star g}
    \qquad
    \forall f, g \in \Functions_{pol}(T^*\Base)\,,
\end{equation}
where $\star$ is the star-product defined
in \eqref{eq:star-product}. Here again, this is
simply a consequence of the fact that $(\Fock_n,\rho)$
is a representation of $(\WeylAlg_{2D},\ast)$,
i.e. we sort of `pullback' the algebra and representation
structure from the fiber to the base manifold $\Base$.
Note that this approach was already outlined in
\cite[App. A]{Grigoriev:2016bzl}.

\subsection{Covariant Action for CHS/Parent Formulation}

We now have all the ingredients needed to express the action of CHS in a covariant manner. The CHS fields are encoded in a pair of fields,
namely a flat connection $A$ and a covariantly constant 
section $F$ of the Weyl bundle \cite{Grigoriev:2006tt, Grigoriev:2016bzl},
\begin{equation}
    \dR A + \tfrac1{2\hbar}\,[A,A]_\ast = 0\,,
    \qquad
    \dR F + \tfrac1\hbar\,[A,F]_\ast = 0\,,
\end{equation}
which is invariant under the gauge transformations
\begin{equation}
    \delta_\xi A = \dR\xi + \tfrac1\hbar\,[A,\xi]_\ast\,,
    \qquad 
    \delta_{\xi,w} F = \tfrac1\hbar\,[F,\xi]_\ast
    + \{F,w\}_\ast\,,
    \qquad 
    \dR w + \tfrac1\hbar\,[A,w]_\ast {=} 0\,,
\end{equation}
where $\xi,w \in \Gamma(\WeylBundle_\Base)$
are $0$-form valued in the Weyl bundle,
with $w$ required to be covariantly constant,
while $\xi$ is unconstrained. The sum
$\varepsilon=\tfrac1\hbar\xi+w$ corresponds
to the symbol of an arbitrary differential operator,
such as $\widehat{\varepsilon}$ appearing in \eqref{eq:Segal},
and it splits into its Hermitian and anti-Hermitian part,
respectively $w$ and $\xi$ (though both are real, 
the latter is dressed with $\hbar$ that we take as imaginary
in the sense that $\hbar^\dagger=-\hbar$)\footnote{Note that we reuse the symbols \(\xi\), \(\omega\), and \(F\) from Segal’s formulation; here, however, they denote the lifts of the base–space (spacetime) symbols, whereas in Segal’s approach they are defined directly on the base.
}.
Such a description is obtained from an approach
known as the `parent formulation' of gauge theories,
developed in \cite{Barnich:2004cr, Barnich:2010sw, Grigoriev:2010ic, Grigoriev:2012xg} and references therein.
Now using the trace described in the previous section in (\ref{FFS trace}) we can write a covariant CHS action as,
\begin{equation}
    S_{CHS}[A,F]=\Tr_A\big(l_{*}(F)\big) = \int_{x \in \Base} \int_{T^*_x\Base} \dR^Dp\ 
    \mu(l_{*}(F)|\underbrace{A,\cdots,A}_{D\,\text{times}})\,.
\end{equation}
where $\ell_{*}(F)$ is some star-function\footnote{Any such star function can be expanded as a formal power series in $\hbar$, of the form
\[
f^{\ast}(a)
= \sum_{n=0}^{\infty} \hbar^{2n} \sum_{k=2}^{2n}
f^{(k)}(a)\, p_{n,k}(a).
\]
where $f^(k)$ denotes the kth derivative of $f$ and $p_{n,k}(a)$ are monomials of $k$ in the first $4n$ derivatives of $a$ with respect to the variables of the Weyl algebra.} of $F$. Now the gauge variation of this action comes from two sources: the variation of $A$ with respect to $\xi$ and the variation of $F$ under both $\xi$, $w$. As mentioned in (\ref{var of trace}), the trace is already invariant under the gauge transformations $\delta_\xi A = \dR\xi + \tfrac1\hbar\,[A,\xi]_\ast$ of the flat connection $A$. Also, because of the cyclicity of trace, the variation coming from the commutator part of $\delta F$ vanishes, and we are only left with the variation of action as,
\begin{equation}
    \delta_{\xi,w}S=\int_{x}\int_{p-\text{fibre}} 2\mu(l'_{*}(F)\ast F\ast w|\underbrace{A,\cdots,A}_{D\,\text{times}})
\end{equation}
As we have seen in the Segal case, we can set $l_{*}(F)=\Theta_{\ast}(F)$ to ensure the invariance of the action under higher spin Weyl transformation.
Note that to get the action on the spacetime (base) we have set $y=0$ and we get the spacetime action in the form 
\begin{equation}
    S^{CHS}_{spacetime}=\int_{x \in \Base} \int_{T^*_x\Base} \dR^Dp\ 
    \mu(l_{*}(F)|\underbrace{A,\cdots,A}_{D\,\text{times}})|_{y=0}
\end{equation}
Indeed if we take flat space then $e^{a}_{\mu}=\delta^{a}_{\mu}$ and our flat connection becomes $A=dx^{\mu}p_{\mu}$ globally and with $F|_{y=0}=f$, the action reduces to Segal action 
\[
S[f]=\int_{flat}\Theta_{\star}(f)
\]

In the next sections, we will see how matter can be coupled with CHS in a covariant manner using similar tools as we used in this section.
\section{Wigner function and quadratic actions}
\label{sec:Wigner}

As usual, when dealing with gauge theories, matter fields
consist of sections of vector bundles associated with a
representation of the gauge algebra (meaning here,
the algebra in which gauge fields take values).
Accordingly, we add the scalar field $\phi$ 
to the previous system in the guise of its lift
as a covariantly constant section of the Fock bundle,
\begin{equation}
    \dR\Phi + \tfrac1\hbar\,\rho(A)\Phi = 0\,,
\end{equation}
which transforms in the corresponding representation, 
\begin{equation}
    \delta_{\xi,w} \Phi = -\rho(\tfrac1\hbar\,\xi + w)\Phi\,,
\end{equation}
thereby preserving the covariant constancy condition.
Now all we need is an action functional implementing
the coupling of $\phi$ to the higher spin background
in a gauge-invariant manner.

Around flat space, Segal's approach consisted in
considering a quadratic action for a \emph{complex}
scalar field $\phi$ in flat spacetime, 
\begin{equation}
    S[\phi] = \int_{\R^D} \dR^Dx\ \phi^*(x)\,
                            (\widehat H\phi)(x)\,,
\end{equation}
for some differential operator $\widehat{H}$
which encodes the coupling of $\phi$ to a background 
of gauge fields, the latter being related
to the `coefficients' of this operator. For instance,
in the case of the conformally-coupled scalar,
$\widehat{H}$ would be the conformal Laplacian
whose expression depends on a metric $g$ (via its inverse
contracting two covariant derivatives,
and via the Ricci scalar term), and which implements
the coupling of $\phi$ to conformal gravity.
The above action can formally be written as
\begin{equation}
    S[\phi] = \bra{\phi} \widehat{H} \ket{\phi}\,,
\end{equation}
so that it becomes relatively simple to see that
it is invariant under the following
infinitesimal transformations
\begin{equation}
    \delta_{\varepsilon}\ket{\phi}
    = -\widehat \varepsilon\ket{\phi}\,,
    \qquad
    \delta_{\varepsilon} \widehat H
    = \widehat{\varepsilon}^{\,\dagger} \circ \widehat{H}
    + \widehat{H} \circ \widehat{\varepsilon}\,,
\end{equation}
where $\varepsilon$ is another, arbitrary,
differential operator. Assuming that the space of operators
we are working with possesses a \emph{trace},
we can further re-write the action as
\begin{equation}
    S[\phi] = \Tr(\widehat{H} \circ \ket{\phi}\!\bra{\phi})\,,
\end{equation}
that is the trace of the operator $\widehat H$ composed
with the projector $\ket{\phi}\!\bra{\phi}$. In this form,
the action can be more easily translated in terms
of symbols, leading to
\begin{equation}
    S[\phi] = \int_{T^*\R^D} \dR^D p\,\dR^D x\
    \big(H \star W_{\phi}\big)(x,p)
\end{equation}
where $H(x,p)$ and $W_{\phi}(x,p)$ are the symbols
of the kinetic operator $\widehat{H}$ and the projector
$\ket{\phi}\!\bra{\phi}$, also known as the Wigner function,
respectively. The integration over the cotangent bundle
$T^*\R^D$ defines a trace over the space of symbols,
at least those which are compactly supported
or vanish at infinity sufficiently fast.
Indeed, in this case one finds
\begin{equation}
    \Tr(f \star g) = \int_{T^*\R^D} \dR^D x\,\dR^D p\
    (f \star g)(x,p) = \int_{T^*\R^D} \dR^D x\,\dR^D p\
    f(x,p)\,g(x,p) = \Tr(g \star f)\,,
\end{equation}
for any symbols $f$ and $g$, since all higher order terms
in the star product are total derivatives on $T^*\R^D$,
and hence can be ignored for the aforementioned
suitable class of symbols. The transformation rule,
in terms of symbols, becomes 
\begin{equation}
     \delta_{\varepsilon} H
     = \varepsilon^{\dagger} \star H + H \star \varepsilon\,,
    \qquad\text{and}\qquad
    \delta_{\varepsilon} W_{\phi} = -\varepsilon \star W_{\phi}
    - W_{\phi} \star \varepsilon^{\dagger}\,,
\end{equation}
under which the action transform as 
\begin{align}
    \delta_{\varepsilon} S[\phi]
    = -\Tr\big([H \star W_{\phi}\,,\varepsilon^{\dag}]_{\star}\big)
    = 0\,,
\end{align}
i.e. the action is left invariant as a consequence
of the cyclicity of the trace.

We have seen in the previous section how to define
a star-product and construct the associated invariant
trace via the FFS cocycle for any, possibly curved,
manifold $\Base$ so that we only need to find a suitable
generalization of the Wigner function to curved settings.
One can think of the Wigner function as a bilinear map
\begin{equation}
    W: \Fock_D \otimes \Fock_D \longrightarrow \WeylAlg_{2D}\,,
\end{equation}
taking two elements of the Fock representation
and constructing an element of the Weyl algebra
out of them. For our purpose, what matters is that
it possesses the following couple of properties
(whose proofs are recalled in Appendix \ref{app:Weyl_calculus}).
\begin{enumerate}[label=$(\roman*)$]
\item\label{item:left-right_Wigner}
First, it intertwines the left and right multiplication
in the Weyl algebra with the Fock action
\begin{equation}
    F \ast W[\Phi,\Psi] = W[\rho(F)\Phi, \Psi]\,,
    \qquad 
    W[\Phi, \Psi] \ast F^\dagger = W[\Phi, \rho(F)\Psi]\,,
\end{equation}
for any element $F(y,p) \in \WeylAlg_{2D}$
and any pair of Fock space states
$\Phi(y), \Psi(y) \in \Fock_n$.

\item\label{item:integral_Wigner}
Second, integrating it over momenta yields
\begin{equation}
    \int_{\R^D} \dR^Dp\
    \tfrac{\partial^k}{\partial p_{a_1} \dots \partial p_{a_k}}\,
    W[\Phi,\Psi] = \delta_{k,0}\,\Phi(y)\,\Psi(y)\,,
\end{equation}
for any Fock space elements $\Phi,\Psi \in \Fock_n$ 
which are seen as embedded in the Weyl algebra
on the right hand side. 
\end{enumerate}

A first naive guess for a curved version $\Wigner_\phi$
of the Wigner function associated with a scalar field
$\phi \in\Functions(\Base)$ is to simply apply
the above bilinear map to two copies of its lift
as covariantly constant sections of the Fock bundle,
i.e.
\begin{equation}
    \Wigner_\phi(x;y,p) := W[\Phi,\Phi]
    = \int \dR^Du\ e^{\frac1\hbar\,p \cdot u}\,
    \Phi(x;y+\tfrac12\,u)\,\Phi^\dagger(x;y-\tfrac12\,u)\,.
\end{equation}
First of all, let us note that this is a covariantly
constant section of the Weyl bundle. Indeed, upon writing
it as $\Wigner_\phi = W[\Phi,\Phi]$ in order to highlight
the fact that it is bilinear in the covariantly constant
section of the Fock bundle $\Phi$, one finds that it verifies
\begin{equation}
    \tfrac1\hbar\,[A, \Wigner_\phi]_\ast
    = \tfrac1\hbar A \ast W[\Phi,\Phi]
    + W[\Phi,\Phi] \ast (\tfrac1\hbar\,A)^\dagger
    = W[\rho\big(\tfrac1\hbar\,A)\Phi,\Phi]
    + W[\Phi,\rho(\tfrac1\hbar\,A)\Phi]\,,
\end{equation}
where we used the properties in \ref{item:left-right_Wigner}.
We can then use the covariant constancy of $\Phi$,
to show that
\begin{equation}
    \dR\Wigner_\phi + \tfrac1\hbar\,[A,\Wigner_\phi]_\ast = 0\,,
\end{equation}
i.e. our curved version of the Wigner function $\Wigner_\phi$
is a covariantly constant section of the Weyl bundle.
Moreover, property \ref{item:left-right_Wigner}
also ensure that $\Wigner_\phi$ transforms as
\begin{equation}
    \delta_{\xi,w} \Wigner_\phi
    = \tfrac1\hbar\,[\Wigner_\phi, \xi]_\ast
    - \{\Wigner_\phi,w\}_\ast\,.
\end{equation}
which implies that its star-product with the covariantly constant lift $F$ behaves as
\begin{equation}
    \delta_{\xi,w}(F \ast \Wigner_\phi)
    = [F \ast \Wigner_\phi, \tfrac1\hbar\,\xi - w]_\ast\,,
\end{equation}
under the gauge transformations of the system.
As a consequence, the functional%
\footnote{Note that the dependence on conformal higher spin fields
in the action \eqref{main} is a little subtle to read-off:
as explained in \cite{Basile:2022nou}, they can be moved around
between $A$ and $F$ via gauge transformations,
which can therefore encode field redefinitions.}
\begin{equation}\label{main}
    S[\phi] = \Tr_A(F \ast \Wigner_\phi)\,,
\end{equation}
is well-defined, being the trace of the star-product
of two covariantly constant sections of the Weyl bundle,
as well as gauge invariant under all transformations
listed above, thanks to the cyclicity of the FFS trace,
which holds \emph{up to boundary terms}. Let us remark that,
contrary to the action for CHS gravity, which is expressed
as the FFS trace of a symbol that dies off at infinity
\emph{both} in spacetime and in the fiber/momenta directions 
\cite{Basile:2022nou}, this is not necessarily the case here:
the $p$-dependency of the integrand may not allow us to discard
boundary terms for arbitrary gauge parameters. In other words,
we expect that the gauge parameters $\xi$ and $w$ should be
restricted so as to ensure that the boundary terms appearing
when checking the cyclicity/gauge invariance of the FFS trace
(see \cite[App. C]{Basile:2022nou}) can actually be neglected.
Modulo this subtlety, eq. \eqref{main} gives a manifestly covariant 
and higher spin invariant form of a coupling
between the scalar field and a background
of conformal higher spin fields, which is one of the main results
of this chapter. On the other hand, irrespectively of the action principle, the equations of motion $\rho(F) \Phi \big|_{y=0}=0$ are well-defined and, in particular, gauge invariant. As it turns out, in the case where $A$ is linear
in $p$, this expression simplifies to
\begin{equation}
    S[\phi] = \int_\Base \dR^Dx\,|e|\,\int_{T^*_x\Base} \dR^Dp\
    W[\rho(F)\Phi,\Phi]\big|_{y=0}
    = \int_\Base \dR^Dx\,|e|\,\phi^*(x)\,(\widehat f\phi)(x)\,,
\end{equation}
as a consequence of the properties \ref{item:left-right_Wigner}
and \ref{item:integral_Wigner} of the Wigner function,
and the fact that the trace takes the form 
\eqref{eq:trace_linear_case}.

\section{Conformally-coupled scalar and higher spins}
\label{sec:HS}
Let us give two examples to show how the formalism and the action \eqref{main} can reproduce what it has to, e.g. the coupling to low-spin background fields and to higher spin background. The latter problem was studied in $d=4$ for a coupling to a spin-three field in \cite{Beccaria:2017nco}.

\subsection{Conformally Invariant Laplacian}
As an illustration, let us show how we can recover 
the conformally-coupled scalar. This boils down to
identifying the symbol of the conformal Laplacian,
\begin{equation}
    \nabla^2 - \tfrac{D-2}{4(D-1)}\,R\,,
\end{equation}
which we can do in a couple of ways: either by
working out its quantization, or by imposing that
it transforms correctly under the above gauge transformations.

Let us start with the former. Considering that
the quantization map yields $\widehat{p}_a=-\hbar\nabla_a$,
we should consider the Ansatz $f=p^2+\alpha\,R$
for the symbol of the conformal Laplacian, where $\alpha$
is a numerical coefficient to be fixed. It is then enough
to compute the lift of this symbol, up to order $2$ in $y$,
\begin{align}
    F=\tau(p^2 + \alpha\,R)
    = p^2 + \tfrac13\,y^a y^b\,R_a{}^c{}_b{}^d\,p_c p_d
    + \alpha\,\big(R + y^a\,\nabla_a R
    + \tfrac12\,y^a y^b\,\nabla_a\nabla_b R\big) + \dots
\end{align}
as well as that 
of the scalar field $\phi$ at order $2$ in $y$
given previously in \eqref{eq:lift_phi}, 
and use
\begin{equation}\label{eq:q_lem}
    \rho(p^2)\rvert_{y=0} = \hbar^2\,\partial_y^2\,,
    \qquad\qquad
    \rho(y^a y^b p_c p_d)\rvert_{y=0}
    = \tfrac{\hbar^2}{2}\,\delta^{(a}_c\,\delta^{b)}_d\,,
\end{equation}
to find that the quantization of the Ansatz $f$ reads
\begin{equation}
    \widehat{f}\phi = \hbar^2\,\big(\nabla^2
    +[\tfrac{\alpha}{\hbar^2} -\tfrac14]\,R\big)\phi\,,
\end{equation}
which implies
\begin{equation}
    \alpha = \frac{\hbar^2}{4(D-1)}
    \qquad\Longrightarrow\qquad
    f = p^2 + \tfrac{\hbar^2}{4(D-1)}\,R\,,
\end{equation}
upon imposing that it reproduces the conformal Laplacian.
Note that this computation also shows that, perhaps contrary
to one's intuition, the symbol of the ordinary Laplacian
is not $p^2$, but should instead be corrected by a curvature
dependent term $\tfrac{\hbar^2}{4}\,R$.

Let us now turn our attention to the symmetries
of our action, focusing on Weyl symmetry. 
Having constructed the $1$-form connection $A$
from a torsionless and metric connection, its coefficients
when expanded order by order in $y$ are tensors
built out of the vielbein and its derivatives only,
and hence have a definite behavior under Weyl transformations%
\footnote{For instance, recall that the spin-connection transforms as
$\delta_\sigma^{\text{Weyl}} \omega^{a,b}
    = 2\,e^{[a}\,\nabla^{b]}\sigma$, under Weyl rescaling.}
\begin{equation}
    \delta_\sigma^{\text{Weyl}} e^a = \sigma\,e^a\,.
\end{equation}
These Weyl transformations can be realized
as gauge symmetries of $A$, by suitably choosing
the gauge parameters
$\xi_{\text{Weyl}}, w_{\text{Weyl}} \in \Gamma(\WeylBundle_\Base)$.
In other words, we can embed the \emph{geometric}
transformations that are Weyl rescalings,
as \emph{gauge} transformations of the system of fields
$A$, $F$ and $\Phi$ (which are affected by both types
of parameters, $\xi$ and $w$).
To explicitly find the gauge parameter
$\xi_{\text{Weyl}}$, one needs to solve the condition
\begin{equation}
    \dR \xi_{\text{Weyl}} + \tfrac1\hbar\,[A,\xi_{\text{Weyl}}]_\ast
    \overset{!}{=} \delta_\sigma^{\Weyl} A\,,
\end{equation}
for $\xi_{\text{Weyl}}$ in terms of $\sigma$.
This can be done  as before, namely order by order in $y$,
using the contracting $h$. More precisely, 
for $\xi_{\text{Weyl}} = \sum_{k\geq1} \xi_{(k)}$
with $\xi_{(k)}$ of order $k$ in $y$ and linear in $p$,
one finds the recursion
\begin{equation}
    \xi_{(k+1)} = h\big(\nabla\xi_{(k)}
    + \tfrac1\hbar\,\sum_{l=2}^k [A_{(l)}, \xi_{(k+1-l)}]_\ast
    - \delta_\sigma^{\scriptstyle\rm Weyl} A_{(k)}\big)\,,
\end{equation}
which yields
\begin{equation}\label{eq:xi_Weyl}
    \xi_{\text{Weyl}} = -\sigma\,y \cdot p
    - \nabla_a \sigma\,(y^a\,y \cdot p - \tfrac12\,y^2\,p^a)
    -\tfrac13\,\nabla_a \nabla_b \sigma\,\big(y^a y^b\,y \cdot p
    -\tfrac12\,y^2\,y^a\,p^b) + \dots\,,
\end{equation}
where as usual, the dots denote higher order terms in $y$.
Now we can focus on the symbol of our differential operator, 
that we assume to be of the form $p^2+\alpha\,R$
for some coefficient $\alpha$ to be fixed by requiring
that, here again, Weyl transformation can be implemented
as gauge symmetries. In other words, we want to impose
\begin{equation}
    \Big(\tfrac1\hbar\,[F, \xi_{\text{Weyl}}]_\ast
    + \{F,w_{\text{Weyl}}\}_\ast\Big)\big|_{y=0}
    \overset{!}{=} \delta_\sigma^{\text{Weyl}}\big(p^2
    + \alpha\,R\big)\,, 
\end{equation}
with $F=\tau(p^2+\alpha\,R)$ its covariantly constant lift,
and where the gauge parameter $w_{\text{Weyl}}$ is assumed 
to be proportional to the lift of the Weyl parameter $\sigma$,
i.e.
\begin{equation}
    w_{\text{Weyl}} = \beta\,\tau(\sigma)
    \equiv \beta\,\sum_{k\geq0} \tfrac1{k!}\,
    y^{a_1} \dots y^{a_k}\,
    \nabla_{a_1} \dots \nabla_{a_k}\sigma\,,
\end{equation}
with $\beta$ a coefficient to be determined as well.
Note that at this point, the choice of $w_{\text{Weyl}}$
is merely an educated guess: it should be covariantly constant,
and related to the Weyl parameter $\sigma$,
hence this is the simplest option---which turns out to be 
the correct one as we shall see. Using the previous formulae,
one finds on the one hand, 
\begin{equation}
    \Big(\delta_{\xi_{\text{Weyl}},w_{\text{Weyl}}} F\Big)\big|_{y=0}
    = 2\sigma\,(\beta+1)\,p^2
    + \tfrac{\hbar^2}{2}\,\beta\,\Box\sigma
    + 2\sigma\,\alpha\beta\,R\,,
\end{equation}
while on the other hand
\begin{equation}
    \delta^{\text{Weyl}}_\sigma (p^2+\alpha\,R)
    = -2\alpha\,\big(\sigma\,R + (D-1)\,\Box\sigma\big)\,,
\end{equation}
which implies
\begin{equation}
    \beta=-1\,,
    \qquad\text{and}\qquad 
    \alpha = \tfrac{\hbar^2}{4(D-1)}\,,
\end{equation}
thereby fixing the symbol of the conformal Laplacian
in accordance with the previous discussion.

As a final consistency check, one can compute
the gauge transformation of the lift of the scalar $\phi$
generated by the parameter $\xi_{\text{Weyl}}$ and $w_{\text{Weyl}}$ 
identified previously, and recover
\begin{equation}\label{eq:var_F}
    \delta_{\xi_{\text{Weyl}},w_{\text{Weyl}}} \Phi\big|_{y=0} 
    = -\tfrac{D-2}2\,\sigma\,\phi\,,
\end{equation}
as expected for a conformally-coupled scalar field.

\subsection{Higher Spin Background}
Let us recall that $A$ is kept purely gravitational
and background conformal higher spin fields are placed into $F$
as an uplift of\footnote{If one wants to consider
all integer spins, a spin-one has to be included,
which is naively missing above. Alternatively,
it is possible to truncate the system to even spins only.}
\begin{align}\label{hsmetriclike}
    f & = p^2 + \tfrac{\hbar^2}{4(n-1)}\,R
    + \sum_{s>2} h^{a_1 \dots a_s}(x)\,p_{a_1} \dots p_{a_s}\,.
\end{align}
It is instructive to work out the gauge transformations
of this symbol generated by the gauge parameters
\begin{align}
    \xi & = \xi_{\text{Weyl}}
    - \tau\Big(\sum_{s>2}\xi^{a_1 \dots a_{s-1}}(x)\,
    p_{a_1} \dots p_{a_{s-1}}\Big)\,,\\
    w & = w_{\text{Weyl}}
    + \tau\Big(\sum_{s>2} \sigma^{a_1 \dots a_{s-2}}(x)\,
    p_{a_1} \dots p_{a_{s-2}}\Big)\,,
\end{align}
that is, we simply append to the gauge parameters
identified previously the covariantly constant uplift
of arbitrary monomials in $p$. Indeed, in this manner
the gauge variation of $A$ is unaffected by this new term,
\begin{equation}
    \delta_\xi A \equiv \delta_{\xi_{\text{Weyl}}} A\,,
\end{equation}
and thus boils down to a Weyl transformation
of the gravitational sector. It does, however,
affect the gauge transformation of $f$. Computing
$\delta_{\xi,w} F\rvert_{y=0}$ and extracting
the piece of order $s>2$ in $p$, one finds
\begin{align}
    \delta_{\xi,\sigma} h^{a_1 \dots a_s}
    & = 2\,\nabla^{(a_1} \xi^{a_2 \dots a_s)}
    + 2\,\eta^{(a_1 a_2} \sigma^{a_3 \dots a_s)}
    + (s-2)\,\sigma\,h^{a_1 \dots a_s} + \dots 
\end{align}
where the dots denote curvature corrections.
The first two terms correspond to the `naive'
covariantization of the linearized gauge transformations
initially proposed by Fradkin and Tseytlin
for conformal higher spin fields,
i.e. the flat space ones wherein partial derivatives
are replaced by covariant derivatives. The third term
tells us that the Weyl weight of a conformal higher spin
field with spin $s$ is $s-2$, which is also in accordance
with expectations \cite{Fradkin:1985am}.%
\footnote{Note that the Weyl weight of a metric-like field
$\phi_{\mu_1 \dots \mu_s}$ is $2s-2$,
e.g. it is $2$ for metric $g_{\mu\nu}$. Its fiber version,
to which $h^{a_1 \dots a_s}$ should be compared to,
is obtained by contracting it with $s$ inverse vielbeins 
$e^\mu_a$, giving Weyl weight of $s-2$.}
This can be seen as another sign of relevance
for this framework in the problem of formulating
CHS gravity in a manifestly covariant manner.

\paragraph{Higher spin currents.}
As a final application, we can derive the higher spin
currents for an arbitrary curved spacetime. To do so,
let us split the previous symbol \eqref{hsmetriclike}
into that of the conformal Laplacian
and the conformal higher spin fields,
\begin{equation}
    f = p^2 + \tfrac{\hbar^2}{4(D-1)}\,R
    + f_{hs}(x,p)\,,
    \qquad 
    f_{hs}(x,p) := \sum_{s>2} h^{a_1 \dots a_s}\,
    p_{a_1} \dots p_{a_s}\,,
\end{equation}
according to which the action obtained from $f$
is the sum of the conformally-coupled scalar
and a Noether coupling part,
\begin{equation}
    S_{\scriptstyle\rm Noether}[h,\phi]
    = \tfrac12\,\Tr_A(F_{hs} \ast \Wigner_\phi)
    = \tfrac12\,\int_\Base \dR^Dx\,|e|\,\phi^*\,
    \big[\rho(F_{hs})\Phi\big]\rvert_{y=0}\,,
\end{equation}
corresponding to the contribution of the higher spin
currents coupled to higher spin sources/ background fields
$h^{a_1 \dots a_s}$. In other words, we can identify
the higher spin current by putting the above functional
in the form
\begin{equation}
    S_{\scriptstyle\rm Noether}[h,\phi]
    = \tfrac12\,\int_\Base \dR^Dx\,|e|\
    \sum_{s>2} h^{a_1 \dots a_s}\,J_{a_1 \dots a_s}(\phi)\,,
\end{equation}
where the spin $s$ current $J_{a_1 \dots a_s}$
here is by definition bilinear in the scalar field $\phi$.

This computation involves the action of the quantization map
on the lift of $f_{hs}$, which is of arbitrary order in $p$.
As a consequence, the relevant terms to compute in this lift,
meaning those that will contribute to the final result
after applying the quantization of $F_{hs}$ to $\Phi$
and setting $y=0$, are those that are $y$-independent
or contain \emph{exactly} the same number of $y$'s and $p$'s.
Indeed, the quantization map applied to a monomial
of order $l$ in $y$ and $m$ in $p$ reads
\begin{equation}
    \rho(y^{a_1} \dots y^{a_l}\,p_{b_1} \dots p_{b_m})
    = (-\hbar)^m\sum_{k=0}^{\min(l,m)} \tfrac1{2^k}\,
    \tfrac{m!}{(m-k)!}\tfrac{l!}{k!(l-k)!}
    y^{(a_1} \dots y^{a_{l-k}}
    \delta^{a_{l+1-k}}_{(b_1} \dots \delta^{a_l)}_{b_k}
    \tfrac{\partial}{\partial y^{b_{k+1}}}
    \dots \tfrac{\partial}{\partial y^{b_m)}}\,,
\end{equation}
so that when setting $y=0$, only monomials with $l \leq m$,
i.e. less $y$'s than $p$'s, remain. This would be difficult
to compute for arbitrary spin $s>2$, so we will focus 
on the curvature independent part of the current.
The relevant part of the lift of $f_{hs}$
is therefore given by its `covariant Taylor series',
\begin{equation}
    F_{hs} = \sum_{k\geq0} \tfrac1{k!}\,y^{a_1} \dots y^{a_k}\,
    \nabla_{a_1} \dots \nabla_{a_k} f_{hs} + \dots\,,
\end{equation}
where the dots denote curvature corrections.
Applying the quantization map on this (partial) lift,
and evaluating the result at $y=0$, one ends up with
\begin{align}
    \rho(F_{hs})\rvert_{y=0} & = \sum_{s>2} (-\hbar)^s\,
    \sum_{k=0}^s \tfrac1{2^k} \tfrac{s!}{k!(s-k)!}\,
    \nabla_{a_1} \dots \nabla_{a_k}
        h^{a_1 \dots a_k\,a_{k+1} \dots a_s}\,
    \tfrac{\partial}{\partial y^{a_{k+1}}}
        \dots \tfrac{\partial}{\partial y^{a_s}} + \dots\,.
\end{align}
Under the same restrictions, the lift of the scalar field reads
\begin{equation}
    \Phi = \sum_{k\geq0} \tfrac1{k!}\,y^{a_1} \dots y^{a_k}\,
    \nabla_{a_1} \dots \nabla_{a_k} \phi + \dots\,,
\end{equation}
so that,
\begin{align}
    \rho(F_{hs})\Phi\rvert_{y=0} & = \sum_{s>2} (-\hbar)^s\,
    \sum_{k=0}^s \tfrac1{2^k} \tfrac{s!}{k!(s-k)!}\,
    \nabla_{a_1} \dots \nabla_{a_k}
        h^{a_1 \dots a_k\,a_{k+1} \dots a_s}
    \nabla_{a_{k+1}} \dots \nabla_{a_s} \phi
    + \dots\,,
\end{align}
again keeping only curvature independent terms.
Upon integration by parts, one finds
\begin{align}
    J_{a_1 \dots a_s} & = \big(-\tfrac{\hbar}2\big)^s\,
    \sum_{k=0}^s \tfrac{(-1)^k\,s!}{k!(s-k)!}\,
    \nabla_{(a_1} \dots \nabla_{a_k} \phi^*\,
    \nabla_{a_{k+1}} \dots \nabla_{a_{s)}} \phi 
    + \dots\,,
\end{align}
as one may have expected. This is the covariantized version
of the well-known `dipole' generating function
$\phi^*(x-y)\phi(x+y)$ that yields conserved quasi-primary 
(higher spin) currents with an admixture of descendants
in the flat space. The curvature corrections
can systematically be worked out, see \cite{Beccaria:2017nco}
for the spin-three example in the bottom-up approach.
However, it is clear that the higher the spin
the more non-linearities in the Riemann tensor $R$
and its derivatives will enter. Therefore, eq. \eqref{main}
seems to be the most compact way of writing the coupling
of the free scalar field to a higher spin background.

\paragraph{First order correction in curvature.}
If we focus on the spin-$3$ case, then we only need
to compute the lift of $f_{s=3} = h^{abc}\,p_a p_b p_c$
to order $3$. Pushing the computation of the lift
of any symbol $f(x,p)$ presented in \eqref{eq:lift_y2}
to the next order, thanks to the recursion \eqref{eq:rec_F},
yields
\begin{align}
    \tau(f) & = \Big(1 + y^a \nabla_a
    + \tfrac12\,y^a y^b\,\big[\nabla_a \nabla_b 
    + \tfrac13\,R_{da}{}^c{}_b\,p_c\,
    \tfrac{\partial}{\partial p_d}\big] \\
    & \hspace{50pt} + \tfrac16\,y^a y^b y^c\,
    \big[\nabla_a \nabla_b \nabla_c
    + \tfrac12\,\nabla_a R_{db}{}^e{}_c\,
    p_e\,\tfrac{\partial}{\partial p_d}
    + R_{da}{}^e{}_b\,p_e\,
    \tfrac{\partial}{\partial p_d}\,\nabla_c\big] \\
    & \hspace{100pt} + \tfrac{\hbar^2}{12}\,
    y^a\,\big[R_{ab}{}^d{}_c\,
    \tfrac{\partial^2}{\partial p_b \partial p_c}\nabla_d
    - \tfrac14\,\nabla_b R_{ac}{}^e{}_d\,p_e\,
    \tfrac{\partial^3}{\partial p_b \partial p_c \partial p_d}\big]
    + \dots \Big)\,f\,,
\end{align}
and applying it to $f_{s=3}$, one finds that
$F_{s=3} = \tau(f_{s=3})$ is given by
\begin{align}
    F_{s=3} & = \Big(h^{abc} + y^i\,\nabla_i h^{abc}
    + \tfrac12\,y^i y^j\,\big[\nabla_i \nabla_j h^{abc}
    + R_{di}{}^a{}_j h^{bcd}\big] \\
    & \hspace{50pt} + \tfrac16\,y^i y^j y^k\,
    \big[\nabla_i \nabla_j \nabla_k h^{abc} 
    + \tfrac32\,\nabla_i R_{dj}{}^a{}_k\,h^{bcd}
    + 3\,R_{di}{}^a{}_j\,\nabla_k h^{bcd}\big] + \dots\Big)\,p_a p_b p_c \\
    & \hspace{100pt} + \tfrac{\hbar^2}{2}\,y^i\,
    \big[R_{ib}{}^d{}_c\,\nabla_d h^{abc} 
    - \tfrac14\,\nabla_b R_{ic}{}^a{}_d\,h^{bcd}\big]\,p_a
    + \dots\,,
\end{align}
where the dots denote terms of order $4$ or higher,
and its quantization evaluated at $y=0$ reads
\begin{align}
    -\tfrac1{\hbar^3}\,\rho(F_{s=3})\rvert_{y=0}
    & = h^{abc}\,\tfrac{\partial^3}{\partial y^a \partial y^b \partial y^c}
    + \tfrac32\,\nabla_a h^{abc}\,
    \tfrac{\partial^2}{\partial y^b \partial y^c}
    + \tfrac14\,\big[3\,\nabla_a \nabla_b h^{abc}
    -R_{ab}\,h^{abc}\big]\,\tfrac{\partial}{\partial y^c} \\
    & \hspace{100pt}
    + \tfrac18\,\big[\nabla_a \nabla_b \nabla_c h^{abc}
    - \nabla_a R_{bc}\,h^{abc} - R_{ab}\,\nabla_c h^{abc}\big]\,.
\end{align}
Similarly, we can use \eqref{eq:rec_phi} to compute
the lift of the scalar field $\phi$ to order $3$,
\begin{equation}
    \Phi = \phi + y^a\nabla_a\phi
    + \tfrac12\,y^a y^b\,\big(\nabla_a\nabla_b
    - \tfrac16\,R_{ab}\big)\phi
    + \tfrac16\,y^a y^b y^c\,\big(\nabla_a \nabla_b \nabla_c
    -\tfrac14\,\nabla_a R_{bc}
    -\tfrac12\,R_{ab}\,\nabla_c\big)\,\phi + \dots\,,
\end{equation}
so that one finds
\begin{align}
    -\tfrac1{\hbar^3}\,\rho(F_{s=3})\Phi\rvert_{y=0}
    & = h^{abc}\,\big(\nabla_a \nabla_b \nabla_c
    -\tfrac38\,\nabla_a R_{bc}
    -\tfrac34\,R_{ab}\,\nabla_c\big)\,\phi
    + \tfrac32\,\nabla_c h^{abc}\,
    \big(\nabla_a\nabla_b- \tfrac14\,R_{ab}\big)\,\phi \\
    & \quad + \tfrac34\,\nabla_a \nabla_b h^{abc}\,\nabla_c\phi
    + \tfrac18\,\nabla_a \nabla_b \nabla_c h^{abc}\,\phi\,.
    \nonumber
\end{align}
which leads to the following expression
\begin{align}
    J_{abc} & = -\tfrac{\hbar^3}{8}\,
    \Big(\nabla_{(a} \nabla_b \nabla_{c)} \phi\,\phi^*
    - 3\,\nabla_{(a} \nabla_b \phi\,\nabla_{c)} \phi^*
    + 3\,\nabla_{(a} \phi\,\nabla_b \nabla_{c)} \phi^*
    - \phi\,\nabla_{(a} \nabla_b \nabla_{c)} \phi^* \\
    & \hspace{250pt} -3\,R_{(ab}\,\nabla_{c)} \phi\,\phi^*
    + 3\,R_{(ab}\,\phi\,\nabla_{c)} \phi^*\Big)\,,
\end{align}
for the spin-$3$ current. More generally, the currents
up to first order in the curvature tensor (and its derivatives)
are obtained from the generating function
\begin{equation}
    {\cal J}(x|u) = e^{-\frac\hbar2\,u \cdot[\nabla_1-\nabla_2]}\,
    \Big(1 - \tfrac{\hbar^2}8\,
    {\rm sinhc}(\tfrac\hbar4\,u \cdot \nabla_3)\,
    R_{ab}(x_3)\,u^a u^b + {\cal O}(R^2)\Big)\,\phi(x_1)\,\phi^*(x_2)\big|_{x_i=x}\,,
\end{equation}
where
\begin{equation}
    {\cal J}(x|u) := \sum_{s \geq 0} \tfrac{(-\hbar)^s}{2^s\,s!}\,
    J_{a_1 \dots a_s}(x)\,u^{a_1} \dots u^{a_s}\,,
    \qquad 
    {\rm sinhc}(z) := \frac{\sinh(z)}{z}\,,
\end{equation}
and $\nabla_i$ denotes the covariant derivative
with respect to $x_i$ (see Appendix \ref{app:curvature}
for the derivation of this formula).
Let us remark here that, in order to make contact 
with, say the computation of \cite{Beccaria:2017nco}
for the spin-three current or even the standard computation
of the energy-momentum tensor, one should find the correct field
redefinition bringing the components of monomials in $p$
into the appropriate field frame (combination of conformal higher spin fields and derivatives thereof),
see e.g. \cite[App. F]{Segal:2002gd} and \cite{Bekaert:2010ky}
for an instance of the same issue around a flat background.

\section{Discussion and Summary}
\label{sec:discuss}
The work presented in this chapter is a natural continuation of the quest to covariantize the construction of conformal higher spin 
gravities started in \cite{Basile:2022nou}.
Now, both the action for conformal higher spin gravity
$S_{CHS}[h_s]$ and the coupling of the scalar matter
to the higher spin background,
$\bra\Phi\widehat{H}[\phi_s]\ket\Phi$,
can be written in a covariant way. The result completes
the study initiated in \cite{Beccaria:2017nco},
where the mixing between covariant spin-three
and spin-one currents that couple to background fields
have been discussed in $D=4$. In addition, one can consider
the matter coupled conformal higher spin gravity,
see \cite{Joung:2015eny} for some amplitudes
in this theory over flat background. Note, however,
that while the scalar matter can be coupled
to a higher spin background for any $n$ the conformal anomaly 
recipe gives $S_{CHS}[h_s]$ only for $n$ even.

The results open up the possibility of considering more general
matter fields in the relevant higher spin background,
such as the higher-derivative scalar fields (also known as
higher order singletons \cite{Bekaert:2013zya}), 
or spinor (and its higher-derivative counterpart),
see \cite{Grigoriev:2018wrx}.
The latter would in principle require the use
of the supersymmetry version of the FFS cocycle, 
i.e. the representative of the cohomology class
of the Clifford--Weyl algebra dual to the unique
Hochschild homology class of the same algebra 
\cite{Engeli:2008}.

Another possible application of the results
is to conformally-invariant differential operators.
Conformal geometry (in the sense of gauge symmetries
realized by diffeomorphisms and Weyl transformations)
is a part of the higher spin system. As we showed,
one can derive the conformal Laplacian as a particular instance 
of the scalar field coupled to the conformal gravity 
background. Generalizations such as Paneitz \cite{Paneitz:2008} 
or Fradkin--Tseytlin \cite{Fradkin:1982xc} operators
and GJMS operators \cite{GJMS} can also be recovered
by considering $F = (p^2)^k + \dots$ 
that would lead to operators of type $(\nabla^2)^k+\dots$,
i.e. starting with the $k$th power of the Laplacian,
and corrected by curvature terms.

It would also be interesting to apply the deformation 
quantization techniques to the self-dual conformal higher spin 
gravity \cite{Hahnel:2016ihf,Adamo:2016ple} that is natural
to formulate on twistor space. Here, the underlying space
$\mathbb{CP}^3$ is already symplectic. The twistor description
of low-spin fields, $s=1,2$, requires usual (holomorphic) 
connections and vector-valued one-forms, which can be understood 
as differential operators of zeroth and first order.
An extension to higher spin calls for differential operators
of arbitrary order, i.e. to the quantization
of the cotangent bundle again
(see also \cite{Bekaert:2021sfc} for additional discussions
of the quantization of the cotangent in relation with
the definition of higher spin diffeomorphisms).

Let us also note that the results and techniques of this chapter bridges a gap in the phase space approach to quantum mechanics.
Indeed, one can attempt to extend the Fedosov construction
to accommodate all the usual ingredients required
in quantum mechanics. The trace is, obviously, given by
the Feigin--Felder--Shoikhet cocycle; wave functions
can be understood as covariantly constant elements
in the Fock representation obtained via the quantization map. 
Wigner function takes exactly the same form as in the flat space, 
but in the fiber. The basic ingredients above do not rely
on the phase space being a cotangent bundle and should extend to 
arbitrary symplectic manifolds (a polarization is needed
to define the Fock space). This seems to depart
from the usual approach of symbol calculus on curved background, 
e.g. \cite{Hormander:1985, Duistermaat:1994, Widom:1980, Safarov:1997, Pflaum:1998, Pflaum:1999,  Dereziski:2018}
and references therein.

Finally, it would be interesting to construct
the $3d$ matter-coupled conformal higher spin gravity,
where the `dynamics' of conformal higher spin fields
is given by the Chern--Simons action
(as there is no conformal anomaly in $3d$).
Such a theory, namely the one based on fermionic matter,
can be seen to exist with the help of the argument
based on the parity anomaly \cite{Grigoriev:2019xmp}
(see e.g. \cite{Niemi:1983rq, Redlich:1983dv, Redlich:1983kn}
for original papers on the derivation of Chern--Simons
theory from the parity anomaly and \cite{Bonora:2016ida}
for the spin-three case). An alternative idea
along the AdS/CFT correspondence lines
was recently explored in \cite{Diaz:2024kpr, Diaz:2024iuz}.

\newpage
\begin{appendices}
\section{A Brief Review of Weyl Calculus}
\label{app:Weyl_calculus}
Let us give a brief summary of the definition
and construction of the Wigner function in flat space
(following e.g. the textbook \cite{deGosson:2006},
or the papers \cite{Segal:2002gd, Bekaert:2008xfa, Bekaert:2009ud, Bekaert:2010ky}).

\paragraph{Quantization map in flat space.}
The deformation quantization of $\R^{2D} \cong T^*\R^D$,
amounts to defining an isomorphism
\begin{equation}
    \begin{tikzcd}
        \Functions(T^*\R^D) \ar[r, "\sim"]
        & {\cal D}(\R^D)\,,
    \end{tikzcd}
\end{equation}
where ${\cal D}(\R^D)$ stands for the space
of differential operators on $\R^D$. This map
is referred to as a `quantization map' since,
as we will recall shortly, it allows one to define
a star-product on the algebra of functions $T^*\R^D$,
and hence a quantization thereof. To do so, 
we can take advantage of the Fourier transform
in flat space, that we denote by
\begin{equation}
    ({\cal F}f)(u,v) := \int_{\mathbb R^{2D}}
    \tfrac{{\rm d}^Dx\,{\rm d}^Dp}{(2\pi\hbar)^D}\,
    f(x,p)\,e^{-\frac{i}\hbar\,(x \cdot u + p \cdot v)}\,,
\end{equation}
for a symbol $f(x,p)$. Given a choice of quantization 
for the phase space coordinates $x^\mu \to \hat x^\mu$
and $p_\mu \to \hat p_\mu$, where hatted symbols
denote the corresponding operator, we want to associate

Schematically, we want to write something like
``$\widehat f(\hat x, \hat p)
\sim f(x,p)\,\delta(x-\hat x)\,\delta(p-\hat p)$'',
where $f(x,p)$ is the symbol of the operator $\widehat f$.
This sketchy formula can be given a precise sense,
using the Fourier representation of the Dirac distribution, 
leading to
\begin{equation}
    \widehat f(\hat x, \hat p) = \int_{\mathbb R^{2D}}
    \tfrac{{\rm d}^Du\,{\rm d}^Dv}{(2\pi\hbar)^D}\,
    ({\cal F}f)(u,v)\,
    e^{\frac{i}\hbar\,(u \cdot \hat x + v \cdot \hat p)}\,,
\end{equation}
and which is called the \emph{Weyl ordering} of operators.
Note that the exponential operator can be re-written as
\begin{equation}\label{eq:BCH}
    \exp\big(\tfrac{i}{\hbar}\,(u \cdot \hat x
                                + v \cdot \hat p)\big)
    = e^{\frac{i}{2\hbar}\,u \cdot v}\,
    \exp\big(\tfrac{i}{\hbar}\,u \cdot \hat x\big)\,
    \exp\big(\tfrac{i}{\hbar}\,v \cdot \hat p\big)\,,
\end{equation}
since we assume
$[\hat x^\mu, \hat p_\nu]=i\hbar\,\delta^\mu_\nu$.
Choosing the usual coordinate representation,
\begin{equation}
    \hat x^\mu = x^\mu\,,
    \qquad 
    \hat p_\mu = -i\hbar\,\partial_\mu\,,
\end{equation}
the action of this operator on a wave function $\varphi(x)$
is given by
\begin{align}
    (\widehat f\,\varphi)(x)
    & = \int \tfrac{\dR^Du\,\dR^Dv}{(2\pi\hbar)^D}\,
    ({\cal F}f)(u,v)\,e^{\frac{i}{2\hbar}\,u \cdot v}\,
    e^{\frac{i}\hbar\,u \cdot x}\,\varphi(x+v) \\
    & = \int \tfrac{\dR^Du\,\dR^Dv}{(2\pi\hbar)^D}\,
    \tfrac{{\rm d}^nx'\,{\rm d}^np}{(2\pi\hbar)^D}\,
    f(x',p)\,e^{-\frac{i}\hbar\,p \cdot v}\,
    e^{\frac{i}\hbar\,u \cdot (x-x'+\frac{v}2)}\,
    \varphi(x+v) \\
    & = \int \tfrac{\dR^Dv\,\dR^Dp}{(2\pi\hbar)^D}\,
    f(\tfrac{x+v}2,p)\,e^{\frac{i}\hbar\,p \cdot (x-v)}\,
    \varphi(v)\,,
\end{align}
where the first equation is obtained using
\eqref{eq:BCH} and the action of the translation operator,
the second line is merely the definition
of the Fourier transform, and the last one is the result
of integrating over $u$, which gives a Dirac distribution,
and then evaluating it by integrating over $x'$.
Upon Taylor expanding $f$ and integrating by part,
one can put this formula into an operatorial form
\begin{equation}\label{eq:exp_quantization}
    (\widehat f\varphi)(x) = f(x,p)\,
    \exp\Big(\!\!-\!i\hbar\,
    \tfrac{\overleftarrow{\partial}}{\partial p} \cdot
    \big[\tfrac12\,\tfrac{\overleftarrow{\partial}}{\partial x}
    + \tfrac{\overrightarrow{\partial}}{\partial x}\big]\Big)\,
    \varphi(x)\big|_{p=0}\,.
\end{equation}

The Moyal--Weyl star-product can be recovered
from the composition of the two operators associated with
two symbols via the above symbol, or quantization, map.
More precisely, it can be defined as the symbol
of the composition of the quantization of two symbols, i.e.
\begin{equation}\label{eq:quantization=rep}
    \widehat f \circ \widehat g = \widehat{f \star g}\,.
\end{equation}
To do so, let us start by recalling
that the action of a symbol $f$ given above
exhibits the \emph{kernel} of that associated operator, 
namely
\begin{equation}
    (\widehat f\,\phi)(x) = \int_{\R^D} \dR^Dq\,
    K_f(x,q)\,\phi(q)\,,
    \qquad\text{with}\qquad
    K_f(x,q) := \int_{\R^D} \tfrac{\dR^Dp}{(2\pi\hbar)^D}\,
    f(\tfrac{x+q}2,p)\,e^{\frac{i}\hbar\,p \cdot (x-q)}\,.
\end{equation}
The symbol of the operator $\widehat f$ can be extract back
from its kernel, via its inverse transform
\begin{equation}
    f(x,p) = \int_{\R^D} \dR^Dq\,
    K_f(x+\tfrac12\,q,x-\tfrac12\,q)\,
    e^{-\frac{i}\hbar\,p \cdot q}\,,
\end{equation}
and therefore, using this together with the fact
that the integral kernel of the composition
of two operators is given by
\begin{equation}
    K_{\widehat f \circ \widehat g}(x,x')
    = \int_{\R^D} \dR^Dq\,
    K_{\widehat f\,}(x,q)\,K_{\widehat g\,}(q,x')\,,
\end{equation}
one ends up with
\begin{equation}
    \big(f \star g\big)(x,p) = \tfrac1{(\pi\hbar)^{2n}}
    \int \dR^Dv_1\,\dR^Dv_2\,\dR^Dw_1\,\dR^Dw_2\
    e^{\frac{2i}\hbar\,(v_1 \cdot w_2 - v_2 \cdot w_1)}\,
    f(x+v_1,p+w_1)\,g(x+v_2,p+w_2)\,.
\end{equation}
Upon Taylor expanding the two functions $f$ and $g$
around $(x,p)$, and integrating by part, 
one 
\begin{equation}\label{eq:exp_star}
    \big(f \star g\big)(x,p) = f(x,p)\,
    \exp\Big(\tfrac{i\hbar}2\,
    \big[\tfrac{\overleftarrow{\partial}}{\partial x}
    \cdot \tfrac{\overrightarrow{\partial}}{\partial p}
    -\tfrac{\overleftarrow{\partial}}{\partial p}
    \cdot \tfrac{\overrightarrow{\partial}}{\partial x}\big]\Big)\,g(x,p)\,,
\end{equation}
Note that the Moyal--Weyl star-product is \emph{Hermitian},
meaning that it satisfies
\begin{equation}
    (f \star g)^* = g^* \star f^*\,,
\end{equation}
where $(-)^*$ denotes the complex conjugation,
i.e. the latter is an anti-involution of the Weyl algebra.

One can think of the quantization map as providing
a \emph{representation} of the Weyl algebra: identifying
the latter as the subalgebra of \emph{polynomial}
functions on $T^*\R^D$, wave functions which are nothing but 
functions on $\R^D$, the base of the cotangent bundle $T^*\R^D$,
are acted upon by the former via the quantization map.
This subspace can be thought of as a Fock space,
which carries a representation of the Weyl algebra
as can be seen from the defining relation 
\eqref{eq:quantization=rep}.

The integration over the cotangent bundle
$T^*\R^D$ defines a trace over the space of symbols,
at least those which are compactly supported
or vanish at infinity sufficiently fast.
Indeed, in this case one finds
\begin{equation}
    \Tr(f \star g) = \int_{T^*\R^D} \dR^D x\,\dR^D p\
    (f \star g)(x,p) = \int_{T^*\R^D} \dR^D x\,\dR^D p\
    f(x,p)\,g(x,p) = \Tr(g \star f)\,,
\end{equation}
for any symbols $f$ and $g$, since all higher order terms
in the star product are total derivatives on $T^*\R^D$,
and hence can be ignored for the aforementioned
suitable class of symbols.

\paragraph{Wigner function in flat space.}
Having worked out how to translate the action
and the composition of differential operators in terms
of their symbol, as well as their trace, we can now
turn our attention to the computation of matrix elements
for these operators, expressing the transition probability
from one state to another. Since the latter can be expressed as
\begin{equation}\label{eq:transition}
    \bra\psi\widehat{H}\ket\phi
        = \Tr(\widehat{H} \circ \ket\phi\!\bra\psi)\,,
\end{equation}
we have everything we need to derive such quantities
using symbols, provided that we know that of the projector
$\ket\phi\!\bra\psi$. In light of the relation
between the symbol of an operator and its integral kernel,
we may first focus on that of the projector.
This integral kernel is easily computed,
\begin{equation}
    \big(\ket{\phi}\!\langle{\psi}\ket{\varphi}\big)(x)
    \overset{!}{=} \phi(x)\,\int_{\R^D} \dR^Dq\,
    \psi^*(q)\,\varphi(q)
    \qquad\Longrightarrow\qquad 
    K_{\ket{\phi}\!\bra{\psi}}(x,q)
    \equiv \phi(x)\,\psi^*(q)\,,
\end{equation}
which leads to 
\begin{equation}
    W[\phi,\psi](x,p) = \int_{\R^D} \dR^Dq\
    \phi(x+\tfrac{q}2)\,\psi^*(x-\tfrac{q}2)\,
    e^{-\frac{i}\hbar\,p \cdot q}\,,
\end{equation}
for its symbol. It obey the following useful properties
\begin{equation}\label{eq:prop_Wigner}
    \xi \star W[\phi,\psi]
        = W[\widehat\xi\,\phi,\psi]\,,
    \qquad 
    W[\phi,\psi] \star \xi^\dagger
        = W[\phi,\widehat\xi\,\psi]\,,
\end{equation}
in accordance with the fact that it is the symbol
of the projector $\ket\phi\!\bra\psi$, and
\begin{equation}\label{eq:integral_Wigner}
    \int_{\R^D} \dR^Dp\ W[\phi,\psi](x,p)
    = \phi(x)\,\psi^*(x)\,.
\end{equation}
Now we can replace the right hand side of \eqref{eq:transition} with its symbol counterpart, leading to
\begin{equation}
    \Tr(H \star W[\phi,\psi]) = \Tr(W[\widehat{H}\phi,\psi])
    = \int_{T^*\R^D} \dR^Dx\,\dR^Dp\ W[\widehat{H}\phi,\psi]
    = \int_{\R^D} \dR^Dx\ \psi^*(x)\,(\widehat{H}\phi)(x)\,,
\end{equation}
upon using the previously listed properties of $W[\phi,\psi]$,
thereby reproducing the expected result for the quantity
$\bra\psi\widehat{H}\ket\phi$ from a quantum mechanical
point of view. The \emph{Wigner function} $W_\phi$
associated with a wave function $\phi$ is the symbol 
of the projector $\ket\phi\!\bra\phi$, i.e. 
\begin{equation}
    W_\phi(x,p) := W[\phi,\phi](x,p)
    \equiv \int_{\R^D} \dR^Dq\ e^{-\frac{i}\hbar\,q \cdot p}\,
    \phi(x+\tfrac{q}2)\,\phi(x-\tfrac{q}2)\,,
\end{equation}
whose integral over $p$ is nothing but the probability
density defined by $\phi$.

To conclude this appendix, let us prove the identity
\eqref{eq:integral_Wigner} and a small variation on it
(the intertwining property \eqref{eq:prop_Wigner}
can be proved by direct computation using the integral formulae 
for the star-product and the quantization map),
by expressing the Wigner function in terms of star-product.
To achieve this, recall that the star-product
of a phase factor $e^{\frac{i}\hbar\,q \cdot p}$,
where $q$ is a fixed parameter, with any symbol $f(x,p)$
yields
\begin{equation}\label{eq:translation}
    e^{\frac{i}\hbar\,q \cdot p} \star f(x,p)
    = e^{\frac{i}\hbar\,q \cdot p} f(x+\tfrac{q}2,p)\,,
    \qquad 
    f(x,p) \star e^{\frac{i}\hbar\,q \cdot p}
    = e^{\frac{i}\hbar\,q \cdot p} f(x-\tfrac{q}2,p)\,,
\end{equation}
i.e. it implements translations in $x$ up to a phase.%
\footnote{To be more precise, the action of translation
on elements depending on $x$ \emph{only} is generated by 
\[
    e^{\frac{i}\hbar\,q \cdot p} \star \phi(x)
    \star e^{-\frac{i}\hbar\,q \cdot p} = \phi(x+q)\,,
\]
which can be recovered from the formulae \eqref{eq:translation}.}
Integrating these formulae over $q$ yields
\begin{equation}
    \big(f \star \delta_p\big)(x,p)
    = \int_{\R^D} \dR^Dq\,e^{-\frac{i}\hbar\,q \cdot p}\,
    f(x+\tfrac{q}2,p)\,, 
    \qquad 
    \big(\delta_p \star f\big)(x,p)
    = \int_{\R^D} \dR^Dq\,e^{-\frac{i}\hbar\,q \cdot p}\,
    f(x-\tfrac{q}2,p)\,,
\end{equation}
where $\delta_p$ is the Dirac distribution
in the space of momenta $p_a$. With these simple identities
at hand, one finds
\begin{align}
    \phi \star \delta_p \star \psi^*
    & = \int_{\R^D} \dR^Dq\, \phi(x)
    \star \big[e^{-\frac{i}\hbar\,q \cdot p}
    \psi^*(x-\tfrac12\,q)\big] \\
    & = \int_{\R^D} \dR^Dq\,\big[\phi(x)
    \star e^{-\frac{i}\hbar\, q \cdot p}]\,
    \psi^*(x-\tfrac12\,q) \\
    & = \int_{\R^D} \dR^Dq\
    \phi(x+\tfrac{q}2)\,\psi^*(x-\tfrac{q}2)\,
    e^{-\frac{i}\hbar\,p \cdot q} = W[\phi,\psi]\,,
\end{align}
where to pass from the first to the second line,
one should notice that since $\phi$ only depends on $x$,
its star-product with any other Weyl algebra element
will produce only derivatives with respect to $p$ 
on the latter.

Now this expression makes it  relatively easy
to evaluate the integral over momenta of the Wigner function
and its derivatives with respect to $p$. Indeed,
since the only term of this star-product that depends
on momenta is the Dirac distribution, the result is of the form
\begin{equation}
    \phi \star \delta_p \star \psi^*
    \sim \sum_{k,l\geq0} \partial_x^k\phi \times
    \partial_p^{k+l} \delta(p) \times \partial_x^l \psi^*\,,
\end{equation}
so that the integral over $p$ schematically reads
\begin{equation}
    \int \dR^Dp\ W[\phi,\psi] \sim \sum_{k,l\geq0}
    \int \dR^Dp\ \partial_p^{k+l} \delta(p)
    \times (\partial_x^k\phi\,\partial_x^l\psi^*)\,,
\end{equation}
which identically vanishes for $k+l>0$ since both
$\phi$ and $\psi$ do not depend on $p$, and yields
\eqref{eq:integral_Wigner} for $k=0=l$. On top of that,
since taking partial derivative with respect to $x$ or $p$
commutes with the star-product, the derivatives
of the Wigner function with respect to $p$ are of the form
$\partial_p^k W[\phi,\phi]
    \sim \phi \star \partial^k_p\delta_p \star \psi^*$,
and hence the same argument shows that the integral
over the momenta identically vanishes, 
\begin{equation}\label{eq:der_Wigner}
    \int_{\R^D} \dR^Dp\ \tfrac{\partial^k}{\partial p_{a_1} \dots \partial p_{a_k}}\,W[\phi,\psi] = 0\,,
    \qquad \forall k>0\,.
\end{equation}

\section{More on Weyl Transformations}
\label{app:Weyl_transfo}
In this appendix, we provide more details concerning
the computation of the gauge variation of the symbol
$p^2 + \alpha\,R$. For convenience, let us introduce 
the tensor
\begin{equation}
    \mathcal{P}_{ab}{}^{cd} := \delta_{(a}^c \delta_{b)}^d
    - \tfrac12\,\eta_\AlgInd{AB}\,\eta^{cd}\,,
\end{equation}
with which the gauge parameter $\xi_{\text{Weyl}}$,
identified in \eqref{eq:xi_Weyl} as the one generating
Weyl transformations for the components of
the $1$-form connection $A$, is given by
\begin{equation}
    \xi_{\text{Weyl}} = -\sigma\,y \cdot p
    - \mathcal{P}_{bc}{}^{ad}\,y^b y^c p_d\,\nabla_a \sigma
    -\tfrac13\,y^{(a}\,\mathcal{P}_{cd}{}^{b)e} y^c y^d p_e\,
    \nabla_a \nabla_b \sigma + \dots\,,
\end{equation}
plus terms of order $3$ and higher in $y$, but all linear in $p$.
In order to compute the gauge transformation
of $F = \tau(p^2+\alpha\,R)$ generated by 
$\xi_{\text{Weyl}}$, and $w_{\text{Weyl}}$ given by
\begin{equation}
    w_{\text{Weyl}} = \beta\,\tau(\sigma)
    = \beta\,\big(1 + y^a\,\nabla_a
    + \tfrac12\,\nabla_a \nabla_b + \dots\big)\,\sigma\,,
\end{equation}
one needs to compute the star-product between elements
of the Weyl algebra which are at most quadratic in $p$.
For our purpose, it will be enough to compute
neglecting terms with less, or as many, $y$'s than $p$'s.
We therefore only need the lift of $p^2$ and $R$
up to order $2$ in $y$,
\begin{equation}
    \tau(p^2) = p^2 + \tfrac13\,y^a y^b\,
    R_a{}^c{}_b{}^d\,p_c p_d + \dots\,,
    \qquad 
    \tau(R) = R + y^a\,\nabla_a R
    + \tfrac12\,y^a y^b\,\nabla_a \nabla_b R + \dots\,,
\end{equation}
which yields
\begin{align}
    \tfrac1\hbar\,\big[p^2, \xi_{\text{Weyl}}\big]_\ast
    & = 2\sigma\,p^2 + 4\,\mathcal{P}_{bc}{}^{da}\,
    y^b\,p^c p_d\,\nabla_a \sigma
    + \tfrac23\,y^c y^d\,\mathcal{P}_{cd}{}^{a(\bullet}\,p^{b)}\,
    p_a\,\nabla_b \nabla_\bullet \sigma \\ & \hspace{150pt}
    + \tfrac43\,y^{(b}\,\mathcal{P}_{de}{}^{c)a}\,
    y^e\,p^d p_a\,\nabla_b \nabla_c \sigma + \dots \\
    & = 2\,p^2\,\big(\sigma + y^a\,\nabla_a \sigma
    + \tfrac13\,y^a y^b\,\nabla_a \nabla_b \sigma\big)
    + \tfrac23\,(y \cdot p\,y^a - \tfrac12\,y^2\,p^a)\,p^b\,
    \nabla_a \nabla_b \sigma + \dots\,, 
\end{align}
while the commutator of $\xi_{\text{Weyl}}$ with other terms
in the lift of $p^2$ or $R$ do not contribute terms
with less $y$'s than $p$'s, and
\begin{align}
    \big\{\tau(p^2), w_{\text{Weyl}}\big\}_\ast
    & = \Big(2\,\big[p^2 + \tfrac13\,R_{a}{}^{c}{}_{b}{}^{d}\,
    y^a y^b\,p_c p_d\big] + \tfrac{\hbar^2}2\,\big[\eta^{ab}
    + \tfrac13\,R_{a}{}^{c}{}_{b}{}^{d}\,y^a y^b\big]\,
    \tfrac{\partial^2}{\partial y^a \partial y^b}\big)
    w_{\text{Weyl}} + \dots \\
    & = 2\beta\,p^2\,\big(\sigma + y^a\,\nabla_a \sigma
    + \tfrac12\,y^a y^b\,\nabla_a \nabla_b \sigma\big)
    + \tfrac{2\beta}3\,\sigma\,R_{a}{}^{c}{}_{b}{}^{d}\,
    y^a y^b\,p_c p_d + \beta\,\tfrac{\hbar^2}2\,\Box\sigma
    + \dots \\
    \big\{\tau(R), w_{\text{Weyl}}\big\}_\ast
    & = 2\,\tau(R)\,w_{\text{Weyl}} = 2\beta\,\sigma\,R + \dots 
\end{align}
where again the dots denote terms of order $3$
or higher in $y$.
Putting everything together, we end up with
\begin{align}
    \delta_{\xi_{\text{Weyl}},w_{\text{Weyl}}} F
    & = 2\sigma\,\big[(\beta+1)\,p^2 + \alpha\beta\,R\big]
        + \beta\,\tfrac{\hbar^2}2\,\Box\sigma 
        + 2(\beta+1)\,p^2\,y^a\,\nabla_a \sigma \\
    & \qquad + y^a y^b\,p_c p_d\,\big([\beta+\tfrac23]\,
    \eta^{cd}\,\delta^\times_a \delta^\bullet_b
    + \tfrac23\,\eta^{\times c}\,\delta^d_a \delta^\bullet_b
    -\tfrac13\,\eta_\AlgInd{AB}\,\eta^{\times c}\eta^{\bullet d}\big)\,
    \nabla_\times \nabla_\bullet \sigma + \dots
\end{align}
whose value at $y=0$, which we gave earlier in \eqref{eq:var_F},
can be compared to the Weyl variation of $p^2+\alpha\,R$
and imposing that the two agree implies
\begin{equation}
    \alpha = \tfrac{\hbar^2}{4(D-1)}\,,
    \qquad 
    \beta = -1\,.
\end{equation}
From now on, we will fix these values, 
and will denote the gauge transformations
generated by $\xi_{\text{Weyl}}$ and $w_{\text{Weyl}}$ 
with the same symbol as for a Weyl transformation
generated by $\sigma$,
\begin{equation}
    \delta_{\xi_{\text{Weyl}},w_{\text{Weyl}}} \equiv \delta^{\text{Weyl}}_\sigma\,,
\end{equation}
since the two agree with the aforementioned values
of $\alpha$ and $\beta$.
As a final cross-check, let us compute
the Weyl transformation of the equation of motion
\begin{equation}
    \widehat{f}\phi
    = \big(\Box - \tfrac{D-2}{4(D-1)}\,R\big)\,\phi\,,
\end{equation}
which, in our formalism, is obtained by evaluating
\begin{equation}
    \delta^{\text{Weyl}}_\sigma\big(\rho(F)\,\Phi\big)
    = \rho(\delta^{\text{Weyl}}_\sigma F)\,\Phi
    + \rho(F)\,\delta^{\text{Weyl}}_\sigma\Phi\,,
\end{equation}
at $y=0$, where recall that $\Phi=\tau(\phi)$
is the lift of $\phi$ as a flat section of the Fock bundle,
whose first order in $y$ are given in \eqref{eq:lift_phi}.
To compute the first term, we only need to use
the simple quantization formula \eqref{eq:q_lem},
to find
\begin{equation}\label{eq:interm1}
    \rho(\delta^{\text{Weyl}}_\sigma F)\,\Phi\rvert_{y=0}
    = \hbar^2\,\tfrac{D-4}6\,\big(\Box\sigma
    + \tfrac1{D-1}\,R\,\sigma\big)\,\phi\,.
\end{equation}
To compute the second term, we should also use
\begin{equation}
    \rho(F)\rvert_{y=0} = \hbar^2\,\eta^{ab}\,
    \tfrac{\partial^2}{\partial y^a \partial y^b}
    - \tfrac{\hbar^2}{12}\,\tfrac{D-4}{D-1}\,R\,,
\end{equation}
as well as
\begin{equation}
    \tfrac1\hbar\,\rho(\xi_{\text{Weyl}})
    = -\big(\sigma + y^a\,\nabla_a \sigma\big)\,
    \big(y \cdot \tfrac{\partial}{\partial y} + \tfrac{D}2\big)
    + \tfrac12\,y^2\,\nabla^a\,\tfrac{\partial}{\partial y^a}
    - \tfrac{D+1}3\,y^a y^b\,\nabla_a \nabla_b \sigma
    + \tfrac16\,y^2\,\Box\sigma + \dots\,,
\end{equation}
and
\begin{equation}
    \rho(w_{\text{Weyl}})= -\big(1 + y^a\,\nabla_a
    + \tfrac12\,[\nabla_a \nabla_b - \tfrac16\,R_{ab}]
    + \dots\big)\sigma\,.
\end{equation}
Acting with the last two operators on the lift
of $\phi$, one finds
\begin{align}
    \delta^{\text{Weyl}}_\sigma\Phi
    & = -\rho(\tfrac1\hbar\,\xi_{\text{Weyl}} + w_{\text{Weyl}})\,\Phi \\
    & = -\tfrac{D-2}2\,\sigma\,\phi
    - \tfrac{D}2\,y^a \nabla_a(\sigma\,\phi)
    - \tfrac{D+2}4\,y^a y^b\,\sigma\,
    \big(\nabla_a \nabla_b - \tfrac16\,R\big)\,\phi \\
    & \quad - \tfrac{D}2\,y^a y^b\,\nabla_a \sigma \nabla_b \phi
    -\tfrac{D-2}6\,y^a y^b\,\phi\,\nabla_a \nabla_b \sigma
    + \tfrac12\,y^2\,\big(\nabla \sigma \cdot \nabla\phi
    + \tfrac16\,\phi\,\Box\sigma\big) + \dots\,,
\end{align}
which leads to
\begin{equation}\label{eq:interm2}
    \rho(F)\,\delta^{\text{Weyl}}_\sigma\Phi\rvert_{y=0}
    = -\hbar^2\,\tfrac{D+2}2\,\sigma\,\Box\phi
    - \hbar^2\,\tfrac{D-4}6\,\phi\,\Box\sigma
    + \hbar^2\,\tfrac{D(3D-4)+4}{24(D-1)}\,\sigma\,R\,\phi\,.
\end{equation}
Collecting the two terms \eqref{eq:interm1}
and \eqref{eq:interm2}, we finally obtain the action
of a Weyl transformation on the equation of motion,
\begin{equation}
   \delta^{\text{Weyl}}_\sigma\big(\widehat{f}\phi)
    = -\tfrac{D+2}2\,\sigma\,
    \big(\Box - \tfrac{D-4}{4(D-1)}\,R\big)\,\phi\,,
\end{equation}
as expected: we recover the fact that the conformal Laplacian sends functions of Weyl weight
$-\tfrac{D-2}2$ to functions of Weyl weight $-\tfrac{D+2}2$.

\section{Feigin--Felder--Shoikhet Invariant Trace}
\label{app:FFS}
The Hochschild cohomology of the Weyl algebra $\WeylAlg_{2n}$
with values in its linear dual $\WeylAlg^*_{2n}$
is known to be concentrated in degree $2n$ and to be 
one-dimensional \cite{Feigin:1989}.
A representative for this cohomology class,
that we will denote by $\Phi$ hereafter,
was given explicitly by Feigin, Felder and Shoikhet 
\cite{Feigin:2005}, and reads as follows:
\begin{align}
    \Phi(a_0|a_1,\dots,a_{2n}) & = \int_{u\in\Delta_{2n}}
    \exp\Big[\hbar\sum_{0 \leq i<j \leq 2n}
    \big(\tfrac12+u_i-u_j\big)\,\pi_{ij}\Big]\,
    \det\big|\tfrac{\partial}{\partial p^I_a},
    \tfrac{\partial}{\partial y^a_I}\big|_{I=1,\dots,2n} \\
    \nonumber & \hspace{150pt} \times\ 
    a_0(y_0,p_0)\,a_1(y_1,p_1) \dots a_{2n}(y_{2n},p_{2n})\rvert_{y_k=0}\,,
\end{align}
where $\Delta_{2n}$ is the standard $2n$-simplex
which can be defined as
\begin{equation}
    \Delta_{2n} = \big\{(u_1,\dots,u_{2n}) \in \R^{2n}
    \mid u_0 \equiv 0 \leq u_1 \leq u_2 \leq
        \dots \leq u_{2n} \leq 1\big\}\,,
\end{equation}
and 
\begin{equation}
    \pi_{ij} := \frac{\partial}{\partial y^a_i}\,
    \frac{\partial}{\partial p_a^j}
    - \frac{\partial}{\partial p_a^i}\,
    \frac{\partial}{\partial y^a_j}\,,
\end{equation}
and the determinant is taken over the $2n \times 2n$ matrix
whose entries are the operators
$\tfrac{\partial}{\partial p^I_a}$
and $\tfrac{\partial}{\partial y^a_I}$
where the index $I$ runs over $1$ to $2n$,
so that the argument $a_0$ remains unaffected
by this determinant operator.

In practice, we need only the Chevalley--Eilenberg cocycle 
obtained from $\Phi$ by skew-symmetrisation of its arguments,%
\footnote{Recall that the skew-symmetrisation map
is a morphism of complexes between the Hochschild
complex of an associative algebra,
and the Chevalley--Eilenberg of its commutator Lie algebra.}
which we will denote by,
\begin{equation}
    [\Phi](a_0|a_1,\dots,a_{2n})
    = \sum_{\sigma\in\mathcal{S}_{2n}} (-1)^\sigma\,
    \Phi(a_0|a_{\sigma_1},\dots,a_{\sigma_{2n}})\,,
\end{equation}
where $(-1)^\sigma$ denotes the signature
of the permutation $\sigma$. The $n$-cochain defined by
\begin{equation}
    \mu(a_0|a_1,\dots,a_n) := \tfrac1{n!}\,
    \epsilon_{b_1 \dots b_n}\,
    [\Phi](a_0|y^{b_1}, \dots, y^{b_n},a_1,\dots,a_n)\,,
\end{equation}
is {\it almost} a Chevalley--Eilenberg cocycle,
in the sense that it satisfies
\begin{equation}
    \sum_{i=0}^n (-1)^i\,
        \mu\big([a_{-1},a_i]_\ast | a_0, \dots,a_n\big)
    + \sum_{0 \leq i < j \leq n} (-1)^{i+j}\,
        \mu\big(a_{-1} | [a_i, a_j]_\ast, a_0,\dots,a_n\big)
    = \tfrac{\partial}{\partial p_a}\,
        \varphi_a(a_{-1}|a_0, \dots, a_n)\,,
\end{equation}
where
\begin{equation}
    \varphi_a(a_{-1} | a_0, \dots, a_n)
    = \tfrac1{(n-1)!}\,\epsilon_{a b_1\dots b_{n-1}}\,
    [\Phi](a_{-1}|y^{b_1}, \dots, y^{b_{n-1}},a_0,\dots,a_n)\,,
\end{equation}
i.e. it verifies the cocycle condition modulo
a total derivative in $p$. As a first step towards
simplifying the expression of $\mu$, let us note that
\begin{equation}
    \det\big|\tfrac{\partial}{\partial y^a_I},
    \tfrac{\partial}{\partial p_a^I}\big|
    = \sum_{\sigma \in \mathcal{S}_{n|n}}
    (-1)^\sigma\,\epsilon^{a_1 \dots a_n}\,
    \frac{\partial}{\partial y^{a_1}_{\sigma_1}}
    \dots \frac{\partial}{\partial y^{a_n}_{\sigma_n}}\,
    \epsilon_{b_1 \dots b_n}\,
    \frac{\partial}{\partial p_{b_1}^{\sigma_{n+1}}}
    \dots \frac{\partial}{\partial p_{b_n}^{\sigma_{2n}}}\,
\end{equation}
where $\mathcal{S}_{n|n}$ denotes the set of permutations
of $2n$ elements which preserve the order of the first $n$
and the last $n$ elements separately,
i.e. $\sigma_1 < \sigma_2 < \dots < \sigma_n$
and $\sigma_{n+1} < \sigma_{n+2} < \dots < \sigma_{2n}$, and
\begin{align}
    \tfrac1{n!}\,\epsilon_{a_1 \dots a_n}\,&
    \sum_{\sigma\in\mathcal{S}_{2n}} (-1)^\sigma\,
    y^{a_1}_{\sigma_1} \dots y^{a_n}_{\sigma_n}\,
    a_1(y_{\sigma_{n+1}}, p_{\sigma_{n+1}})
    \dots a_n(y_{\sigma_{2n}},p_{\sigma_{2n}})\\
    & = \epsilon_{a_1 \dots a_n}\,
    \sum_{\sigma\in\mathcal{S}_{n|n}} (-1)^\sigma\,
    y^{a_1}_{\sigma_1} \dots y^{a_n}_{\sigma_n}\,
    \sum_{\tau\in\mathcal{S}_n} (-1)^\tau\,
    a_{\tau_1}(y_{\sigma_{n+1}}, p_{\sigma_{n+1}})
    \dots a_{\tau_n}(y_{\sigma_{2n}},p_{\sigma_{2n}})\,,
    \nonumber
\end{align}
so that, put together, these two formulae yield
\begin{align}
    \det\big|\tfrac{\partial}{\partial y^a_I},
    \tfrac{\partial}{\partial p_a^I}\big|
    & \Big(\tfrac1{n!}\,\epsilon_{a_1 \dots a_n}\,
    \sum_{\sigma\in\mathcal{S}_{2n}} (-1)^\sigma\,
    y^{a_1}_{\sigma_1} \dots y^{a_n}_{\sigma_n}
    a_1(y_{\sigma_{n+1}}, p_{\sigma_{n+1}})
    \dots a_n(y_{\sigma_{2n}},p_{\sigma_{2n}})\Big) \\
    & = (2n)!\,\sum_{\substack{\{i_1<\dots<i_n\}\\
                    \subset \{1,\dots,2n\}}}\
    \sum_{\sigma\in\mathcal{S}_n} (-1)^\sigma\,
    \epsilon_{a_1 \dots a_n}\,
    \frac{\partial a_{\sigma_1}}{\partial p_{a_1}}
    (y_{i_1},p_{i_1}) 
    \dots \frac{\partial a_{\sigma_n}}{\partial p_{a_n}}
     (y_{i_n},p_{i_n})\,,
\end{align}
where the first sum is taken over all \emph{ordered} subsets
of $n$ integers in the set $\{1,\dots,2n\}$.
We are now in position of writing down the cochain $\mu$:
for any $a_0,a_1,\dots,a_n \in \WeylAlg_{2n}$,
it is given explicitly by 
\begin{equation}
    \mu(a_0|a_1,\dots,a_n)
    = (2n)!\,\sum_{\sigma\in\mathcal{S}_n} (-1)^\sigma\,
    \int_{u\in\Delta_{2n}} \mathscr{D}(u;a_0,a_{\sigma_1},
    \dots,a_{\sigma_n})\rvert_{y=0}\,,
\end{equation}
where
\begin{equation}
    \mathscr{D}(u;-) = \sum_{f\in\Delta([n],[2n])}\
    \exp\Big[\hbar\sum_{0 \leq i<j \leq n}
    \big(\tfrac12+u_{f(i)}-u_{f(j)}\big)\,\pi_{ij}\Big]
    \epsilon_{a_1 \dots a_n}\,\big(1 \otimes
    \tfrac{\partial}{\partial p_{a_1}} \otimes
    \dots \otimes \tfrac{\partial}{\partial p_{a_n}}\big)\,,
\end{equation}
and 
\begin{equation}
    \Delta([k],[l]) := \big\{f: \{1,2,\dots,k\}
    \longrightarrow \{1,2,\dots,l\} \mid f(i) < f(j)\,,\ 
    1 \leq i < j \leq k\big\}
\end{equation}
denotes the set of order-preserving maps from the set
$[k]$ of the first $k$ integers, to the set $[l]$
of the first $l$ integers. Note that by convention,
we put $f(0)=0$ and $u_0=0$.

\paragraph{Trace on the deformed algebra of functions.}
Suppose that $a_1, \dots, a_n$ are linear in $p$,
and write $\tfrac{\partial a}{\partial p_b} = a^b(y)$
for their derivative with respect to $p$.
Then the above operator collapses to
\begin{equation}
    \mathscr{D}(u;a_0,a_1,\cdots,a_n)
    = \sum_{f\in\Delta([n],[2n])} \exp\Big[\hbar\sum_{i=1}^n
    \big(u_{f(i)}-\tfrac12\big)\,\tfrac{\partial}{\partial p_a}\,
    \tfrac{\partial}{\partial y_i^a}\Big]\,a_0(y,p)
    \times \epsilon_{b_1 \dots b_n}\,
    a^{b_1}_1(y_1) \dots a^{b_n}_n(y_n)\,,
\end{equation}
thereby exhibiting a clear distinction between
the arguments: the zeroth one will only receive
derivative with respect to $p$, while the remaining
$n$ arguments will only receive derivatives
with respect to $y$.
Now consider the case where $a_0=F(y,p)$,
and all other arguments are equal to the Fedosov connection,
$a_1=\dots=a_n=A$.
Since $A$ is linear in $p$ we can write it as
\begin{equation}
    A(y,p) = \dR x^\mu\,e_\mu^a\,A_a{}^b(y)\,p_b\,,
\end{equation}
and introducing the notation
\begin{equation}\label{eq:alt}
    \mathbb{A}(y_1, \dots, y_n)
    := \epsilon^{a_1 \dots a_n}\,\epsilon_{b_1 \dots b_n}\,
    A_{a_1}{}^{b_1}(y_1) \cdots A_{a_n}{}^{b_n}(y_n)\,,
\end{equation}
we end up with
\begin{equation}
    \mathscr{D}(u;F,A,\dots,A) = \dR^Dx\,|e|\,
    \sum_{f\in\Delta([n],[2n])} \exp\Big[\hbar\sum_{i=1}^n
    \big(u_{f(i)}-\tfrac12\big)\,\tfrac{\partial}{\partial p_a}\,
    \tfrac{\partial}{\partial y_i^a}\Big]\,
    F(y,p) \times \mathbb{A}(y_1, \dots, y_n)\,,
\end{equation}
where $|e|$ is the determinant of the vielbein.
This formula exhibits a couple of properties:
\begin{itemize}
\item First, as we noticed earlier, the argument $F(y,p)$
is the only one to receive derivatives with respect to $p$.
This means that in order to compute $\mu(F|A,\dots,A)$,
one only needs to know $F\rvert_{y=0}$, the $y$-independent
part of the symbol $F$.

\item Second, the integral over the simplex will produce
some combinatorial coefficients
\begin{equation}
    \sum_{f\in\Delta([n],[2n])} \int_{\Delta_{2n}}
    \big(u_{f(\ell_1)}-\tfrac12\big)\,\cdots\,
    \big(u_{f(\ell_k)}-\tfrac12\big)\,,
\end{equation}
which depends on a $k$-tuple of integers
$(\ell_1, \dots, \ell_k)$ comprised between $1$ and $n$.
In fact, one can refine this dependency a little bit
by remarking that if two $k$-tuple are related by
a permutation $\tau \in \mathcal{S}_k$, the associated
coefficients are equal,
so that these coefficients may as well be labeled
by partitions of $k$.
\end{itemize}

Putting this together, one ends up with
\begin{align}
    \mu(F|A,\dots,A) & = \dR^Dx\,|e|\,
    \sum_{k\geq0} \mu^{\scriptscriptstyle\nabla}_{a_1 \dots a_k}(x)
    \frac{\partial^k}{\partial p_{a_1} \cdots \partial p_{a_k}}
    F(y,p)\big|_{y=0}\,,
\end{align}
where $\mu^{\scriptscriptstyle\nabla}_{a_1 \dots a_k}(x)$
are polynomials in the (covariant derivatives of the) curvature
of $\nabla$, which is obtained by computing the term
of order $\hbar^k$ in
\begin{equation}
    \sum_{f\in\Delta([n],[2n])} \int_{\Delta_{2n}}
    \exp\Big[\hbar\sum_{i=1}^n \big(u_{f(i)}-\tfrac12\big)\,
    \tfrac{\partial}{\partial y_i^a}
    \otimes \tfrac{\partial}{\partial p_a}\Big]\,
    \mathbb{A}(y_1, \dots, y_n) \otimes F(y,p)\big|_{y_i=0}\,.
\end{equation}

\section{Curvature expansion}
\label{app:curvature}
Since the components of $\Completion$ are constructed
from the curvature tensor of $\nabla$, its covariant
derivatives and contractions thereof, we can rearrange
its expansion in the power of the curvature, which appears through 
\begin{equation}
    \Curvature \equiv -\tfrac13\,\dR x^\mu\,R_{\mu a}{}^c{}_b\,
    y^a y^b\,p_c\,,
\end{equation}
namely we write $\Completion=\sum_{k\geq1} \Completion^{(k)}$ 
where $\Completion^{(k)}$ is of order $k$ in $\Curvature$
and its derivatives. Let us now evaluate
its defining equation \eqref{eq:rec_A}
at order $n$ in $\Curvature$, 
\begin{equation}
    \Completion^{(k)} = \Curvature\,\delta_{k,1}
    + \SymDer \Completion^{(k)}
    + \sum_{l=1}^{k-1} \tfrac{1}{2\hbar}\,
    h\big[\Completion^{(l)}, \Completion^{(k-l)}\big]_\ast\,,
\end{equation}
which we can re-write, for $k>1$, as
\begin{equation}\label{eq:rec_curv}
    \Completion^{(k)} = \tfrac1{2\hbar}\,
    \sum_{l=1}^{k-1} \Taylor h\,
    \big[\Completion^{(l)}, \Completion^{(k-l)}\big]_\ast\,,
    \qquad\text{with}\qquad 
    \Taylor := \frac1{1-\SymDer} = \sum_{m=0}^\infty \SymDer^m\,.
\end{equation}
The first few orders in this curvature expansion
\begin{equation}
    \Completion^{(1)} = \Taylor\Curvature\,,
    \qquad 
    \Completion^{(2)} = \tfrac1{2\hbar}\,\Taylor
        \langlecket{\Curvature, \Curvature}\,,
    \qquad 
    \Completion^{(3)}
    = \tfrac1{2\hbar^2}\,\Taylor\langlecket{\Curvature,
        \langlecket{\Curvature,\Curvature}}\,,
\end{equation}
\begin{equation}
    \Completion^{(4)} = \tfrac1{2\hbar^3}\,\Taylor
    \langlecket{\Curvature,\langlecket{\Curvature,
        \langlecket{\Curvature,\Curvature}}}
    +\tfrac1{8\hbar^3}\,\Taylor
        \langlecket{\langlecket{\Curvature,\Curvature},
            \langlecket{\Curvature,\Curvature}}\,,
\end{equation}
where we introduce the bracket
\begin{equation}
    \langlecket{-,-}
        := h\,\big[\Taylor(-),\Taylor(-)\big]_\ast\,,
\end{equation}
as a shorthand notation. This approach has a couple
of advantages: first, it allows us to access in one go
whole pieces of $\Completion$ at arbitrary order in $y$,
and second, the recursion in order of curvature
exhibits an interesting structure, namely it appears
that it is controlled by the grafting (non-planar)
binary trees. Indeed, denoting the operator $\Taylor$
by an edge, and the composition of the star-commutator
with the contracting homotopy by a vertex, i.e.
\begin{equation}
    \Taylor(X) = 
    \begin{tikzpicture}[baseline=(m1), scale=0.75]
        \node[above] (in) at (0,0.5) {$X$};
        \node (m1) at (0,0) {};
        \node (out) at (0,-0.5) {};
        \draw[thick] (in) -- (out);
    \end{tikzpicture}
    \hspace{50pt}
    \tfrac1{2\hbar}\,h[X, Y]_\ast = 
   \tikzset{
    vertex/.style={circle, draw, fill=black, inner sep=0pt, minimum size=2pt}
} 
    \begin{tikzpicture}[baseline=(m2), scale=0.75]
        \node[above] (1st) at (-1,0.5) {$X$};
        \node[above] (2nd) at (1,0.5) {$Y$};
        \node (out) at (0,-1) {};
        \node[vertex] (m2) at (0,0) {};
        \draw[dashed] (1st) -- (m2);
        \draw[dashed] (2nd) -- (m2);
        \draw[dashed] (m2) -- (out);
    \end{tikzpicture}
\end{equation}
where the diagrams should be read from top to bottom,
so that for instance
\tikzset{
    vertex/.style={circle, draw, fill=black, inner sep=0pt, minimum size=2pt}
}
\begin{equation}
    \Taylor\langlecket{X,Y}\ =
    \begin{tikzpicture}[baseline=(m2), scale=0.75]
        \node[above] (1st) at (-1,0.5) {$X$};
        \node[above] (2nd) at (1,0.5) {$Y$};
        \node (out) at (0,-0.75) {};
        \node[vertex] (m2) at (0,0) {};
        \draw[thick] (1st) -- (m2) -- (out);
        \draw[thick] (2nd) -- (m2);
    \end{tikzpicture}
\end{equation}
one can re-write the recursion relation \eqref{eq:rec_curv} as
\begin{equation}
    \Completion^{(k)} = \sum_{l=1}^{k-1}\
    \begin{tikzpicture}[baseline=(m2), scale=0.75]
        \node[above] (1st) at (-1,0.5) {$\Completion^{(l)}$};
        \node[above] (2nd) at (1,0.5) {$\Completion^{(k-l)}$};
        \node (out) at (0,-0.75) {};
        \node[vertex] (m2) at (0,0) {};
        \draw[dashed] (1st) -- (m2);
        \draw[dashed] (2nd) -- (m2);
        \draw[thick] (m2) -- (out);
    \end{tikzpicture}
\end{equation}
for all $k > 1$. Now it becomes relatively easy
to see that the result of this recursion relation
is to express $\Completion^{(k)}$ as a sum 
over all \emph{rooted planar binary trees with $k$ leaves.}
Indeed, the latter can be obtained by successively
\emph{grafting}---meaning summing over all trees
resulting from attaching the root of a tree to the leaves
of another one---the rooted binary tree with a single vertex
to itself, $k$ times, which is exactly what the above relation
produces. Taking into account the fact that $\langlecket{-,-}$
is antisymmetric amounts to identifying any two rooted
planar binary trees which can be related by permuting
the two leaves at each node, that is, one should 
sum over rooted \emph{non-planar} binary trees
and take into account the number of planar ones
that it is equivalent to as a multiplicity.

Having worked out the recursion formula for $\Completion$
in order of the curvature, we can do the same
for the lift of symbols. Indeed, resumming the defining
relation \eqref{eq:rec_F} for the lift $F(x;y,p)$
of a symbol $f(x,p)$ yields,
\begin{equation}
    F = \SymDer F + \tfrac1\hbar\,h\,[\Completion,F]_\ast
\end{equation}
from which we can extract the order $k$ piece via
\begin{equation}
    F^{(0)} = \Taylor f\,,
    \qquad 
    F^{(k>0)} = \tfrac1\hbar\,\sum_{l=0}^{k-1} \Taylor h\,
    [\Completion^{(k-l)}, F^{(l)}]_\ast\,.
\end{equation}
The first few orders are given by
\begin{equation}
    F^{(1)} = \tfrac1\hbar\,\Taylor\,\langlecket{\Curvature,f}\,,
    \qquad 
    F^{(2)} = \tfrac1{2\hbar^2}\,\Taylor
    \Big(\langlecket{\langlecket{\Curvature,\Curvature},f}
    + 2\,\langlecket{\Curvature,\langlecket{\Curvature,f}}\Big)\,,
\end{equation}
and
\begin{equation}
    F^{(3)} = \tfrac1{2\hbar^3}\,\Taylor
    \Big(\langlecket{\langlecket{\Curvature,
            \langlecket{\Curvature,\Curvature}},f}
    + \langlecket{\langlecket{\Curvature,
        \Curvature},\langlecket{\Curvature,f}}
    + \langlecket{\Curvature,
        \langlecket{\langlecket{\Curvature,\Curvature},f}}
    + 2\,\langlecket{\Curvature,
        \langlecket{\Curvature,\langlecket{\Curvature,f}}}\Big)\,.
\end{equation}

Finally, the same can be done for the lift of a function
$\phi(x)$ to a covariantly constant section $\Phi(x;y)$
of the Fock bundle: the re-summed for of the recursion
relation \eqref{eq:rec_phi} reads
\begin{equation}
    \Phi = \SymDer\Phi
    + \tfrac1\hbar\,h\,\rho(\Completion)\Phi\,.
\end{equation}
which, when evaluated at order $n$ in $\Curvature$ 
gives us
\begin{equation}
    \Phi^{(0)} = \Taylor\,\phi\,,
    \qquad 
    \Phi^{(k>0)} = \tfrac1\hbar\,\sum_{l=0}^{k-1} \Taylor h\,
    \rho(\Completion^{(k-l)})\Phi^{(l)}\,.
\end{equation}
The first few orders read
\begin{equation}
    \Phi^{(1)} = \tfrac1\hbar\,\Taylor\big(\Rep{\Curvature}{\phi}\big)\,,
    \qquad 
    \Phi^{(2)} = \tfrac1{2\hbar^2}\,\Taylor
    \Big(\Rep{\langlecket{\Curvature,\Curvature}}{\phi}
    + 2\,\Rep{\Curvature}{\big(\Rep{\Curvature}{\phi}\big)}\Big)\,,
\end{equation}
and
\begin{equation}
    \Phi^{(3)} = \tfrac1{2\hbar^3}\,\Taylor
    \Big(\Rep{\langlecket{\Curvature,
            \langlecket{\Curvature,\Curvature}}}{\phi}
    + \Rep{\langlecket{\Curvature,
        \Curvature}}{(\Rep{\Curvature}{\phi})}
    + \Rep{\Curvature}{\big(\Rep{\langlecket{\Curvature,\Curvature}}{\phi}\big)}
    + 2\,\Rep{\Curvature}{\big(\Rep{\Curvature}{(\Rep{\Curvature}{\phi})}\big)}\Big)\,,
\end{equation}
where we introduced the shorthand notation
\begin{equation}
    \Rep{\bullet}{(-)}
    := h\,\rho\big(\Taylor(\bullet)\big)\,\Taylor(-)\,,
\end{equation}
for the sakes of conciseness. The term of order $k$
will be almost identical to that of $F$,
except for the replacement of $f$ with $\phi$,
and any bracket of the form $\langlecket{-,f}$,
i.e. whose second argument is $f$, with $\Rep{(-)}{\phi}$.
This is of course not surprising since the only 
difference between the two case is the representation
of the Weyl algebra in which the covariantly constant
section that we are solving for sits in: the adjoint
for symbols like $f$ and the Fock one for functions like $\phi$.

\paragraph{Simplifying the elementary operations.}
Let us try to find a concise expression for the operator $\Taylor$.
To do so, first notice that
\begin{equation}
    \SymDer\alpha = h\nabla\alpha = \tfrac1N\,
    \big(y^a \nabla_a \alpha - \nabla\,N\,h\alpha\big)\,,
\end{equation}
so that, on forms valued in the Weyl algebra
which are \emph{annihilated} by the homotopy operator $h$,
one finds
\begin{equation}
    h\alpha=0
    \qquad\Longrightarrow\qquad 
    \SymDer\alpha = \tfrac1N\,y \cdot \nabla\alpha
    = y \cdot \nabla\,\tfrac1{N+1}\,\alpha\,,
\end{equation}
where we used the fact that $y \cdot \nabla$ is obviously
of degree $1$ in $y$, and hence increases the eigenvalue
of the number operator $N$, a fact we should take into account
when moving the latter to the right of the former.
Repeating this operation, we end up with
\begin{equation}
    \Taylor = \sum_{k\geq0} (y \cdot \nabla)^k\,\tfrac1{(N+1)_k}
    = \sum_{k\geq0} \tfrac1{k!}\,(y \cdot \nabla)^k\,
    \tfrac{(1)_k}{(N+1)_k}
    \equiv {}_1F_1\big[1;N+1; y \cdot \nabla\big]\,,
\end{equation}
where $(a)_k=a(a+1)\dots(a+k-1)$ is the raising Pochhammer symbol,
and where we used the fact that $(1)_k=k!$ to recognize
the \emph{confluent hypergeometric function}.
Let us stress that the above expression is valid only
for $\Taylor$ acting on elements in ${\rm Ker}(h)$,
which is enough for us since we are interested in applying it 
to either $\Completion$ or a $0$-form, both annihilated by $h$,
by definition.

We can now use the integral representation of the confluent
hypergeometric function,
\begin{equation}\label{eq:1F1}
    {}_1F_1\big[a;b;x\big] := \sum_{k\geq0} \tfrac{x^k}{k!}\,
    \tfrac{(a)_k}{(b)_k} = \frac{\Gamma(b)}{\Gamma(a)\,\Gamma(b-a)}\,
    \int_0^1 \dR t\,e^{t\,x}\,t^{a-1}\,(1-t)^{b-a-1}\,,
\end{equation}
which holds for $\Re(b) > \Re(a) > 0$. In our case,
both parameters are positive integer, and verify the inequality
except if $N=0$, i.e. if we act on a $y$-independent $0$-form.
This case is particularly simple to treat since the hypergeometric
series collapses to an ordinary exponential, i.e.
\begin{equation}\label{eq:j_0-form}
    \Taylor\rvert_{\Functions(T^*\Base)} = e^{y \cdot \nabla}\,.
\end{equation}
We can therefore exclude this case, that is consider $N>0$,
and use the above integral representation to re-write $\Taylor$ as
\begin{equation}
    \Taylor = N\,\int_0^1 \dR t\,e^{t\,y \cdot \nabla}\,(1-t)^{N-1}
    =\int_0^1\, \dR t\,e^{(1-t)\,y \cdot \nabla}\,
    \tfrac{\dR}{\dR t}\,t^N\,,
\end{equation}
where we used the change of variable $t \to 1-t$
before recognizing the derivative. We therefore find
\begin{align}
    \Taylor(\alpha) & = \int_0^1 \dR t\,e^{(1-t)\,y \cdot \nabla}\,
    \tfrac{\dR}{\dR t}\alpha(x,t\,\dR x, t\,y,p) \\
    & = \alpha + y \cdot \nabla\,\int_0^1 \dR t\,
    e^{(1-t)\,y \cdot \nabla}\,\alpha(x,t\,\dR x, t\,y,p)\,,
\end{align}
for any Weyl algebra-valued form $\alpha$ such that $h\alpha=0$.
Remark that this last form obtained by integrating by part, namely
\begin{equation}
    \Taylor = 1 - \int_0^1 \dR t\,\tfrac{\dR}{\dR t}
    \big(e^{(1-t)\,y\cdot\nabla}\big)\,t^N\,,
\end{equation}
also makes sense for $N=0$, since it reproduces \eqref{eq:j_0-form}.

\paragraph{Lifts at first order in curvature.}
Let us now use this operator to compute the lift 
of symbols and wave functions at first order in curvature.
Starting with the latter, we need to first compute
\begin{align}
    \rho(\Taylor \Curvature)\,\Taylor \phi 
    & = \exp\Big(\!-\hbar\,\partial_p \cdot
    \big[\tfrac12\,\partial_{y_1} + \partial_{y_2}\big]\Big)
    \int_0^1 \dR u\,u^2\,e^{(1-u)\,y_1 \cdot \nabla}\,
    \big(-\dR x^\mu\,R_{\mu a}{}^c{}_b\,y_1^a y_1^b\,p_c\big)\,
    \times\, e^{y_2 \cdot \nabla} \phi\big|_{p=0, y_1=y=y_2} 
    \nonumber \\
    & = -\tfrac\hbar2\,\int_0^1 \dR u\,u^2\,
    e^{(1-u)\,y \cdot \nabla}\,
    \big(\dR x^\mu\,R_{\mu a}\,y^a\big)
    \times e^{y \cdot \nabla} \phi + (\cdots)\,,
\end{align}
where the dots denote terms annihilated by the contracting
homotopy $h$. Applying the latter composed with $\Taylor$,
we end up with
\begin{align}
    \Phi^{(1)} & \equiv \tfrac1\hbar\,
    \Taylor\,h\,\rho(\Taylor \Curvature)\,
    \Taylor \phi = -\tfrac12\,\int_{[0,1]^3} \dR t\,
    \dR s\,\dR u\ e^{(1-t)\,y \cdot [\nabla_1+\nabla_2]} \\
    & \hspace{150pt} \times\ \tfrac{\dR}{\dR t}\,\Big(s\,t^2\,u^2\,
    e^{t\,s\,y \cdot [\nabla_1 + (1-u)\,\nabla_2]}
    \phi(x_1)\,R_{ab}(x_2)\,y^a y^b\Big)\rvert_{x_1=x=x_2}\,,
    \nonumber
\end{align}
where $\nabla_i$ denotes the covariant derivative
with respect to $x_i$. Evaluating this formula up
to third order in $y$ yields
\begin{equation}
    \Phi = \phi + y^a\,\nabla_a\phi + \tfrac12\,y^a y^b\,
    \big(\nabla_a \nabla_b - \tfrac16\,R_{ab}\big)\phi
    + \tfrac16\,y^a y^b y^c\,\big(\nabla_a \nabla_b \nabla_c
    -\tfrac12\,R_{ab}\,\nabla_c -\tfrac14\,\nabla_a R_{bc}\big)\phi
    + (\dots)\,,
\end{equation}
as previously derived from the defining recursion relation
for the lift of the wave function $\phi$ in the Fock bundle.
We can re-write this lift up to first order in curvature as
\begin{equation}
    \Phi(x;y) = \Big(\tau^{(0)}(y \cdot \nabla_1)
    + \tau^{(1)}(y \cdot \nabla_1, y \cdot \nabla_2)\,
    \tfrac12\,y^a y^b\,R_{ab}(x_2)
    + \dots\Big)\phi(x_1)\rvert_{x_i=x}\,,
\end{equation}
with $\tau^{(0)}(z_1) = e^{z_1}$ and
\begin{equation}
    \tau^{(1)}(z_1,z_2)
    = -\int_{[0,1]^3} \dR t\,\dR s\,\dR u\
    e^{(1-t)\,[z_1+z_2]}\ \tfrac{\dR}{\dR t}\,\Big(s\,t^2\,u^2\,
    e^{t\,s\,[z_1 + (1-u)\,z_2]}\Big) 
    = -\tfrac16\,e^{z_1}\,{}_1F_1\big[2;4;z_2\big] 
\end{equation}
which, upon using the integral representation \eqref{eq:1F1},
can be expressed as
\begin{equation}
    \tau^{(1)}(z_1,z_2) = e^{z_1}\,
    \int_0^1 \dR t\ (t-1)\,t\,e^{t\,z_2}\,.
\end{equation}
and hence
\begin{equation}\label{eq:Phi1}
    \Phi(x;y) = \Big(1 +\tfrac12\,y^a y^b\,
    \int_0^1 \dR t\ t\,(t-1)\,e^{t\,y \cdot \nabla} R_{ab} 
    + \dots\Big)\,e^{y \cdot \nabla} \phi\,.
\end{equation}

Now let us turn our attention to the piece of first order
in curvature of the lift of $f$, for which we need to compute
\begin{align}
    [\Taylor \Curvature, \Taylor f]_\ast
    & = \sum_{\sigma=\pm} \sigma\,
    \exp\Big(\tfrac{\sigma\hbar}{2}\,
    \big[\partial_{y_1} \cdot \partial_{p_2}
    - \partial_{p_1} \cdot \partial_{y_2}\big]\Big) \\
    & \hspace{50pt} \times \,\int_0^1 \dR u\,
    e^{(1-u)\,y_1 \cdot \nabla}\,\big(-u^2\,\dR x^\mu\,
    R_{\mu\,a}{}^c{}_b y^a_1 y^b_1\,p_{1\,c}\big)
    \times e^{y_2 \cdot \nabla} f(p_2)
    \big|_{\substack{y_1=y=y_2 \\ p_1=p=p_2}} \\
    & = \sum_{\sigma=\pm} \sigma\,
    \exp\Big(\tfrac{\sigma\hbar}{2}\,
    \partial_{y_1} \cdot \partial_{p_2}\Big)
    \int_0^1 \dR u\,e^{(1-u)\,y_1 \cdot \nabla} 
    \big(u^2\,\dR x^\mu\,R_{\mu\,a}{}^c{}_b\,y_1^a y_1^b\big) \\
    & \hspace{180pt} \times\,
    \big(-p_{1\,c} + \tfrac{\sigma\hbar}{2}\,\nabla_c\big)\,
    e^{y_2 \cdot \nabla}\,f(p_2)
    \big|_{\substack{y_1=y=y_2\\p_1=p=p_2}} \\
    & = \tfrac{\hbar}{2}\,\sum_{\sigma=\pm} 
    \int_0^1 \dR u\ e^{(1-u)
    [\,y + \frac{\sigma\hbar}{2}\,\partial_p] \cdot \nabla}
    \Big(u^2\,\dR x^\mu\,R_{\mu\,a}{}^c{}_b\,
    \big[y^b\,\tfrac{\partial}{\partial p_a}
    + \tfrac{\sigma\hbar}{2}\,
    \tfrac{\partial^2}{\partial p_a \partial p_b}\big]\Big) \\
    & \hspace{150pt} \times\,
    \big(-p_c' + \tfrac{\sigma\hbar}{2}\,\nabla_c\big)\,
    e^{y \cdot \nabla}\,f(p)\rvert_{p'=p} + (\dots)\,, 
\end{align}
where the dots denote terms that are annihilated by $h$.
Applying it followed by $\Taylor$, one finds
\begin{align}
    F^{(1)} & \equiv \tfrac1\hbar\,\Taylor h\,
    [\Taylor\Curvature, \Taylor f]_\ast 
    = \tfrac12\,\sum_{\sigma=\pm} 
    \int_{[0,1]^3} \dR s\,\dR t\,\dR u\
    e^{(1-t)\,y \cdot [\nabla_1 + \nabla_2]} \\
    & \hspace{150pt} \times\ \tfrac{\dR}{\dR t}\,
    \Big(t\,u^2\,e^{s\,t\,y \cdot [(1-u)\,\nabla_1+\nabla_2]
    +\frac{\sigma\hbar}2\,(1-u)\,\partial_p \cdot \nabla_1}\,
    y^d\,R_{da}{}^c{}_b(x_1) \\ & \hspace{150pt}
    \times\,\big[t\,s\,y^b\,\tfrac{\partial}{\partial p_a}
    + \tfrac{\sigma\hbar}{2}\,
    \tfrac{\partial^2}{\partial p_a \partial p_b}\big]\,
    \big(-p_c' + \tfrac{\sigma\hbar}{2}\,\nabla_c\big)\,
    f(x_2,p)\Big)\big|_{\substack{x_1=x=x_2\\p'=p}}\,.
\end{align}
After explicitly performing the integrals as for the lift
of $\phi$, one finds 
\begin{align}
    F^{(1)}(x,p;y) & = \tfrac12\,e^{y \cdot \nabla_1}\,
    \sum_{\substack{k,l\geq0\\\sigma=\pm}}
    \big(\tfrac{\sigma\hbar}{2}\big)^k\,
    \tfrac{(\partial_p \cdot \nabla_2)^k}{k!}\,
    \tfrac{(y \cdot \nabla_2)^l}{l!}\,y^a\,R_{da}{}^c{}_b(x_2)\,
    \tfrac{1}{(k+1)(l+k+2)(l+k+3)} \\
    & \hspace{100pt} \times\,
    \Big[y^b\,\tfrac{\partial}{\partial p_d}
    + \tfrac{\sigma\hbar}{2}\,\tfrac{l+2k+4}{(k+2)}\,
    \tfrac{\partial^2}{\partial p_b \partial p_d}\Big]\,
    \big(p_c' - \tfrac{\sigma\hbar}{2}\,\nabla_c\big)\,
    f(x_1,p)\big|_{\substack{x_i=x,\\p'=p}} \\
    & = \tfrac12\,\sum_{\sigma=\pm}\int_{[0,1]^2} \dR s\,\dR t\
    t\,(1-t)\,e^{y \cdot \nabla_1
    + t\,[y + s\frac{\sigma\hbar}2\,\partial_p] \cdot \nabla_2}\,
    \big[p' - \tfrac{\sigma\hbar}2\,\nabla_1\big]_c 
    \label{eq:F1} \\
    & \hspace{50pt} \times\,y^a\,\Big(y^b
    + \sigma\hbar\,\big[1 + \tfrac{t}2\,(1-s)\,
    y \cdot \nabla_2\big]\,\tfrac{\partial}{\partial p_b}\Big)\,
    R_{da}{}^c{}_b(x_2)\,\tfrac{\partial}{\partial p_d}
    f(x_1,p)\big|_{\substack{x_i=x,\\p'=p}}\,, 
\end{align}
for the piece of first order in curvature of the lift
$\tau(f)$ of any symbol $f(x,p)$.

\paragraph{Generating function.}
Combining the first order lift $\Phi^{(1)}$
of the scalar field given in \eqref{eq:Phi1} with the fact that 
\begin{equation}
    \rho(\Taylor f)\rvert_{y=0}
    = \exp\Big(\!-\hbar\partial_p \cdot \big[\tfrac12\,\nabla
    + \partial_y\big]\Big)f\,\big|_{y=0=p}
\end{equation}
for a symbol $f=f(x,p)$, i.e. $y$-independent,
one ends up with
\begin{align}
    \rho(F^{(0)})\Phi^{(1)}\rvert_{y=0}
    & = \tfrac{\hbar^2}{2}\,\int_0^1 \dR t\ (t-1)t\,
    \times\, e^{-\hbar\,\partial_p \cdot [\nabla_1 + t\,\nabla_2
    +\frac12\,\nabla_3]}\,\phi(x_1)\,R_{ab}(x_2)\,
    \tfrac{\partial^2}{\partial p_a \partial p_b} f(x_3,p)
    \big|_{x_i=x, p=0}\,,
\end{align}
upon using the BCH formula. Note that the commutator appearing
in these manipulations can be discarded as a consequence
of the fact that we are working at first order in curvature.
Now applying the quantization map to the first order lift
$F^{(1)}$ of a symbol given in \eqref{eq:F1},
we end up with,
\begin{align}
    \rho(F^{(1)})\Phi^{(0)}\rvert_{y=0} & = \tfrac{\hbar^2}{8}\,
    \sum_{\sigma=\pm}\int_{[0,1]^2} \dR s\,\dR t\ t(t-1)\,
    e^{-\frac{\hbar}2\,\partial_p \cdot
    [2\nabla_1 + t(1-\sigma\,s)\,\nabla_2+\nabla_3]} \\
    & \hspace{50pt} \times\,
    \Big[(1-2\sigma) + \tfrac{\sigma\hbar}2\,t(1-s)\,
    \partial_p \cdot \nabla_2\Big]\,\phi(x_1)\,R_{ab}(x_2)\,
    \tfrac{\partial^2}{\partial p_a \partial p_b}
    f(x_3,p)\big|_{\substack{x_i=x,\\p=0}}\,, \nonumber
\end{align}
after using the BCH formula a few times as before.
Performing the integrals, multiplying the result by $\phi^*$,
and integrating by part so that all derivatives on $f$
are re-distributed on $\phi$, $\phi^*$ and the curvature,
one ends up with
\begin{equation}
    {\cal J}(x|u) = e^{-\frac\hbar2\,u \cdot[\nabla_1-\nabla_2]}\,
    \Big(1 - \tfrac{\hbar^2}8\,
    {\rm sinhc}(\tfrac\hbar4\,u \cdot \nabla_3)\,
    R_{ab}(x_3)\,u^a u^b + {\cal O}(R^2)\Big)\,\phi(x_1)\,\phi^\dagger(x_2)\big|_{x_i=x}\,,
\end{equation}
where
\begin{equation}
    {\cal J}(x|u) := \sum_{s \geq 0} \tfrac{(-\hbar)^s}{2^s\,s!}\,
    J_{a_1 \dots a_s}(x)\,u^{a_1} \dots u^{a_s}\,,
\end{equation}
is the generating function for the higher spin currents, and 
\begin{equation}
    {\rm sinhc}(z) := \frac{\sinh(z)}{z}\,.
\end{equation}
is the hyperbolic version of the $\rm sinc$ function.
 
\end{appendices}

\end{document}